
\input epsf
\input rotate
\input pstricks
\input pst-node
\psset{linewidth=.3pt, unit=1mm, arrowsize=5pt 3}
\def\ov(#1)#2#3{\rput(#1){\ovalnode{#2}{#3}}} 
\def\bx(#1)#2#3{\rput(#1){\rnode{#2}{\psframebox{#3}}}}
\def\fig#1#2{{\epsfysize #1\epsfbox{#2}}} 
\def\figr#1#2{\setbox\figurarot=\hbox{\epsfxsize#1\epsfbox{#2}}\rotr\figurarot}

\long\def\figura#1#2#3#4{\midinsert
\ps 3\centerline{\fig {#1}{#2}}\ps 2
\noindent{\mayp #3 }{\peqfont #4\ps0}\ps 3
\endinsert} 

\newbox\figurarot
\long\def\figrot#1#2#3#4{\midinsert
\setbox\figurarot=\hbox{\epsfxsize#1\epsfbox{#2}}
\ps 3\centerline{\rotr\figurarot}
\noindent{\mayp #3 }{\peqfont #4\ps0}\ps 3
\endinsert}

\newrgbcolor{verdeosc}{0 0.5 0}
\newrgbcolor{rojoosc}{0.5 0 0}
\newrgbcolor{azulosc}{0 0 0.5}

\font\ninerm=cmr9     \font\sixrm=cmr6
\font\ninei=cmmi9     \font\sixi=cmmi6
\font\ninesy=cmsy9   \font\sixsy=cmsy6
\font\ninebf=cmbx9   \font\sixbf=cmbx6
\font\ninett=cmtt9 
\font\nineit=cmti9 
\font\ninesl=cmsl9 
\font\thesisface=cmr10 at 11pt
\font\thesisit=cmti10 at 11pt
\font\thesistt=cmtt10 at 11pt
\font\thesisbf=cmbx10 at 11pt
\font\thesissl=cmsl10 at 11pt
\font\thesispeq=cmr8 at 9pt

\font\mathtext=cmss10      at 11pt
\font\mathsubtext=cmss7    at 8pt
\font\mathsubsubtext=cmss5 at 6pt
\font\mathlet=eurm10       at 11 pt
 \font\mathsublet=eurm7    at 8pt
 \font\mathsubsublet=eurm5 at 6pt
\font\mathsym=cmsy10       at 11pt
 \font\mathsubsym=cmsy7    at 8pt
 \font\mathsubsubsym=cmsy5 at 6pt
\font\mathext=cmex10       at 11pt
 \font\mathsubext=cmex10   at 11pt
 \font\mathsubsubext=cmex10 at 11pt
\font\mathscr=euxm10       at 11pt

\font\amsmath=msbm10       at 11pt 


\font\grandota=cmssdc13 at 20pt  
\font\mediota=cmssdc13  at 15 pt 
\font\pequenota=cmss13 
\font\maymuypeq=cmcsc7
\font\maypeq=cmcsc8      at 9pt
\font\maynormal=cmcsc10  at 11pt


\def\normal{
\def\rm{\thesisface}%
\def\it{\thesisit}\def\sl{\thesissl}\def\tt{\fam\ttfam\thesistt}%
\def\bf{\thesisbf}%
\def\may{\maynormal}\def\mayp{\maypeq}
\textfont0=\mathtext\scriptfont0=\mathsubtext\scriptscriptfont0=\mathsubsubtext
\textfont1=\mathlet\scriptfont1=\mathsublet\scriptscriptfont1=\mathsubsublet
\textfont2=\mathsym\scriptfont2=\mathsubsym\scriptscriptfont2=\mathsubsubsym
\textfont3=\mathext\scriptfont3=\mathsubext\scriptscriptfont3=\mathsubsubext
\normalbaselineskip=13.5pt
\baselineskip=13.5pt
\lineskip=2pt
}

\newskip\ttglue
\def\peqfont{
\def\rm{\thesispeq}%
\textfont0=\ninerm \scriptfont0=\sevenrm \scriptscriptfont0=\sixrm%
\textfont1=\ninei \scriptfont1=\seveni \scriptscriptfont1=\sixi%
\textfont2=\ninesy \scriptfont2=\sevensy \scriptscriptfont2=\sixsy%
\textfont3=\tenex \scriptfont3=\tenex \scriptscriptfont3=\tenex%
\textfont\itfam=\nineit \def\it{\fam\itfam\nineit}%
\textfont\slfam=\ninesl \def\sl{\fam\slfam\ninesl}%
\textfont\ttfam=\ninett \def\tt{\fam\ttfam\ninett}%
\textfont\bffam=\ninebf \scriptfont\bffam=\sevenbf%
\scriptscriptfont\bffam=\sixbf \def\bf{\fam\bffam\ninebf}%
\def\oldstyle{\fam1 \ninei}%
\def\may{\maypeq}%
\def\mayp{\maymuypeq}%
\tt \ttglue=.6em plus .25em minus .15em%
\normalbaselineskip=12pt%
\normallineskip=1pt%
\setbox\strutbox=\hbox{\vrule height 9pt depth 2pt width 0pt}%
\let\sc=\sevenrm \normalbaselines\rm}

\newcount\veces
\def\defr#1(#2)#3{\veces=0
\def#1{#3}
\loop
  \edef#1{#1#3}
  \advance\veces by 1
  \ifnum\veces<#2
\repeat}

\long\def\boxit#1#2{\hbox{\vrule
\vtop{%
\vbox{\hrule\kern#1%
\hbox{\kern#1#2\kern#1}}%
\kern#1\hrule}%
\vrule}}


\long\def\btabla(#1){
\defr\tablerule(#1){&&}
\edef\tablerule{height3pt\tablerule\cr}
\vbox\bgroup \offinterlineskip\halign\bgroup
\vrule ##&&\ \hfil##\hfil\ &\vrule ##\cr
\noalign{\hrule}}

\def\lx#1\cr{\tablerule &#1&\cr \tablerule\noalign{\hrule}}%

\def\etabla{\egroup\egroup}

\newdimen\mtabla
\mtabla=3pt 

\long\def\sbtabla(#1){
\defr\ands(#1){&}
\defr\rulefills(#1){&\omit{\hrulefill}}
\def\tablerulepeq{&height 2\mtabla\ands\cr}
\vbox\bgroup \offinterlineskip\halign\bgroup
\quad##\hfill\ &\vrule ##&&\quad\hfill##\hfill\ \cr
\noalign{\hrule height 1pt \vskip\mtabla}}

\def\li#1&&#2\cr{\noalign{\vskip\mtabla} 
#1&\omit&#2\cr
\noalign{\vskip\mtabla}
&\rulefills\cr}

\def\lt#1\cr{\tablerulepeq#1\cr}
\def\setabla{\tablerulepeq 
\noalign{\hrule height 1pt}\egroup\egroup}

\def\ps#1{\vskip #1mm}     
\def\bul{\item{$\bullet$}} 
\def\nind{\noindent} 
\def\ot{\leftarrow}
\def\pl{\partial}


\newif\iftitular 
\newif\ifblanco
\newtoks\temaizq  \temaizq={\hfil}
\newtoks\temader  \temader={\hfil}
\def\makefootline{
\ifblanco\blancofalse\else
\iftitular\vbox to 0pt{\vskip 3mm\global\titularfalse %
\hfil--- {\oldstyle \folio} ---\hfil\vss}
\fi\fi}
\def\makeheadline{\ifblanco\blancofalse\else\vbox to 0pt{%
\vskip -22.5pt\hbox to \hsize{
\peqfont
\iftitular \hfil 
\else
  \ifodd\pageno \the\temader\hfil\folio 
  \else \folio\hfil\the\temaizq
  \fi
\psline(-\hsize,-1)(0,-1) 
\fi}\vss}\fi}
\def\plainoutput{
\ifodd\pageno \hoffset=0cm \else \hoffset=-5true mm \fi
\shipout\vbox{\makeheadline
\pagebody
\makefootline}
\advancepageno
\ifnum\outputpenalty>-20000\else\dosupereject\fi}

\nopagenumbers
\def\pss{\vskip 4mm}
\def\<{\left<}
\def\>{\right>}
\def\[{\left[}
\def\]{\right]}
\def\({\left(}
\def\){\right)}
\def\bra<#1|{\left<#1\right|}
\def\ket|#1>{\left|#1\right>}
\def\elem<#1|#2|#3>{\left<{#1}\right|{#2}\left|{#3}\right>}

\def\tr{\hbox{tr}\!}
\def\Sb{{\hbox{\mathsubtext Sb}}}
\def\lin#1{{\hbox{\mathtext #1}}}
\def\linp#1{{\hbox{\mathsubtext #1}}}

\def\pasapagina{\vskip 2cm\penalty-9000\vskip -2cm}

\def\chapterp#1;#2.{\titulartrue
\vbox{\ps{10}\vbox{\hrule height 0.5mm 
\ps3\hbox{\verdeosc\grandota #1}}
\ps3\hrule height 0.5mm\ps3
\rightline{\vbox{\hbox{\verdeosc \grandota #2}\ps3\hrule height
0.5mm}}\ps{35}}
\global\temaizq={#1#2}}

\def\chapter#1{\titulartrue\denuevas
\vbox{\ps{10}\hrule height 0.5mm
\global\temaizq={#1}
\ps 3\hbox{\verdeosc\grandota #1}\ps 3\hrule height 0.5mm\ps{45}}}

\def\section#1{\pasapagina\temader={#1}%
\ps7\vbox{\ps7\hbox{\azulosc\mediota#1}}\ps5}

\def\part#1{\pasapagina\ps7\vbox{\ps7\line{\rojoosc\mediota\hfil #1}}\ps8}

\def\subsectionp#1{{\may #1}\pss}
\def\subsection#1{\pasapagina\ps7{\may #1}\ps4}
\def\sang#1{\item{\hbox to 4mm{#1\hfil}}}
\def\ct[#1-#2]{{\mayp [#1\hskip2pt{\oldstyle #2}]}} 
\def\denuevas{\global\eqnumber=0\global\footnotenumber=0}

\newcount\eqnumber
\eqnumber=0
\def\eq#1{\global\advance\eqnumber by 1
\global\edef#1{\the\eqnumber}\eqno{[#1]}}

\newcount\bibnumber
\long\def\bibliography{\pss}
\bibnumber=0
\long\def\bib#1#2{\global\advance\bibnumber by 1
\global\edef#1{(\the\bibnumber)} 
\global\edef\bibliography{\bibliography\item{#1}#2\ps2}}

\def\cx[#1-#2] #3\par{\hbox to \hsize{\peqfont\hbox to 2.5cm{\rojoosc
[{\mayp #1}\hskip 2pt{\oldstyle #2}]\hfil}\hfil
\vtop{\advance\hsize by -3cm \nind #3}}\ps2}

\newcount\footnotenumber
\footnotenumber=0
\long\def\footn#1{\global\advance\footnotenumber by 1
\footnote{${}^{\rm \the\footnotenumber}$}{\peqfont #1 \ps{-3}}}

\def\pasacap{\global\footnotenumber=0\global\eqnumber=0\vfill\eject}

\normal\rm

\blancotrue
\vbox{
\vskip 2.5cm
\centerline{\grandota\verdeosc Real Space Renormalization Group}
\vskip 1cm
\centerline{\grandota\verdeosc Techniques and Applications.}}
\vskip 3cm
\centerline{\pequenota Ph.D. memory of}
\vskip 5mm
\centerline{\mediota\rojoosc Javier Rodr\'{\i}guez Laguna}
\vskip 1cm
\centerline{\pequenota supervised by the doctors}
\vskip 5mm
\centerline{\mediota\azulosc Miguel \'Angel Mart\'{\i}n-Delgado Alc\'antara}
\vskip 5mm
\centerline{\pequenota and}
\vskip 5mm
\centerline{\mediota\azulosc Germ\'an Sierra Rodero.}
\vskip 2cm
\centerline{\fig{3cm}{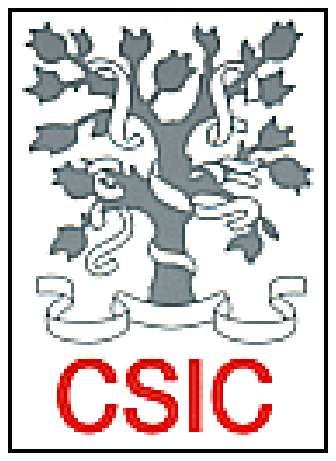}\hskip 5cm\fig{3cm}{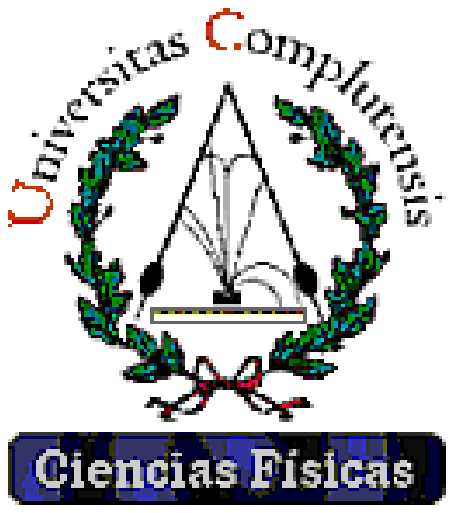}}

\vfill\eject
\vbox{}
\vfill\eject
\vbox{
\vskip 2cm
\font\gotica=yswab at 14pt
\gotica
{\rojoosc
\rightline{\hbox to 2cm{\fig{7mm}{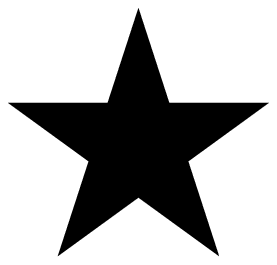}\hfil}}
\ps5
\rightline{Verum sine mendacio, certum et verissimum:}
\ps2
\rightline{quod est inferius est sicut quod est superius,}
\ps2
\rightline{et quod est superius est sicut quod est inferius,}
\ps2
\rightline{ad perpetranda miracula rei unius.}}
\ps6
\rightline{\raise3pt\hbox{\vrule height .5pt width 20pt} Hermes Trismegistus}
\ps5
\rightline{\hbox to 2cm{\fig{7mm}{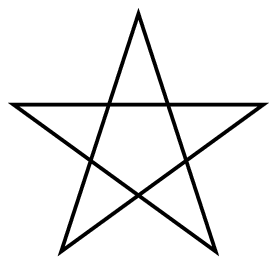}\hfil}}
}
\ps{50}
\rput(40,0){\fig{8cm}{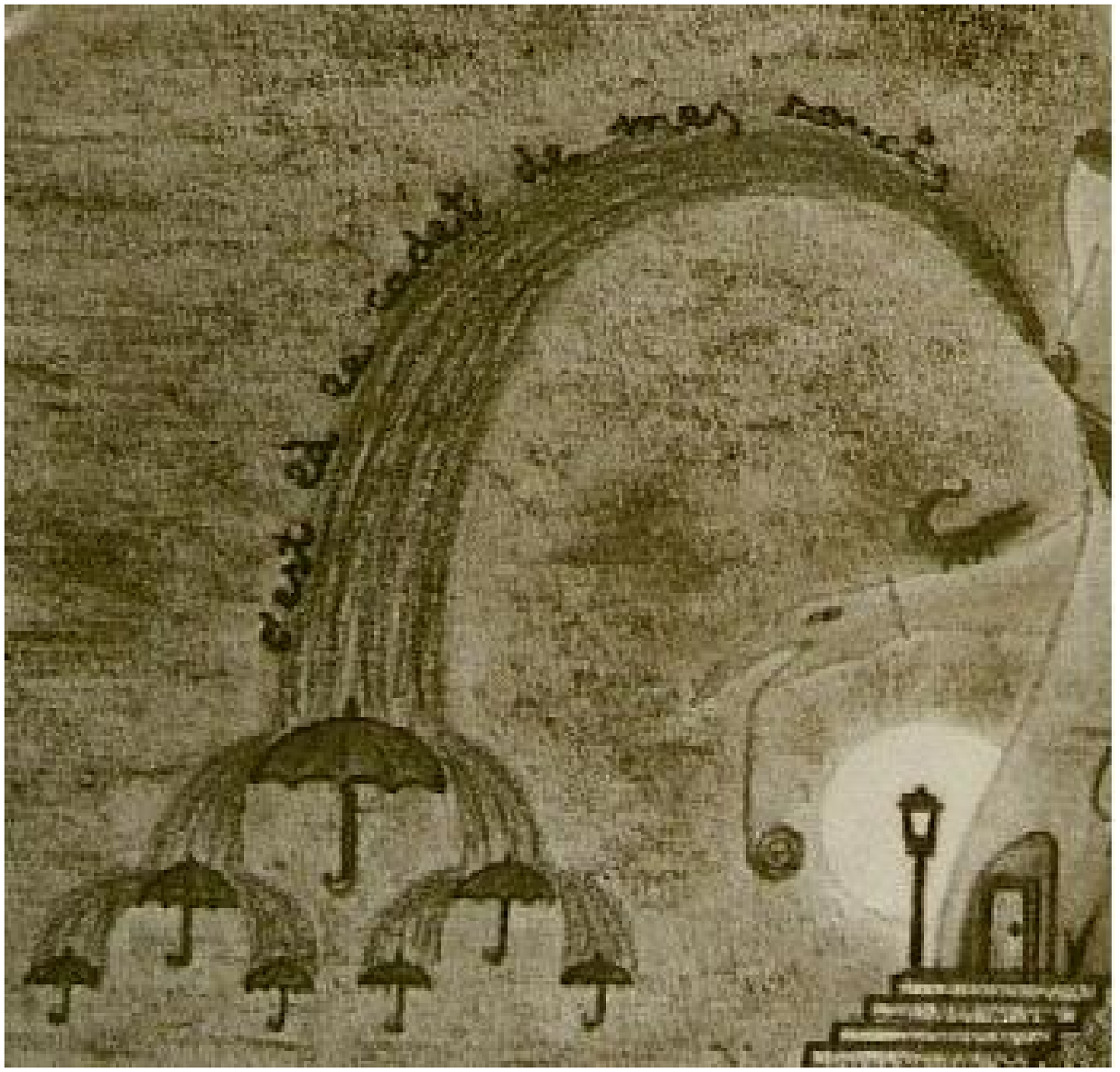}}
\ps{48}
{\it C'est l\`a le cadet de mes soucis.} (detail)

Daniel Omar Romero, private collection.

\vfill\eject
\vbox{}
\vfill\eject
\section{Index.}
\newdimen\oldparindent \oldparindent=\parindent
\parindent=0pt
\def\enx#1-#2.{\hbox to \hsize{{\may #1}\tenrm\dotfill\rm {\oldstyle #2}}\ps2}

{\may Preface.}
\ps2

\enx1. Introduction -1.

1.1. New techniques for new questions.

1.2. The scale axis and the renormalization group.

1.3. Blocking physics.

1.4. Brief history of the renormalization group.

1.5. Plan of this work.

1.6. Bibliography.
\pss

\enx2. The Correlated Blocks Renormalization Group -17.

Part I. The Blocks Renormalization Group.

2.1 BRG for a quantum magnetism model.

2.2 A simple test model: the particle in a box.
\ps2

Part II. The Correlated Blocks Renormalization Group.

2.3 Correlated blocks in 1D.

2.4 CBRG in practice.

2.5 Wave--function reconstruction.

2.6 Bidimensional CBRG.

2.7 Self--replicability.

2.8 Bibliography.
\pss

\enx3. Density Matrix Renormalization Group Algorithms -51.

Part I. DMRG for Quantum Mechanics.

3.1. The density matrix renormalization group.

3.2. DMRG algorithms for particles in a 1D potential.
\ps2

Part II. DMRG, Trees and Dendrimers.

3.3. DMRG for trees.

3.4. Physics and chemistry of dendrimers.

3.5. DMRG algorithm for exciton dynamics in dendrimers.

3.6. Bibliography.
\pss

\enx4. Multidimensional formulations of DMRG -83.

4.1. Long--range formulation of DMRG.

4.2. Puncture renormalization group.

4.3. 1D Implementation and numerical results.

4.4. PRG analysis for 2D and 3D lattices.

4.5. Warmups for PRG.

4.6. Blocks algebra.

4.7. Towards a many--body PRG.

4.8. Application of PRG to an exciton model with disorder.

4.9. Bibliography.
\pss

\enx 5. A RSRG approach to field evolution equations -119.

5.1. Introduction.

5.2. RSRG formalism for field evolution prescriptions.

5.3. Cells--overlapping truncators.

5.4. Applications and numerical results.

5.5. Towards a physical criterion for truncation.

5.6. Bibliography.
\pss

\enx Conclusions and Future Work -137.
\pss

\enx Appendices -139.

A. The laplacian on a graph.

B. Computational issues.

C. Asymptotics of the discrete heat equation.

D. Fast Gram-Schmidt procedure.

E. Scaling dimension associated to the energy.
\pss

\enx Glossary -151.

\ps{7}
\rput(77,0){
\ov(0,0){A}{Introduction to RG.}
\ov(0,-20){B}{\vbox{\hbox{RG variational}\hbox{techniques for QM.}}}
\ov(-30,-40){C1}{Without sweep}
\ov(20,-40){C2}{\vbox{\hbox{With sweep}\hbox{\peqfont Convergence to exact.}}}
\ncline{->}{A}{B}
\ncline{->}{B}{C1}
\ncline{->}{B}{C2}
\ov(-55,-65){D1}{\vbox{\hbox{Election criteria}\hbox{for freedom degrees.}}}
\ov(-20,-60){D2}{CBRG}
\ov(3,-55){D3}{DMRG}
\ov(40,-60){D4}{PRG}
\bx(-55,-90){F1}{Application to PDE}
\ncline{->}{D1}{F1}
\ncline{->}{C1}{D1}
\ncline{->}{C1}{D2}
\ncline{->}{C2}{D3}
\ncline{->}{C2}{D4}
\ncline{->}{E1}{F1}
\ncline{->}{E2}{F1}
\ncline{->}{E3}{F1}
\bx(-13,-75){E4}{Implicit}
\bx(15,-72){E5}{\vbox{\hbox{Explicit}\hbox{\peqfont Long range or $>$1D}}}
\ncline{->}{D3}{E4}
\ncline{->}{D3}{E5}
\ov(-13,-90){F2}{Dendrimers}
\ncline{->}{E4}{F2}
\bx(26,-92){E6}{\vbox{\hbox{Graphs}\hbox{\peqfont Short range.}}}
\bx(60,-80){E7}{Long range.}
\ncline{->}{D4}{E6}
\ncline{->}{D4}{E7}
\ov(60,-95){F3}{\vbox{\hbox{Excitons with}\hbox{disorder}}}
\ncline{->}{E7}{F3}
\ov(40,-110){G1}{PRG for many-body}
\ncline{->}{D4}{G1}
}

\parindent=\oldparindent

\vfill\eject

\vbox{
\ps{10}
\centerline{\grandota\verdeosc Preface.}
\ps{20}}

\moveleft5mm\hbox{\may Scientific Aspects.}
\pss

The ``real space renormalization group'' concept unites many of the
reasons for which I like mathematical physics. It is a key tool (and
shall remain so) in the explanation both of fundamental laws of Nature
and human scale physics. The basic idea is accessible to high school
students and, at the same time, deep enough to put forward some of the
hardest conceptual problems we are facing.
\pss

The bet for the real space renormalization group is at the heart of a
certain world view as something fundamentally intelligible. This new
physics aspires to be a {\it constructions game\/}: blocks, cells,
graphs... discrete geometry and analysis as the basis of the
explanation of magnetism, elementary particle physics, chemistry,
cosmology or turbulence. This view would result too naive if the
scales interaction and fractal geometry were not taken into account.
They endow this view with the necessary complexity.
\pss

Computers are the new tool, and {\sl programming} the new algebra. But
intelligence must stay as the guide, and we should not allow the tools
to become the crutches of thought. I.e.: the ideas must be clear,
every step in a computation must be deeply understood and meaningful
(``never make a calculation whose result you do not know''). When we
are able to write an efficient algorithm to predict a physical
phenomenon, the reason is that we understand it well. When a problem
is solved through computational ``brute force'', physics have failed.
\pss\pss

\moveleft5mm\hbox{\may Methodological aspects.}
\pss

All computations in this thesis have been carried out with our own
programs or with {\it free software\/}. The usage of propietary
programs is, for the present author, incompatible with scientific
work. It is equivalent to the usage of mathematical formulae whose
proof is not accessible to us. Programs have been written under LINUX
in the C and C++ languages (with Emacs) for the GNU
compilers. Libraries were written (see appendix B) for the usage of
matrices, graphs, screen (X11) and printer ({\may PostScript})
graphics managers... which shall be released to the public domain in
due time. This thesis is written in pure \TeX, and its format was
designed by the author.
\pss

Internet is changing the social aspects of physics. It has already
removed many spatial barriers to scientific communication, and is now
weakening the most powerful institutional burden: scientific
journals. As a tiny contribution, all our papers have been sent to the
Los \'Alamos, including this thesis.
\vfill\eject

\moveleft5mm\hbox{\may Acknowledgements.}
\pss

I have made most of this thesis while working as a mathematics,
physics, computer science, biology, geology, chemistry, technology and
astronomy teacher in public high schools (IES Ignacio Ellacur\'{\i}a
(Alcorc\'on), IES Galileo Galilei (Alcorc\'on), IB Cardenal Cisneros
(Madrid), IES Rayuela (M\'ostoles) and IES Arquitecto Peridis
(Legan\'es)). I can't help reminding some of my fellow teachers for
encouraging me not to leave research, and my students for their
confidence and their ease to get illusioned. On the other hand, I must
remark the tremendous bureaucratic barriers in the (luckily few) times
I have required support from the Spanish Academic Administration.
\pss

This thesis was written between the {\it Instituto de Matem\'aticas y
F\'{\i}sica Fundamental\/} of the {\it CSIC\/} and the {\it Dpto.~de
F\'{\i}sica Te\'orica\/} of the {\it UCM\/} in Madrid. I must say that
they both have been kinder to me than it was their strict duty, and
have made my work a lot easier despite the lack of any contractual
relation.
\pss

Panic to forget important names prevents me from writing a list of all
the friends which have contributed with their encouragement and time
to this work. I owe to my family my affinity for mathematics, critical
thought and confidence on reason (special mention to my grandfather),
along with an overload of intellectual honestity which shall do no
good to me.
\pss

In the scientific domain, I would like to thank my supervisors
(Germ\'an Sierra and Miguel \'Angel Mart\'{\i}n-Delgado) for their
tolerance with my dispersion and heterodoxy. I also thank Rodolfo
Cuerno for his patience in long discussions and Andreas Degenhard (the
fighter) for his confidence. An special position, both in personal and
scientific issues, is for Silvia Santalla: without her enthusiasm and
ingenuity this thesis would have never been completed.
\pss\pss

The point is that, although it is alredy written, I do not yet know
what is exactly a thesis and what is expected from it. Between the
alternative of putting together the papers with a staple along with a
brief introduction, and writing the text I would like to read if I were
a recent graduate getting into the field, I have chosen the second
alternative. It has taken more time and efforts than I intended, but I
hope it has been worth. I beg the examination board not to feel
offended when reading of too obvious things...
\pss\pss\pss

\rightline{Madrid, November 2001.}
\rightline{Javier Rodr\'{\i}guez Laguna.}
\rightline{Instituto de Matem\'aticas y F\'{\i}sica Fundamental
(CSIC).}
\rightline{C/ Serrano, 123. 28006 Madrid.}
\rightline{E-mail: {\tt javirl@sisifo.imaff.csic.es}}

\pasacap
\blancofalse
\pageno=1

\chapter{1. Introduction}

\subsection{Synopsis.}


1.1. New techniques for new questions.

1.2. The scale axis and the renormalization group.

1.3. Blocking physics.

1.4. Brief history of the renormalization group.

1.5. Plan of this work.

1.6. Bibliography.
\pss


The present chapter consists of a non-technical introduction to the
Real Space Renormalization Group (RSRG) and the associated
mathematical-physics concepts. There are no new results, although the
exposition is original. The purpose is merely to make the whole work
as self-contained and accessible to the non-specialist as possible.


\section{1.1. New Techniques for New Questions.}

It may be argued that the biggest renewal movement in theoretical
physics since the middle of the past century has been the relevance of
the analysis of complex systems whose basic rules are known. Nowadays,
almost all resarch fields involve the study of systems with many
(maybe infinite) degrees of freedom. A fast survey of the key problems
range from the large-scale structure of the Universe \ct[KT-90],
\ct[PEB-80], high critical temperature superconductivity \ct[GMSV-95],
hydrodynamical turbulence \ct[FR-95], \ct[MY-65], surface growing
phenomena \ct[BS-95] or quark-gluon plasma physics \ct[DGMP-97].

It is probably excessive to assert that complex phenomena physics
involves a new scientific {\it Weltanschauung}\footn{From German:
world view.}. But, nevertheless, it is important to acknowledge the
truth contained in the proposition:
\pss

\bul There are new methods: doing physics involves different jobs than
before. Maybe the most important one is the extensive usage of computers.

\bul There is a new series of ``paradigmatic examples'' about what is an
appropriate answer to a problem. We may cite the solution to the 2D
Ising model by L.~Onsager \ct[FEY-72], the renormalization group
analysis of the Kondo problem by K.G.~Wilson \ct[WIL-75], the
turbulence theory of A.N.~Kolmogorov \ct[K-41] or the application of
the density matrix renormalization group to the problem of the
particle in a box \ct[WHI-98].

\bul The questions considered to be relevant have also changed. Thus,
e.g., the precise prediction of the power law according to which the
specific heat of a ferromagnet diverges as it approaches Curie's point
is considered to be a great success even if the position of that
temperature is not quite accurate \ct[BDFN-93].
\pss

The massive usage of computers has had a great influence. A problem is
considered to be well solved when a fast and correct computational
algorithm has been provided which allows the precise calculation of
physical observables. A greater importance is being assigned to
discrete methods in physics, which were almost completely
forlorn. Although traditional mathematical analysis was developed in
parallel with discrete analysis\footn{Newton's interpolation formula
is considered the discrete analogue of Taylor's expansion, and both
are not too distant in time \ct[KLI-72].}, it was abandoned during the
XVIII and XIX centuries but for a few geniuses such as Leonhard Euler
or George Boole \ct[GKP-89].

Some of the most important physicists of the era previous to this
stage, to which they greatly contributed (such as, e.g. L.D.~Landau),
were against the expansion in the usage of computers. In a certain
sense, they were right. During the technological explosion which,
since the 70's, took computers to the physics departments all over
the world, the physical and mathematical insight was abandoned {\it to
some extent\/} for the mere computer--aided exploration of solutions
to problems. The incredible boost in the capacity of the computers
allowed many researchers to neglect their algorithms. Many problems
were out of reach of their computational power, but the number of
those which might be tackled was so big that this did not constitute
a serious problem.

But a parallel movement was being started. The idea that the search
for computer algorithms to solve problems in physics was as
respectable as it had been the search for solution methods for 
differential equations, was getting mature.


\section{1.2. The Scale--Axis and the Renormalization Group.}

\subsection{The Scale--Axis.}

The new problems in physics are characterized by the great number of
degrees of freedom. Their objects of study are extended systems,
whether spatial and/or temporally, which have a {\sl basic scale} of
actuation at which its dynamical rules are simply determined
\footn{The contents of this section can be checked in
any of the basic books on RSRG \ct[BDFN-93], \ct[GOL-93],
\ct[CFP-92]. Nevertheless, the presentation is new and
the terms {\sl basic scale} and {\sl observational frame} are
original.}.

Unfortunately (or maybe not so) the observation scale is always much
greater than the basic scale, and an amount of non-trivial phenomena
appear which are believed to be {\it direct consequence\/} of the
physics at the basic scale. The search for these basic rules belongs
to another ``scientific program'', which was preponderant in the first
half of the past century.
\pss

No observational device can have infinite resolution or infinite
range. Therefore, each description of a physical system must always be
labelled by a minimum scale (the grain-size of the film) and a maximum
scale (the plate-size). In the quantum field theory nomenclature,
these parameters are respectively called ``ultraviolet-cutoff'' (UV)
and ``infrared-cutoff'' (IR).

Let us, therefore, define an {\sl observational frame} to be a
reference frame (whether Lorentz or Galilei), along with the
specification of the ranged scales. The analogy with a magnifying
glass may be interesting \ct[GAW-96].

\midinsert
\ps3
\rput(20,-80){\peqfont
\psline(0,0)(0,80)\psline(3,0)(3,80)
\rput[l](7,5){Planck scale}
\rput[l](7,15){Quarks}
\rput[l](7,25){Atoms}
\rput[l](7,35){Bacteria}
\rput[l](7,45){Cats}
\rput[l](7,55){Stars}
\rput[l](7,65){Galaxies}
\rput[l](7,75){Galaxy clusters}
\psframe[fillstyle=solid,fillcolor=gray](-3,15)(6,24)
\psframe[fillstyle=solid,fillcolor=white](-2,16)(5,23)
\rput[r](-4,15){UV-cutoff}
\rput[r](-4,24){IR-cutoff}
\rput[r](-10,20){Obs. frame}
\psline{->}(-9,20)(-4,20)}
\ps{83}
\noindent{\mayp Figure 1.} {\peqfont An observational frame which is
carefully placed for the examination of atomic nuclei.\par}
\endinsert

The fundamental frame shall be the one which contains the scales
immediately above and below the basic scale. A translation in the
scale--axis is, therefore, a movement of the observational frame as a
whole. This operation is usually termed a {\sl Renormalization Group
Transformation} (RGT).

When the basic scale has long been surpassed, no new physics is
supposed to appear. Thus, the dynamics is simply ``transported'' along
the scale--axis from its focus by a set of propagators.

\subsection{The Renormalization Group.}

We might characterize the dynamics seen at an observational frame
different from the fundamental one as an ``effective'' or ``{\sl
renormalized}'' dynamics. This new dynamics might have the same form
as the original one, and differ only in the values of some
parameters. In general, it is illicit to make this supposition.

Even so, we shall describe one of those textbook examples in which
everything works properly: the Ising model \ct[LEN-20], \ct[ISI-25]
which explains ferromagnetism in uniaxial magnets. It only contains
classical spins which are either parallel or anti-parallel to a given
fixed axis. It is a model with only two parameters at the basic scale:
the coupling strength between neighbour spins (attempting to order the
system) and temperature (attempting to disorder it)\footn{As a matter
of fact, there is {\it only one\/} parameter, which is the ratio of
these two. Nevertheless, it is preferrable to state it in these
terms.}.
\pss

Any physical magnitude defined at the basic scale may be ``translated''
along the scale--axis by the RGT. During its displacement, the
relevance of the given element is either diminished or boosted. In the
first case it shall be termed an {\sl irrelevant} element, while in
the second case it shall be said to be {\sl relevant}\footn{Of course,
{\sl marginal} elements which neither increase nor decrease may also
exist, and are usually rather difficult to analyze.}.

Physics observed with the naked eye is so far away from the basic
scale that the RGT are usually allowed to reach a {\sl fixed
point}. I.e.: advance until the relative importance of the different
elements of the dynamics is {\sl invariant} under the
transformation\footn{This really means to take the thermodynamic
limit.}.

The fixed point may yield a trivial dynamics if only one of the
elements is relevant. Thus, in the Ising model of ferromagnetism, if
only the coupling between spins is relevant, the {\sl phase} shall be
magnetized and homogeneous. On the other hand, if only temperature is
relevant, then the phase shall be disordered.

\subsection{Critical points.}

The fixed point can be much more interesting. Being invariant under
scale changes, they may have non-trivial {\sl fractal} properties
\ct[DUP-89], \ct[KR\"O-00], and in that case the system is said to be at
a {\sl critical point}. No deterministic fractals are involved (such
as the Mandelbrot set or the Sierpi\'nski triangle), but statistical
ones. In other words: the probabilities for different configurations
have scale invariance properties.

The term ``fractal'' was coined by Beno\^{\i}t Mandelbrot \ct[MAN-82]
so as to unify a big number of explanations to physical phenomena
which were characterized by a wide range of scales for which either
dynamics or geometry was {\it self-similar\/}. Certainly, the study of
fixed points of the RGT is a powerful tool in the explanation of the
origin of fractal structures in Nature.
\pss

In the Ising model example, the magnet at Curie temperature (its
critical point) presents {\it domains\/} with either up or down
magnetization, which are themselves splashed with smaller domains
which themselves... and so on {\it ad scalam basicam\/}. The system
has a hyerarchy of domains, which makes it highly susceptible. Being
all scales strongly coupled, local changes are greatly
magnified. Thus, a small external field may impose a global
magnetization on the system quite easily.
\pss

Critical phenomena, i.e., phenomena with a certain scale-invariance,
are ubiquitous in the real world. Although the discussed example has
been a phase transition in an equilibrium statistical model, there are
many models far from equilibrium which present self-organized
crisis\footn{The word ``criticality'', albeit widely used, is not
correct. The term {\sl crisis} is more appropriate.}. Beyond
continuous phase transitions (without latent heat), the
self-gravitating media \ct[PEB-80], turbulence \ct[FRI-95], granular
media movements \ct[BTW-88], polymers in solution \ct[GEN-72], surface
roughening \ct[BS-95], etc.

Moreover, it should not be surprising that the mathematics of scale
invariance is an essential tool for the theories on the fundamental
constituents of matter \ct[COL-85], \ct[KOG-79]. In fact, as it is
explained in 1.4.II., it was in that field where the term
``Renormalization Group'' was born.

\subsection{Critical Exponents and Universality.}

At the critical point there are many physical magnitudes which {\it
diverge\/}. For a phase transition those might be e.g. susceptibility,
specific heat, etc. These are physically meaningful divergences, and
in many cases are related to the inability of euclidean measures to
capture a fractal phenomenon (see \ct[DUP-89] and \ct[KR\"O-00]). In
other terms, they diverge for the same reason as the length of a coast
does, and in the a similar way: because a 1D measure is not
appropriate. The length of a coast only diverges when there is either
no highest or no lowest scale, and similarly the magnitudes at a
critical point only diverge for an infinite system.

In order to characterize the magnitude of these divergences, power
laws are appropriate: $F\propto t^{-\alpha}$, where $F$ is the
observable, $t$ is a variable which is zero at the phase transition
(e.g.: reduced temperature) and $\alpha$ is what we shall call a {\sl
critical exponent}.

Critical exponents may be obtained from the RGT. The procedure starts
with the obtention of the fixed points of the transformation for a
given physical system. Afterwards, the neighborhood of the fixed
points is carefully studied. Obviously, only the {\sl relevant}
elements of the dynamics shall be important in the computation.

Universality is almost trivial after the previous remarks. If two
physical systems share all the relevant variables and differ only in
the irrelevant ones, they must share critical exponents and they will
enter the same {\sl universality class}. Thus, in the ferromagnet
example, the crystallization type of the lattice, the chemical aspects
of the atoms, etc. happen to be irrelevant. Therefore, all
ferromagnets have the same critical exponents. Moreover, models as
different beforehand as the ``mixing--non-mixing'' phase transition
between $CH_3OH$ and $C_6H_{14}$, the ``superfluid--normal fluid''
transition for ${}^3$He and many others share the {\it essentials\/}
of the ferromagnet transition and fall into the same universality
class.

Experience shows that the ``essentials'' for a phase transition (i.e.:
the variables labelling universality classes) are merely the
dimensionality of the configurational space and that of the {\sl order
parameter} (i.e.: the field which is zero at the disordered phase and
non-zero at the ordered one).


\section{1.3. Blocking Physics.}

\subsection{Spin Blocks.}

The basic idea of Kadanoff \ct[KAD-66] is the image which is kept in
mind by any practicioner of RG techniques: the spin blocks. Figure 2
shows a square lattice of spins with three levels of block composition
on it.

\midinsert
\ps 3
\newgray{gris}{.8}
\rput(30,0){
\multirput(0,0)(0,-5){8}{
\multirput(0,0)(5,0){8}{\pscircle[fillstyle=solid,fillcolor=gris](0,0){1}}}
\multirput(0,0)(0,-10){4}{
\multirput(0,0)(10,0){4}{\psframe(-1.5,1.5)(6.5,-6.5)}}
\multirput(0,0)(0,-20){2}{
\multirput(0,0)(20,0){2}{\psframe(-2.5,2.5)(17.5,-17.5)}}
\psframe(-4,5)(40,-40)}
\ps{42}
\nind{\mayp Figure 2. }{\peqfont A square lattice of $8\times 8$
spins with three levels of composition of blocks superimposed. Each
level corresponds to a RG step in Kadanoff's approach.\par}
\ps2
\endinsert

The fundamental operation is the composition of spins to form blocks,
and the fundamental idea is that ``a spin block behaves, {\it to some
extent\/}, in the same way as a single spin does''\footn{Hence the
quotation for this work.}. A description of the system in terms of
blocks instead of spins only requires an appropriate change in the
coupling constants.

Unfortunately (or perhaps not so) that bucolic image is impossible to
implement in practice. In many cases the effective description of a
system in terms of spin blocks is possible, but at the expense of
introducing new interactions between them which were not present
beforehand. These new interactions, at the next step, might require
even more interactions and so on {\it ad nauseam\/}. Some theories are
exactly renormalizable (among them all the interactions considered to
be fundamental except gravitation), and many of them are {\it
approximately\/} renormalizable. I.e.: they allow the application of
the RG to a good precision with the inclusion of a small number of
extra interactions.
\pss

Once the set of interactions which is {\it approximately\/} closed
under RG has been delimited, it is only needed to find which is the
appropriate change in the coupling constants. A {\sl parameter space}
is established with them all, and the RG has a ``realization'' on it
as a {\sl flow}. This flow has probably some fixed points which, as it
was discussed in the previous section, determine the macroscopic
physics of the system.

In this text we shall not enter the rather interesting theory of fixed
points of the RGT and the determination of the critical exponents from
the study of the flow around them. This work is heading in other
direction.

\subsection{How to Make up a Block.}

In spite of the conclusions which might be drawn from the previous
section, the formation of a block from its constituent spins is not a
straight-forward procedure. In physical terms, we must determine
which are the variables which will characterize the block.

Let us consider an Ising-like model with two states per spin ($+1$ and
$-1$) and the formation of blocks, as in figure 2, through the
composition of four neighboring spins making up a square. If the block
spin must be an Ising spin again, it must take a value $+1$ or
$-1$. How should that value be chosen?
\pss

\bul Through decimation: the value of the upper-left spin is chosen
(for example). It seems to be an arbitrary method, but there are
suitable techniques for its realization.

\bul Through majority: the sign of the sum of all the spins is
chosen. It has one inconvenient: it does not provide a rule when the
block magnetization is null. In that case, adopt a random value.
\pss

The inconvenient of the majority rule is solved if the restriction of
Ising spins is lifted, and the adoption of arbitrary values for the
spin is allowed. In that case, the rule shall also be simple: take the
average value of the spins of the block (or the total spin, which is
equivalent modulo a scaling factor).

A different way to focus the problem may provide some insight. From
the four original degrees of freedom (see again fig. 2), RG must
choose a single one. According to the majority rule, it shall be the
{\sl mode} which is homogeneous on all four spins. Therefore, the
system is {\it projected\/} on the degrees of freedom which are
homogeneous for scales below the block size (which becomes a new UV
cutoff).

Is it necessary to choose precisely that degree of freedom for the
block? By no means. An interesting option, which shall appear to have
outmost importance for the rest of this work, is the adoption of
different degrees of freedom for each block, depending on their
circumstances (e.g.: distance to the border, values of some
inhomogeneous parameters...). Moreover, each block may be represented
by {\it more\/} than one degree of freedom, allowing a more refined
analysis.


\section{1.4. Brief History of the Real Space Renormalization Group.}

\subsection{I. Prehistory}

The idea of scale invariance is old and venerable in geometry. Scaling
arguments were commonplace among the pithagorean school/sect (the
golden section and its realizations), culminating with Euclid's proofs
(see, e.g. \ct[KLI-72]). These ideas, held alive through the
Middle Ages in a quasi-mystical disguise, had a great
influence on the re-foundation of physics in the XVII. For example, 
Galileo \ct[GAL-1638] makes a beautiful discussion on the scaling
properties of animal sizes\footn{E.g., if an animal doubles it size,
its weight is multiplied by eight, while the section of its bones is
only multiplied by four, rendering it unstable.}. Notwithstanding, the
mathematical tools of the rationalist era, early algebra and analysis,
were not well suited to deal with problems with many scales. The
appropriate tools may have appeared only recently.
\pss

Until the end of the XIX century, scaling arguments were not common in
physics. Fluid mechanics (\ct[MY-65], \ct[FRI-95]) is a classical
source of successful scaling hypothesis. The idea of Osborne Reynolds
about considering velocity fluctuations in a flow as the generators of
an ``effective viscosity'' is probably one of the first examples of a
parameter renormalization in the modern sense. Moreover, the
qualitative image of Lewis F. Richardson (1922) about
turbulence\footn{Lewis F. Richardson was a peculiar meteorologist with
a strong intuition and unorthodox expressive means. He paraphrased a
poem by Jonathan Swift \ct[MY-65]:\ps1 {\it Bigwhorls have little
whorls,\ps0 which feed on their velocity;\ps0 and little whorls have
lesser whorls,\ps0 and so on to viscosity (in the molecular sense).}}
was converted by Andrei N. Kolmogorov (1941) \ct[K-41] into a
quantitative theory of a suprising success.

By the beginning of the past century, acoustics and biophysics had
already generated a great amount of such kind of arguments, although
most of them were merely empirical \ct[SCH-91].
\pss

The last decades of the XIX century saw the rebirth of the attempts to
explain human scale phenomena on the basis of the movement and
interactions of atoms\footn{We have said rebirth because yet Plato's
``Timaeus'' was the first textbook on that subject
\ct[PLA-360$\;${\mayp B.C.}].}: kinetic theory and statistical
physics. On the other hand, the predominant scientific philosophy was
{\it positivism\/}, which proclaimed that deep understanding was not a
main objective of physics, and discarded atoms as unnecessary\footn{It
is said that Ludwig Boltzmann commited suicide partly because of the
reject by Ernst Mach and his school.}. Albert Einstein was partly
guided by scaling arguments in his search for a diffusion theory
\ct[EIN-05], whose importance is similar to his other two 1905 papers,
since it destroyed the opposition to the atomic hypothesis and
statistical physics.
\pss

Statistical physics opened up a new mathematical problem: the study of
many similar elements (molecules, spins, electrons...) which create a
coercive field on their neighbours. The most immediate ancestor of the
renormalization group is probably {\sl Mean Field Theory}
\ct[BDFN-93], \ct[GOL-93], which is an attempt of general applicability
to understand such systems by assuming the same behaviour for all the
items and searching a {\sl self-consistent solution}. In other words,
each element must act in such a way that its behaviour gets explained
by the interaction with neighboring elements behaving as it does
\footn{Act in such a way that your behaviour may serve as a general
rule: the kantian {\it cathegorical imperative\/} provides an
interesting analogy within ethics \ct[KAN-1788].}.

Some of the prototypical examples (of great success) are Van der Waals
theory of the liquid-vapour transition, Weiss theory of ferromagnetism
and Hartree-Fock theory of atomic spectra \ct[HAR-57] (which was
extended to interacting electrons in solids \ct[MAT-67]).

Mean field theory predicted a phase transition for the Ising model of
ferromagnetism in any dimension. Since the theory presupposes the
absence of dissent among the spins (there are no {\sl fluctuations}),
the higher the number of neighbours of each item is, the better it
works. Specifically, in dimension $\geq 4$ mean field yields {\sl
exact} results for many theories. But for unidimensional problems it
is unable to predict that, due to geometrical scaling
reasons\footn{The frontier of a unidimensional region has a magnitude
which is independent of its size. Thus, if a fluctuation is created
with downwards spins in the middle of a sea of upwards spins, the cost
is the same whether it is small or big. These fluctuations destroy any
possibility of ferromagnetism (long range order) at any finite
temperature.}, {\it there are no\/} phase transition.

Mean field theory reached its most highly sophisticated expression
with the ideas of L.D.~Landau, who employed it to analyze the first
great statistical theory of continuous fields which served to model a
big number of phase transitions with a mere change of its
parameters. V.L.~Ginzburg proceeded to analyze the influence of the
fluctuations and to give an applicability criterion for mean field on
Landau theory. But the mathematical theory was so intrincate that it
was impossible to improve the approximations for the critical
exponents given by mean field until the introduction of the RG.
\pss

Since the end of the XIX century a series of concepts which were later
known as {\sl fractal geometry} were getting mature \ct[MAN-92],
\ct[EDG-90]. The monster collection of classical analysis (Peano
curves, Weierstra\ss\ functions...) along with the insightful comments
of Henri Poincar\'e on the nature of non-integrability, the measure
theory of Felix Hausdorff (which includes non-integer dimensions)
joined the powerful intuition of Beno\^{\i}t Mandelbrot to create a
geometric frame which has proved to be better suited than the old
euclidean scheme for the understanding of phenomena with many
scales\footn{Perhaps it is worth to notice Mandelbrot's comment
\ct[MAN-82] on the fact that it was UV catastrophes which killed both
classical physics (\dag1900) and classical (smooth) mathematics
(\dag1875).}.

\subsection{II. The Archaic Age.}

RG made its appearance in a rather different disguise as it is known
today. An article of E.C.G.~Stueckelberg and A.~Peterman in 1953 and
another one by M.~Gell-Mann and F.E.~Low in 1954 \ct[SP-53],
\ct[GML-54] opened the field, reaching maturity with the text of
N.N.~Bogoliubov and D.V.~Shirkov in 1959 \ct[BOG-59].

These original works were dedicated to the elimination of infinites in
quantum field theories through the ``renormalization'' (change,
adaptation) of the physical parameters of the system. They were
posteriorly accused of lack of insight. In words of Kenneth Wilson
\ct[WIL-75], ``the worse feature of the standard renormalization
procedure is that it is a purely mathematical technique for
substracting out the divergent parts of integrals in the continuum
limit. It gives no insight into the physics of the statistical
continuum limit''.

The interpretation of the RG for quantum field theories in terms of
scale invariance (or its lack thereof) was carried out by C.G.~Callan
\ct[CAL-70], K.~Symanzik \ct[SYM-70] and K.G.~Wilson himself, around
1970, opening the path to the new era.

The archaic era provided us with a great amount of notation, beginning
with the very term ``renormalization group''. Despite the general
consensus on the inappropriateness of the term, it is sure that there
is no general agreement about the necessity of a change or the
direction it should take.

\subsection{III. The Golden Age.}

The main actors appear into stage around the middle sixties. In 1966,
in a short-lived journal called {\sl Physics}, Leo P.~Kadanoff
\ct[KAD-66] proposed the transformation of {\sl spins} into
{\sl block spins} for Ising-like models and proved that this
transformation explains some (so far) empirical relations among the
scaling exponents. 

The honour of converting the idea into an efficient computation method
corresponds to Kenneth G.~Wilson. In 1971 he published two consecutive
papers \ct[WIL-71{\mayp A}] and \ct[WIL-71{\mayp B}]. The first one
recasts Kadanoff's theory in differential form, rendering it more
suitable for mathematical analysis. The second one anticipated {\sl
wavelets} \ct[LEM-89] in more than a decade, and used them to analyze
Ginzburg-Landau's model. It constituted an intermediate step between
RSRG and the more developed Momentum Space RG.

The early RSRG techniques were successfully applied to the 2D Ising
model (see, e.g., Niemeyer and Van Leeuwen \ct[NvL-73]). In 1974
Wilson solved the Kondo problem, which dealt with the effect of a
magnetic impurity on the conduction band electrons of a metal
\ct[WIL-75]. He remarked that ``it was the first example where the
renormalization group program had been carried out in full''. His
solution was based on the division into {\sl shells} of the whole
lattice around the impurity, in such a way that the further a shell
was from the center, the nearer to the Fermi surface the electrons
were supposed to be\footn{The usage of mixed real space and momentum
space techniques is one of Wilson's great ideas. Electrons which are
far away from the impurity only contributed to the magnetic
susceptibility (his main target) if they were so near to the Fermi
surface that their ability to be excited could compensate.}. Shells
were integrated in an iterative way, starting from the center, so the
external ones only saw the impurity sheltered by the inner shells
(reminding somewhat of Gauss' theorem). It is important to remark that
the solution required a strong amount of numerical computations (in
1974 terms) and it was considered to be a great success with an error
in the observables of a few percent.
\pss

It was also in 1974 when Leo Kadanoff and his collaborators worked out
the first results for one of the most significative problems: the 2D
and 3D Ising model \ct[KAD-75] \ct[KH-75]. An advance of an
extraordinary significance for the applicability of the RG to the
computation of the critical exponents of phase transitions was the
{\sl Migdal--Kadanoff transformation}. It is a variational relation
between the free energies of two lattices which differ only in the
position of some bonds, as it can be seen in figure 3.

\figura{4cm}{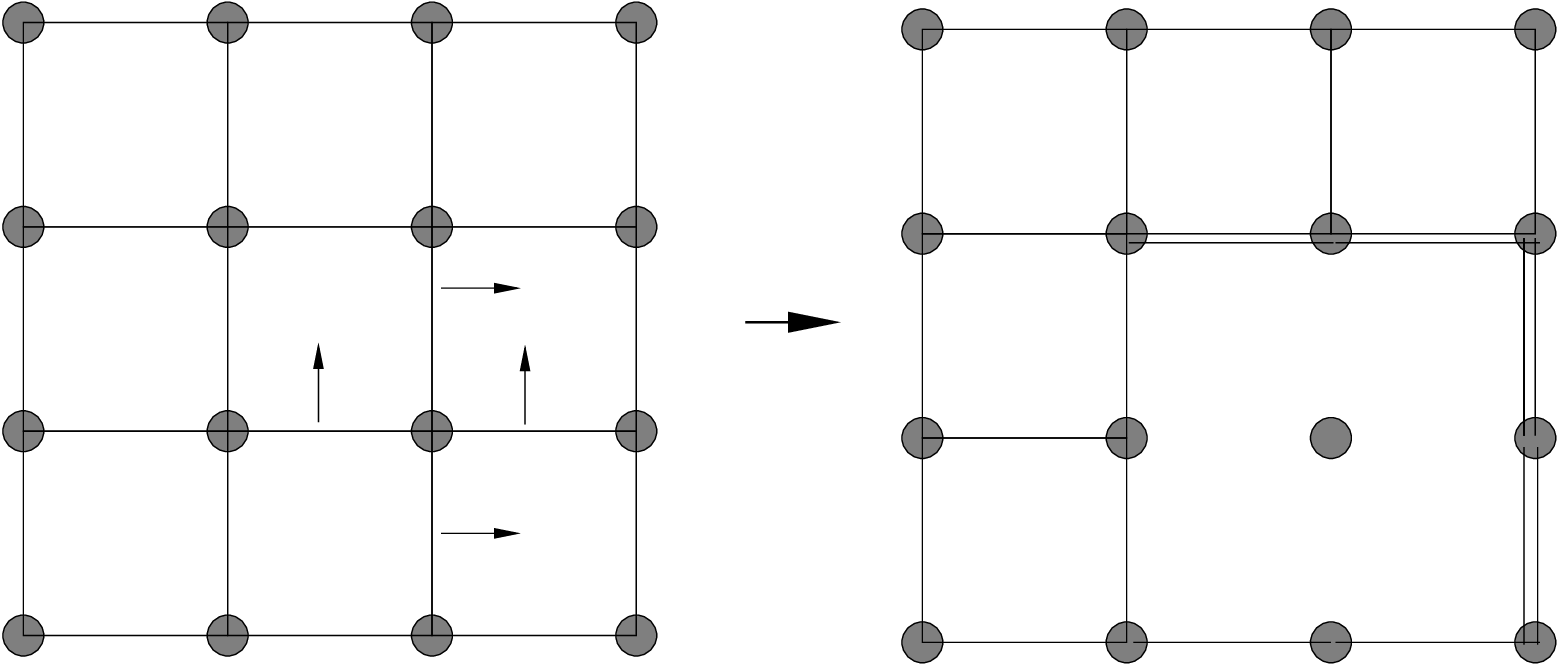}{Figure 3.}{Migdal--Kadanoff
transformation. Moving the marked bonds in the given directions leave
an unlinked site.}

The right side lattice in figure 3 has a spin site which is completely
isolated, and hence its contribution is trivial. The free energy of
the right lattice is a lower bound for the free energy of the
primitive lattice. Although the transformation was initially developed
by A.A.~Migdal in 1976 \ct[MIG-76], it was the interpretation given by
L.P.~Kadanoff in terms of moving bonds which made it popular
\ct[KAD-76].
\pss

By the end of the 70's the idea of developing the RSRG as a
non-perturbative computational method to obtain results for quantum
field theories was getting mature. Some groups working in parallel,
one in the United States (\ct[DWY-76], \ct[DWY-77], \ct[DSW-78],
\ct[DW-78], \ct[SDQW-80]) and others in the Netherlands and Paris
(\ct[BCLW-77], \ct[PJP-82]) developed similar techniques. The first
one had as their target the low energy properties of quantum
chromodynamics (QCD), meanwhile the others focused principally in
condensed matter many-body theories.

The method was baptized as the {\sl Blocks Renormalization Group}
(BRG). A detailed explanation may be found in chapter 2, but the
general idea deserves to be exposed here succintly. The objective is
to obtain {\sl variationally} the lower energy states of bigger and
bigger blocks using as {\it Ansatz\/} the lower energy states of the
smaller ones.
\pss

In 1982 Kenneth G. Wilson was awarded (in solitary) the Nobel prize
for his RG solution to the Kondo problem. It is interesting to notice
that by that time RSRG was going through an {\it impasse\/}, since the
results of the previously stated applications to QCD and condensed
matter physics were not as successful as it was hoped. On the other
hand, big computer facilities allowed the implementation of massive
Quantum Monte-Carlo \ct[BH-92] algorithms to solve some of the same
problems. Thus, RSRG was slowly abandoned.
\pss

By the same time, the ideas of scale invariance had already invaded
the theoretical analysis of quantum field theories (Callan-Symanzyk
equations, scaling anomalies...) \ct[COL-85], \ct[ITZD-89]. Practical
Monte-Carlo calculations for lattice field theories required the
conceptual apparatus of RG to take the ``statistical continuum
limit'', as Wilson had foreseen.

\subsection{IV. The Scholastic Age.}

The (relative) abandonment by the groups which were mostly interested
in the obtention of numerical results left the field ready for deeper
theoretical analysis. Through the 80's a series of algebraic and
conceptual techniques were developed which helped to understand the
RSRG process.

Touching briefly a topic which surpasses this brief introduction we
shall remark the analysis of Belavin, Polyakov and Zamolodchikov in
1984 \ct[BPZ-84], \ct[ITZD-89], which issued the idea that 2D
universality classes might be thought to be linked to representations
of an algebra derived from RG\footn{Scaling exponents would be
universal in the same sense as spins for different particles are,
since $J^2$ is a Casimir invariant of $SU(2)$. The appropriate algebra
is called after {\sl Virasoro}.} in their work on 2D Conformal Field
Theories (CFT)\footn{Conformal transformations are all those which
conserve angles locally, though not necessarily distances. Scale
invariance along with translational, rotational and under inversion
yields full conformal invariance. In 2D, conformal invariance is
special because, since any complex analytic function determines a
conformal transformation, the group has infinite dimension.}. This
work triggered an avalanche of activity, mainly due to the interest
which in a few years was unleashed by string theory\footn{Quantum
String Theory's \ct[GSW-87] main job is to ``count'' all possible
surfaces which start and end at given curves. The necessity to avoid
repetition forces to take that sum {\it modulo\/}
re-parametrizations. That led the practicioners to the study of
CFT.}. The application of CFT and, afterwards, of quantum groups to
the problems of statistical physics was a task which, although
originated at this time, was matured only through the decade of the
90's \ct[GRS-96].

\subsection{V. The Industrial Age.}

Numerical methods and theoretical ideas which had ``something'' in
common with RSRG had been developing in parallel with it. This proves
that the main idea was mature and needed, both technical and
theoretically. Among the most applications--oriented ideas we may cite
multi-grid methods \ct[PTVF-97], among the most theoretical, quantum
groups or conformal field theories \ct[GRS-96] and among the mixed
ones, {\it wavelets\/} theory \ct[LEM-89] and multi-fractal analysis
\ct[FRI-95].

Quantum Monte-Carlo (QMC) method, which had displaced RSRG by the
beginning of the 80's, had serious problems when applied to critical
phenomena. Beyond the strong requirements on CPU time, memory and the
so-called {\sl critical slow-down} (which made the convergence of the
observables despairingly slow in the vicinity of a critical point), it
suffered from the {\sl sign problem} \ct[BH-92]. 

QMC is based on the conversion of a quantum problem into an equivalent
classical equilibrium one through use of the Trotter-Suzuki formula
(Feynman and Hibbs \ct[FH-65] established the analogy between a
partition function and a path integral, while Suzuki \ct[SUZ-76]
converted it into a useful formula\footn{The relation between quantum
systems in dimension $d$ and classical equilibrium systems in $d+1$
seems to require a more fundamental explanation.}). The sign problem
is related to this conversion in the case of fermions: the method
yields absurd results, such as negative ``ghost-like''
probabilities. Although some tricks were developed, very scarce
progress was made during the 80's.

These disappointing results made some numerically inclined researchers
turn their eyes again to RSRG.
\pss

The success of Wilson's computation had never been repeated. Kondo's
problem (and, in general, impurity problems) is, in some sense,
special: it consists of a single source of interaction; each added
shell is further away from it and, therefore, interacts more weakly.

Wilson himself had proposed, in an informal talk in 1986, to try a
very simple problem in order to understand the failure of the RSRG:
the problem of a spinless particle in a 1D box studied with quantum
mechanics \ct[WHI-99]. A few people from the public, among them Steve
R.~White, undertook the project.

This problem had the advantage of its simplicity and, moreover, it
seemed to include all the elements which made the RSRG fail for more
sophisticated problems. In effect, the blocks method (also known as
BRG) obtained variationally the lowest energy states of the system
using as {\it Ansatz\/} for the wave-functions the approximations
which were obtained for the smaller blocks. Unfortunately (or maybe
not so) the energies were incorrect by {\it some orders of
magnitude\/} for big blocks, and did not scale properly.

In 1992 S.R.~White, after some unfruitful attempts\footn{There {\it
was\/} a successful method among the ones tried by Steve White and
Reinhard Noack, but it was impossible to generalize. It used as
``bricks'' states from the blocks created with different boundary
conditions.}, arrived at the satisfactory solution. The reason for
which BRG did not work correctly with the particle in a box was that
the block was {\sl isolated} from its neighbours. White observed that the
low energy states from the small blocks are, in many cases, quite bad
{\it bricks\/} for building the low energy states of bigger blocks
(see figure 5 in chapter 2).

The best states to be chosen may not necessarily correspond to the
lowest energy states. Let us consider, e.g., two blocks and two sites
in 1D as are depicted in figure 4. The ground state is found for the
full system and it is projected on the left and right parts (left
block + site, right block + site), providing us with new blocks (which
have one more site) and which constitute good bricks to continue
growing.

\ps3
\rput(47,0){
\multirput{0}(0,0)(5,0){4}{\pscircle*(0,0){1.5}\psline(1.5,0)(3.5,0)}
\psline(20,0)(22,0)\pscircle(23.5,0){1.5}
\psline(25,0)(27,0)\pscircle(28.5,0){1.5}\psline(30,0)(32,0)
\multirput{0}(33.5,0)(5,0){4}{\psline(0,0)(2,0)\pscircle*(3.5,0){1.5}}
\rput(1,-5){\peqfont Left}
\rput(50,-5){\peqfont Right}}
\ps8
\nind{\mayp Figure 4. }{\peqfont Two blocks of several spins, and two
intermediate sites which shall be swallowed by the blocks.\par}
\pss

The projection is carried out with the help of a density matrix, which
provides the name of {\sl Density Matrix Renormalization Group
(DMRG)}, profusely described at chapter 3.

DMRG has reached numerical precision which is not only much higher
than any one obtained with BRG, but also surpasses Wilson's solution
to the Kondo problem. It has some handicaps, since it is based upon
the left-right distinction. This makes precision diminish drastically
in higher dimensions.

At this point the historical exposition comes to an end in order to
leave the stage for the explanations on our own work on this field.


\section{1.5. Plan of this Work.}

Chapter 2 exposes a modification of the Blocks Renormalization Group
(BRG) which allows to analyze correctly quantum-mechanical problems
both in 1D and 2D. This technique is called Correlated Blocks
Renormalization Group (CBRG). Most part of the material of this
chapter is taken from the paper of M.A.~Mart\'{\i}n-Delgado,
J.~Rodr\'{\i}guez-Laguna and G.~Sierra \ct[MRS-95], albeit some ideas
are exposed here for the first time.

The third chapter describes in detail the application of the DMRG for
problems in quantum mechanics, both for 1D systems and for any kind of
tree-like structure. As an application with technical interest, the
excitonic spectrum of some macromolecules with fractal properties
called {\sl dendrimers} is obtained. Most part of the contents of this
chapter is published in \ct[MRS-00{\mayp B}], from the same authors,
and the references therein.

The standard formulation of the DMRG must necessarily distinguish
between left and right, which is inappropriate for multidimensional
systems. In order to analyze these, the Punctures Renormalization
Group (PRG) was introduced, which is extensively described in chapter
4. The exposition is based on that of \ct[MRS-00{\mayp A}], from the
former authors, but an application to the problem of localization in
disordered excitonic systems is provided, taken from \ct[DLMRS-01],
along with some reflections on the possibility to extend the method to
many-body problems.

The formerly exposed methods are suitable for application to the
analysis of numerical methods for partial differential equations. Two
RSRG methods for the improvement of the efficiency of numerical
algorithms, one based on the overlapping of cells and the other on
more physical criteria, are developed in chapter 5. The source is the
papers of A.~Degenhard and J.~Rodr\'{\i}guez-Laguna \ct[DRL-01{\mayp
A}] and \ct[DRL-01{\mayp B}].

Each chapter has a bibliography attached. This work ends with six
appendices and a glossary of important terms, either classical in the
literature or introduced in our work.


\section{1.6. Bibliography.}

\cx[BCLW-77] {\may H.P. van de Braak, W.J. Caspers, C. de Lange,
M.W.M. Willemse}, {\sl The determination of the ground-state energy
of an antiferromagnetic lattice by means of a renormalization
procedure}, Physica A, {\bf 87}, 354-368 (1977).

\cx[BDFN-93] {\may J.J. Binney, N.J. Dowrick, A.J. Fisher, M.E.J. Newman},
{\sl The theory of critical phenomena.} Clarendon Press (1993).

\cx[BH-92] {\may K. Binder, D.W. Heermann}, {\sl Monte Carlo simulation
in statistical physics}, Springer (1992).

\cx[BOG-59] {\may N.N. Bogoliubov, D.V. Shirkov}, {\sl The theory of
quantized fields}, Interscience (1959).

\cx[BPZ-84] {\may A.A. Belavin, A.M. Polyakov, A.B. Zamolodchikov},
{\sl Infinite conformal symmetry in two-dimensional quantum field
theory}, Nucl. Phys. B {\bf 241}, 333-380 (1984).

\cx[BS-95] {\may A.L. Barab\'asi, H.E. Stanley}, {\sl Fractal concepts in
surface growth}, Cambridge U.P. (1995).

\cx[BTW-88] {\may P. Bak, C. Tang, K. Wiesenfeld}, {\sl Self-organized
criticality}, Phys.~Rev. {\bf 38}, 364-374 (1988).

\cx[CAL-70] {\may C.G. Callan}, {\sl Broken scale invariance in scalar
field theory}, Phys.~Rev.~D {\bf 2}, 1541-47 (1970).

\cx[CFP-92] {\may R.J. Creswick, H.A. Farach, C.P. Poole Jr.}, {\sl
Introduction to renormalization group methods in physics}, John Wiley
\& Sons, (1992).

\cx[COL-85] {\may S. Coleman}, {\sl Aspects of symmetry}, Cambridge
U.P. (1985).

\cx[DGMP-97] {\may A. Dobado, A. G\'omez-N\'{\i}cola, A. Maroto,
J.R. Pel\'aez}, {\sl Effective lagragians for the standard model},
Springer (1997).

\cx[DLMRS-01] {\may F. Dom\'{\i}nguez-Adame, J.P. Lemaistre,
V.A. Malyshev, M.A. Mart\'{\i}n-Delgado, A. Rodr\'{\i}guez,
J.~Ro\-dr\'{\i}\-guez-Laguna, G.~Sierra}, {\sl Absence of weak
localization in two-dimen\-sional disordered Frenkel lattices},
available at {\tt cond-mat/0201535} and J.~Lumin. {\bf 94-95}, 359-363
(2001).

\cx[DRL-01{\mayp A}] {\may A. Degenhard, J. Rodr\'{\i}guez-Laguna}, 
{\sl A real space renormalization group approach to field evolution
equations}, {\tt cond-mat/0106155} and Phys.~Rev.~E {\bf 65}
036703 (2002). 

\cx[DRL-01{\mayp B}] {\may A. Degenhard, J. Rodr\'{\i}guez-Laguna}, {\sl 
Towards the evaluation of the relevant degrees of freedom in nonlinear
partial differential equations}, {\tt cond-mat/0106156} and
J.~Stat.~Phys. {\bf 106} 5/6, 1093 (2002).

\cx[DSW-78] {\may S.D. Drell, B. Svetitsky, M. Weinstein}, {\sl
Fermion field theory on a lattice: variational analysis of the
Thirring model}, Phys.~Rev.~D {\bf 17}, 2, 523-36 (1978).

\cx[DUP-89] {\may B. Duplantier}, {\sl Fractals in two dimensions and
conformal invariance}, Physica D {\bf 38}, 71-87 (1989).

\cx[DW-78] {\may S.D. Drell, M. Weinstein}, {\sl Field theory on a
lattice: absence of Goldstone bosons in the U(1) model in two
dimensions}, Phys.~Rev.~D {\bf 17}, 12, 3203-11 (1978).

\cx[DWY-76] {\may S.D. Drell, M. Weinstein, S. Yankielowicz}, {\sl
Strong coupling field theory I. Variational approach to $\phi^4$
theory}, Phys.~Rev.~D {\bf 14}, 2, 487-516 (1976).

\cx[DWY-77] {\may S.D. Drell, M. Weinstein, S. Yankielowicz}, {\sl Quantum
field theories on a lattice: variational methods for arbitrary coupling
strengths and the Ising model in a transverse magnetic field}, Phys.~Rev.~D
{\bf 16}, 1769-81 (1977).

\cx[EDG-90] {\may G.A. Edgar}, {\sl Measure, topology and fractal
geometry}, Springer (1990).

\cx[EIN-05] {\may A. Einstein}, {\sl \"Uber die von der
molekularkinetischen Theorie der W\"arme gefordete Bewegung von in
ruhenden Fl\"ussigkeiten suspendierten Teilchen}, Annalen der Physik
{\bf 17}, 549-560 (1905).

\cx[FEY-72] {\may R.P. Feynman}, {\sl Statistical Mechanics: 
A Set of Lectures}, Addison-Wesley (1972).

\cx[FH-65] {\may R.P. Feynman, A.R. Hibbs}, {\sl Quantum Mechanics and
Path Integrals}, McGraw-Hill (1965).

\cx[FRI-95] {\may U. Frisch}, {\sl Turbulence. The legacy of
A.N. Kolmogorov}, Cambridge U.P. (1995).

\cx[GAL-1638] {\may G. Galilei}, {\sl Discorsi e dimostrazioni
matematiche, intorno \`a due nuove scienze atteneti alla Mecanica ed
ai muovimenti locali}, freely available at {\tt
http://www.liberliber.it} (1638).

\cx[GAW-96] {\may K. Gaw\c edzki}, {\sl Turbulence under a magnifying
glass}, talk given at the Carg\`ese Summer Institute in
1996, available at {\tt chao-dyn/9610003} (1996).

\cx[GEN-79] {\may P.G. de Gennes}, {\sl Scaling concepts in polymer
physics}, Cornell U.P. (1979).

\cx[GKP-89] {\may R.L. Graham, D.E. Knuth, O. Patashnik}, {\sl
Concrete mathematics}, Addison-Wesley Publ. Co. (1994) (First ed.:
1989).

\cx[GML-54] {\may M. Gell-Mann, F.E. Low}, {\sl Quantum
electrodynamics at small distances}, Phys.~Rev. {\bf 95}, 5,
1300-12 (1954).

\cx[GMSV-95] {\may J. Gonz\'alez, M.A. Mart\'{\i}n-Delgado, G. Sierra,
A.H. Vozmediano}, {\sl Quantum electron liquids and High-$T_c$
superconductivity}, Springer (1995).

\cx[GOL-93] {\may N. Goldenfeld}, {\sl Lectures on phase transitions and
the renormalization group}, Addison-Wesley (1993).

\cx[GRS-96] {\may C. G\'omez, M. Ruiz-Altaba, G. Sierra}, {\sl Quantum
groups in two-dimensional physics}, Cambridge U.P. (1996).

\cx[GSW-87] {\may M.B. Green, J.H. Schwarz, E. Witten}, {\sl
Superstring theory}, Cambridge U.P. (1987).

\cx[HAR-57] {\may D.R. Hartree}, {\sl The calculation of atomic
structure}, John Wiley \& sons (1957).

\cx[ISI-25] {\may E. Ising} Z. der Physik {\bf 31}, 253 (1925).

\cx[ITZD-89] {\may C. Itzykson, J.M. Drouffe}, {\sl Statistical Field Theory},
Cambridge U.P. (1989).

\cx[K-41] {\may A.N. Kolmogorov}, {\sl The local structure of
turbulence in incompressible viscous fluid for very large Reynolds
numbers}, Dokl.~Akad.~Nauk SSSR {\bf 30}, 4 (1941). Reprinted in
Proc.~Roy.~Soc. A {\bf 434}, 9 (1991).

\cx[KAD-66] {\may L.P. Kadanoff}, {\sl Scaling laws for Ising models near
$T_c$}, Physica {\bf 2}, 263 (1966). 

\cx[KAD-75] {\may L.P. Kadanoff}, {\sl Variational principles and
approximate renormalization group calculations}, Phys.~Rev.~Lett. {\bf
34}, 16, 1005-8, (1975).

\cx[KAD-76] {\may L.P. Kadanoff}, Ann.~Phys.~(N.Y.) {\bf 100}, 359 (1976).

\cx[KAN-1788] {\may I. Kant}, {\sl Kritik der praktischen Vernunft},
freely available at {\tt http://gutenberg.aol.de/kant/\break
kritikpr/kritikpr.htm} (1788).

\cx[KH-75] {\may L.P. Kadanoff, A. Houghton}, {\sl Numerical
evaluation of the critical properties of the two-dimensional Ising
model}, Phys.~Rev.~B {\bf 11}, 1, 377-386 (1975).

\cx[KLI-72] {\may M. Kline}, {\sl Mathematical thought from ancient to
modern times}, Oxford U.P. (1972).

\cx[KOG-79] {\may J.B. Kogut}, {\sl An introduction to lattice gauge theory
and spin systems}, Rev.~Mod.~Phys. {\bf 51}, 4, 659-713 (1979).

\cx[KR\"O-00] {\may H. Kr\"oger}, {\sl Fractal geometry in quantum
mechanics, field theory and spin systems}, Phys.~Rep. 323, 81-181
(2000).

\cx[KT-90] {\may E.W. Kolb, M.S. Turner}, {\sl The Early Universe},
Addison-Wesley (1990).

\cx[LEM-89] {\may  P.G. Lemari\'e} (ed.), {\sl Les Ondelettes en 1989},
Springer (1989).

\cx[LEN-20] {\may W. Lenz}, Phys. Zeitschrift {\bf 21}, 613 (1920). 

\cx[MAN-82] {\may B.B. Mandelbrot}, {\sl The fractal geometry of
nature}, Freeman \& Co. (1982).

\cx[MAT-67] {\may R.D. Mattuck}, {\sl A guide to Feynman diagrams in
the many-body problem}, Dover (1992). First ed. by McGraw-Hill in
1967.

\cx[MIG-76] {\may A.A. Migdal}, Sov. Phys. JETP {\bf 42}, 743 (1976).

\cx[MRS-95] {\may M.A.~Mart\'{\i}n-Delgado, J.~Rodr\'{\i}guez-Laguna, 
G.~Sierra}, {\sl The correlated block renormalization group},
available at {\tt cond-mat/9512130} and Nucl.~Phys.~B {\bf 473},
685-706 (1996).

\cx[MRS-00{\mayp A}] {\may M.A.~Mart\'{\i}n-Delgado,
J.~Rodr\'{\i}guez-Laguna, G.~Sierra}, {\sl Single block formulation of
the DMRG in several dimensions: quantum mechanical problems},
available at {\tt cond-mat/0009474}, and Nucl.~Phys.~B {\bf 601},
569-590 (2001).

\cx[MRS-00{\mayp B}] {\may M.A.~Mart\'{\i}n-Delgado,
J.~Rodr\'{\i}guez-Laguna, G.~Sierra}, {\sl A density matrix
renormalization group study of excitons in dendrimers}, available at
{\tt cond-mat/0012382}, and Phys.~Rev.~B {\bf 65}, 155116 (2002).

\cx[MY-65] {\may A.S. Monin, A.M. Yaglom}, {\sl Statistical fluid 
mechanics}, The MIT press (1971). First ed. in russian by Nauka (1965).

\cx[NvL-73] {\may T. Niemeyer, J.M.J. van Leeuwen}, {\sl Wilson theory
for spin systems on a triangular lattice}, Phys.~Rev.~Lett {\bf 31},
1411 (1973).

\cx[PEB-80] {\may P.J.E. Peebles}, {\sl The Large-Scale Structure of the
Universe}, Princeton U.P. (1980).

\cx[PJP-82] {\may P. Pfeuty, R. Jullien, K.A. Penson}, {\sl
Renormalization for quantum systems}, in {\sl Real Space
Renormalization}, ed. by T.W. Burk\-hardt and J.M.J. van Leeuwen
pages 119-147, Springer (1982).

\cx[PLA-360$\;${\mayp B.C.}] {\may Plato}, {\sl Timaeus}, freely
available at {\tt http://perseus.mpiwg-berlin.mpg.de} both in
classical Greek and English (360 B.C.).

\cx[PTVF-97] {\may W.H. Press, S.A. Teukolsky, W.T. Vetterling,
B.P. Flannery}, {\sl Numerical Recipes in C}, Cambridge U. P. (1997),
and freely available at {\tt http://www.nr.com}.

\cx[SCH-91] {\may M. Schroeder}, {\sl Fractals, chaos, power laws},
W.H. Freeman \& Co. (1991).

\cx[SDQW-80] {\may B. Svetitsky, S.D. Drell, H.R. Quinn,
M. Weinstein}, {\sl Dynamical breaking of chiral symmetry in lattice
gauge theories}, Phys.~Rev.~D {\bf 22}, 2, 490-504 (1980).

\cx[SP-53] {\may E.C.G. Stuckelberg, A. Peterman}, Helv.~Phys.~Acta, {\bf
26}, 499 (1953). 

\cx[SUZ-76] {\may M. Suzuki}, {\sl Relationship between $d$-dimensional
quantal spin systems and $(d+1)$-dimensional Ising systems},
Prog.~Theor.~Phys. {\bf 56}, 5, 1454-69 (1976).

\cx[SYM-70] {\may K. Symanzik}, Comm.~Math.~Phys. {\bf 18}, 227 (1970).

\cx[WHI-98] {\may S.R. White}, {\sl Strongly correlated electron
systems and the density matrix renormalization group}, Phys.~Rep. {\bf
301}, 187-204 (1998).

\cx[WHI-99] {\may S.R. White}, {\sl How it all began}, en
{\sl Density-Matrix Renormalization} ed. por I. Peschel et
al. (Dresden conference in 1998), Springer (1999).

\cx[WIL-71{\mayp A}] {\may K.G. Wilson}, {\sl Renormalization group and 
critical phenomena I. Renormalization group and the Kadanoff scaling
picture}, Phys.~Rev.~B {\bf 4}, 9, 3174-3183 (1971).

\cx[WIL-71{\mayp B}] {\may K.G. Wilson}, {\sl Renormalization group
and critical phenomena II. Phase-cell analysis of critical behaviour},
Phys.~Rev. {\bf B}, 9, 3184-3205 (1971).

\cx[WIL-75] {\may K.G. Wilson}, {\sl The renormalization group: critical
phenomena and the Kondo problem}, Rev. Mod. Phys. {\bf 47}, 4, 773-840 
(1975).

\pasacap

\chapter{2. The Correlated Blocks Renormalization Group.}

\subsection{Synopsis.}

\bul Part I. The Blocks Renormalization Group.

2.1 The BRG for a quantum magnetism model.

2.2 A simple test model: the particle in a box.
\ps2

\bul Part II. The Correlated Blocks Renormalization Group.

2.3 Correlated blocks in 1D.

2.4 CBRG in practice.

2.5 Wave--function reconstruction.

2.6 Bidimensional CBRG.

2.7 Self--replicability.
\ps2

2.8 Bibliography.
\pss\pss

This chapter develops in its first part the structure of technique
called the {\it Blocks Renormalization Group method (BRG)\/} in some
detail, along with the origin of its poor quantitative (and sometimes
qualitative) results. The second part contains the original work of
this chapter, which is the method known as {\it Correlated Blocks
Renormalization Group (CBRG)\/}, introduced by our group in 1996
\ct[MRS-96]. It is an extension of the BRG algorithm
which repairs its deficiencies and, in some sense, respects its
spirit.

\part{Part I. The Blocks Renormalization Group.}

The name Blocks Renormalization Group (BRG) is associated with 
one of the realizations of the blocking idea of Leo Kadanoff
\ct[KAD-66]: the obtention of the low energy spectrum of a quantum 
system. The main advantage of the technique is that it is {\it
non-perturbative\/}\footn{It is non-perturbative on the coupling
constants, but it admits a perturbative expansion on the inter-blocks
interaction. See \ct[GMSV-95].\ps{-3}}. Thus, states with topologically
non-trivial configurations and structures quite far away from the
non-interacting states are equally available to the method.

It was developed by some independent groups. One of them, working at
SLAC, had as their main target the low energy properties of the
theories composing the standard model, specially QCD \ct[DWY-76],
\ct[DWY-77], \ct[DW-78], \ct[DSW-78], \ct[SDQW-80]. Others,
working from the Netherlands and Paris, focused on models inspired in
condensed matter physics, such as the Heisenberg model \ct[BCLW-77],
\ct[JUL-81], \ct[PJP-82].

The information contained in this first part of this chapter is not
original to our work. It is included merely for completeness and to
make the text as self-contained as possible. Classical reviews on
these topics include \ct[JUL-81], \ct[PJP-82] or \ct[GMSV-95].

\section{2.1. BRG for a Quantum Magnetism Model.}

\subsection{The Ising Model in a Transverse Field.}

The 1D Ising model in a transverse field (ITF) is possibly the
simplest model presenting a {\sl quantum phase transition}
\ct[GMSV-95]. The hamiltonian is a summation of terms like this one

$$h_{12}=-\Gamma(\sigma_1^x+\sigma_2^x) - J\sigma_1^z\sigma_2^z \qquad
\Gamma,J>0 \eq{\itfterm}$$

\nind with $\sigma^x$ and $\sigma^z$ two of the Pauli matrices. The
parameters $\Gamma$ and $J$ represent, respectively, an external
magnetic field applied in the $x$--direction and the uniaxial coupling
constant between neighbour spins along the $z$--direction. We shall
assume both constants to be positive.

The problem is non-trivial because of the competition of two effects:
the magnetic field attempts to align all the spins in the $x$
direction, meanwhile the coupling constant forces them to align in the
$z$ direction. The compromise is complicated because both components
of the spin {\it do not commute\/}. Therefore, if the
$\sigma^z$--eigenstates basis is assumed, the states which are aligned
along the $x$ axis (due to the external field) appear to be
``disordered''.

The known relation between quantum models at dimension $d$ and
equilibrium classical models at dimension $d+1$ is exemplified in the
correspondence which may be established between the 1D ITF and the 2D
classical Ising model. Within it, the external magnetic field plays
the r\^ole of a temperature (see the text by Feynman and Hibbs
\ct[FH-65] for general considerations, and the article by Suzuki
\ct[SUZ-76] for the explicit construction of that correspondence).

The development of the application of the BRG to the Ising problem in
a transverse field was carried out in \ct[MDS-96], though we shall
follow closely the exposition made at \ct[GMSV-95].

\subsection{The BRG applied the ITF Model.}

The main idea of the BRG procedure may be quite succintly described in
a single paragraph. Let us consider two neighboring $1/2$ spins
interacting and let $\ket|+>$ and $\ket|->$ denote the eigenstates of
$\sigma^x$. The total system may be in four states $\ket|-->$,
$\ket|-+>$, $\ket|+->$ and $\ket|++>$. A hamiltonian matrix may be
written for them and diagonalized. Let us suppose that only the low
energy spectrum is our target. Then, the two lowest energy states may
be retained and supposed to be the only possible states for the spins
pair. To complete the analogy, we may label these states with the
signs ``$\ket|-'>$'' and ``$\ket|+'>$''. Now the interaction between
{\it four\/} real $1/2$ spins may be studied as the interaction
between {\it two\/} such ``block spins''.
\pss

If the system has only two sites, the eigenstates are given by:

$$\eqalign{
\ket|G>={1\over \sqrt{1+a^2}}( \ket|--> + a\ket|++> ) & \qquad
E=-\sqrt{J^2+4\Gamma^2}\cr
\ket|E_1>={1\over\sqrt{2}}(\ket|-+>+\ket|+->) & \qquad E=-J \cr
\ket|E_2>={1\over\sqrt{2}}(\ket|-+>-\ket|+->) & \qquad E=J \cr
\ket|E_3>={1\over\sqrt{1+a^2}}(-a\ket|-->+\ket|++>) &
\qquad E=\sqrt{J^2+4\Gamma^2}\cr}\eq{\itfstates}$$

\noindent where $a$ is a parameter which depends on the coupling constant
$g\equiv J/2\Gamma$,

$$a(g)\equiv {-1+\sqrt{1+g^2}\over g}\eq{\itfdefa}$$

\nind Notice that the parameter $a$ ranges from $a(0)=0$
to $a(\infty)=1$. The function $a(g)$ interpolates between these to
extreme situations: only magnetic field when $a=0$ and only spins
coupling when $a=1$.

The first two states are now going to be assimilated as new ``block
spin'' states $\ket|-'>\ot\ket|G>$ and $\ket|+'>\ot\ket|E_1>$. This
operation is equivalent to a mapping from a Hilbert space of {\sl two
spins} (dimension 4) into a Hilbert space of a {\sl single block spin}
(dimension 2). Let us define the operators:
\ps{-4}

$$\displaylines{T: \hbox{\amsmath C}\,{}^4 \mapsto 
\hbox{\amsmath C}\,{}^2 \cr
T^\dagger: \hbox{\amsmath C}\,{}^2 \mapsto 
\hbox{\amsmath C}\,{}^4 \cr}$$

\nind such that

$$\eqalign{ 
T\ &=\ \ket|-'>\bra<G|\ +\ \ket|+'>\bra<E_1| \cr
T^\dagger\ &=\ \ket|G>\bra<-'|\ +\ \ket|E_1>\bra<+'| \cr }$$ 

\nind The first of these operators, $T$, is known as {\sl truncation
operator}, which takes a ``two spins'' state and projects it on a
``block spin'' state. The second, $T^\dagger$, called {\sl embedding
operator}, takes a block spin state and returns a real ``two spins''
state. It is straightforward to observe that (since $\<-'|+'\>=0$)

$$TT^\dagger=I \qquad \hbox{Identity on ``block spins'' space}$$

\nind i.e.: the successive application of the embedding and
the truncation operators yields the identity on the block states
space. But this relation may not be inverted:

$$T^\dagger T \neq I \qquad \hbox{Identity on real ``two spins'' space}$$

\nind So, the operator $T^\dagger T$, which first truncates a real
``two spins'' state into a block state and then tries to reconstruct
it back, is {\sl not} the identity. It is the {\sl projector} on the
subspace of the {\sl retained degrees of freedom}. This shall be a
recurrent idea throughout this thesis.
\pss

The next step is to write the hamiltonian for the system formed by
{\sl two block spins}. This hamiltonian has (hopefully) the same form
as the previous one, so as the RG step may be iterated and reach
arbitrarily long chains.

Two blocks are placed side by side and allowed to interact via the
same ITF hamiltonian. The ``real'' spins have indices $1$ to $4$,
meanwhile the block spins have indices $1'$ and $2'$. If $j$ is a real
spin index, $j'$ shall denote the index of its associated block spin:
$j'=\lfloor j/2 \rfloor$. Figure 1 represents the numbering of sites
and blocks graphically.

\midinsert
\pss\pss
\rput(40,0){
\psline(0,0)(20,0)\psline(40,0)(60,0)
\multirput(0,0)(20,0){4}{\pscircle*(0,0){1.5}}
\rput(0,6){$j=1$}\rput(20,6){$j=2$}\rput(40,6){$j=3$}\rput(60,6){$j=4$}
\rput(10,-8){$j'=1$}\rput(50,-8){$j'=2$}
\psframe(-3,-3)(23,3)\psframe(37,-3)(63,3)
\psline[linestyle=dashed](20,0)(40,0)}
\ps{10}
\nind{\mayp Figure 1. }{\peqfont Numbering of the sites and blocks for
the BRG application on the 1D-ITF. The continuous line corresponds to
the intra--block links, and the dashed one to the inter--blocks
link.\par}
\pss
\endinsert

Now let us define the block spin operators so as they fulfill
relations which are {\it equivalent\/} to the ones between the old
$\sigma$'s:

$$\displaylines{
\sigma^z_{j'}\ket|-'>_{j'}=-\ket|-'>_{j'} \qquad 
\sigma^z_{j'}\ket|+'>_{j'}=\ket|+'>_{j'} \cr
\sigma^x_{j'}\ket|-'>_{j'}=\ket|+'>_{j'} \qquad 
\sigma^x_{j'}\ket|+'>_{j'}=\ket|-'>_{j'} \cr}$$

An easy algebraic manipulation leads us to

$$\displaylines{ 
T\sigma^x_j T^\dagger={1-a^2\over 2 (1+a^2)} ( I + \sigma^x_{j'})\cr 
T\sigma^z_j T^\dagger={1+a^2\over\sqrt{2(1+a^2)}} \sigma^z_{j'}\cr 
T\sigma^z_{2j-1}\sigma^z_{2j}T^\dagger={(1+a)^2\over
2(1+a^2)}I + {(1-a)^2\over 2(1+a)^2}\sigma^x_{j'}
\rlap{\qquad (Intra--block link)}\cr}$$

These transformed operators shall enable us to re-write the
hamiltonian for each block in its own terms. But the inter-blocks part
(the dashed line in figure 1) requires a new operator:

$$T \sigma^z_{2j}\sigma^z_{2j+1} T^\dagger={(1+a)^2\over
2(1+a^2)}\sigma^z_{j'}\sigma^z_{j'+1}
\rlap{\qquad (Inter--blocks link)}$$

The most interesting conclusion of these equations is that, in fact,
the hamiltonian for the renormalized states has {\sl the same form} as
the original one:

$$T H_N(\Gamma,J) T^\dagger=\Delta E+H_{N/2}(\Gamma',J')$$

\nind with $\Delta E$ as a change in the ``zero energy level'' and two
renormalized values for $\Gamma$ and $J$. The three parameters are
given by:

$$\displaylines{
\Delta E=-{N\over 2}\[ \Gamma{1-a^2\over (1+a)^2} + {J\over
2}{(1+a)^2\over (1+a^2)} \]\cr
\Gamma'=\Gamma {1-a^2\over (1+a)^2}- J{(1+a)^2\over 2(1+a^2)} \qquad
J'=J{(1+a)^2\over 2(1+a^2)}}$$

We should remark that the explicit form of the function $a(g)$ was not
used in the derivation of the renormalization group
transformation. Thus, the whole function may be used as a {\sl
variational parameter} with respect to which the energy may be
minimized. The conditions which must be imposed to $a(g)$ on physical
grounds are its positiveness and the ``boundary conditions'' $a(0)=0$
and $a(\infty)=1$ (check [\itfstates] and the discussion
immediately below it). We shall say that each function $a(g)$
fulfilling these relations constitute an {\sl RG-prescription}.

The properties of the system depend only on the parameter $g$. Thus,
it is appropriate to ask about its flow. From the previous equations
it is easy to obtain

$$g'=R(g)\equiv {J'\over 2\Gamma'}={1\over 2}{g(1+a(g))^2\over
1-a(g)^2-a(1-a(g))^2} \eq{\runningg}$$

For any function $a(g)$ the previous transformation has three fixed
points: $g=0$, $g=\infty$ and a nontrivial intermediate one, which we
shall denote by $g_c$. The two trivial fixed points correspond to the
ordered and disordered phases, respectively, and are both
attractive. On the other hand, the nontrivial $g_c$ is repulsive. The
{\sl exact} value of $g_c$ is $1/2$, and corresponds to the same
universality class as the 2D classical Ising model, if $(g-g_c)$
replaces the reduced temperature $(T-T_c)/T_c$ (i.e.: as the variable
which measures the ``distance'' to the critical point).

From equation [\runningg] it may be deduced that the value of
$a_c\equiv a(g_c)$ is a {\sl universal function}. Whatever the
function $a(g)$, 

$$a_c=a(g_c)={2\sqrt{1-2g_c}+2g_c-1\over 3+2g_c}$$

Since $a_c\in[0,1]$, this equation implies that $g_c\leq 1/2$,
inequality which saturates at the exact value. 

The value of $g_c$ for a given prescription (i.e.: an specification
for $a(g)$) can be found by seeking the intersection between $a(g)$
and the function

$$f(g)=\cases{ \displaystyle{2\sqrt{1-2g}+2g-1\over 3+2g} & 
	if $0\leq g \leq 1/2$\cr
		0 & if $g>1/2$\cr}$$

This analysis is usually performed numerically. As an example, figure
2 shows both $f(g)$ and $a(g)$ for the prescription we have found
previously. Numerically, we find $g_c\approx 0.39$.

\figrot{5cm}{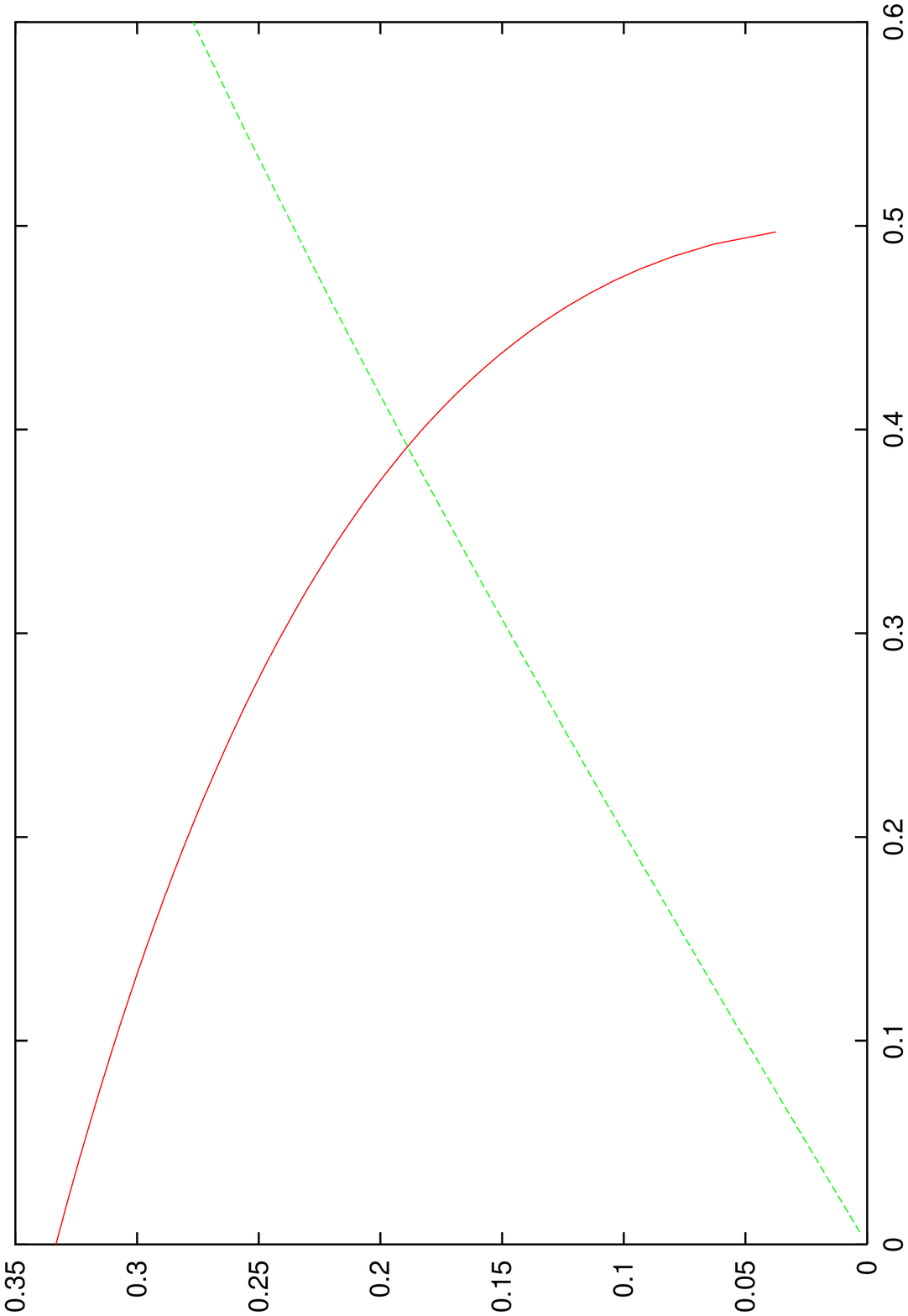}{Figure 2.}
{RG prescription function (green line) $a(g)$ and universal function
$f(g)$ (red one). Their intersection yields the value of $g_c$
for this prescription, which is approximately $0.39$.}

Despite the numerical nature of the final part of the procedure, much
can be said from merely qualitative considerations. Let us consider,
as it is a common practice in quantum field theory, the function
$\beta(g)\equiv R(g)-g$. If this function is positive, the coupling
constant increases. The zeroes of this function, shown in figure 3,
signal fixed points of the RG transformation.

\figrot{5cm}{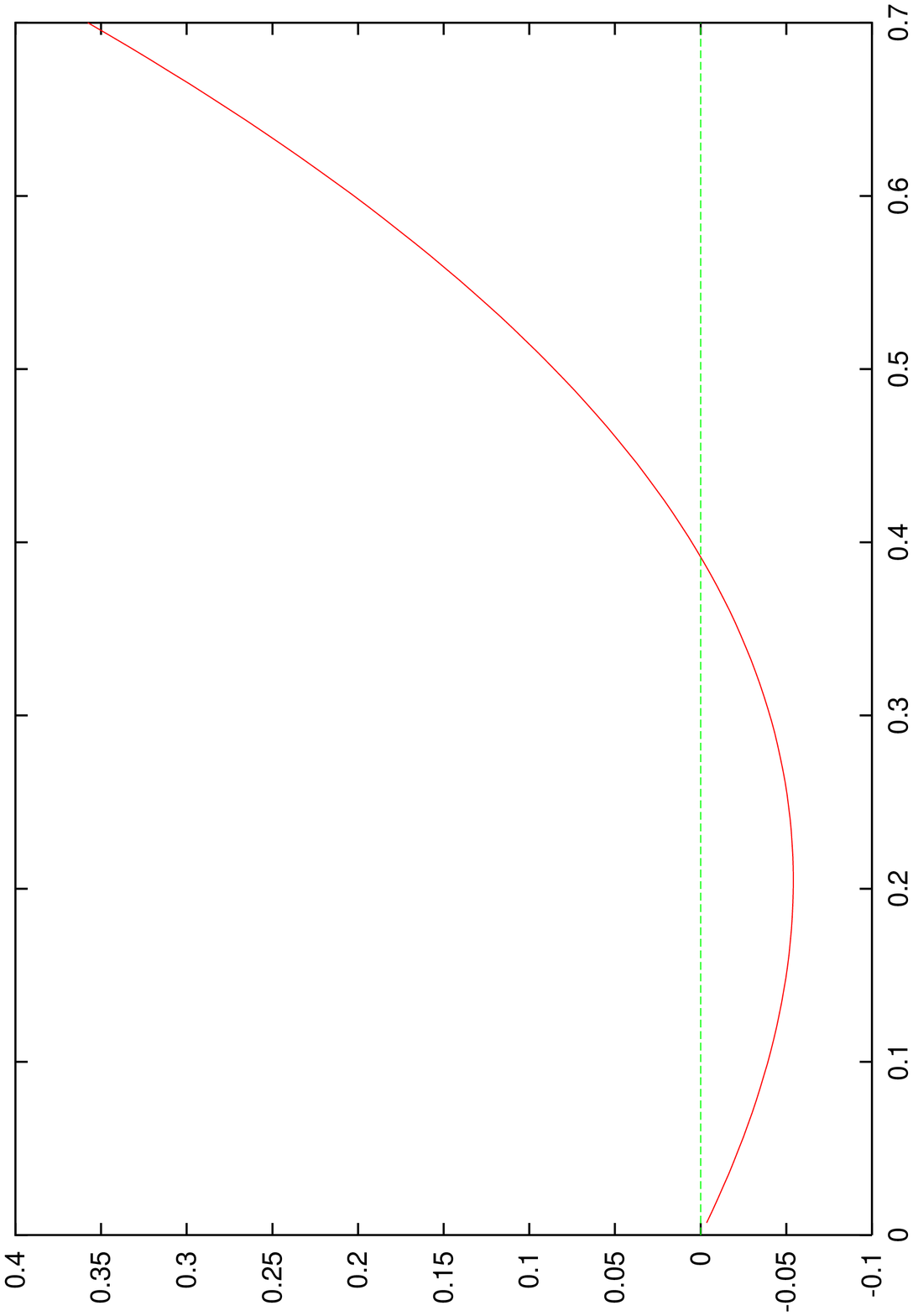}{Figure 3.}{Function $\beta(g)$, whose zero
and infinite values correspond to fixed points. Concretely, the zero
around $g\approx 0.39$ is the critical point of the theory.}

The function $\beta(g)$ fully characterizes the RG-flow of the
theory. In this case its description is rather straightforward, since
the parameter space is 1D. As it was stated above, the positivity of
the function states the growth of $g$. Following that rule, the
RG-flow is shown in figure 4.

\midinsert
\pss
\rput(5,0){
\def\fl(#1,#2){\psline*(#1,-1.2)(#2,0)(#1,1.2)}
\psline{->}(0,0)(125,0)\rput[l](127,0){$g$}
\multirput(0,0)(20,0){7}{\psline(0,0)(0,-1)}
\rput[t](0,-3){$0$}
\rput[t](60,-3){$0.3$}
\rput[t](100,-3){$0.5$}
\rput[t](76,-3){$g_c$}\psline[linewidth=1pt](76,-2)(76,4)
\fl(14,13)
\fl(32,31)
\fl(46,45)
\fl(65,64)
\fl(86,87)
\fl(110,111)}
\ps{10}
\nind{\mayp Figure 4. }{\peqfont RG-flow chart of the Ising model in a
transverse field according to our BRG prescription. Notice the
critical point $g_c\approx 0.39$, necessarily repulsive.\ps0}
\endinsert

\subsection{How to Compute Critical Exponents.}

The {\sl correlation length}, $\xi$, is rigorously defined through the
correlation function (in quantum field theory, the two-points Green's
function). For the present discussion it is enough to consider $\xi$
as the size of the biggest structures in the system in units of the
lattice spacing (effective UV-cutoff). Thus, if the RG transformation
really yields the long scales physics invariant, it must be true that

$$\xi \to {\xi\over 2}$$

\nind {\it whichever\/} the prescription may be, as long as the
blocking parameter is $2$. At the fixed points the correlation length
may only be zero (complete order or complete disorder) or infinite
(nontrivial scale invariance, fractal structures).

The correlation length is infinite only for the $g=g_c$ case. Its
divergence as we approach that critical point fulfills a power law:

$$\xi\approx (g-g_c)^{-\nu} \eq{\divlongcorr}$$

Let us consider a point $g$ quite near $g_c$, so we may consider the
$R(g)$ function approximable by a straight line with slope
$R'(g_c)$. Defining the reduced coupling constant $\tilde g\equiv
(g-g_c)$, we may say that $R(\tilde g)\approx R'(g_c)\tilde g$ if
$\tilde g$ is small.

Thus, when $\xi\to\xi/2$, the coupling constant $\tilde g\to
R'(g_c)\tilde g$. Merging both relations with the expected form
[\divlongcorr] we obtain:

$$\nu={\ln 2\over \ln R'(g_c)}$$

These computations yield $\nu\approx 1.482$, while the exact value
is just $1$.
\pss

Other critical exponents may be computed in analogous ways. We have
calculated $\nu$ merely for illustrating purposes, and the full
computations would lead us too much astray. We refer the interested
readers to \ct[GMSV-95] and the rest of the general works on RG, such
as \ct[PJP-82] and \ct[JUL-81].

The conclusion to be drawn, in any case, is that, despite the
qualitative general agreement, critical exponents are not as precise
{\it as it was expected\/}.


\section{2.2. A Simple Test Model: the Particle in a Box}

As it was stated in the introduction, despite the theoretical
appealing of the RSRG techniques and their impressive success on
impurity problems, their numerical application was slightly
disappointing and was discarded as a practical numerical technique
during the 80's. An informal talk by Kenneth G.~Wilson in 1986 has
become legendary \ct[WHI-99]. In it he proposed to focus on a simple
problem which seemed to contain the core difficulties: a quantum
spinless particle moving freely in a box.

Once a physical scenario has been settled, a set of robust observables
should be specified as an appropriate target. In our case we shall
focus on the {\sl energy spectrum} and its associated {\sl scaling
exponents} which shall be defined in due time. But in order to gain a
deep physical understanding (which is the true aim of this work) we
shall attempt to depict the full wave--functions. In fact, very
interesting conclusions shall be drawn from geometrical considerations
on these pictures.
\pss

The mathematical formulation of this problem is rather simple. The
configuration space where the particle dwells is discretized into a
graph, where vertices represent cells in real space and edges take
into account the topology (i.e.: connectivity). The Hilbert space gets
finite--dimensional and, if no potential is present, the hamiltonian
shall be proportional to the laplacian on the graph\footn{For further
explanations the reader is referred to appendix A.}.

The boundary conditions (b.c.) which are imposed on the problem shall
prove to be of outmost importance. If fixed b.c. are held
($\psi(0)=\psi(L)=0$), then the hamiltonian is proportional to the
matrix:

$$H_f = rI -A$$ 

\nind where $r$ is the connectivity of the bulk vertices, $I$ is the
identity matrix and $A$ is the adjacency matrix of the graph (i.e.:
element $(i,j)$ takes value $+1$ when vertex $i$ is linked to vertex
$j$ and zero otherwise). For example, for a one--dimensional finite
box split into $N$ cells we have:

$$H_f=\pmatrix{ 2 & -1   &      &      &  \cr
	       -1 &  2   &\ddots&      &  \cr
	          &\ddots&\ddots&\ddots&  \cr
                  &      &\ddots&  2   &-1\cr 
                  &      &      & -1   & 2\cr }$$	          

This matrix may be exactly diagonalized. The eigenvalues and
eigenvectors are given by:

$$\ket|\psi^{(n)}>_j \propto \sin\left({\pi(n+1)\over N+1}j\right) \qquad
E^{(n)}=4\sin^2\left({\pi(n+1)\over 2(N+1)}\right) \ \lin{with}\ 
j\in[1\ldots N],\ n\in[0\ldots (N-1)]$$

On the other hand, if the boundary conditions are free
($\psi'(0)=\psi'(L)=0$), the hamiltonian is directly proportional to
the combinatorial laplacian on the graph\footn{Of course, a
differential condition may not be imposed directly on a discrete
structure. See appendix A for the accurate meaning of the analogy.}:

$$H_l=\pmatrix{1 & -1   &      &      &  \cr
	       -1 &  2   &\ddots&      &  \cr
	          &\ddots&\ddots&\ddots&  \cr
                  &      &\ddots&  2   &-1\cr 
                  &      &      & -1   & 1\cr }$$	    

This operator has a {\sl zero mode}: a spatially uniform wave--function
is eigenstate of the $H_l$ matrix with eigenvalue $E_0=0$. The rest of
the spectrum may be exactly obtained too:

$$\ket|\psi^{(n)}>_j\propto \cos \left({\pi n\over N} (j-1/2)\right)
\qquad E_n=4\sin^2\left({\pi n\over 2N}\right) \ \lin{with}\ 
j\in[1\ldots N],\ n\in[0\ldots (N-1)]$$

There are many different physical problems which lead to the same
mathematical formulation. It is a good technique to have in mind the
alternative physical realizations of a given mathematical model when
trying to solve it. In our case, a vibrating string or a tightly bound
electron in a lattice of similar atoms are some of the available
possibilities.

Such Tight Binding Model (TBM) is a consistent analogy. Let us
consider a molecule or solid (i.e.: the lattice), composed of either
atoms or (small) molecules (i.e.: the sites). An electron may only
occupy a single orbital per site, whose energies are the diagonal
elements of the hamiltonian or {\sl self-energies}. But orbitals at
nearby sites overlap, thus providing {\sl hopping terms} (non-diagonal
elements). Because of them, the ground state of the electron gets
dispersed throughout the lattice\footn{Unless some sites might have a
rather low self-energy and act as deep wells, as it shall be explored
in section 4.7.}. It is remarkable the exposition of \ct[FEY-65],
where the TBM is used to introduce the wave--function concept.


\subsection{BRG for the Particle in a Box.}

Let us consider a 1D lattice of $N$ sites: $L_1\equiv[1\ldots N]$ and
a new lattice which stands at the right of this one:
$L_2\equiv[(N+1),\ldots,(2N)]$. The ground states for each of these
lattices are known. Now we are asked whether we might {\sl
variationally} build up the ground state of the composite lattice
$L\equiv [1\ldots 2N]$ {\it using the small block ground states as an
Ansatz\/}.

According to the standard BRG technique, we should write an effective
hamiltonian for the subspace spanned by the two subsystem ground
states, extended appropriately to the whole lattice by writing
zeroes outside its domain.

The problem reduces to the diagonalization of a $2\times 2$ effective
hamiltonian matrix. Let $\ket|\psi_L>$ be the left ground state and
$\ket|\psi_R>$ the right one. Thus, $H_t$ shall denote the total
hamiltonian (for the full system):

$$\pmatrix{ \elem<\psi_L|H_t|\psi_L> & \elem<\psi_L|H_t|\psi_R> \cr
	    \elem<\psi_R|H_t|\psi_L> & \elem<\psi_R|H_t|\psi_R> \cr}$$

Variational approaches of physical problems are always highly
dependent on the quality of the {\it Ansatz\/}. In this case, it
proves to be surprisingly {\it inappropriate\/}.

Checking the numbers for a definite size of the system ($N_t=40$ sites
for the composite system): 

\pss
\centerline{
\btabla(3)
\lx $E^{(0)}_{20}$ && $\elem<\psi_L|H_t|\psi_R>$ && $E^{(0)}_{20+20}$ &&
$E^{(0)}_{40}$ \cr
\lx $0.0223$ && $-0.00212$ && $0.0202$ && $0.00587$ \cr
\etabla}
\ps2
\nind{\mayp Table 1. }{\peqfont Results obtained with the BRG for the
free particle in a box. $E^{(0)}_N$ is the exact ground state energy
for a system of $N$ sites, and $E^{(0),BRG}_{N_1+N_2}$ means the
approximate energy using a $BRG$ algorithm with blocks of sizes $N_1$
and $N_2$.\ps0}
\pss

\nind The best way to show the cause of the failure of the
BRG ($400\%$!) is to plot the wave--functions of the exact ground
states for the small blocks and the resulting approximation, as it is
done in figure 5.

\figura{6cm}{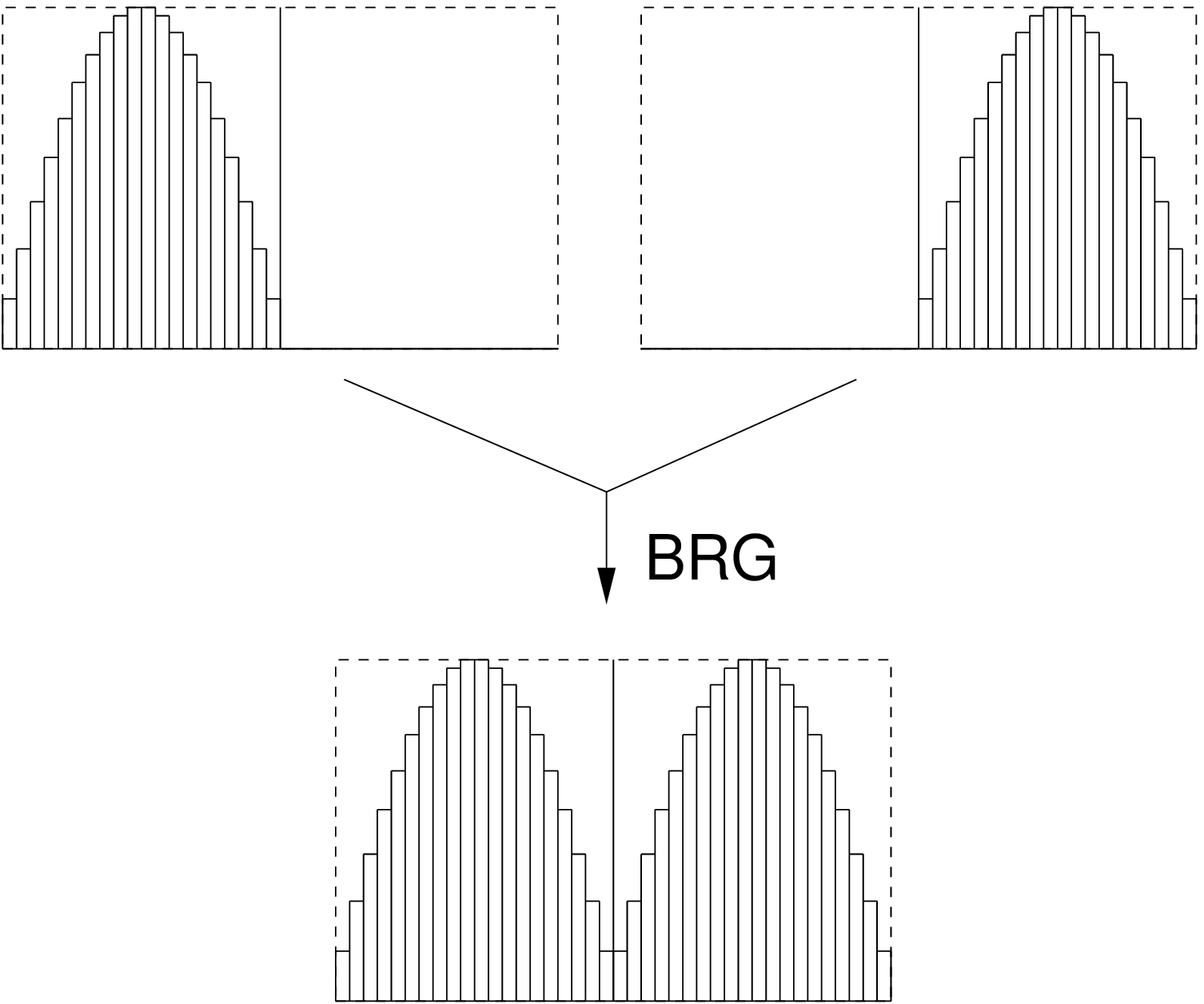}{Figure 5.}{Above: the exact ground states of 
the small boxes are depicted. Below: the BRG prediction for the ground 
state of the full system.}

The boundary conditions force the wave--functions to take the value
zero at the borders of each block, thus making a spurious ``kink''
appear in the center of the complete system.

Therefore, the first lesson to be obtained is that ``{\it Boundary
conditions may be determinative for the failure of a RG--scheme\/}''
\pss

New RSRG techniques were developed which dealt with these problems
successfully by the early 90's. These works pointed out the {\sl
correlation} between the blocks as the reason of the failure of the
BRG and designed a thoroughly new RG algorithm, called the Density
Matrix Renormalization Group (DMRG)\footn{It is interesting to remark
that S.R.~White and R.M.~Noack, creators of the DMRG, had formerly
invented the ``combination of boundary conditions'', which used as
bricks to build the ground state of the fixed b.c. laplacian the
states of small blocks created with both types of b.c. The algorithm
worked, but the method was rejected since it was not generalizable to
interacting systems \ct[WHI-99].}. A more insightful analysis of the
failure of BRG appeared in 1995 \ct[MDS-95], which proved its
correctness in the design of a modification which, while conservating
the old block constructions idea, yielded the energies precisely.

The original approach used {\it free\/} boundary conditions because of
the homogeneity of the ground state (which is flat). This fact makes
the RG prescription recover {\it exactly\/} the ground state of the
composite block. The energy of the ground state and the first excited
state are obtained with the BRG technique with a good precision. At
the end of the process, the wave--functions were reconstructed and
depicted.

The techniques which appeared in \ct[MDS-95] were further developed
and expanded in our work, which shall be exposed in the second part of
this chapter.
\vfill\eject

\part{Part II. The Correlated Blocks Renormalization Group.}
\pss

Among the successful RSRG techniques, the one known as {\sl Correlated
Blocks Renormalization Group (CBRG)} is probably the one which
resembles the old BRG more closely. The seeds which grew to make up
the technique were published in \ct[MDS-95] and became a full fledged
algorithm in \ct[MRS-96], as a part of the present work.

Along this second part of the chapter, the simple problem exposed at
the end of the previous section is addressed with a technique which
derivates from old BRG, but which takes the correlation between blocks
explicitly into account. Despite the simplicity of the problem under
study, it is possible to obtain conclusions which are generalizable to
more complicated ones.

The essential part of the material of this section is published in
\ct[MRS-96], but many results and discussions appear in this thesis
for the first time.


\section{2.3. Correlated Blocks in 1D.}

Let us consider a linear chain of $N$ sites. This chain is split into
small blocks of $m$ sites, denoted by ${\bf b}_p$ with $p\in[1\ldots
N_b]$ ($N_b\equiv N/m$), as depicted in figure 6.

\midinsert
\newcount\vecesq
\def\bloque#1{%
\multirput{0}(0,0)(7,0){#1}{\pscircle*(0,0){1}}
\vecesq=#1
\advance\vecesq by -1
\multiply\vecesq by 7
\psline(0,0)(\the\vecesq,0)}

\ps3
\rput(50,0){%
\bloque{9}
\multirput{0}(0,-8)(21,0){3}{\bloque{3}}}
\ps{10}

\nind{\mayp Figure 6. }{\peqfont Splitting a $9$-sites chain into 
three blocks ($m=3$). Notice that some links are missing.\ps0}
\pss
\endinsert

The hamiltonian for each of the {\it fully isolated\/} blocks shall be
denoted by $A_p$. Due to their isolation, they shall always have {\it
free\/} boundary conditions at both extremes. These matrices shall be
denoted as {\sl self--energy} operators.

In order to re-compose the full chain, the whole set $\{A_p\}_1^{N_b}$
is not enough. The resulting hamiltonian matrix would then be (setting
$m=3$ for definiteness):


\newtoks\matA
\newtoks\matH

\matA={\hbox{$\matrix{ 1 & -1 & 0 \cr
	              -1 &  2 & -1 \cr
	               0 & -1 & 1 \cr}$}}

\matH={\vbox{\offinterlineskip
\halign{&#\cr
\the\matA & \vrule & & &\cr
&\vrule height 3pt & & &\cr
\hrulefill & &\hrulefill & &\cr
&\vrule height 3pt & &\vrule &\cr
&\vrule & \the\matA &\vrule &\cr
&\vrule height 3pt & &\vrule \cr
& &\hrulefill & &\hrulefill\cr
& & &\vrule height 3pt&\cr
& & &\vrule &\hbox to 7mm{\hfil$\ddots$}\cr
}}}

$$\pmatrix{ A_1 &    &   &   \cr
	        &A_2 &   &   \cr
                &    &\ddots&\cr
	        &    &   &A_{N_b}}=
\pmatrix{\the\matH}$$


It is therefore necessary to {\it complete\/} the $A_p$ matrices with
new kinds of matrices which are related to the {\it missing
links\/}. Let us remark that the needed completion amounts to $(+1)$'s
added along the diagonal in the block border vertices and some
non--diagonal $(-1)$'s, corresponding to the removed links.

Denoting by $\<pR\>\equiv\<p\ot p+1\>$ the (directed) edge linking the
block $p$ to $p+1$ (or, equivalently, $\<pL\>\equiv\<p\ot p-1\>$) we
may define:

$$\displaylines{
B_{\<p \ot p+1\>}=B_{\<pR\>}=
\pmatrix{0 & 0 & 0\cr 0 & 0 & 0 \cr 0 & 0 & 1}
\qquad \hbox{if $p<N_b$} \cr
B_{\<p \ot p-1\>}=B_{\<pL\>}=
\pmatrix{ 1 & 0 & 0\cr 0 & 0 & 0 \cr 0 & 0 & 0}
\qquad \hbox{if $p>1$}\cr}
$$ 

\nind which shall be called {\sl influence matrices}, since
the blocks receive through them {\it influence\/} from their
neighbours. The physical meaning of the $B_{\<pR\>}$ matrix is the
energetic perturbation which the block $p$ receives from the block which
stands at its right side. Standard BRG, therefore, would have set the
intra--block hamiltonian to

$$(H_B)_p=A_p + B_{\<pL\>} + B_{\<pR\>}$$

\nind for bulk blocks $p\in [2\ldots (N-1)]$. But the non--diagonal
elements corresponding to the missing links are still out of our
matrix. We introduce the matrices $C_{\<p,p+1\>}$ for (undirected)
edges:

$$C_{\<p,p+1\>}=\pmatrix { 0 & 0 & 0 \cr 0 & 0 & 0 \cr -1 & 0 & 0}$$

\nind These matrices shall be called {\sl interaction matrices}, because
they are not to be added to the block hamiltonians, but to {\it stay
between\/} two blocks. The full hamiltonian may be reconstructed in
this way:

\matH={\vbox{\offinterlineskip \halign{&#\cr
$A_1+B_{\<1R\>}$\ & \vrule &\ $C_{\<1,2\>}$ &\vrule \cr
                 & \vrule height 3pt&       &\vrule \cr
\hrulefill       &        & \hrulefill    & &\hrulefill\cr
&\vrule height 3pt& &\vrule & &\vrule\cr
\ $C^\dagger_{\<1,2\>}$\ &\vrule &\ $A_2+B_{\<2L\>}+B_{\<2R\>}$\ & \vrule
&\ $C_{\<2,3\>}$\ &\vrule \cr
&\vrule height 3pt&& \vrule && \vrule \cr
\hrulefill &&\hrulefill && \hrulefill \cr
&\vrule height 3pt&& \vrule && \vrule \cr
&\vrule&\ $C^\dagger_{\<2,3\>}$\ &\vrule&\ $A_3+B_{\<3L\>} +
B_{\<3R\>}$ \ &\vrule &\hfil$\ddots$\hfil\cr
&\vrule height 3pt&& \vrule && \vrule \cr
&&\hrulefill &&\hrulefill \cr
\noalign{\vskip 3pt}
&&&&&&&&\hrulefill\cr
&&&&\hfil$\ddots$\hfil&&\hfil$\ddots$\hfil
&\vrule&\ $C_{\<N_b-1,N_b\>}$\ \cr
&&&&&&&\vrule height 3pt\cr
&&&&&&\hrulefill&&\hrulefill\cr
&&&&&\vrule height 3pt&&\vrule \cr
&&&&&\vrule&\ $C^\dagger_{\<N_b-1,N_b\>}$\ &\vrule&\
$A_{N_b}+B_{\<N_bL\>}$\cr
}}}

$$H_l = \pmatrix{\the\matH}$$

\noindent The structure of this matrix may be represented pictorially,
as it is done in figure 7:
\ps{18}

\rput(20,0){
\cnodeput(0,0){A}{1}
\cnodeput(30,0){B}{2}
\cnodeput(60,0){C}{3}
\cnodeput(90,0){D}{4}
\psset{nodesep=3pt}
\ncline{<->}{A}{B}
\lput*{:U}{$C_{\<1,2\>}$}
\ncline{<->}{B}{C}
\lput*{:U}{$C_{\<2,3\>}$}
\ncline{<->}{C}{D}
\lput*{:U}{$C_{\<3,4\>}$}
\pscurve(0,2.5)(-2,10)(0,12)(2,10)(0,2.5)
\rput[c](0,14){$A_1$}
\rput(30,0){\pscurve(0,2.5)(-2,10)(0,12)(2,10)(0,2.5)
\rput[c](0,14){$A_2$}}
\rput(60,0){\pscurve(0,2.5)(-2,10)(0,12)(2,10)(0,2.5)
\rput[c](0,14){$A_3$}}
\rput(90,0){\pscurve(0,2.5)(-2,10)(0,12)(2,10)(0,2.5)
\rput[c](0,14){$A_4$}}
\nccurve[angleA=45,angleB=135]{->}{A}{B}
\lput{:U}{\rput[b](0,1){$B_{\<2L\>}$}}
\nccurve[angleA=-45,angleB=-135]{<-}{A}{B}
\lput{:U}{\rput[t](0,-1){$B_{\<1R\>}$}}
\nccurve[angleA=45,angleB=135]{->}{B}{C}
\lput{:U}{\rput[b](0,1){$B_{\<3L\>}$}}
\nccurve[angleA=-45,angleB=-135]{<-}{B}{C}
\lput{:U}{\rput[t](0,-1){$B_{\<2R\>}$}}
\nccurve[angleA=45,angleB=135]{->}{C}{D}
\lput{:U}{\rput[b](0,1){$B_{\<4L\>}$}}
\nccurve[angleA=-45,angleB=-135]{<-}{C}{D}
\lput{:U}{\rput[t](0,-1){$B_{\<3R\>}$}}
}
\ps{14}
\nind{\mayp Figure 7. }{\peqfont Pictorial representation of the
decomposition of the full free b.c. hamiltonian.\par}
\pss

The CBRG step proceeds now by forming {\sl superblocks} by joining the
blocks into pairs. A superblock hamiltonian is specified by the
matrix:

$$H^\Sb_{[p,p+1]}=\pmatrix{ A_p + B_{\<pR\>} & C_{\<p,p+1\>} \cr
                 C^\dagger_{\<p,p+1\>} & A_{p+1}+B_{\<p+1,L\>}}$$

\nind The key point is to ``renormalize'' it into a single block with
its corresponding $A$ matrix. This is graphically represented in
figure 8:

\midinsert
\ps{20}
\rput(20,0){
\cnodeput(0,0){A}{1}
\cnodeput(30,0){B}{2}
\psset{nodesep=3pt}
\ncline{<->}{A}{B}
\lput*{:U}{$C_{\<1,2\>}$}
\pscurve(0,2.5)(-2,10)(0,12)(2,10)(0,2.5)
\rput[c](0,14){$A_1$}
\rput(30,0){\pscurve(0,2.5)(-2,10)(0,12)(2,10)(0,2.5)
\rput[c](0,14){$A_2$}}
\nccurve[angleA=45,angleB=135]{->}{A}{B}
\lput{:U}{\rput[b](0,1){$B_{\<2L\>}$}}
\nccurve[angleA=-45,angleB=-135]{<-}{A}{B}
\lput{:U}{\rput[t](0,-1){$B_{\<1R\>}$}}
\pcline{->}(40,0)(60,0)
\rput[b](50,2){RG step}
\cnodeput(70,0){C}{1'}
\rput(70,0){\pscurve[linewidth=1pt](0,3)(-2,10)(0,12)(2,10)(0,3)
\rput[c](0,14){$A_{1'}$}}
}
\ps{15}
\nind{\mayp Figure 8. }{\peqfont The superblock is formed with the 
elements depicted in the left graph. It shall be renormalized into a 
single new block along with its self--energy matrix.\par}
\pss
\endinsert

As $m$ is the number of states for each block, the superblock has
dimension $2m\times 2m$. That matrix is exactly diagonalized, and the
$m$ lowest energy eigenstates $\ket|\psi_1>$...  $\ket|\psi_m>$ are
retained and written down as the rows of a matrix $T$, which is called
the {\sl truncation operator}:

$$T_{[p,p+1]} \equiv \sum_{i=1}^{m} \ket|\delta_i>\bra<\psi_i|
=\pmatrix{\psi_1 \cr \vdots\cr \psi_{m}}$$ 

\nind where $\ket|\delta_i>$ means the $m$ components vector 
which has a single $1$ at position $i$-th.

Therefore, $T_{[p,p+1]}$ is an operator which takes from the direct
sum of the vector spaces of both blocks to the $m$-dimensional space
where each dimension ``represents'' a different eigenstate of the
superblock hamiltonian. Its matricial representation, thus, has
dimension $(2m)\times m$. Its action, in physical terms, is to take a
vector which dwells on both blocks and return its $m$ {\sl weights} on
each of the retained states.

Its adjoint operator, $T^\dagger_{[p,p+1]}$ is called the {\sl
embedding operator}, which takes a state expressed by its weights on
the set of retained states ($m$ numbers) and returns its full form
($2m$ numbers). We have:

$$T^\dagger_{[p,p+1]}=\sum_{i=1}^{m} \ket|\psi_i>\bra<\delta_i|$$
\pss

Now the superblock shall be renormalized to be a simple block in the
following RG step. Therefore, it shall have an own self--energy
operator $A'_{p'}$ with $p'={p+1\over 2}$ which shall be defined as
the reduction of $H^\Sb$ to the linear space spanned by the retained
states:

$$A'_{p'}\equiv T H^{\Sb}_{[p,p+1]} T^\dagger$$

$$A'_{p'}=\sum_{i,j=1}^{m} \ket|\delta_i>\bra<\psi_i|
H^{\Sb}_{[p,p+1]} \ket|\psi_j>\bra<\delta_j|$$

So, in components:

$$(A'_{p'})_{ij}= \elem<\psi_i|H^{\Sb}_{[p,p+1]}|\psi_j>$$

\nind This transformatin is just a ``lossy'' basis change. 
Performing this procedure with all the $N_b/2$ superblocks we get a
new set of $A$ matrices, which shall be denoted by
$\{A^{(1)}_p\}_1^{N_b/2}$.
\pss

The power of the RG stems from its {\it iterativity\/}, i.e.: the
possibility to proceed recursively. In our case, we miss a set of
renormalized influence ($B$) and interaction ($C$) operators in order
to proceed.

The renormalized influence and interaction operators are obtained by
building a ($4m\times 4m$) {\sl inter--superblocks} effective
hamiltonian, which contains the links which were missed when building
the superblock:

\matH={\vcenter{\offinterlineskip\halign{&#\cr
\ $0$ &&\vrule& & \cr
&&\vrule height 3pt&&\cr
&\ $B_{\<(p+1)R\>}$\ &\vrule&\ $C_{\<p+1,p+2\>}$\ &\cr
&&\vrule height 3pt&&\cr
\noalign{\hrule}
&&\vrule height 3pt&&\cr
&\ $C^\dagger_{\<p+1,p+2\>}$\ &\vrule&\ $B_{\<(p+2)L\>}$\ &\cr
&&\vrule height 3pt&&\cr
&&\vrule&&\ $0$ \cr}}}

$$H^{\Sb-\Sb}_{[p\ldots p+3]}=\left(\the\matH\right)$$

It is intuitively clear that the $B$ and $C$ matrices should be
extracted from this inter--superblocks hamiltonian. But it is
necessary to take care in this process, since we have {\it two\/}
different truncation matrices. Thus, we define the $2m\times 4m$ combined
truncation matrices:

$$T_{[p\ldots p+3]}\equiv T_{[p,p+1]}\oplus I_{[p+2,p+3]} + 
I_{[p,p+1]}\oplus T_{[p+2,p+3]}$$

\noindent I.e.: they act on a field with values on the sites of the
four blocks and returns $2m$ numbers. The first $m$ are the weights on
the left part, and the last $m$ are the weights on the right part.

The inter--superblocks matrix is truncated using these combined
operators:

$$T_{[p\ldots p+3]}\  H^{\Sb-\Sb}_{[p\ldots p+3]}\  T^\dagger_{[p\ldots p+3]} =
\pmatrix{ B'_{\<p'R\>} & C'_{\<p',p'+1\>} \cr
	  C'^\dagger_{\<p',p'+1\>} & B'_{\<(p'+1)L\>} }$$

\nind So we may read the new sets of matrices:

$$\{B'_{\<pL\>}\}_2^{N_b/2} \qquad 
\{B'_{\<pR\>}\}_1^{N_b/2-1} \qquad
\{C'_{\<p,p+1\>}\}_1^{N_b/2-1}$$
\pss

From this point on, the CBRG iteration is easily established. The main
idea is to recur the procedure until the whole system is of the size
of a superblock, rendering therefore the approximate diagonalization
of the full hamiltonian feasible.

The details for a concrete numerical implementation are left
for the next paragraph.


\section{2.4. The CBRG Algorithm in Practice.}

Throughout this section, the technical details of the main applications
of the CBRG are discussed.

\subsection{Implementation of the Homogeneous System.}

It is usual to discard the computational aspects when discussing a
mathematical-physics algorithm. It must be remarked that without the
concrete application within a computer language, the above discussion is
almost void. For a general discussion on the numerical
implementations see appendix B.
\pss

Since the system is {\sl homogeneous}, all the $A_i$ matrices are
equal, and the same holds for the $B_{\<iL\>}$, $B_{\<iR\>}$ and
$C_{\<p,q\>}$. Therefore, the data structure is rather simple: the
system just consists of a ``pack'' $\{A,B_L,B_R,C\}$. Let us remind
that $m$ is the number of sites of the original blocks and $N_b=N/m$
is the number of such blocks. Therefore, matrices $A$, $B_L$, $B_R$
and $C$ shall have dimension $m\times m$.

\def\composition{\hbox{\tt Composition}\,}
A procedure called $\composition$ is of outmost importance. It builds a
$2n\times 2n$ matrix out of four $n\times n$ matrices. It works like
this:

$${\tt \composition(A1,A2,A3,A4)}= \pmatrix{ A_1 & A_2 \cr A_3 & A_4 }$$

The superblock hamiltonian $2m\times 2m$ is written as ({\tt Ct} means
the transpose of {\tt C}):

$$\hbox{\tt Hsb}={\tt \composition(A+BR,C,Ct,A+BL) }=
\pmatrix{ A+B_R & C \cr C^\dagger & A+B_L }$$

This superblock hamiltonian is diagonalized. The $m$ lowest
eigenstates are retained and the truncation matrix {\tt T} ($m\times
2m$) is written down with these vectors as rows.

The easiest renormalization procedure is that for the $A$ matrix. If
{\tt Tt} denotes the transpose of {\tt T} (i.e.: the embedding
operator), then\footn{We shall employ the notation ${\tt x}\ot {\tt
A}$ to denote the assignment of the value {\tt A} to the variable {\tt
x}, as it is common practice in computational sciences.}:

$${\tt A'} \ot {\tt T * Hsb * Tt \hfil }\eq{\rghom}$$

The rest of the matrices renormalize in this way ({\tt Z} denotes a
null $m\times m$ matrix):

$$\matrix{ {\tt H00=\composition( Z, Z, Z, BR)} &
           {\tt H01=\composition( Z, Z, C, Z)} \cr 
	   {\tt H10=\composition( Z, Ct, Z, Z)} &
	   {\tt H11=\composition( BL, Z, Z, Z)}}$$

$$\eqalign{
{\tt BR'} &\ot {\tt T * H00 * Tt } \cr
{\tt C'}  &\ot {\tt T * H01 * Tt } \cr
{\tt BL'} &\ot {\tt T * H11 * Tt } \cr }\eqno{[\rghom']}$$

(Of course, {\tt H10} is not needed since it would renormalize to {\tt
Ct}, but it has been written down so as the pattern is more easily
recognized).

The number of superblocks to be diagonalized is just the number of
RG-steps which must be taken before the superblock contains the whole
system, i.e.: $\log_2(N_b/2)$.

\subsection{Numerical Results for the Free Homogeneous System.}

In the homogeneous chain with free boundary conditions the results are
satisfactory. Table 2 shows the numbers for a chain with free b.c. at
both ends with $768$ sites split into blocks of $6$ sites each
(Therefore, $N_b=128$ and $6$ RG-steps are required).

\midinsert
\centerline{
\sbtabla(2)
\li Energy && Exact & CBRG \cr
\lt \hfill$E_0$ && $0$ & $1.1340 \times 10^{-14}$ \cr
\lt \hfill$E_1$ && $1.6733\times 10^{-5}$ & $1.9752\times 10^{-5}$\cr
\lt \hfill$E_2$ && $6.6932\times 10^{-5}$ & $7.6552\times 10^{-5}$\cr
\lt \hfill$E_3$ && $1.5060\times 10^{-4}$ & $1.8041\times 10^{-5}$\cr
\lt \hfill$E_4$ && $2.6772\times 10^{-4}$ & $2.9681\times 10^{-4}$\cr
\lt \hfill$E_5$ && $4.1831\times 10^{-4}$ & $5.1078\times 10^{-4}$\cr
\setabla}
\nind{\mayp Table 2. }{\peqfont Numerical results for a free chain
of $6\times 2^7=768$ sites.\par}
\pss 
\endinsert

The results for the excited states, with errors in the range
$15-25\%$, are quite correct quantitatively. We shall obtain more
robust tests, such as scaling exponents of the energies.

The ground state energy is always zero, which is reproduced within
machine precision. In the large $N$ regime, the excited states fulfill
a simple relation:

$$E^n_N\approx \pi^2{n^2\over N^2} \qquad N\to\infty$$

A least squares fit of the data to a power law of the type

$$E^n_N\approx K {n^\alpha\over N^\beta}$$

\nind yields $\alpha=2.09 \pm 0.03$ and $\beta=1.95 \pm 0.03$ using
$n\in[0\ldots 6]$ and $N\in[96\ldots 3072]$. The biggest source of
error lies in the prefactor, which fits to $K=7\pm 1$ (the exact value
is $\approx 9.86$).

\ps7
\subsectionp{Inhomogeneous System.}

The data structure for the system when it is inhomogeneous is slightly
more intrincate. Every $A_i$, $B_{\<iL\>}$, etc. matrix is
different. At each RG-step, memory should be reserved for a different
number of matrices: $N_b$ matrices of size $m\times m$ are stored at
the beginning, but this number is divided by $2$ after each step.

The only step which is slightly dangerous and should be kept in mind
is that the $T$ matrices employed in equations [\rghom] and [\rghom']
differ.
\pss

For definiteness, let {\tt A(i)} be the matrix corresponding to
the $i$-th block at a given RG step, while {\tt BL(i)}, {\tt BR(i)}
and {\tt C(i,i+1)} denote the corresponding influence and interaction
matrices. Let $N_b$ be the number of blocks at this RG
step. Therefore, $N_b/2$ superblock hamiltonians should be written
down. For any $i\in[1\ldots N_b/2]$,

$$\displaylines{\hbox{\tt Hsb(i)}=
{\tt \composition(A(2i-1)+BR(2i-1), C(2i-1,2i),
Ct(2i-1,2i),A(2i)+BL(2i)) }= \cr
=\pmatrix{ A_{2i-1}+B_{\<(2i-1)R\>} & C_{\<(2i-1),2i\>} \cr 
C_{\<(2i-1),2i\>}^\dagger & A_{2i}+B_{\<(2i)L\>}}\cr}$$

Diagonalizing this superblock hamiltonian and discarding the highest
energy eigenvectors we obtain a truncation matrix {\tt T(i)}. The
renormalization of the {\tt A} matrices is straightforward:

$${\tt A'(i)} \ot {\tt T(i) * Hsb(i) * Tt(i)}$$

The inter--superblocks hamiltonian is written this way:

$$\eqalign{
{\tt BR'(i)} &\ot 
{\tt T(i) * \composition( Z, Z, Z, BR(2i)) * Tt(i)} \cr
{\tt BL'(i)} &\ot 
{\tt T(i+1) * \composition( BL(2i+1), Z, Z, Z) * Tt(i+1)}\cr
{\tt C'(i,i+1)} &\ot 
{\tt T(i) * \composition( Z, Z, C(2i,2i+1), Z) * Tt(i+1)}\cr}$$

The number of RG steps given before the superblock hamiltonian covers
the whole range is still\footn{If a RG-step index is added to the
matrices (e.g. $A^{(k)}(i)$, $C^{(k)}(i,i+1)$...) and all the levels
are stored, the resulting data structure may remind strongly of
wavelets \ct[LEM-89]. Of course, this fact is not casual: see
1.4.III.} $\log_2(N_b/2)$. The number of superblocks to be
diagonalized is now given by:

$$1+2+4+\cdots+N_b/2=\sum_{i=0}^{\log_2 N_b/2} 2^i= 2N_b-1$$

According to \ct[PTVF-97], typical diagonalization of a tridiagonal
matrix takes $O(n^2)$ steps (without eigenvectors). Our technique,
which is {\it totally general for tridiagonal matrices\/}, gives a
good numerical estimate in only $O(n)$ steps ($O(\log_2(n))$ if the
homogeneous algorithm may be applied).

\subsection{Fixed boundary conditions.}

The case of a homogeneous chain with fixed (or mixed fixed--free)
boundary conditions may be included under the title of inhomogeneous
system. As it is discussed in appendix A, the {\sl natural} system is
the free chain (isolated system), while the fixed chain implies a
trivial yet existent exterior space, which makes sites near to the
border ``really'' different.

This issue is made more clear in our approach. In order to solve the
fixed boundary conditions system it is necessary to include
$B_{\<1L\>}$ and $B_{\<N_bR\>}$ matrices, which account for the
influence of the (trivial) sites outside the chain (the so-called
``tack'' sites).

Of course, it is not required to compute all the superblock
matrices. The system is {\it almost\/} homogeneous and therefore it is
enough to treat differently the border blocks.

\subsection{Numerical results for Inhomogeneous Systems.}

Any potential may be used to test the CBRG in a non-homogeneous 1D
system. Many of them have been checked, and a few results are shown in
table 3.

\midinsert
\pss
\centerline{
\sbtabla(1)
\li Potential && Error range\cr
\lt $V(x)=0$ with free b.c.                && 0 -- 22     \%\cr
\lt $V(x)=0$ with fixed b.c.              && 17 -- 28    \%\cr
\lt $V(x)=0$ with fixed-free b.c.          && 17 -- 22    \%\cr
\lt $V(x)=\sin(8x)+\cos(8\phi x)$        && 0.02 -- 0.2 \%\cr
\lt $V(x)=1/2\,(x-0.5)^2$                && 3 -- 5      \%\cr
\lt $V(x)\in {\cal R}[0,1]$              && $10^{-4}$ -- 0.4 \%\cr
\setabla}
\noindent{\mayp Table 3.} {\peqfont Some results for inhomogeneous 1D
systems, with $768$ sites and $6$ states. In all cases, the variable
$x=(i-1)/N_t$, i.e.: the site index over the number of sites.
$\phi=0.5(1+\sqrt{5})$ is the golden section and $V\in{\cal R}[0,1]$
means that $V(x)$ is a random variable equally distributed on $[0,1]$,
drawn independently for each $x$.\par}
\pss
\endinsert

A more robust check shall be performed on the fixed boundary 
conditions and the harmonic oscillator potential $V(x)=1/2\,x^2$. In
the first case, a $(n+1)^2/N^2$ law may be obtained. The second is
characterized by the equally spaced spectrum.
\pss

\bul Fixed boundary conditions.

Fitting to a functional form $E_N^n=K(n+1)^\alpha/N^\beta$ we obtain
$K=7\pm1$, $\alpha=2.05\pm0.02 $ and $\beta=1.94 \pm 0.02$. The $N$
and $m$ values are taken from the same range as in the free case, and
the results may be seen to be similar in both cases. The exact values
in the asymptotic regime are, of course, $\alpha=\beta=2$ and
$K=\pi^2$.
\pss

\bul Harmonic oscillator.

We fit the energies to the law $E_N^n=K(n+1/2)^\alpha/N^\beta$. The
asymptotic exact values are, in this case, $\alpha=\beta=1$. The
numerical calculations for the same range ($6$ retained states and
size of the system from $192$ to $3072$) yields $\alpha=1.011\pm
0.008$ and $\beta=0.988\pm 0.007$.
\pss

Many questions related to the relative success of CBRG in so different
systems need an answer. We shall postpone the due explanations until
the wave-functions have been depicted.

\subsection{Fixed points of the CBRG.}

Let us consider the 1D homogeneous case with free b.c. The parameter
space for the RG procedure is formed in this case by the matrix
elements of $A$, $B_L$, $B_R$ and $C$. Which are the fixed points for
these parameters?

Of course, there is a trivial fixed point corresponding to old BRG:
$B_L=B_R=C=0$. The evolution of matrix $A$ towards this fixed point
is interesting. Each RG-step the $m/2$ lowest eigenvalues of matrix
$A$ are duplicated and the rest is discarded. Thus, matrix $A$ reaches
the null value in a finite number of steps even though its eigenvalues
never decrease in magnitude.

The non-trivial fixed point is more interesting. Let us take, e.g.,
$m=3$. Then all matrices $A$, $B_L$, $B_R$ and $C$ tend to zero
asymptotically according to power law:

$$A^*=N^{-z'}a^*, \quad B_L^*=N^{-z}b_L^*,\quad
B_R^*=N^{-z}b_R^*, \quad C^*=N^{-z}c^*$$

And the matrices in the fixed point are

$$\displaylines{a^*=\pmatrix{0&0&0\cr 0&e_1&0\cr 0&0&e_2\cr},
\quad b_L^*=\pmatrix{1&s&-s\cr s&t&-t\cr -s&-t&t\cr} \cr
b_R^*=\pmatrix{1&-s&-s\cr -s&t&t\cr -s&t&t\cr},
\quad c^*=\pmatrix{-1&s&s\cr-s&t&t\cr s&-t&-t\cr}}$$

With $e_1\approx 2.3$, $e_2\approx 9$, $s\approx 1.3$ and $t\approx
1.8$. The fit for $z$ and $z'$ yields numbers which are compatible
with $1$ and $2$ respectively. Thus, starting with the matrices
$a^*$,... and applying $2^z$ or $2^{z'}$ after each RG step, they stay
the same.


\section{2.3. Wave--function Reconstruction.}

One of the most interesting features of the CBRG technique is the
possibility of wave--function reconstruction. It is not only an
interesting property by itself, but it shall also lead us to a deeper
insight into its mechanisms.

The key to the procedure is the storage of the truncation operators at
each step for all superblock hamiltonians. They contain the full
information for the reconstruction of the wave--function, even if it
is stored in an intrincate recursive fashion.

The first RG step is the only one to take place in {\it real
space\/}. All other steps take place on a space where only the
difference between left and right is real, but where the states for
each part do not represent real sites. They stand for the {\it
weights\/} of the smaller scale wave--functions on its side.

\subsection{An illustrating example.}

We shall exemplify the procedure with the most simple non-trivial case
available. Let us diagonalize the block hamiltonian with $4$
sites. The two lowest energy eigenvectors are interpreted as two real
space functions which are shown in figure 9A.

\figura{3cm}{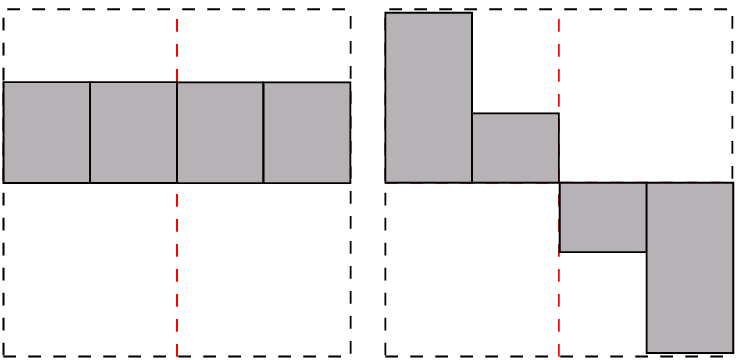}{Figure 9A.}{The two lowest energy states for a
$4$ sites free-free chain.}

Now we form the truncation operator using the two shown eigenstates as
rows. The following RG step proceeds {\it by joining\/} two systems
similar to the former one. The new system has a left block and a right
block, but its ``sites'' only represent the {\it weights\/} of the old
states on each of the sides.

Let us form, thus, the effective superblock hamiltonian for the four
mentioned states (two for the left side and two for the right
side). Such a superblock, even though it represents $8$ real sites, is
only a $4\times 4$ matrix. Its $2$ lowest energy eigenstates are shown
in figure 9B.
\ps6

\def\psfr{\psframe[fillstyle=solid,fillcolor=lightgray]}
\midinsert
\rput(20,-3){
\rput(25,3){\it New ground state}
\psframe(0,0)(50,-40)
\psline(0,-20)(50,-20)
\psline(25,0)(25,-40)
\rput(12.5,-3){\peqfont Left side}
\rput(37.5,-3){\peqfont Right side}
\psfr(5,-20)(8,-8)
\psfr(30,-20)(33,-8)
\rput*(7,-23){\peqfont Ground}
\rput*(18,-23){\peqfont Exc.}
\rput*(32,-23){\peqfont Ground}
\rput*(43,-23){\peqfont Exc.}}
\rput(80,-3){
\rput(25,3){\it New first excited}
\psframe(0,0)(50,-40)
\psline(0,-20)(50,-20)
\psline(25,0)(25,-40)
\rput(12.5,-3){\peqfont Left side}
\rput(37.5,-3){\peqfont Right side}
\psfr(5,-20)(8,-12)
\psfr(16,-20)(19,-15)
\psfr(30,-20)(33,-28)
\psfr(41,-20)(44,-15)
\rput*(7,-23){\peqfont Ground}
\rput*(18,-23){\peqfont Exc.}
\rput*(32,-31){\peqfont Ground}
\rput*(43,-23){\peqfont Exc.}
}
\ps{44}
\nind{\mayp Figure 9B. }{\peqfont The two lowest energy eigenstates of
the superblock. The left graph represents the formation of the new
ground state, and the right one that for the new excited state. Bars
represent the {\it weights\/} of the old block states.\par}
\endinsert

The depicted states show that only the left and right ground states
contribute to the ground state of the bigger system. Both states are
flat, and so is the global state.

The first excited state is more interesting. The left side receives a
positive contribution from the small ground state and another from the
small excited state. This implies that it shall have the same shape as
the small excited state, but ``{\sl raised}'' due to the constant
term. On the right side, the contribution from the ground state is
negative, which implies a uniform ``{\sl descent}'' to the other
contribution, which is from the old right excited state. Figure 9C
shows the full wave--functions at the end of the last step.

\figura{3cm}{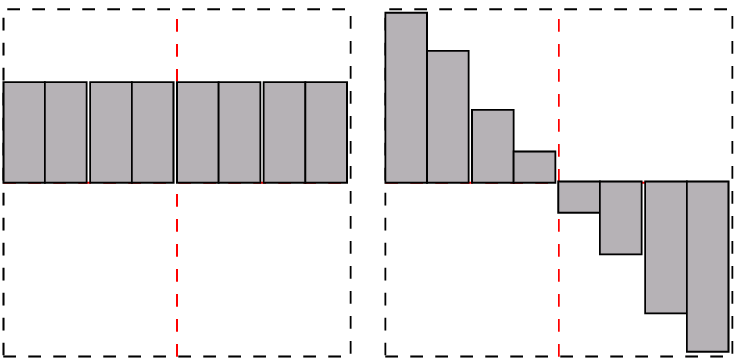}{Figure 9C.}{Full wave--functions at the end of
the RG step.}

\vfill\eject
\subsection{The Reconstruction Algorithm.}

If all embedding $T^\dagger_i$ matrices are stored, then the complete
wave--functions may be reconstructed. The $T^\dagger_i$ matrices take
a vector of $m$ components (which represent the weight of each of the
$m$ lowest energy states of the superblock), and return a $2m$
components vector. The first $m$ components of this vector represent
the contribution from each of the left states, meanwhile the last $m$
represent the contribution from the right ones.

The reconstruction algorithm works downwards. In a certain sense, it
reminds of the application of more and more powerful magnifying
glasses. Starting from the eigenvectors of the last superblock
hamiltonian (which represent the whole system), the previous step
$T^\dagger$ is applied to them. Since that operator doubles the number
of components, we may introduce two operators called $L$ and $R$, which
represent respectively the restriction to the left and right halves of
the system. Matricially,

$$T^\dagger = \pmatrix{ L \cr R }$$

All operators $L$ and $R$ are $m\times m$ matrices.

\def\ov(#1)#2#3{\rput(#1){\ovalnode{#2}{#3}}}

\midinsert
\ps{10}
\rput(80,0){\ov(0,0){A}{\hbox to 1cm{\hfil$\ket|\psi_0>$\hfil}}
\ov(-20,-20){B1}{$L_1\ket|\psi_0>$}
\ov(20,-20){B2}{$R_1\ket|\psi_0>$}
\psset{nodesep=3pt}
\ncline{->}{A}{B1}
\ncline{->}{A}{B2}
\ov(-40,-40){C1}{$L_2L_1\ket|\psi_0>$}
\ov(-15,-40){C2}{$R_2L_1\ket|\psi_0>$}
\ov(15,-40){C3}{$L_2R_1\ket|\psi_0>$}
\ov(40,-40){C4}{$R_2R_1\ket|\psi_0>$}
\ncline{->}{B1}{C1}
\ncline{->}{B1}{C2}
\ncline{->}{B2}{C3}
\ncline{->}{B2}{C4}}
\ps{44}
\noindent{\mayp Figure 10. }{\peqfont This tree shows the reconstruction
algorithm. An initial superblock eigenstate taken from the last
RG-step is subsequently split into left and right parts until the basic
level (real space) is reached.\par}
\pss
\endinsert

\def\lr{\leftrightarrow}
Figure 10 shows how the reconstruction tree develops. Its root is the
highest level and it grows until its branches reach ground: real
space. If each $R$ is written as a $0$ and is $L$ as a $1$, the
prescription to obtain a given block may be easily given for the
homogeneous case.

\bul Write the block index in base $2$. The number must have $N_l$
binary digits (i.e.: bits), the highest of which may be zero.

\bul Use the dictionary $1\lr L$, $0\lr R$ to obtain a ``word'' in the
$\{L,R\}$ alphabet.

\bul Multiply all matrices, taking each $L$ or $R$ from the
appropriate level. The columns of the resulting matrix are the real
space components of the global states on the sites corresponding to
the given block.
\pss

In the inhomogeneous case the only difference is that at each level
there are more than a pair of $\{L,R\}$ matrices. A general formula
may be provided. Let $i_0\in[0\ldots N-1]$ be a site in real
space. Thus, $i_1\equiv \lfloor i/m\rfloor$ is the index of the block
it belongs to, and $j_0\equiv i_0-mi_1$ is its position index {\it
within\/} that block: $j_0\in [0\ldots m-1]$ and $i_1\in [0\ldots
N_b-1]$. Following this line, we may define a whole set of indices:

$$i_k \equiv \cases{\lfloor i_{k-1}/2 \rfloor & if $k\geq2$ \cr
	            \lfloor i_0/m \rfloor & if $k=1$ \cr}$$

$$j_k \equiv \cases{i_k-2i_{k-1} & if $k\geq 1$ \cr
	            i_0 - m i_1 & if $k=0$ \cr}$$

Thus, $i_k$ is the index of the $k$-th level block in which the site
is included, and $j_k$ is the sub-block within that block. At levels
$>1$, $j_k$ may only take the values $0$ and $1$, standing for right
and left respectively.

The last convention we shall adopt shall be to define $B(l,i,0)$ to be
the right part of the embedding operator at the $l$-th RG step for the
$i$-th block. Consequently, $B(l,i,1)$ accounts for the
left part.

With this notation, if there are $N_l$ levels ($m\times 2^{N_l} = N$),
the $i_0$-th component of the $l$-th global wave--function is given by:

$$\psi^l_{i_0}= \[ B(0,i_1,j_1) \]_{j_0,k_1}
\[B(1,i_2,j_2)\]_{k_1,k_2} \cdots
\[B(N_l-1,i_{N_l-1},j_{N_l-1})\]_{k_{N_l-1},l} \eq{\wfrecons}$$ 

\nind where the summation convention is implicit on all the $k_i$ indices.

\subsection{Graphical results.}

Once the reconstruction algorithm has been described, we shall
reproduce some of the wave--functions which have been obtained with
this technique. Figure 11A shows the most simple case, the 
homogeneous algorithm for a free chain of $64$ sites.

\figura{5.4cm}{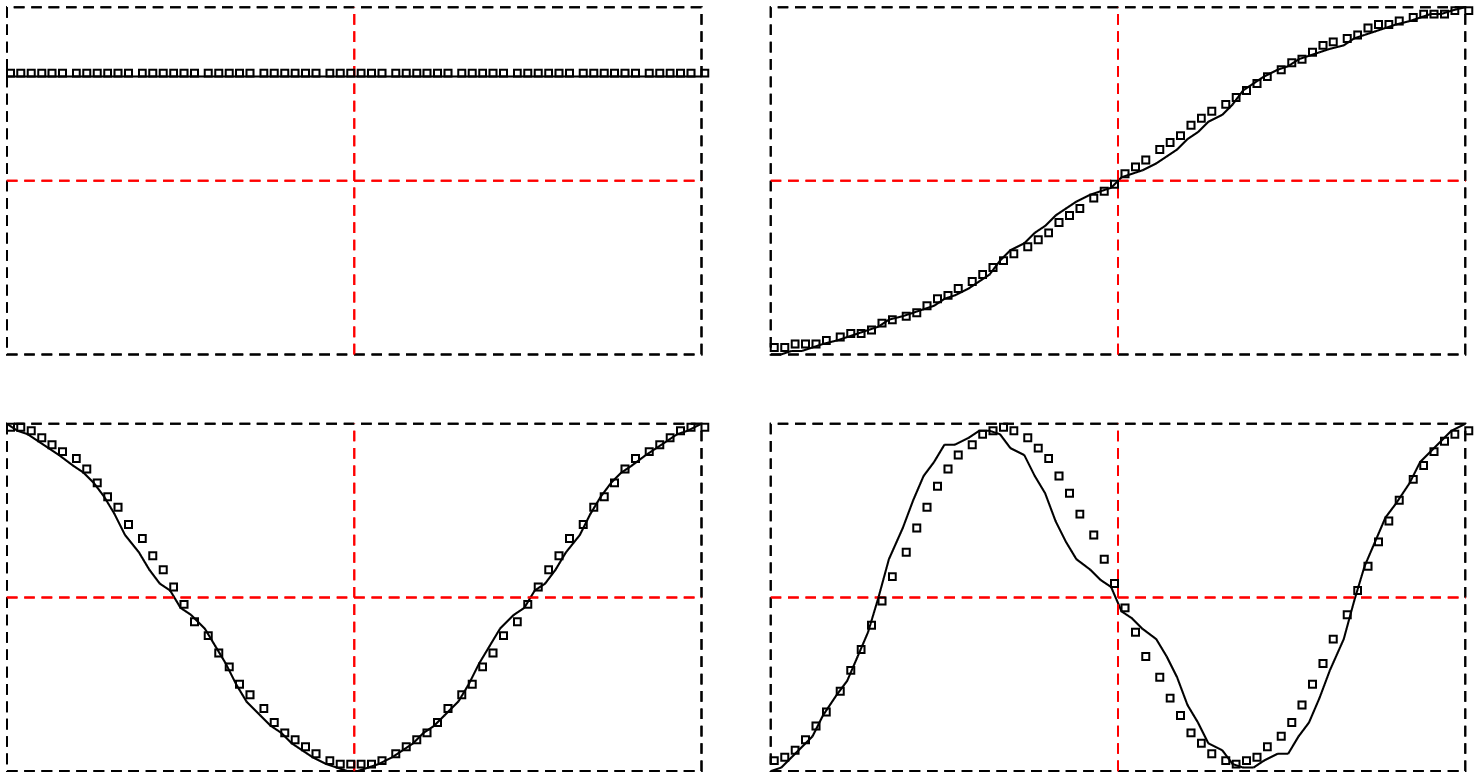}{Figure 11A.}{In continuous line, the CBRG
wave--functions for the free system with $64$ sites. The dots
mark the exact solutions.}

The ground state of the system is exactly reproduced, as it might have
been foreseen. The excited states may be seen to ``degrade
gracefully'' as the energy grows. The number of nodes is conserved,
which is the key for the accuracy of the results. It is always true
that states with even symmetry stand a better reproduction (and,
therefore, smaller errors in energy), because the zero slope is
respected at the origin.
\pss

Figure 11B shows the full wavefunctions for the system with fixed
boundary conditions (without potential). Figure 11C shows the lowest
energy states for a system with free b.c. but in presence of a harmonic
potential $V(x)=(x-1/2)^2$ with $x\in [0,1]$.

\figura{5.4cm}{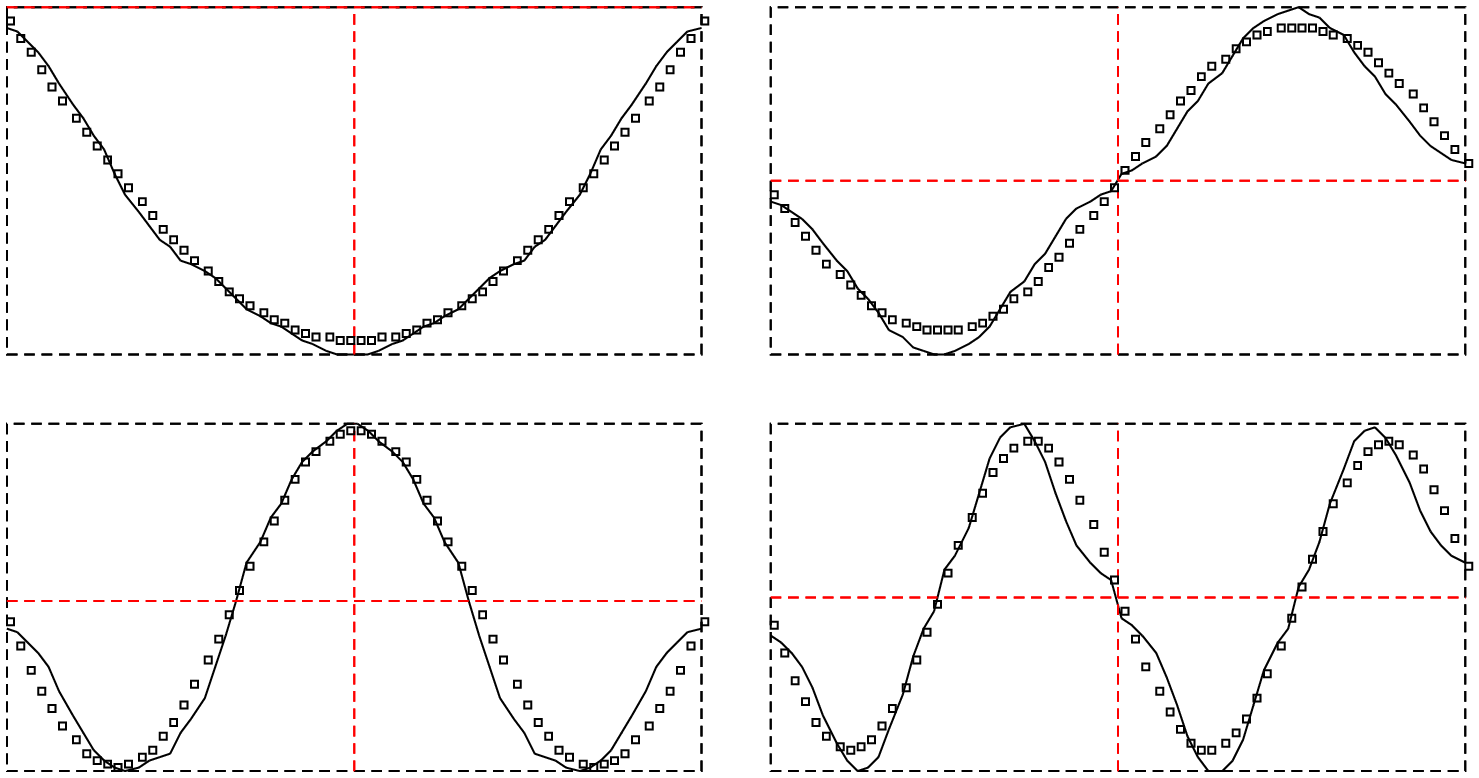}{Figure 11B.}{The fixed b.c.
wave--functions with the same plotting conditions as in figure 11A.}

\figura{6cm}{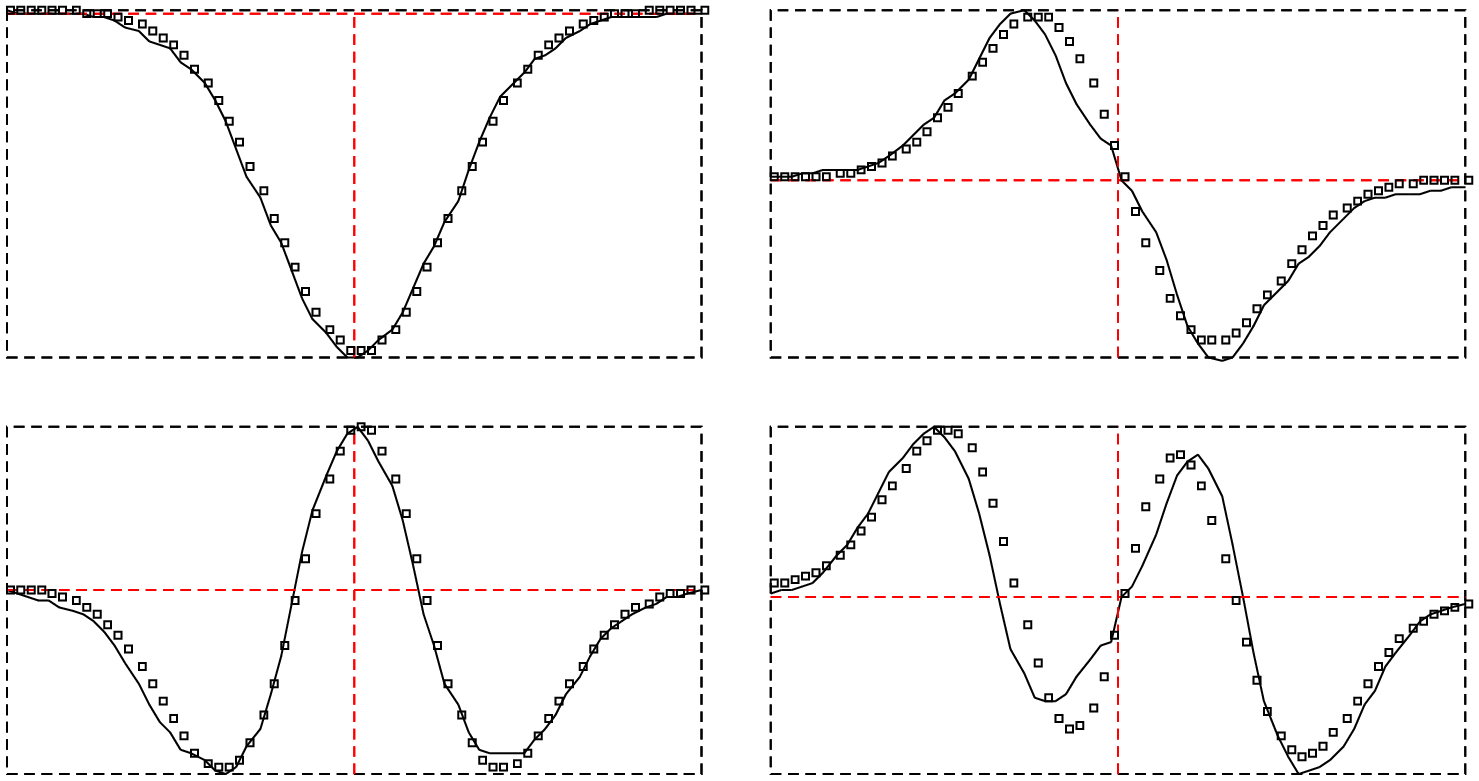}{Figure 10.}{Lowest energy states for the
harmonic oscillator according to CBRG.}

\vfill\eject

\section{2.6. Two--dimensional CBRG.}

The CBRG algorithm is by no means limited to work in one dimension.
Influence and interaction matrices are easily generalized to work in
higher dimensional systems. In this work we shall only expose the 2D
case.

Let us consider a $2\times 2$ block with free boundary conditions. The
laplacian for such a graph is:

\ps{5}
\hbox to \hsize{\hfil$\displaystyle A=\pmatrix{ 
 2 & -1 & -1 &  0 \cr
-1 &  2 &  0 & -1 \cr
-1 &  0 &  2 & -1 \cr
 0 & -1 & -1 &  2 \cr}$\hfil}
\rput(24,4){
\cnodeput(0,0){A}{2}
\cnodeput(0,10){B}{0}
\cnodeput(10,0){C}{3}
\cnodeput(10,10){D}{1}
\ncline{-}{A}{B}
\ncline{-}{B}{D}
\ncline{-}{C}{D}
\ncline{-}{C}{A}}

Influence and interaction are related to neighbouring blocks, which
are four in our case. From the graph theory viewpoint, influence $B$
matrices are associated to directed links, and interaction $C$
matrices to undirected links.

Each block must now have four $C$ matrices ($C_{DU}$, $C_{UD}$,
$C_{LR}$ and $C_{RL}$) and four $B$ matrices ($B_U$, $B_D$, $B_R$ and
$B_L$), whose r\^ole is shown in the the following graph, along with
their numerical values:

\rput(40,-36){\peqfont
\cnodeput(0,0){A}{Block}
\cnodeput(-30,0){B1}{Block}
\cnodeput(30,0){B2}{Block}
\cnodeput(0,-30){B3}{Block}
\cnodeput(0,30){B4}{Block}
\ncline{<->}{B1}{A}
\lput*{:U}{$C_{RL}$}
\ncline{<->}{A}{B2}
\lput*{:U}{$C_{LR}$}
\ncline{<->}{B3}{A}
\lput*{:U}{$C_{DU}$}
\ncline{<->}{A}{B4}
\lput*{:U}{$C_{UD}$}
\nccurve[angleA=30,angleB=150]{->}{B1}{A}
\lput{:U}{\rput[b](0,1){$B_L$}}
\nccurve[angleA=-30,angleB=-150]{<-}{A}{B2}
\lput{:U}{\rput[t](0,-1){$B_R$}}
\nccurve[angleA=60,angleB=-60]{<-}{A}{B4}
\lput{:U}{\rput[t](0,-1){$B_U$}}
\nccurve[angleA=120,angleB=-120]{->}{B3}{A}
\lput{:U}{\rput[b](0,1){$B_D$}}
}

\vbox{\halign{\hbox to \hsize{\hfil #}\cr
$\displaystyle B_U=\pmatrix{1&0&0&0\cr 0&1&0&0\cr 0&0&0&0\cr
0&0&0&0\cr}
B_D=\pmatrix{0&0&0&0\cr 0&0&0&0\cr 0&0&1&0\cr 0&0&0&1\cr}$\cr
\noalign{\ps2}
$\displaystyle B_L=\pmatrix{1&0&0&0\cr0&0&0&0\cr0&0&1&0\cr0&0&0&0\cr}
B_R=\pmatrix{0&0&0&0\cr0&1&0&0\cr0&0&0&0\cr0&0&0&1\cr}$\cr
\noalign{\ps2}
$\displaystyle C_{DU}=\pmatrix{0&0&-1&0\cr 0&0&0&-1\cr 0&0&0&0\cr
0&0&0&0\cr}
C_{UD}=\pmatrix{0&0&0&0\cr 0&0&0&0\cr -1&0&0&0\cr
0&-1&0&0\cr}$\cr
\noalign{\ps2}
$\displaystyle C_{RL}=\pmatrix{0&-1&0&0\cr 0&0&0&0\cr 0&0&0&-1\cr
0&0&0&0\cr}
C_{LR}=\pmatrix{0&0&0&0\cr -1&0&0&0\cr 0&0&0&0\cr
0&0&-1&0\cr}$\cr}}
\pss\pss

It is easy to convince oneself that $C_{LR}=C_{RL}^\dagger$ and
$C_{DU}=C_{UD}^\dagger$. Under fixed boundary conditions, the
laplacian would have been:

$$L_{\rm fixed}=A+B_L+B_R+B_U+B_D=
\pmatrix{ 4 & -1 & -1 & 0 \cr
	 -1 &  4 &  0 & -1 \cr
	 -1 &  0 &  4 & -1 \cr
	  0 & -1 & -1 & 4 \cr}$$

The superblock is formed by four blocks, and its hamiltonian is a
$16\times 16$ matrix in our case (in general, $4m\times 4m$). Its
graphical representation coincides with the graph for the small $2\times
2$ system so its matrix expression is:

$$H_\Sb=\pmatrix{ A+B_R+B_D & C_{LR} & C_{UD} & 0\cr
	          C_{RL} & A+B_L+B_D & 0      & C_{UD}\cr
	          C_{DU} & 0         & A+B_R+B_U & C_{LR}\cr
	          0      & C_{DU}    & C_{RL} & A+B_L+B_U\cr}$$

The superblock hamiltonian is diagonalized and the truncation and
embedding operators, $T$ and $T^\dagger$, are obtained using the same
operation as before. With these $4\times 16$ and $16\times 4$ matrices
we start the renormalization of all operators:

$$A'=T H_\Sb T^\dagger$$

In order to renormalize the rest of them we need to write
down the inter--superblocks hamiltonian. This should contain $2\times
2$ superblocks, but it is easier to write it down in parts: a {\sl
vertical} and a {\sl horizontal} inter-superblock. The first one links
two superblocks on a column:

\rput(30,-5){\peqfont
\ov(0,0){A}{Sb.}
\ov(0,-25){B}{Sb.}
\ncline{<->}{A}{B}
\lput*{:L}{$H_{\Sb-\Sb}^{\rm vert}$}}

\def\tpaso{&&&&height 4pt&&&&\cr}
\def\trule{\tpaso\noalign{\hrule}\tpaso\cr}
\hbox to \hsize{\hfil
$\displaystyle
H_{\Sb-\Sb}^{\rm vert}=\(
\vcenter{\offinterlineskip
\halign{\quad$#$\quad&$#$\quad&$#$\quad&$#$\quad&\vrule#\quad&&$#$\quad\cr
0 & 0 & 0 & 0 && 0 & 0 & 0 & 0 \cr
\tpaso
0 & 0 & 0 & 0 && 0 & 0 & 0 & 0 \cr
\tpaso
0 & 0 & B_D & 0 && C_{UD} & 0 & 0 & 0\cr
\tpaso
0 & 0 & 0 & B_D && 0 & C_{UD} & 0 & 0\cr 
\trule
0 & 0 & C_{DU} & 0 && B_U & 0 & 0 & 0\cr
\tpaso
0 & 0 & 0 & C_{DU} && 0 & B_U & 0 & 0\cr
\tpaso
0 & 0 & 0 & 0 && 0 & 0 & 0 & 0 \cr
\tpaso
0 & 0 & 0 & 0 && 0 & 0 & 0 & 0 \cr}}\)$}
\pss

The horizontal inter--superblocks hamiltonian is calculated in the
same way:

\rput(10,-17){\peqfont
\ov(0,0){A}{Sb.}
\ov(35,0){B}{Sb.}
\ncline{<->}{A}{B}
\lput*{:U}{$H_{\Sb-\Sb}^{\rm hor}$}}

\hbox to \hsize{\hfil
$\displaystyle
H_{\Sb-\Sb}^{\rm hor}=\(
\vcenter{\offinterlineskip
\halign{\quad$#$\quad&$#$\quad&$#$\quad&$#$\quad&\vrule#\quad&&$#$\quad\cr
0 & 0 & 0 & 0   && 0 & 0 & 0 & 0 \cr
\tpaso
0 & B_R & 0 & 0 && C_{LR} & 0 & 0 & 0 \cr
\tpaso
0 & 0 & 0 & 0   && 0 & 0 & 0 & 0\cr
\tpaso
0 & 0 & 0 & B_R && 0 & 0 & C_{LR} & 0\cr 
\trule
0 & C_{RL} & 0 & 0 && B_L & 0 & 0 & 0\cr
\tpaso
0 & 0 & 0 & 0      && 0 & 0 & 0 & 0\cr
\tpaso
0 & 0 & 0 & C_{RL} && 0 & 0 & B_L & 0 \cr
\tpaso
0 & 0 & 0 & 0 && 0 & 0 & 0 & 0 \cr}}\)$}
\pss
From these $32\times 32$ matrices we may read the new influence and
interaction operators. If we denote, e.g., $H^{\rm hor}(DL)$
the down-left quarter of the $H_{\Sb-\Sb}^{\rm hor}$ matrix, we have:

$$\eqalign{
B_D'=TH^{\rm vert}(UL)T^\dagger\qquad & 
C_{UD}'=TH^{\rm vert}(UR)T^\dagger \cr
C_{DU}'=TH^{\rm vert}(DL)T^\dagger\qquad & 
B_U'=TH^{\rm vert}(DR)T^\dagger \cr
B_R'=TH^{\rm hor}(UL)T^\dagger \qquad & 
C_{LR}'=TH^{\rm hor}(UR)T^\dagger \cr
C_{RL}'=TH^{\rm hor}(DL)T^\dagger \qquad &
B_L'=TH^{\rm hor}(DR)T^\dagger \cr}$$

This way, the RG cycle has been closed.

\subsection{Wave--Function Reconstrunction.}

The reconstruction of the wave--function is also possible in the 2D
case, but it is certainly more involved. The global idea is the same,
but the details are more cumbersome.

An auxiliary idea which may help to understand the procedure is that
of a {\sl quad--tree}. It is a usual concept in fractal image
compression \ct[FIS-95] and consists in a data structure which is able
to store a 2D image in such a way that:

\bul Information is local.

\bul It is easy to establish a UV--cutoff in the image (so as to compress
it).

\bul This UV--cutoff may depend on position.
\pss

The first condition is important so as to obtain high compression
rates, but usual Fourier--transform algorithms (such as JPEG), storing
only {\sl global} information, suffer from Gibb's phenomenon
(wave-like distorsion near edges) and other undesirable effects: {\sl
aliasing}\footn{The term {\sl aliasing} refers to the set of
distortions which are introduced in a digital image when some of its
spatial frequencies with a high weight are (almost) commensurable with
the sampling frequency.}.

Figure 12 shows a typical quad--tree structure spanning a square
region. The outmost node is split into four links, each of which is
also split into four minor ones and so on.

\figura{4cm}{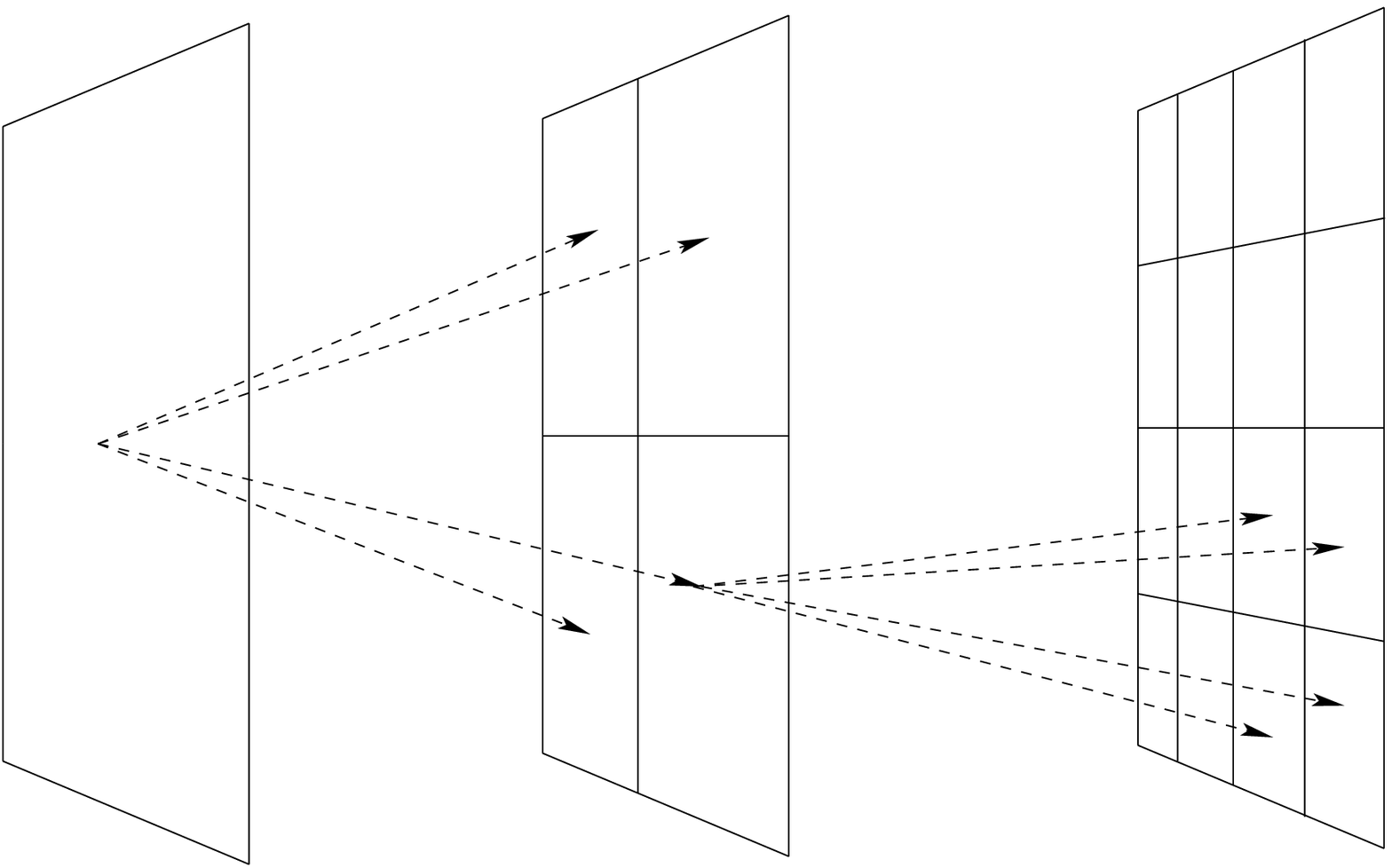}{Figure 12.}{A graphical representation of a
quad--tree structure.}

If nodes in a block are numbered as in the first picture of this
section, the quad--tree structure allows us to substitute the binary
notation used in the 1D algorithm with a base $4$ notation. Each block
is determined by a word in the alphabet $\{0,1,2,3\}$, where each
letter is to be read according to the following dictionary:

\pss
\centerline{\vbox{\halign{&#\qquad\cr
$0$ $\lr$ Upper--Left & $1$ $\lr$ Upper--Right. \cr
$2$ $\lr$ Lower--Left & $3$ $\lr$ Lower--Right. \cr}}}
\pss

The embedding matrices $T^\dagger$ may be split into four matrices,
corresponding to upper-left, upper-right, lower-left and lower-right
regions of the superblock. Let them receive names from $B_k(0)$ to
$B_k(3)$ (with $k$ denoting the RG-step). An equation analogous to
[\wfrecons] may be written down for this case.

Let $x$ and $y$ be the coordinates of a site in real space,
$x,y\in[0\ldots L-1]$. Let us also consider the smallest blocks to be of
size $l_b\times l_b$, and the number of RG-steps to be $N_l$. We define the
following sets of integer indices:

$$x_i'=\cases{ \lfloor x/l_b \rfloor & if $i=0$\cr 
	\lfloor x_{i-1}'/2 \rfloor & if $i>0$\cr}$$ 

\nind I.e.: the $x$-coordinate of the $i$-th level block to which the
site belongs. The same definition goes for the $y_i'$. These $x_i'$
and $y_i'$ indices are only needed so as to define the following:

$$x_i=\cases{ x-x_0l_b & if $i=0$\cr 
	x_i-2x_{i-1} & if $i>0$\cr}$$

\nind This is to say: the position inside each $i$-th level block of
the lower level block to which the site belongs. The $y_i$ are
defined in an analogous way. Now the set of indices $j_i$ for $i\geq
0$ is defined:

$$j_i=2y_i+x_i$$

These $j_i$ indices specify the quad--tree path to find the site
from the root node. In effect, the analogue of [\wfrecons] is:

$$\psi^l_i = \(B_0(j_0)\)_{ik_1} \(B_1(j_1)\)_{k_1k_2} \cdots
\(B_{N_l-1}(j_{N_l-1})\)_{k_{N_l}l}$$

\subsection{Numerical Results of the 2D Algorithm.}

The procedure described in the previous section is the most simple
2D-CBRG algorithm possible. It can be easily generalized to include a
potential or non-trivial boundary conditions of any kind.

For the $16\to 4$ RG scheme proposed above, energy stays within $10\%$
of the exact value for all states in $6$ RG-steps ($128\times 128$
sites). The scaling relation is almost perfect: the energy fulfills
$E^n_N\propto N^\alpha$ with $\alpha\approx 1.99999981$.

Figure 13A shows the four lowest energy states for the $128\times 128$
lattice. The third excited state is also represented in figure 13B
within an axonometric projection with contour lines depicted on the
basis plane.

\figura{3.7cm}{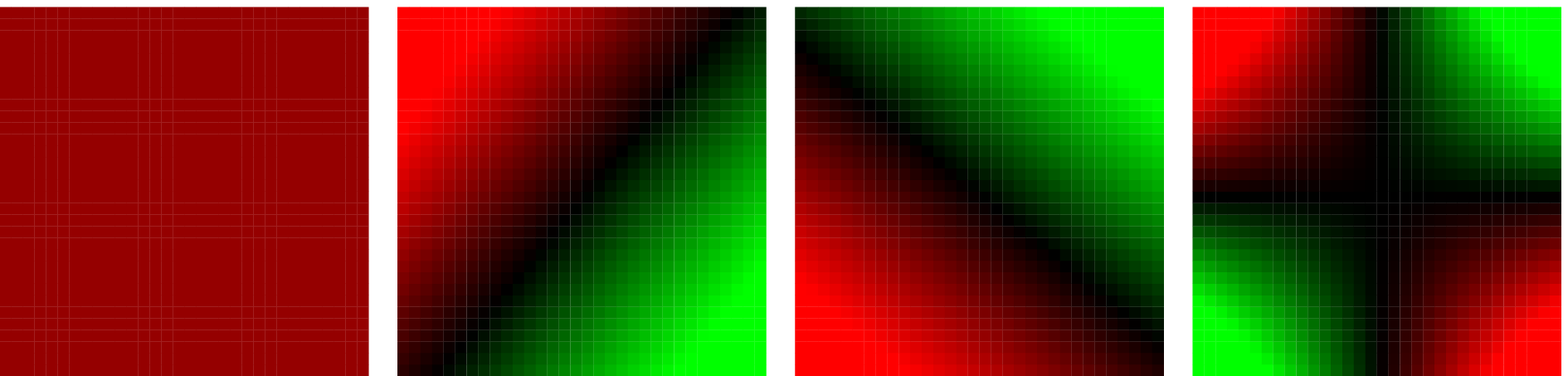}{Figure 12A.}{Density plot for the lowest
energy states for the free b.c. laplacian on a square lattice of
$128\times 128$ sites.}

\midinsert
\rput{270}(80,-30){\epsfysize 12cm\epsfbox{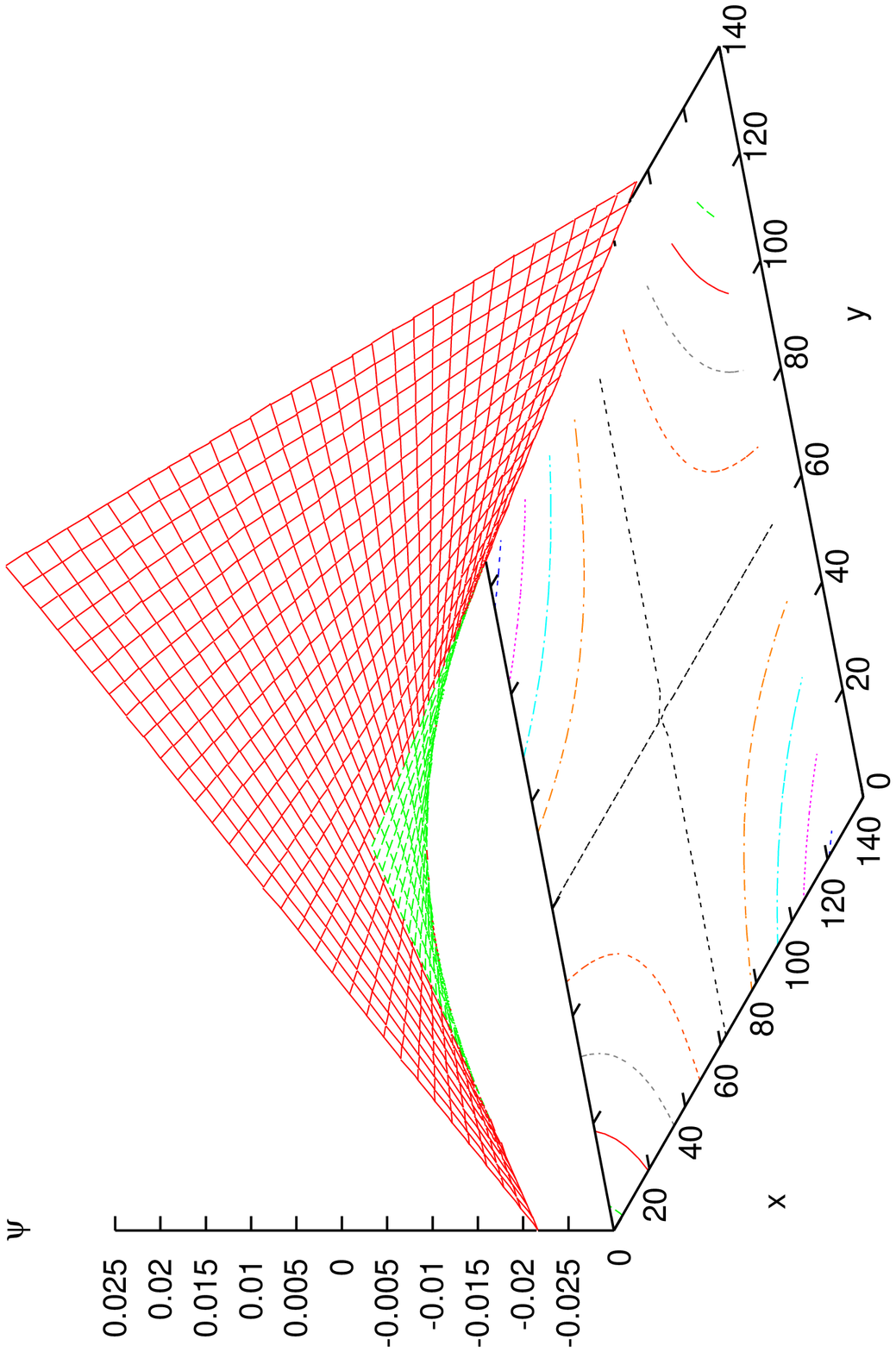}}
\ps{65}
\nind{\mayp Figure 13B. }{\peqfont Axonometric plot of the third
excited state for the free b.c. laplacian on the $128\times 128$
lattice, obtained with CBRG.\par}
\endinsert


\section{2.7. Self--Replicability and the CBRG.}

This section might also have been entitled ``Why do free boundary
conditions work for homogeneous CBRG while the fixed ones do not?''
The answer shall provide us with an interesting insight on the full
CBRG procedure.

The correct point of view in order to find the answer is the
``bricks'' image. Let us focus again on the homogeneous 1D case. Each
RG step works in the following manner: ``{\sl build the best
approximation to the lowest energy eigenfunctions for the whole system
using as bricks the lowest energy eigenfunctions for the left and
right halves}''.

\subsection{Self--replicability.}

We might have started the research the other way round. Given a
function $\phi\in {\cal C}[0,a]$, let us define the operators
$L_{[0,a]}$ and $R_{[0,a]}$ as:

$$(L_{[0,a]}\phi)(x)=
\cases{\phi(2x) & if $x<a/2$ \cr
              0 & if $x\geq a/2$\cr}$$

$$(R_{[0,a]}\phi)(x)=\cases{0 & if $x<a/2$\cr
	            \phi(2(x-a/2)) & if $x\geq a/2$\cr}$$

\nind I.e.: they provide us with reduced copies which are similar to
the original function for each of the parts (left and right).

Starting from a function $\phi$ we obtain a pair of them, $L\phi$ and
$R\phi$. We shall try to reproduce the original $\phi$ within our
subspace spanned by $L\phi$ and $R\phi$. Is this possible?

Let us consider all functions to be ${\cal L}^2$ normalized and let us
define the {\sl replica} transformation:

$${\cal R}\phi=\phi_{\hbox{\peqfont approx}}=\<\phi|L\phi\>\,L\phi +
\<\phi|R\phi\>\,R\phi$$

\nind which is sure to be the best approximation within this
subspace. Its accuracy shall be given by the parameter

$$S\equiv \<\phi_{\hbox{\peqfont approx}}|\phi\>$$

\nind where the symbol $S$ stands for {\sl self--replicability}. The value
$1$ means ``perfect''.
\pss

In practice, of course, we are working with discrete functions stored
in a computer. If the new functions must have the same number of sites
(i.e.: they belong to the same discrete functional space), then two
values of the old function must enter a single site for each of the
parts (left and right). The most symmetric solution is to take the
average of both values. Thus,

$$(L\phi)_i =\cases{ {1\over 2}(\phi_{2i-1}+\phi_{2i}) & if $i\leq
N/2$ \cr 0 & otherwise \cr}$$

\nind along with an equivalent formula for the right side. This is the
computational expression we shall assume. All the examples shall use
this implementation on a 1D lattice with $256$ sites.

\subsection{Numerical Experiments in 1D.}

Let us apply the process on the ground state of the laplacian with
fixed b.c. on a 1D lattice. The best approximation is given in figure
14A.

\figura{3cm}{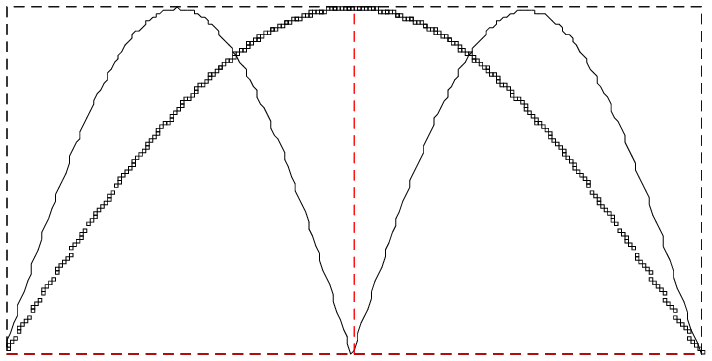}{Figure 14A.}{The ground state of a 1D laplacian
with fixed b.c. is not self--replicable.}

Here $S$ takes the value $0.84883$, which does not seem to be too low
when one observes the obvious differences. But the replica procedure
may be iterated. Figure 14B shows us the second, third and fifteenth
iterations.

\midinsert
\pss
\hbox{\epsfysize=2.2cm\epsfbox{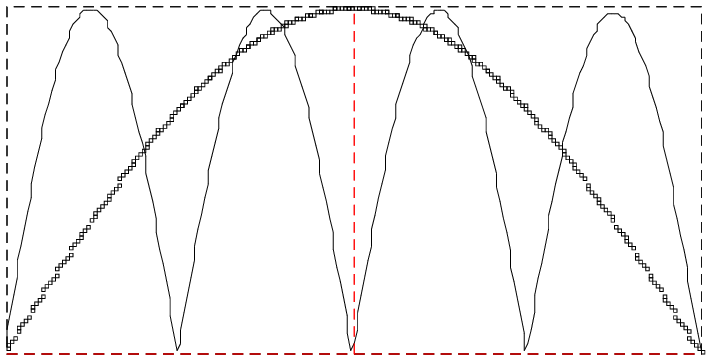}\hskip 1cm%
\epsfysize=2.2cm\epsfbox{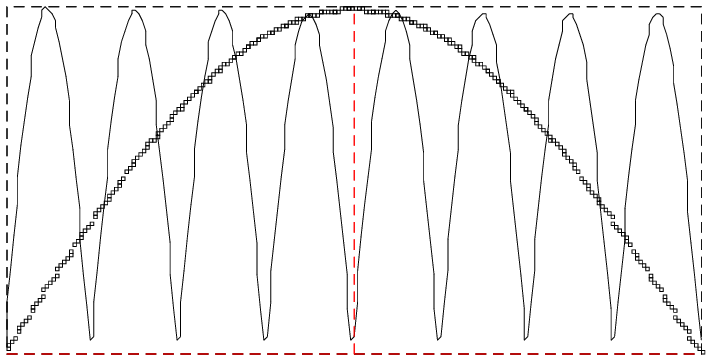}\hskip 7mm%
\raise 1cm\hbox{$\cdots$}\hskip5mm%
\vbox{\epsfysize=2.2cm\epsfbox{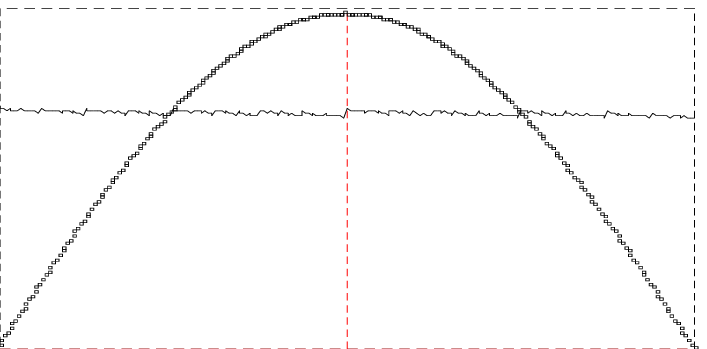}}}
\ps 2
\nind{\mayp Figure 14B. }{\peqfont The procedure is iterated. The last
box represents the fifteenth iteration.\par}
\endinsert

After some iterations the finite resolution of the computer yields a
quasi--constant function as approximation. This function {\it is\/}
exactly {\sl self--replicable}.
\pss

The key idea is that, when a function is self--replicable, it is {\it
exactly\/} attainable by a homogeneous CBRG procedure (which only
requires $\approx \log_2(N)$ operations!).
\pss

The procedure is easily extended to sets of functions. Let us denote
any such set, with $m$ functions, as $\{\phi_i\}_{i=1}^m$. These
functions must be approximated within the subspace spanned by the $2m$
functions $\{L\phi_i,R\phi_i\}_{i=1}^m$. All such functions may
contribute to the reproduction of their ``sisters''.

For example, the low energy spectrum of the 1D laplacian with free b.c.
is not exactly self--replicable, but it is to a good approximation, as
it is shown in figure 15A.

\figura{5cm}{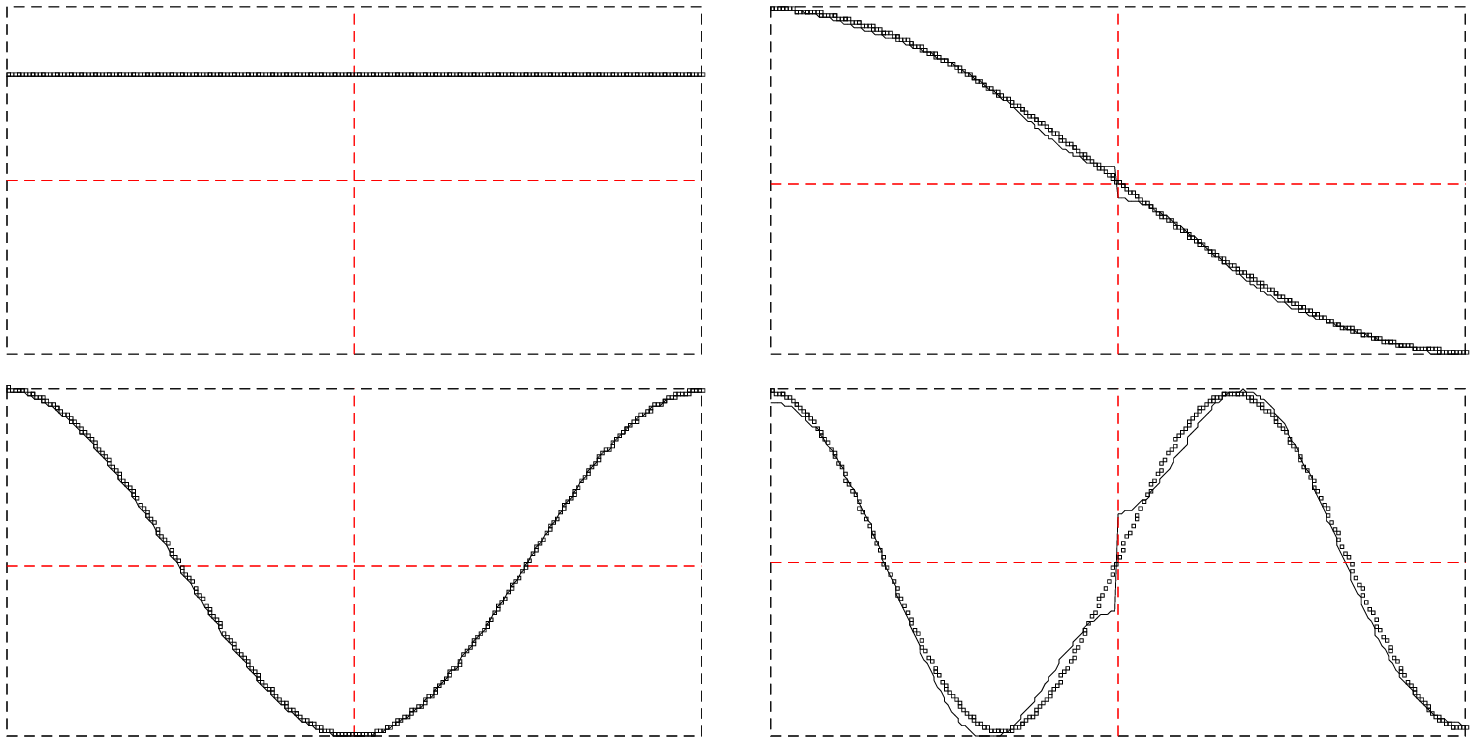}{Figure 15A.}{A single step of the replica
transformation on the lowest energy states of the free b.c. laplacian.}

The $S$ parameters are $1$, $0.999613$, $1$ and $0.995715$. This means
that the ground state (flat) and the third states are exactly
reproduced. The weights show that:
\ps2

\bul The first state is absolutely self-replicable by itself.

\bul The second consist of two copies of itself, the first one raised
(using the first state) and the second one lowered (also using the
first state). The finite slope at the origin is not correctly
represented.

\bul The third state only requires the second one.

\bul The fourth state is even more interesting. Both the left and the
right parts are a combination of the second and third states. The
finite slope at the origin is again incorrectly represented.
\ps3

The procedure may be easily iterated without excessive distortion. The
results are shown in figure 15B. These functions have a rough look,
but factors $S$ are not too different: $1$, $0.99958$, $0.999589$ and
$0.99492$.

\figura{5cm}{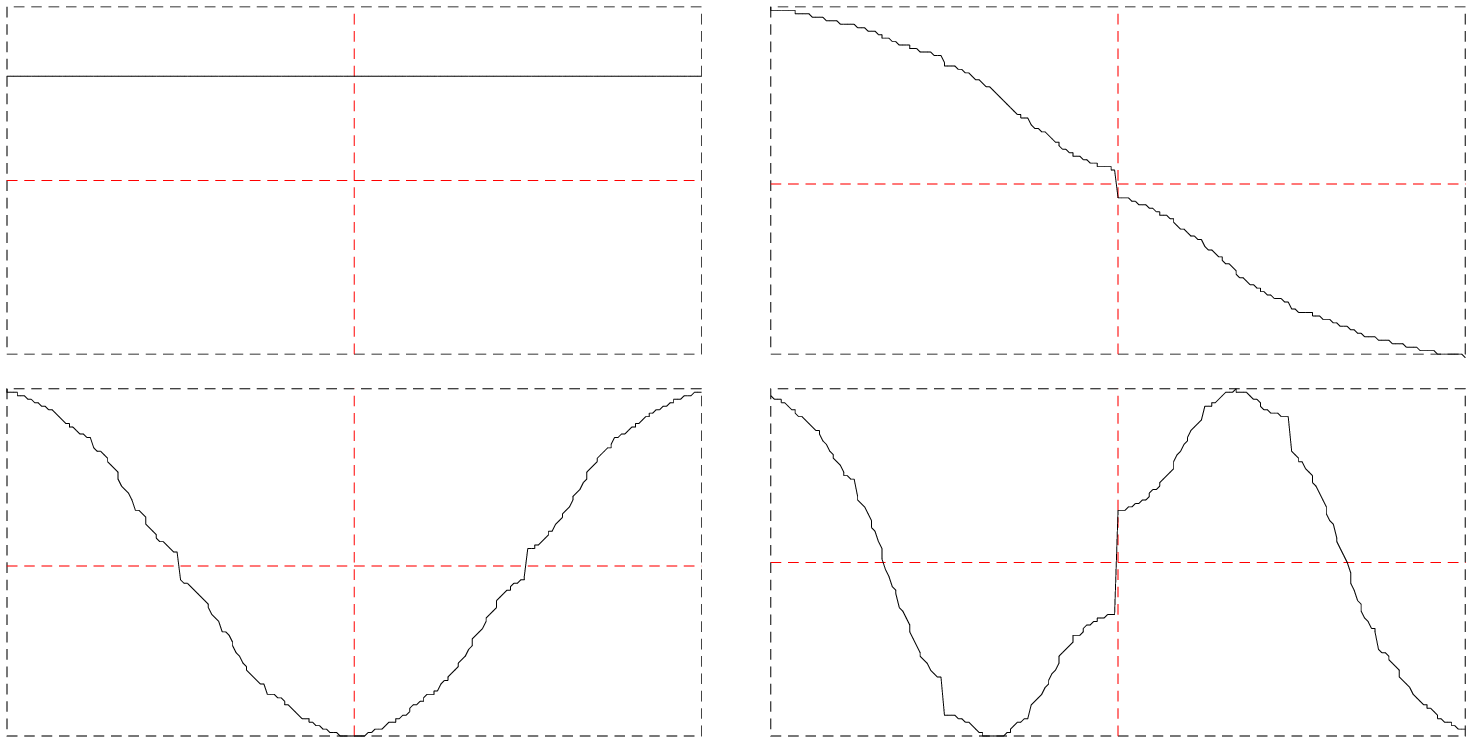}{Figure 15B.}{The same functions after 15
iterations (once the fixed point has been reached).}

It is perhaps more illuminating to perform the same experiment on the
four lowest energy states of the fixed b.c. laplacian (see figures 16A
and 16B).

\figura{5cm}{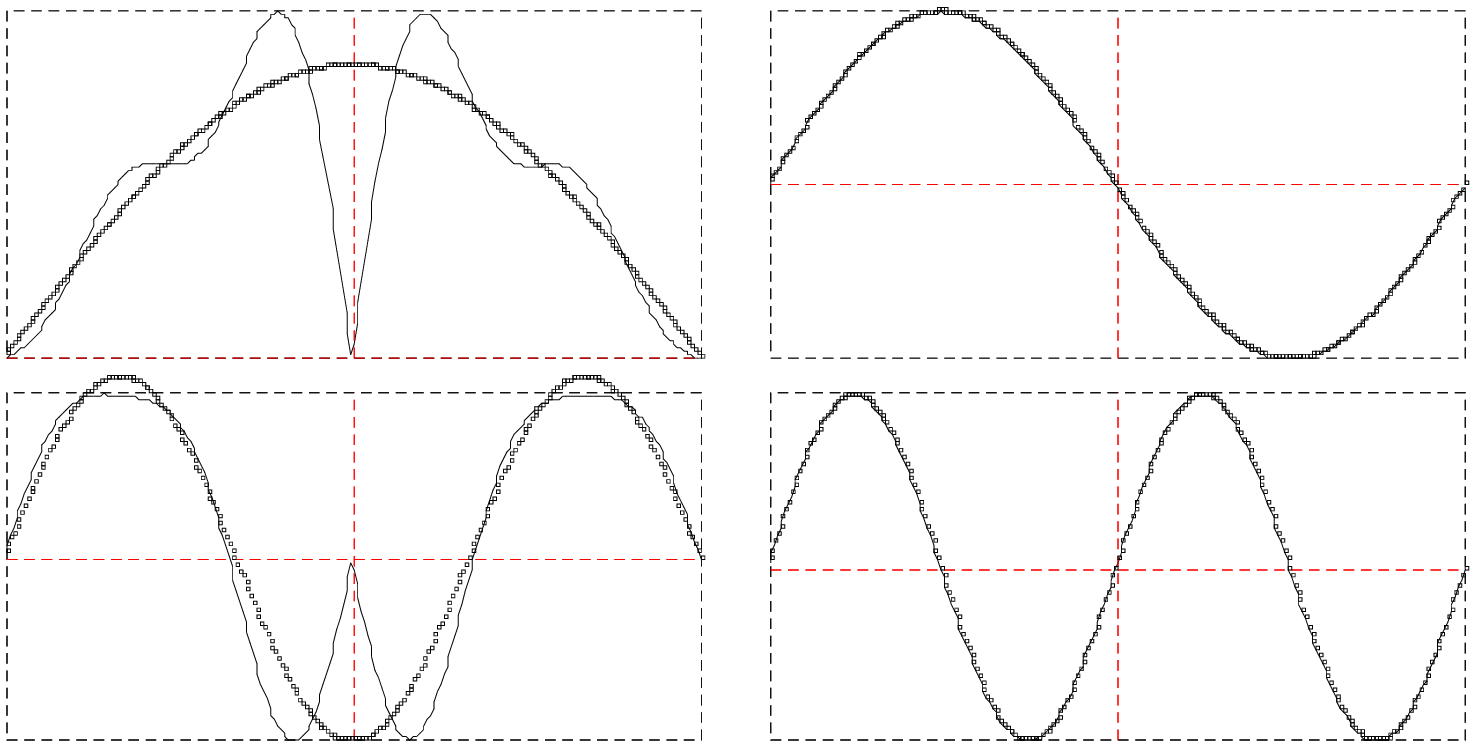}{Figure 16A.}{The lowest energy states of the
fixed b.c. laplacian after 15 iterations. The fixed point has not yet
been reached.}

\figura{5cm}{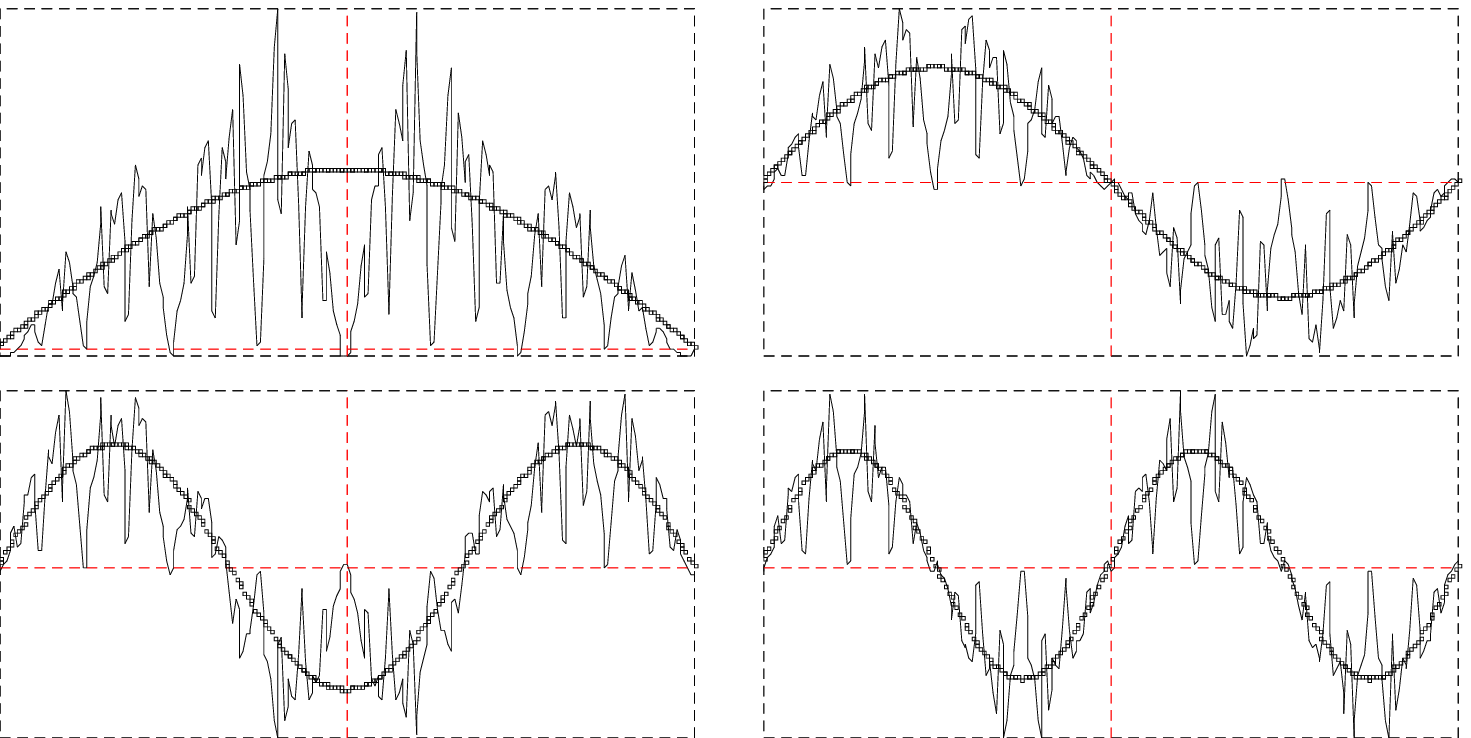}{Figure 16B.}{The same functions as figure 16A,
once the fixed point has been reached (100 iterations).}

The values of the $S$ parameters at the first iteration are not
excessively bad\footn{Sometimes functions which are quite different
may be separated by a short distance in ${\cal L}^2$. It might be more
appropriate to use a discrete Sobolev space (see section 5.4), in
order to take derivatives into account. This way, an overimposed
roughness might be detected.}: $0.953725$, $0.999997$, $0.945245$ and
$0.99998$. But the fixed point yields very different numbers:
$0.843138$, $0.872436$, $0.851632$ and $0.80208$.
\pss

The aspect of the fixed points of the replica transformation is often
quite rough. Appendix E describes the meaning and calculation method
for a ``quasi--fractal'' dimension which corresponds to the enery
scaling exponent under RG transformations. The value for smooth
functions is approximately $2$. Values around zero might be physically
interesting. The fixed points for the eigenstates of the free and
fixed b.c. laplacian yield the results of table 4.

\midinsert
\pss
\line{\sbtabla(4)
\li && Ground state & 1st exc. & 2nd Exc. & 3rd Exc.\cr
\lt Fixed b.c. && $0.034 \pm 0.027$ & $0.21 \pm 0.08$ & $0.33 \pm
0.08$ & $0.42 \pm 0.11$ \cr
\lt Free b.c. && --- & $1.55\pm 0.06$ & $1.65 \pm 0.05$ & $1.4\pm
0.05$\cr
\setabla\hfil}
\ps2
\nind{\mayp Table 4. }{\peqfont Scaling dimensions of the energy under
RG for the fixed point reached with free and fixed b.c.
wave-functions. The free ground state is missing since its calculation
is meaningless: its energy is exactly zero.\par}
\endinsert

\vfill\eject

\subsection{Generalization to 2D.}

The process may be easily generalized to 2D, if instead of splitting
the interval into two regions we part it into four. The case of the
eigenfunctions of the free b.c. laplacian yields a fixed point which is
much smoother than in the 1D case, as it is shown in figure 17.

\midinsert
\pss
\vbox{
\centerline{\fig{2.5cm}{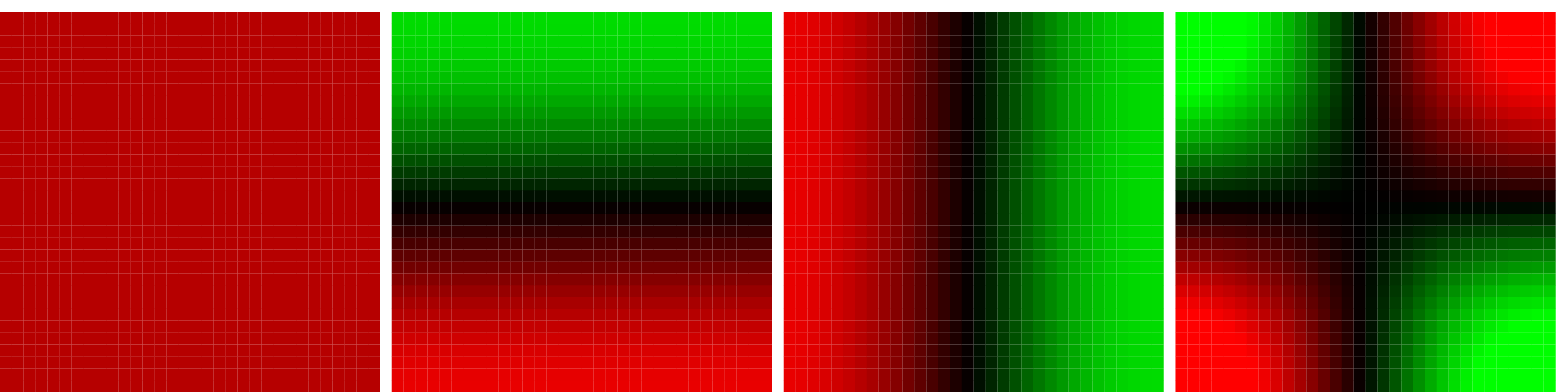}}
\centerline{\fig{2.5cm}{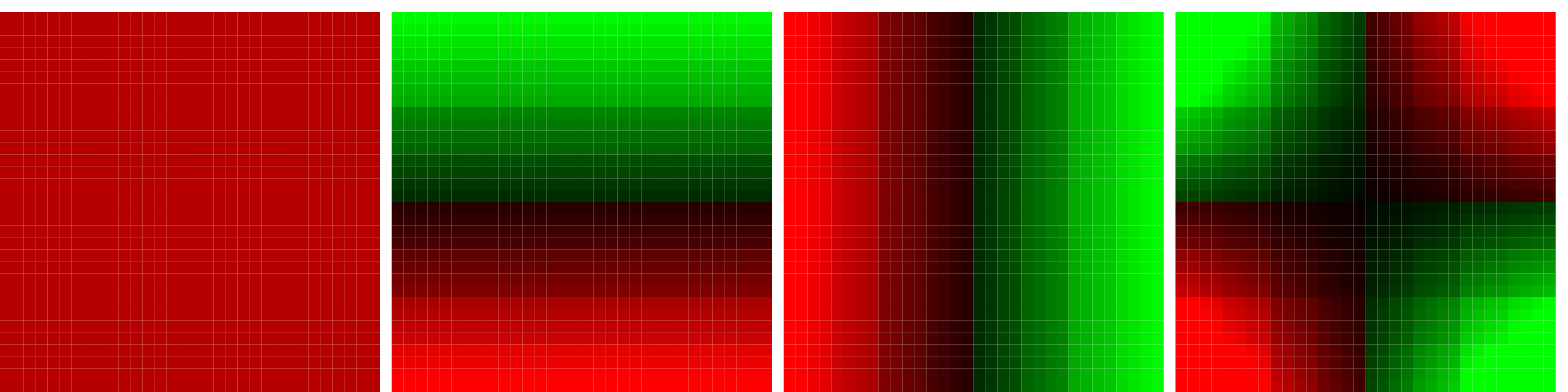}}
\ps2
\nind{\mayp Figure 17. }{\peqfont Above, wave-functions for the 2D
laplacian with free b.c. for a $32\times 32$ system. Below, the fixed
point we reach. These last functions are also smooth.\par}}
\pss
\endinsert

In the fixed b.c. case, we obtain a rather different fixed point, as it
is shown in figure 18.

\midinsert
\pss
\vbox{
\centerline{\fig{2.5cm}{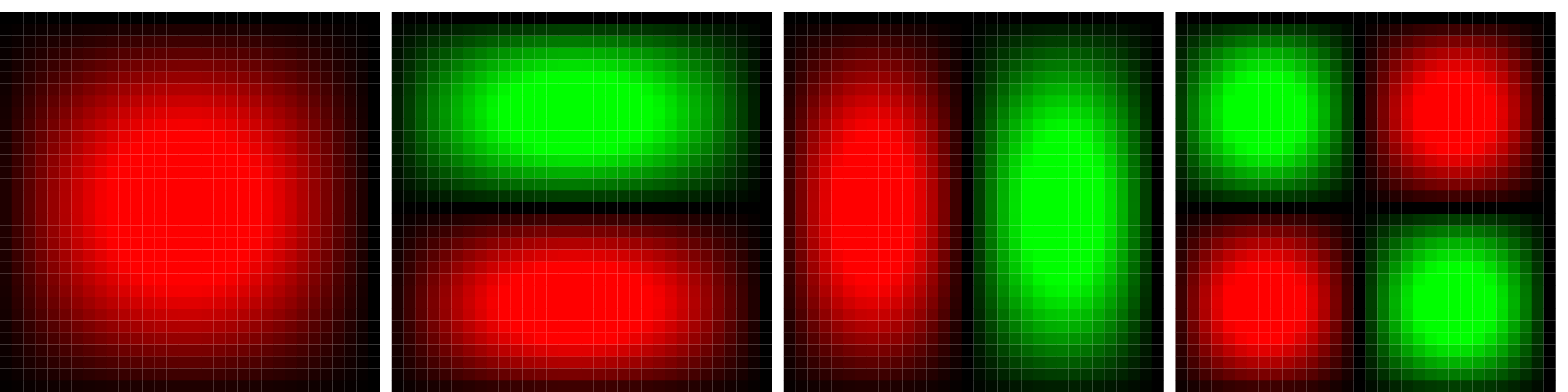}}
\centerline{\fig{2.5cm}{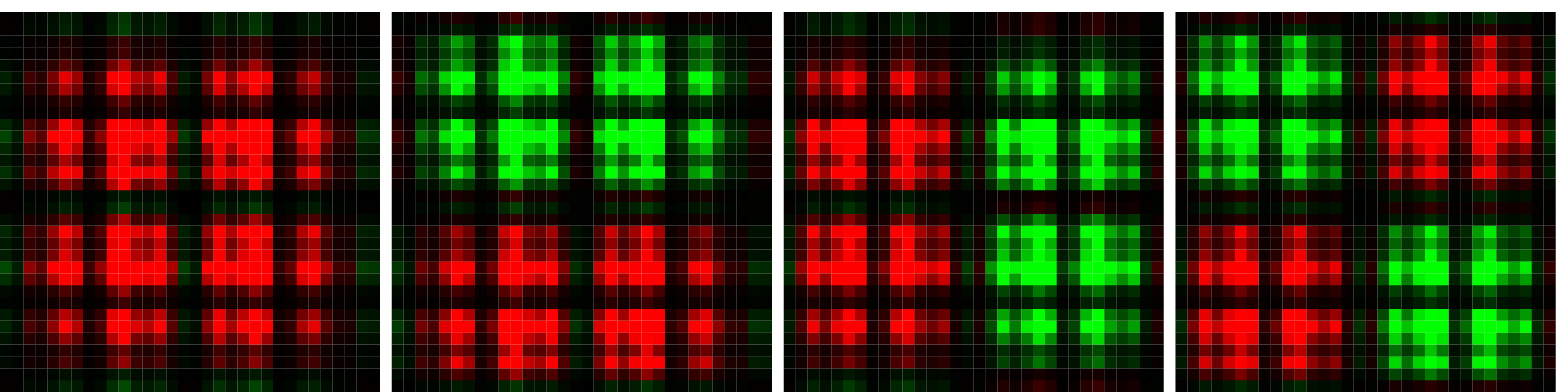}}
\ps2
\nind{\mayp Figure 18. }{\peqfont Same concept as in figure 17, but
for initial wave-functions with fixed b.c. The fixed point is
thoroughly different.\par}}
\pss
\endinsert

Among the fixed points we have found a great richness of
structures. Figure 19 shows a pattern which may result familiar.

\midinsert
\pss
\vbox{
\centerline{\fig{2.5cm}{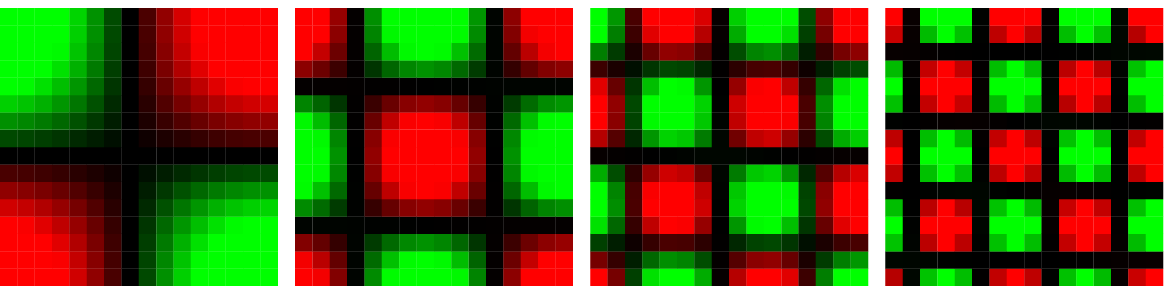}}
\centerline{\fig{2.5cm}{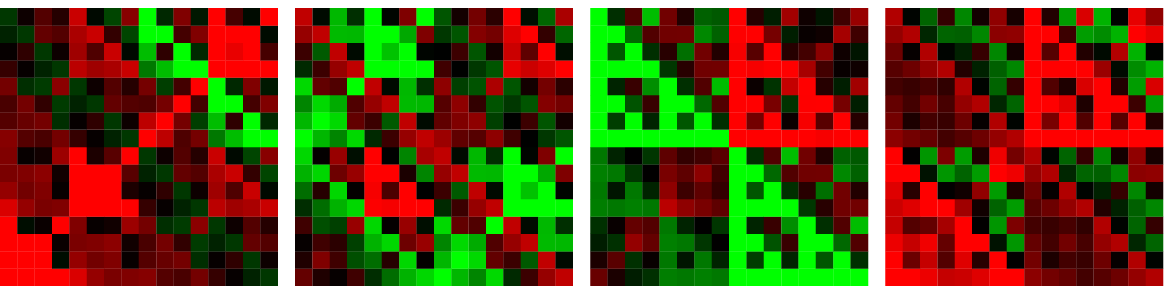}}
\ps2
\nind{\mayp Figure 19. }{\peqfont Wave--functions $16\times 16$ of
higher energy than the previous ones, which yield a repeated pattern
based on the Pascal or Sierpi\'nski triangle. It must be remarked that
this curious structure is not stable, and that it is due to a slight
asymmetry of numerical origin in the initial states.\par}}
\pss
\endinsert

\subsection{Self--replicability and the CBRG.}

Once the question around the free b.c. has been settled (they work
because they are approximately self-replicable), there appears a
second question: why are free b.c. states approximately
self-replicable? The answer is: because they resemble a fixed point of
smooth functions, towards which they do not converge.

There is a great number of fixed points for this replica
transformation\footn{The present author has not been able to find
references to this transformation in the literature on
fractals. Nevertheless, many of the fixed points have non-trivial
fractal structure, even on a discrete functional space.}, but most of
them are non-smooth. The only such set to the knowledge of the author
is really simple: the polynomials.

It may be proved that the set $\{1,x,x^2\ldots x^n\}$ is exactly
self-replicable for any $n$ and interval. The reason is that
translations and scaling transformations of the set return as {\sl
garbage\/} a contribution from lower order polynomials.

Therefore, in a certain sense, we might say that the success of free
b.c. with CBRG is due to the fact that their eigenstates look like the
polynomials. Fixed b.c. states also resemble them, but they lack the
two first ones: $1$ and $x$. This guarantees their failure.


\section{2.8. Bibliography.}

\cx[BCLW-77] {\may H.P. van der Braak, W.J. Caspers, C. de Lange,
M.W.M. Willemse}, {\sl The determination of the ground-state energy of
an antiferromagnetic lattice by means of a renormalization procedure},
Physica A, {\bf 87}, 354-368 (1977).

\cx[DSW-78] {\may S.D. Drell, B. Svetitsky, M. Weinstein}, {\sl
Fermion field theory on a lattice: variational analysis of the
Thirring model}, Phys. Rev. D {\bf 17}, 2, 523-36 (1978).

\cx[DW-78] {\may S.D. Drell, M. Weinstein}, {\sl Field theory on a
lattice: absence of Goldstone bosons in the U(1) model in two
dimensions}, Phys. Rev. D {\bf 17}, 12, 3203-11 (1978).

\cx[DWY-76] {\may S.D. Drell, M. Weinstein, S. Yankielowicz}, {\sl
Strong coupling field theory I. Variational approach to $\phi^4$
theory}, Phys. Rev. D {\bf 14}, 2, 487-516 (1976).

\cx[DWY-77] {\may S.D. Drell, M. Weinstein, S. Yankielowicz}, {\sl Quantum
field theories on a lattice: variational methods for arbitrary coupling
strengths and the Ising model in a transverse magnetic field}, Phys. Rev. D
{\bf 16}, 1769-81 (1977).

\cx[FEY-65] {\may R.P. Feynman, R.B. Leighton, M. Sands}, {\sl The
Feynman Lectures on Physics}, Volume III (Quantum Mechanics),
Addison-Wesley (1965).

\cx[FH-65] {\may R.P. Feynman, A.R. Hibbs}, {\sl Quantum Mechanics and
Path Integrals}, McGraw-Hill (1965).

\cx[FIS-95] {\may Y. Fisher} (ed.), {\sl Fractal image compression},
Springer (1995).

\cx[GMSV-95] {\may J. Gonz\'alez, M.A. Mart\'{\i}n-Delgado, G. Sierra,
A.H. Vozmediano}, {\sl Quantum electron liquids and High-$T_c$
superconductivity}, Springer (1995).

\cx[JUL-81] {\may R. Jullien}, {\sl Transitions de phases \`a $T=0$ dans
les syst\`emes quantiques par le groupe de renormalisation dans l'espace
r\'eel}, Can. J. Phys. {\bf 59}, 605-31 (1981).

\cx[KAD-66] {\may L.P. Kadanoff}, {\sl Scaling laws for Ising models near
$T_c$}, Physica {\bf 2}, 263 (1966). 

\cx[LEM-89] {\may P.G. Lemari\'e} (ed.), {\sl Les Ondelettes en 1989},
Springer (1989).

\cx[MDS-95] {\may M.A. Mart\'{\i}n-Delgado, G. Sierra}, {\sl The
role of boundary conditions in the real-space renormalization group},
Phys.~Lett.~B {\bf 364}, 41 (1995).

\cx[MDS-96] {\may M.A. Mart\'{\i}n-Delgado, G. Sierra}, {\sl Analytic
formulations of the density matrix renormalization group},
Int.~J.~Mod.~Phys. A {\bf 11}, 3145 (1996).

\cx[MRS-96] {\may M.A. Mart\'{\i}n-Delgado, J. Rodr\'{\i}guez-Laguna, 
G. Sierra}, {\sl The correlated block renorma\-lization group}, 
{\tt cond-mat/9512130} and Nucl. Phys. B {\bf 473}, 685 (1996).

\cx[PJP-82] {\may P. Pfeuty, R. Jullien, K.A. Penson}, {\sl
Renormalization for quantum systems}, in {\sl Real
Space Renormalization}, ed. by T.W. Burk\-hardt and J.M.J. van
Leeuwen p\'ag. 119-147. Springer (1982).

\cx[SDQW-80] {\may B. Svetitsky, S.D. Drell, H.R. Quinn,
M. Weinstein}, {\sl Dynamical breaking of chiral symmetry in lattice
gauge theories}, Phys. Rev. D {\bf 22}, 2, 490-504 (1980).

\cx[WHI-99] {\may S.R. White}, {\sl How it all began}, in
{\sl Density-Matrix Renormalization} ed. by I. Peschel et al. (Dresden
workshop in 1998), Springer (1999).

\cx[WIL-71b] {\may K.G. Wilson}, {\sl Renormalization group and critical
phenomena II. Phase-cell analysis of critical behaviour},
Phys. Rev. {\bf B}, 9, 3184-3205 (1971).

\pasacap

\chapter{3. Density matrix renormalization group algorithms.}

\subsection{Synopsis.}

\bul Part I. DMRG for Quantum Mechanics.

3.1. The Density Matrix Renormalization Group.

3.2. DMRG Algorithms for Particles in a 1D Potential.
\ps2

\bul Part II. DMRG, Trees and Dendrimers.

3.3. DMRG for trees.

3.4. Physics and Chemistry of Dendrimers.

3.5. DMRG Algorithms for Exciton Dynamics on Dendrimers.
\ps2

3.6. Bibliography.


\part{Part I. DMRG for Quantum Mechanics.}

The ``particle in a box'' problem, in which K.G.~Wilson had found a
synthesis of the difficulties of the BRG approach\footn{See section
1.4.V and the first part of the previous chapter.}, was solved for the
first time using RSRG techniques by Steve~R.~White in 1992
\ct[WHI-92]. The technique became known as {\it Density Matrix
Renormalization Group\/} (DMRG). With its appearance on stage, RSRG
received in a few years full acknowledgement as a high precision
numerical method. Since Wilson's solution of the Kondo problem no RSRG
method had ever achieved comparable results. In a sense, DMRG holds a
certain ``magical'' aura: its success is still surprising both to the
sporadic practicioner and the specialist.
\pss

Our work on the DMRG has focused on the less intrincate of its
applications so as to reach a deeper understanding of its inner
workings: quantum mechanics of a single particle.

The first two sections of this chapter contain a detailed introduction
to DMRG applied to quantum mechanics. The basic aspects of the DMRG
are taken from the original works of Steven R.~White and Reinhard
M.~Noack (\ct[WHI-92], \ct[NWH-92], \ct[NWH-93] and \ct[WHI-93]) and
from \ct[GMSV-95], which contains its first appearance in a
``textbook''.  The application to quantum mechanical problems is based
on the work of M.A.~Mart\'{\i}n-Delgado, R.M.~Noack and G.~Sierra
\ct[MSN-99]. We would like to remark that all the numerical results
exposed in this work have been obtained with our own programs.

Afterwards, an extension of the method is developed which was carried
out by our group \ct[MRS-00]. Initially, the DMRG was only applied to
unidimensional problems, but it is also possible to employ it to
analyze systems whose configurational space has a tree-like
topology. In the second part of this chapter a suitable extension is
applied to the study of the dynamics of excitons in a kind of
polymeric molecules which have been recently synthesized, known as
{\sl dendrimers}.

The literature on the DMRG is rather wide and quickly growing (see the
article by Karen Hallberg \ct[HAL-99] for a global point of view). In
this thesis we have only cited the works which were most closely
related to ours, and no importance sampling may be assumed.


\section{3.1. The Density Matrix Renormalization Group.}

The global idea of DMRG may be stated in a single sentence: ``when
splitting a system into blocks, these should not remain isolated''.

S.R.~White realized that the lowest energy states of the isolated
blocks {\it need not\/} be the best building bricks to construct the
big block. It is necessary for the block to be related to its
environment so as it may choose the best states.

The solution starts with the {\sl superblock} idea: a block which is
made of some smaller blocks. Its dynamics is studied and the lowest
energy state is found. From that state, the block states which ``fit''
better may be obtained, i.e.: the best bricks. This objective is
accomplished by introducing the {\sl density matrix}.

The development of this section is freely based on the text of
R.P.~Feynman \ct[FEY-72] for the analysis of density matrices, and the
work of S.R.~White \ct[WHI-98] for the general DMRG algorithm.

\subsection{Why Density Matrices?}

The DMRG algorithm receives its name from one of the tools
incorporated into it. It is arguable whether the density matrix is the
key ingredient of the method, but history has made it be considered to
be so.

The reason for which the density matrix appears is the necessity of
``fitting'': it is necessary to find out which are the block states
which reproduce most accurately a chosen global state, which shall be
termed the {\it target state\/}. The manner in which this state for
the superblock is found is a problem which shall be addressed
later.

The states which fit better shall be considered to be the {\it most
probable ones\/}, in which the block may be found, even though they
have nothing to do with the low energy states for an isolated block
(see an example in figure 2).
\pss

Density matrices are, according to some authors, the most fundamental
way to describe quantum mechanical systems. In our description we
shall adopt the more ``practical'' view of R.P.~Feynman \ct[FEY-72],
according to which these matrices are necessary because systems must
always be separated from their environment in order to be analyzed.
This comment was the one which took S.~White to use them as the
foundation of his RG formulation.

Even though we shall follow \ct[FEY-72], the topic is discussed in
similar terms in \ct[LL-71] and in \ct[CDL-73]. At the end of the
section we shall briefly comment on the other possible view.
\pss

Figure 1 shows the structure of the entire system (the superblock)
which is split into {\sl block} and {\sl environment}.

\midinsert
\rput(50,-10){
\multirput{0}(0,0)(5,0){10}{\pscircle*(0,0){1}}
\qline(0,0)(45,0)
\rput(5,-3.5){Block}
\psframe(-2,-6)(17,3)
\psframe(-3,-7)(47,4)
\rput(35,-3.5){Environment}
\rput(25,6){Superblock}
}
\ps{18}
\noindent{\mayp Figure 1.} {\peqfont Superblock split into block and 
environment.\par}
\endinsert

Let the ground state of the full superblock be given by the (pure
state!) ket $\ket|\psi_0>$. In a typical many-body situation, this ket
contains {\it entangled\/} information for block and environment. The
state is expressed as a sum of tensor products:

$$\ket|\psi_0>=\sum_{i,j} \Psi_{i,j} \ket|\alpha_i>\otimes\ket|\beta_j>$$

\nind where $\ket|\alpha_i>$ make up a full basis for the block and
$\ket|\beta_j>$ one for the environment. Let us consider an
operator $A^B$ which only acts on the block variables. Then,

$$\<A^B\>_{\psi_0}=\sum_{i,i',j,j'} \Psi^*_{i,j} \Psi_{i',j'} 
\elem<\beta_j|{\otimes\elem<\alpha_i|A^B|\alpha_{i'}>\otimes}|\beta_{j'}>$$

\nind so, since  $\<\beta_i|\beta_j\>=\delta_{ij}$, we have

$$\<A^B\>_{\psi_0}=\sum_{i,i'}\sum_j \Psi^*_{i,j} \Psi_{i',j}
\elem<\alpha_i|A^B|\alpha_{i'}>$$

Now we may define the {\sl block density matrix} through its matrix
elements in a complete basis

$$\elem<\alpha_{i'}|\rho^B|\alpha_i> = \sum_j \Psi^*_{i,j} \Psi_{i',j}$$

\nind i.e.: the environmental components are ``traced
out''. With this definition our expectation values may be written as:

$$\<A^B\>_{\psi_0} = \sum_{i,i'} \elem<\alpha_i|A|\alpha_{i'}>
\elem<\alpha_{i'}|\rho^B|\alpha_i> = \tr\left(\rho^B A\right)$$

The block density matrix is positive-defined, self-adjoint and has
unit trace (see \ct[FEY-72]). The interpretation, nevertheless, varies
with the author to be considered. The original formulation was given
by L.D.~Landau and J.~von~Neumann \ct[OMN-94] in order to introduce
thermal averages in quantum mechanics. The target was to deal with any
``{\sl second level of uncertainty}'' beyond the one coming from
quantum mechanics. Density matrices are a fundamental tool to study
thermodynamically a quantum system or to take into account any lack of
information concerning the exact way in which the state was
prepared. In other terms:
\ps2

``{\it Whether the environment is not relevant for the computation or
if it is out of reach, the block (system) must not be considered to be
in a pure state, but in a {\sl mixed} state. I.e.: an {\sl statistical
ensemble} of pure states with given probabilities.\/}''
\ps2

The states we are referring to do not need to be orthogonal, but each
one of them must of course correspond to a different state of the
environment. We remark {\sl Gleason's theorem}, which asserts that,
under very general conditions, every mixed state may be described by a
self-adjoint positive-defined and unit-trace density matrix
\ct[GP-89].
\pss

It must be noticed, even though briefly, that the theoretical
physicists which have considered seriously the measurement problem in
quantum mechanics tend to give a different interpretation: ``{\it A
density matrix is the most complete possible description of an
ensemble of identically prepared physical systems\/}''. According to
this second picture, we should have started our discussion with a
density matrix for the complete system. For a defence of this position
the reader is referred to, e.g., \ct[GP-89] and \ct[BAL-90].

\subsection{DMRG: The Basic Idea.}

The objective of the DMRG is to choose for each block the states
which are the best possible building blocks. Thus, an eigenstate of
the full system is sought, which is called the {\sl target
state}. Afterwards we build the set of states of the block which are most
appropriate to reproduce that state.

This ``fit'' is performed by building a density matrix for the block
from the target state, by ``tracing out'' the variables which belong
to the environment. This density matrix is diagonalized. Its
eigenvalues are the {\sl probabilities} for each of the different
states of the block. Therefore, it is necessary to choose the
eigenstates of highest eigenvalues of the density matrix.

Figure 2 shows graphically the difference between choosing an
eigenstate of the density matrix (with highest eigenvalue) and an
eigenstate of the block hamiltonian (with lowest eigenvalue).

\midinsert
\rput(30,-35){
\psline{->}(0,0)(0,30)\rput(-3,30){$\psi$}
\psline{->}(0,0)(60,0)
\pscurve(0,0)(25,20)(50,0)
\pcline{<->}(0,-3)(50,-3)
\lput*{:U}{Superblock}
\pscurve[linewidth=1pt]{-}(30,18.7)(35,15.8)(40,11.5)
\psline[linestyle=dashed]{-}(30,0)(30,30)
\psline[linestyle=dashed]{-}(40,0)(40,30)
\pcline{<->}(30,30)(40,30)
\rput(35,33){Block}
\rput(60,25){\it Eigenstate of $\rho^B$}
\psline{<-}(36,16)(47,24)
\pscurve[linewidth=1pt]{-}(30,0)(35,10)(40,0)
\rput(67,13){\it Eigenstate of $H^B$}
\psline{<-}(37.5,9.5)(53,13)}
\ps{43}

\nind{\mayp Figure 2. }{\peqfont A QM example showing how the 
$\rho^B$ eigenstate fits perfectly the target state at the block,
while the $H^B$ eigenstate, which has to boundary conditions, does
not.\par}
\endinsert

Even though the details of the procedure are strongly dependent on the
particular implementation, the main lines of work are essentially the
ones that follow. The basic DMRG algorithm consists of a {\sl warmup},
which establishes a consistent set of variables, and the {\sl
sweeping}, which iterates the process on the full system until
convergence is reached.

The following discussion is rather formal, and all the details on
concrete implementations are left open.

\subsection{DMRG: The Warmup.}

The system, which we shall consider to be 1D, starts with a small chain.

\ps 3
\rput(50,0){
\multirput{0}(0,0)(5,0){9}{\pscircle*(0,0){1.5}\psline(1,0)(4,0)}
\pscircle*(45,0){1.5}}
\ps{5}

This chain is split into a left and a right part, with two sites between
them:

\ps{3}
\rput(47,0){
\multirput{0}(0,0)(5,0){4}{\pscircle*(0,0){1.5}\psline(1.5,0)(3.5,0)}
\psline(20,0)(22,0)\pscircle(23.5,0){1.5}
\psline(25,0)(27,0)\pscircle(28.5,0){1.5}\psline(30,0)(32,0)
\multirput{0}(33.5,0)(5,0){4}{\psline(0,0)(2,0)\pscircle*(3.5,0){1.5}}
\rput[l](-1,-5){\peqfont Left}
\rput[r](53,-5){\peqfont Right}}
\ps{8}

This system shall be schematically represented as follows:

\ps{8}
\rput(60,0){%
\psframe[fillstyle=solid,fillcolor=lightgray](0,0)(8,8)
\psline(8,4)(10,4)\pscircle(11.5,4){1.5}
\psline(13,4)(15,4)\pscircle(16.5,4){1.5}\psline(18,4)(20,4)
\psframe[fillstyle=solid,fillcolor=lightgray](20,0)(28,8)}
\ps{5}

In this figure two types of objects are to be seen: {\sl blocks}
(whether big, which contain many sites, or small, with a single site)
and links between them, which we shall call {\sl hooks}. Both shall be
represented by certain operators.
\pss

The DMRG growth loop works in this way:
\pss

1.- Using the block and hook operators we compose an effective matrix
    which represents the superblock hamiltonian.
\ps3

2.- The ground state of this superblock hamiltonian is found.
\ps3

3.- With that state, density matrices are found for the left and right
    halves of the system:

\ps{10}
\rput(60,0){%
\psframe[fillstyle=solid,fillcolor=lightgray](0,0)(8,8)
\psline(8,4)(10,4)\pscircle(11.5,4){1.5}
\psline(13,4)(15,4)\pscircle(16.5,4){1.5}\psline(18,4)(20,4)
\psframe[fillstyle=solid,fillcolor=lightgray](20,0)(28,8)
\psframe(-2,-2)(14,10)
\rput[l](-10,-5){\peqfont Left part}
\psframe(14,-2)(30,10)
\rput[r](40,-5){\peqfont Right part}
}
\ps{10}

4.- We retain the $m$ eigenvalues of the density matrix with highest
    eigenvalues on both sides. With them, appropriate truncation
    operators are built.
\ps3

5.- Using these truncation operators, the left and right parts are
    ``renormalized'' to yield new blocks and hooks:

\ps{8}
\rput(62.5,0){%
\psframe[fillstyle=solid,fillcolor=lightgray](0,0)(8,8)
\psline(8,4)(10,4)
\psline(12,4)(14,4)
\psframe[fillstyle=solid,fillcolor=lightgray](14,0)(22,8)}
\ps{5}

6.- The hooks which are dangling may not be tied up together, since
    they represent links to individual sites. We introduce two {\sl
    new sites} between both blocks and we have closed the RG loop.

\ps{4}
\rput(60,-6){%
\psframe[fillstyle=solid,fillcolor=lightgray](0,0)(8,8)
\psline(8,4)(10,4)\pscircle(11.5,4){1.5}
\psline(13,4)(15,4)\pscircle(16.5,4){1.5}\psline(18,4)(20,4)
\psframe[fillstyle=solid,fillcolor=lightgray](20,0)(28,8)}
\ps{8}

After this process the system has increased the number of sites: $N\to
N+2$.

\subsection{DMRG: The Sweep.}

The warmup by itself may yield rather satisfactory results: the
diagonalization of the superblock hamiltonian provides at each step an
approximation to the lowest energy states of the system. But, when
combined with the sweeping process, the results of the DMRG may reach
arbitrarily high precision\footn{This is true for QM. For interacting
systems, the number $m$ of conserved states must be high enough.}.

The warmup provides a series of numerical values for the matrix
elements of the block and hook operators which are {\sl coherent} (in
a sense which shall be now explained). {\it All\/} the operators which
have been obtained through the warmup process must be stored.

Let us call $L(p)$ and $R(p)$ the set of block and hook operators
representing the $p$ leftmost and rightmost sites respectively. In
this way we may write the formal equation:

$$L(p) + \bullet + \bullet + R(N-p-2) = \Omega$$

\nind where $\Omega$ represents the whole system, the symbols
$\bullet$ represent individual sites and the addition symbol is
suitably ``extended'' (see section 4.6).

This very equation may be graphically represented:

\ps{10}
\rput(60,0){
\psframe[fillstyle=solid,fillcolor=lightgray](2,2)(6,6)
\rput(3,-1){\peqfont $L(p)$}
\psline(6,4)(8,4)\pscircle(9.5,4){1.5}
\psline(11,4)(13,4)\pscircle(14.5,4){1.5}\psline(16,4)(18,4)
\psframe[fillstyle=solid,fillcolor=lightgray](18,-2)(30,10)
\rput(26,-5){\peqfont $R(N-p-2)$}}
\ps4

Taking $p=2$, we have

\ps{5}
\rput(30,0){
\multirput{0}(0,0)(5,0){9}{\pscircle*(0,0){1.5}\psline(1,0)(4,0)}
\pscircle*(45,0){1.5}
\psframe(-2.5,-2.5)(7.5,2.5)
\rput(0,-6){\peqfont Left block}
\psframe(18,-2.5)(47,2.5)
\rput(37,-6){\peqfont Right block}}
\rput(110,0){$L(2)+\bullet+\bullet+R(6)$}
\ps{10}

Since we have stored all the $R(p)$ and $L(p)$ during the warmup, it
is possible to build up the superblock hamiltonian which represents
the whole system. Once this has been done, the ground state of the
superblock (target state) is found and the density matrix is built for
one side. In our example, we take the left one:

\vbox{
\ps{13}
\rput(60,0){
\psframe(-2,-4)(12,8)
\psframe[fillstyle=solid,fillcolor=lightgray](2,2)(6,6)
\rput(3,-1){\peqfont $L(p)$}
\psline(6,4)(8,4)\pscircle(9.5,4){1.5}
\psline(11,4)(13,4)\pscircle(14.5,4){1.5}\psline(16,4)(18,4)
\psframe[fillstyle=solid,fillcolor=lightgray](18,-2)(30,10)
\rput(27,-5.5){\peqfont $R(N-p-2)$}
}
\ps 8}

We obtain the most probable states (i.e.: the eigenstates of the
density matrix with highest eigenvalues) and ``renormalize'' the pack
$L(p)+\bullet$ so as to make $L(p+1)$ up. Afterwards, we search in
the ``warehouse'' the pack $R(N-p-3)$, which ``matches'' the $L(p+1)$
we have just built to make up another superblock:

$$L(p+1)+\bullet+\bullet+R(N-p-3)=\Omega$$

In our concrete example, $p$ takes now the value $3$. Then $4$, $5$...
\vfill\eject
\vbox{}
\ps{5}
\rput(30,0){
\multirput{0}(0,0)(5,0){9}{\pscircle*(0,0){1.5}\psline(1,0)(4,0)}
\pscircle*(45,0){1.5}
\psframe(-2.5,-2.5)(12.5,2.5)
\rput(-2,7){\peqfont Left Block}
\psframe(22.5,-2.5)(47,2.5)
\rput(40,7){\peqfont Right Block}}
\rput(110,0){$L(3)+\bullet+\bullet+R(5)$}
\ps{5}
\rput(30,0){
\multirput{0}(0,0)(5,0){9}{\pscircle*(0,0){1.5}\psline(1,0)(4,0)}
\pscircle*(45,0){1.5}
\psframe(-2.5,-2.5)(17.5,2.5)
\psframe(27.5,-2.5)(47,2.5)}
\rput(110,0){$L(4)+\bullet+\bullet+R(4)$}
\ps{5}
\rput(30,0){
\multirput{0}(0,0)(5,0){9}{\pscircle*(0,0){1.5}\psline(1,0)(4,0)}
\pscircle*(45,0){1.5}
\psframe(-2.5,-2.5)(22.5,2.5)
\psframe(32.5,-2.5)(47,2.5)}
\rput(110,0){$L(5)+\bullet+\bullet+R(3)$}
\ps{5}
\rput(30,0){
\multirput{0}(0,0)(5,0){9}{\pscircle*(0,0){1.5}\psline(1,0)(4,0)}
\pscircle*(45,0){1.5}
\psframe(-2.5,-2.5)(27.5,2.5)
\psframe(37.5,-2.5)(47,2.5)}
\rput(110,0){$L(6)+\bullet+\bullet+R(2)$}
\ps{10}

This process goes on, making the left block {\it grow\/} and the right
one {\it shrink\/} until the later is as small as possible. Once this
point is achieved, we invert the sense of the sweep and the left block
starts decreasing (and the $R$ blocks are the ones which get
modified).

Once the initial point is regained, it is said that a {\sl full sweep}
has been completed. Energies usually converge in a few sweeps up to
machine precision.
\pss

Some details were removed from the former description for the sake of
clarity.
\ps2

\bul At each step only a given number $m$ of states are conserved. But the
packs $L(p)$ and $R(p)$ only contain the matrix elements of the needed
operators between those states.

\bul There is a minimum block size, which we shall term $p_0$, which
depends on the number of conserved states $m$. In other words,
$p_0$ sites should contain exactly $m$ states.

\bul When the warmup procedure finishes, values for all $L(p)$ and
$R(p)$ have not been obtained, but only for

$$\{L(p_0),L(p_0+1),\ldots,L(N/2-1)\} \qquad
\{R(p),R(p_0+1),\ldots,R(N/2-1)\}$$

\nind The reason is that the two last blocks of each series, along
with two free sites, make up the whole system. Therefore, the sweep
always starts with this configuration, making the left block grow
(e.g.) until it reaches it maximum size $L(N-p_0-2)$.

\subsection{The Key Features of DMRG.}

The DMRG is a high precision method. In the next section we shall
describe its application to quantum mechanical problems and we shall
discuss its performance. Before providing numerical results we would
like to remark some of its basic features:
\pss

\bul It is a {\sl variational} method. There are no perturbative
series: the system may be as strongly coupled as desired. Meanwhile
not all implementations guarantee the lowering of the energy of all
states at every step, it is sure that the accuracy of the
approximation of the target state shall increase.
\ps2

\bul It is an {\sl implicit} method: the states we are working with
are only known through their matrix elements of the operators making
up the packs $L(p)$ and $R(p)$. Hopefully, the number of such values
is much lower than the dimension of the Hilbert space.


\section{3.2. DMRG Algoritms for Particles in a 1D Potential.}

Now we shall describe a concrete non-trivial implementation of the
DMRG algorithm: the solution to quantum mechanical problems in 1D.
This application was developed by M.A.~Mart\'{\i}n-Delgado, R.M.~Noack
and G.~Sierra in 1999 \ct[MSN-99].
\pss

Let us consider a similar problem to the one we found in section 2.2,
i.e.: the problem of a particle in a discretized box with a
potential. We write the matrix elements of the total hamiltonian:

$$H_{ij} = \cases{ 2/h^2 + V_i & if $i=j$ \cr
	           -1/h^2 & if $|i-j|=1$\cr
	           0 & otherwise\cr}$$

\nind where $N$ is the number of cells and $h=\Delta x=1/N$, since the
box is considered to be the interval $[0,1]$. The values of $V_i$ are
``sampled'' at each cell, e.g.: $V_i=V((i-1)h)$.
\pss

\bul At each RG step there are some states $\ket|\psi^L_i>$ with
$i\in[1\ldots m]$ for the left block and other $\ket|\psi^R_i>$ for
the right one. These states are not {\it explicitly\/} stored. The
left block embraces $p$ sites and the right one $N-p-2$. The two
central sites are represented by two delta states (concentrated at a
single cell), denoted by ``{\it cl\/}'' (center--left) and ``{\it
cr\/}'' (center--right).

\midinsert
\pss
\rput(50,0){
\multirput{0}(0,0)(5,0){9}{\pscircle(0,0){1.5}}
\psline(0,0)(10,0)\psline(25,0)(40,0)
\psframe(-2.5,-2.5)(12.5,2.5)
\psframe(22.5,-2.5)(42.5,2.5)
\rput(-20,0){$\ket|\psi^L_i>$}
\psline{->}(-16,0)(-4,0)
\rput(3,-8){$\ket|\delta_{cl}>$}
\pscurve{->}(8,-8)(12,-6)(14.5,-2)
\pscurve{->}(27,-8)(23,-6)(20.5,-2)
\rput(33,-8){$\ket|\delta_{cr}>$}
\rput(61,0){$\ket|\psi^R_i>$}
\psline{->}(56,0)(45,0)}
\ps{12}
\nind{\mayp Figure 3. }{\peqfont States making up the superblock:
$\ket|\psi^L_i>$ correspond to the left block, $\ket|\delta_{cl}>$ and
$\ket|\delta_{cr}>$ to the free central sites and $\ket|\psi^R_i>$
corresponds to the right block.\par}
\pss
\endinsert

\bul Let us consider all possible decompositions of the system into a
left part, two sites and a right part. Let us see how many packs we
need:

$$\{L(m), L(m+1)\ldots L(N-m-2)\} \qquad \{R(m), R(m+1)\ldots
R(N-m-2)\}$$

\item{} A block smaller than $m$ sites makes no sense, since it would
not even contain $m$ independent states.
\ps2

\bul Each one of the packs, e.g. $L(p)$ contains a block matrix and a
hook vector. We shall denote the matrix with the same symbol (abusing
of notation): $L(p)_{ij}$ and the vector ones by $T_L(p)_i$. In an
explicit fashion:

$$L(p)_{ij}\equiv \elem<\psi^L_i|H|\psi^L_j> \qquad T_L(p)_i\equiv
\elem<\psi^L_i|H|\delta_{cl}> \eq{\leftpacket}$$

\item{} On the right side the definition is equivalent:

$$R(p)_{ij}\equiv \elem<\psi^R_i|H|\psi^R_j> \qquad T_R(p)_i\equiv
\elem<\psi^R_i|H|\delta_{cr}> \eq{\rightpacket}$$

These are the data we shall work with.

\subsection{Blocks Composition.}

One of the keys of the new renormalization methods is its
generalization of the {\sl blocks fusion} idea. For the DMRG each
block is represented by a ``pack'', which consists of a series of
matrix elements of certain operators. Two packs may get fused and
yield other pack which is a representation of the block including
both of them.

Let us suppose the system to be divided into a block with $p$ sites,
two free sites and another with $q$ sites such that $p+2+q=N$. We
write formally:

$$L(p)+\bullet+\bullet+R(q)=\Omega$$

In this paragraph we shall describe the process

$$L(p)+\bullet \to L(p+1)$$

\nind which provides a representation (a pack) for the block
containing the first $p+1$ sites.
\pss

Foremost, we shall write down a variational {\it Ansatz\/} which
includes all the states of the superblock shown in figure 3.

$$\ket|\phi>=\sum_{i=1}^m a^L_i \ket|\psi^L_i> + 
a^{cl}\ket|\delta_{cl}> + a^{cr}\ket|\delta_{cr}> + 
\sum_{j=1}^m a^R_j \ket|\psi^R_j> \eq{\DMRGansatz}$$

The variational parameters are the $\{a^L_i\}_{i=1}^m$,
$\{a^R_i\}_{i=1}^m$, $a^{cl}$ and $a^{cr}$: in total $2m+2$. Since the
set of states is orthonormal, obtaining the superblock hamiltonian is
equivalent to calculate the matrix elements of the hamiltonian between
each pair.

Using the matrix elements $L(p)_{ij}$, $R(q)_{ij}$, $T_L(p)_i$ and
$T_R(q)_i$ we may make up the superblock hamiltonian like this:

\def\h{\hrulefill}\def\v{\vrule height 3pt}
$$H_\Sb=\(
\vcenter{\offinterlineskip\halign{&#\cr
\strut&&\vrule&\strut&\vrule\cr
$\quad$&$\ L\ $&\vrule&$\ T^\dagger_L\ $&\vrule\cr
&&\v&&\v\cr
\h&\h&\h&\h&\h&\h&\h\cr
&&\v&&\v&&\v\cr
&$\ T_L\ $&\vrule&$\ H_{cl,cl}\ $&\vrule&$\ H_{cl,cr}\ $&\vrule\cr
&&\v&&\v&&\v\cr
\h&\h&\h&\h&\h&\h&\h&\h&\h\cr
&&\v&&\v&&\v\cr
&&\vrule&$\ H_{cr,cl}\ $&\vrule&$\ H_{cr,cr}\ $&\vrule&$\ T_R\ $\cr
&&\v&&\v&&\v\cr
&&\h&\h&\h&\h&\h&\h&\h&\h\cr
&&&&\v&&\v\cr
&&&&\vrule&$\ T^\dagger_R\ $&\vrule&$\ R\ $&$\quad$\cr
&&&&\vrule&\strut&\vrule&\strut\cr
}}\)\eq{\hamsuperblock}$$

The diagonalization of such an effective superblock hamiltonian shall
provide us with an estimate for the lowest energies of the total
system. Its eigenvectors shall yield global states for the total
system.

$$\ket|\phi_i>=\sum_{j=1}^m b^j_i \ket|\psi^L_j> + b^{m+1}_i
\ket|\delta_{cl}> + b^{m+2}_i \ket|\delta_{cr}> + \sum_{j=1}^m
b^{m+2+j}\ket|\psi^R_j>$$

\nind with $b^j_i$ denoting the $j$-th component of the $i$-th
eigenvector of the superblock hamiltonian ($j\in[1\ldots 2m+2]$,
$i\in[1\ldots m]$).
\pss

Now the properly called ``renormalization'' procedure is executed,
which we are denoting with the formal expression $L(p)+\bullet\to
L(p+1)$.

Let us focus on the left half of the system (left block + free
site). We put zeroes on the components with $j$ ranging from $m+2$
upto $2m+2$ of matrix $b$ and we obtain the states

$$\sum_{j=1}^m b^j_i \ket|\psi^L_j> + b^{m+1}_i \ket|\delta_{cl}>$$

These {\it do not\/} make up an orthonormal set. A Gram-Schmidt
process solves the problem and the correct states shall be expressed
by

$$\ket|\hat\psi^L_i>=\sum_{j=1}^m B^j_i\ket|\psi^L_i> +
B^{m+1}_i\ket|\delta_{cl}>$$

With these new states, including one more site than the ones before,
shall be used to build up the new pack $L(p+1)$. The block matrix
elements shall be

$$\eqalign{
\hat L(p+1)_{ij}=\elem<\hat\psi^L_i|H|\hat\psi^L_j> =&
\sum_{i',j'=1}^m \elem<\psi^L_{i'}| B^{i'}_i H B^{j'}_j |\psi^L_{j'}> + 
\sum_{j'=1}^m \elem<\delta_{cl}| B^{m+1}_i H B^{j'}_j |\psi^L_{j'}> \cr
&+ \sum_{i'=1}^m \elem<\psi^L_{i'}|B^{i'}_i H B^{m+1}_j |\delta_{cl}> 
+ \elem<\delta_{cr}|B^{m+1}_i H B^{m+1}_j|\delta_{cr}>\cr}$$

\nind This expression, apparently rather complicated, may be written
as a basis change on the upper left part of the superblock
hamiltonian:

\def\tablerule{&height 3pt&&\cr \noalign{\hrule} &height 3pt&&\cr}
$$\hat L(p+1)=B \left(
\vcenter{\offinterlineskip\halign{&\hfil$\;#\;$\hfil&\vrule#\cr
\hbox to 1cm{\hfil\strut}&&&\cr
L(p) && T^\dagger_L &\cr 
\tablerule
T_L && H_{cl,cl}&\cr
&height3pt&&\cr
\noalign{\hrule}}}\right) B^\dagger \eq{\DMRGrenormalizer}$$
\pss

The renormalization of the hook is even more simple. Since it denotes
the linking of the block to an external site, it is by definition

$$\hat T_L(p+1)_i=\elem<\hat\psi^L_i|H|\delta_{cr}>$$

\nind but the whole block is only linked to the central-right site
{\it through\/} the site which was formerly the central-left one (and
is now part of the block). Thus:

$$\hat T_L(p+1)_i=B^{m+1}_i \elem<\delta_{cl}|H|\delta_{cr}>
=B^{m+1}_i H_{p+1,p+2} \eq{\DMRGrenormalizehook}$$
\pss

To sum up, the process $L(p)+\bullet\to L(p+1)$ is divided in these
steps:
\ps2

\bul We take the left and right packs (equations [\leftpacket] and
[\rightpacket]).

\bul We make up the superblock hamiltonian (equation
[\hamsuperblock]).

\bul It is diagonalized, and we retain the $m$ first states.

\bul We make zero the last $m+1$ components of the vectors and
re-orthonormalize them (Gram-Schmidt), to obtain a basis change matrix
$B$.

\bul We use that basis--change matrix to obtain, from the appropriate
quadrant of the superblock matrix, the new $L(p+1)$ (equations
[\DMRGrenormalizer] and [\DMRGrenormalizehook]).
\pss

Obviously, the scheme for the right side renormalization is fully
analogous.
\pss

N.B.: The density matrix {\it is not necessary\/} for the quantum
mechanical problem. It is possible to introduce it, as it is done in
\ct[WHI-92], but the {\sl same result} is obtained. Despite that, the
method is so similar in spirit that it conserves the name.

\subsection{Warmup Cycle.}

In the first step ({\it ab initio\/}) the system is formally
represented by 

$$\Omega(2m+2)=L(m)+\bullet+\bullet+R(m)$$

The left block states, with $m$ sites, may be (e.g.) the delta states
concentrated on each of the $m$ first sites. {\it Mutatis mutandis\/},
the same may be said about the right side. The ``central left'' and
``central right'' sites shall be respectively the sites $m+1$ and
$N-m$.

Therefore, the packs for this first step are:

$$\eqalign{
L(m)_{ij}=\elem<\delta_i|H|\delta_j> \qquad &
T_L(m)=\elem<\delta_i|H|\delta_{m+1}>\cr
R(m)_{ij}=\elem<\delta_{N+1-i}|H|\delta_{N+1-j}> \qquad &
T_R(m)=\elem<\delta_{N+1-i}|H|\delta_{N-m}>}$$

We renormalize it according to the formerly stated scheme both for the
left and right sides:

$$L(m)+\bullet\to L(m+1) \qquad R(m)+\bullet \to R(m+1)$$

\nind by inserting two new sites between both blocks. The process is
repeated until the total size of the system is reached:

$$L(N/2-1)+\bullet+\bullet+R(N/2-1)=\Omega(N)$$

At this very moment we state that the warmup is finished. All the
packs

$$\{L(m)\ldots L(N/2-1)\} \qquad \{R(m)\ldots R(N/2-1)\}$$

\nind have been initialized. Even though not all the $L(p)$ and $R(p)$
have been initialized, it is possible to start working with the values
we have so far. Notice that, if the hamiltonian is symmetric under
parity, it is possible to assume that, at every step, $L(i)=R(i)$, and
the problem gets simplified.
\pss

N.B.: Obviously, there is {\it no real link\/} between the two sites
``$\bullet+\bullet$'' at every step of the warmup. It is necessary to
include a fictitious one so as the algorithm works. When the hopping elements
between neighbouring sites are equal all along the lattice, this is
not a problem.

\subsection{Sweeping Cycles.}

At a practical level, a sweep cycle starts with the system as the
warmup leaves it. Although we might be tempted to start with the
smallest possible left block ($m$ sites) and the biggest right one
($N-m-2$), the difficulty is obvious: only the blocks with sizes
smaller than half the system are available.

Therefore, the sweep consists of three parts:
\ps2

\bul The left block is made grow against the right one till the
extreme is reached (from $N/2-1$ to $N-m-2$).
\pss
\rput(40,0){\psline[linewidth=1.5pt](0,0)(50,0)\pscircle*(0,0){1.5}%
\pscircle*(50,0){1.5}%
\psline{->}(25,2)(48,2)}
\pss

\bul The right block now grows from its minimum size $m$ to its
maximum $N-m-2$.

\ps6
\rput(40,0){\psline[linewidth=1.5pt](0,0)(50,0)\pscircle*(0,0){1.5}%
\pscircle*(50,0){1.5}\psline{<-}(2,2)(48,2)}
\pss

\bul The left block grows again from $m$ to $N/2-1$.

\ps6
\rput(40,0){\psline[linewidth=1.5pt](0,0)(50,0)\pscircle*(0,0){1.5}%
\pscircle*(50,0){1.5}\psline{->}(2,2)(23,2)}
\pss

Once the third phase has been finished, a sweeping cycle has been
completed. All the blocks which include less than $N/2$ sites have
been refreshed at least once from their creation and the rest of them
have been created in the process.

The DMRG algorithm repeats the sweeping cycle until convergence is
attained. If the values for the superblock hamiltonian eigenvalues are
not modified (within some precision) in a whole cycle, the problem is
assumed to be solved.

\subsection{Numerical Results.}

In \ct[MSN-99] the behaviour of the preceding algorithm is analyzed
for a variety of potentials, and results are obtained whose errors are
smaller than one part in $10^{10}$ for all of them: harmonic
oscillator, anharmonic oscillator and double well.

Maybe the most interesting results were obtained with the double well
potential, where in perturbation theory it is rather difficult to
obtain the modification of the spectrum due to the tunnel effect
between them. This modification is analytically studied perturbatively
using the instantonic approximation. The results which are obtained
are fully consistent with exact diagonalization whenever possible and
with the instantonic approximation when it is not.

Since these numerical results were not obtained originally for this
work and are not relevant for its understanding, we refer the
reader again to \ct[MSN-99] for its careful analysis.
\pss

It is important to remark that the computation is rather fast because
the process is {\it implicit\/}, i.e.: the components of the complete
wave-vectors are never explicitly used, and they are not even
stored. The matrix elements of specific operators between those states
is enough to carry out the RG cycle.

This feature of being implicit is quite desirable for an RG algorithm
based on the block idea, since the number of operations to take an RG
step {\it does not scale\/} with the system size. Unfortunately, as we
shall see in the rest of this work, this is not always possible.


\part{Part II. DMRG, Trees and Dendrimers.}

\section{3.3. DMRG for Trees.}

The implicit DMRG algorithm which was described in the former section
was strongly dependent on the left-right distinction. Is it possible
to find an algorithm with the same features on an space where this
distinction does not exist?

A {\sl tree} is a graph which contains no loops. In a certain sense,
it keeps the unidimensional character which is needed\footn{There {\it
are\/} approaches to tackle 2D systems with DMRG (see next chapter),
but the error range is much higher and the speed much lower than for
the 1D case.} for DMRG and, at the same time, it prepares the terrain
to tackle afterwards multidimensional problems. In \ct[MRS-00]
M.A. Mart\'{\i}n-Delgado, G. Sierra and the author of this thesis
extended the quantum-mechanical DMRG algorithm so as it might work on
a tree-like graph without losing its implicit character. This
technique found a practical application in the study of excitons on
dendrimers, which is developed in the next section.

\subsection{Climbing the Trees.}

In the former chapters we have defined the hamiltonian operators on
discretized spaces based on a graph structure. In this section
graphs acquire a crucial importance.
\pss

Let ${\cal G}$ be a graph, described by a set of sites $S$ and an
associated {\sl neighbourhood structure}: for each $i\in S$ there is a
set $N(i)\subset S$ such that $i\in N(j)\iff j\in N(i)$. As usual, a
{\sl path} shall be an ordered set of sites $\{i_1\ldots i_n\}$ such
that $i_{k+1}\in N(i_k)$ (i.e.: each site is connected to the previous
one). The number of elements of $N(i)$ is called $d(i)$, the {\sl
degree} of the sites or coordination index in physical terminology
(see appendix A and \ct[BOL-98] for more details).
\pss

On a connected graph, the removal of a site $i$ splits the set $S$
into a certain number $k(i)$ of disconnected components: $S(i)_1\ldots
S(i)_{k(i)}$, and it must always be true that $k(i)\leq d(i)$. If the
inequality saturates for all the sites, the graph is called {\sl a
tree} (see figure 4).
\midinsert
\pss
\rput(50,0){
\cnodeput(0,0){n1}{1}
\cnodeput(20,0){n2}{2}
\cnodeput(10,-20){n3}{3}
\cnodeput(-10,-20){n4}{4}
\cnodeput(35,-10){n5}{5}
\ncline{-}{n1}{n2}
\ncline{-}{n2}{n3}
\ncline{-}{n1}{n4}
\ncline{-}{n2}{n5}}
\ps{25}
\noindent{\mayp Figure 4 }{\peqfont A tree: the removal of any site
leaves the graph split in disconnected components such that each
neighbour of the site remains in a different one.\par}
\endinsert

It is important to consider the previous definition in more intuitive
terms (however, this explanation does not make up a proof, and we
refer the interested reader to \ct[BOL-98]).

Let us consider a connected graph in which, when removing a site $p$
which contains three neighbours, there appear three disconnected
components. It must be true that each of the components contains one
of the neighbours. Let us call the three components $S(p)_1$, $S(p)_2$
and $S(p)_3$. If the site $p$ did not exist, it would be impossible to
trace a path from a site in $S(p)_1$ to a site of $S(p)_2$. Thus, the
path between those sites passes necessarily through $p$.

If that is true for any $p$ in the graph, then it is not difficult to
conclude that for any pair of points in ${\cal G}$ there is just one
possible path and, therefore, there are no closed paths (which imply
the presence of two different paths between two points). This is the
classical definition of a tree.
\pss

As a matter of fact, there are various alternative definitions of a
{\sl tree}, and it may be useful to know them:
\ps2

\bul A connected graph for which the removal of any site leaves the
graph split into as many isolated components as neighbours it has.

\bul Connected graph for which the removal of any site leaves its
neighbours disconnected among themselves.

\bul Graph in which for each pair of sites there is a single path
which connects them without repeating sites.

\bul Connected graph without loops (i.e.: there are no closed loops
which do not repeat sites).
\pss

It must be clearly remarked that ``legitimate'' multidimensional
spaces may not be represented as tree-like graphs. In some way, trees
present {\it extended\/} unidimensionality. On the other hand, a
unidimensional space with periodic boundary conditions is {\it not\/}
a tree. The final thesis of this chapter shall be that ``The {\sl
implicit} DMRG technique finds its natural place in the trees''.

\subsection{DMRG on Trees. The Idea.}

Let us see how the DMRG algorithm for 1D chains is adapted to tackle
problems on trees.

It is fundamental to observe the system from the ``point of view'' of
a particular site. Let $p$ be that site. For it, the universe is
reduced to a certain number of blocks ($d(p)$ to be exact) which are
disconnected among themselves, as it is shown in figure 5.

\pss
\rput(70,0){
\cnodeput(0,-10){n1}{$p$}
\ov(-15,-5){n2}{$S(p)_1$}
\ov(15,-5){n3}{$S(p)_2$}
\ov(0,-20){n4}{$S(p)_3$}
\ncline{-}{n1}{n2}\ncline{-}{n1}{n3}\ncline{-}{n1}{n4}
} 
\ps{25} 
\nind{\mayp Figure 5. }{\peqfont The world as it is seen
from site $p$.\par} \pss

If we are endowed with appropriate ``packs'' (block + hook) for each
of these blocks, we may build a superblock hamiltonian which
represents the whole system. This hamiltonian would be diagonalized
and, with the help of its $m$ lowest energy states, we would carry out
the block fusion process. For example, blocks $S(p)_1$ and $S(p)_2$
along with the site $p$ itself may form a block

$$S(p)_1 + S(p)_2 + \bullet_p \to B$$

\nind Graphically this appears in figure 6.

\pss
\rput(70,0){
\cnodeput(0,-10){n1}{$p$}
\ov(-15,-5){n2}{$S(p)_1$}
\ov(15,-5){n3}{$S(p)_2$}
\ov(0,-20){n4}{$S(p)_3$}
\ncline{-}{n1}{n2}\ncline{-}{n1}{n3}\ncline{-}{n1}{n4}
\psccurve(-25,-6)(25,-6)(0,-14)
\rput(-26,-2){$B$}
}
\ps{25}
\nind{\mayp Figure 6. }{\peqfont Blocks $S(p)_1$ and $S(p)_2$ along
with site $p$ get fused to make up block $B$.\par}
\pss

Let us call $q$ the neighbour of $p$ which belongs to the block
$S(p)_3$. Block $B$ {\it is a part\/} of the system seen from
$q$. Thus, there must exist an index $y$ such that

$$B=S(q)_y$$

Thus, the operation is really 

$$S(p)_1 + S(p)_2 + \bullet_p \to S(q)_y$$

Now we may consider site $q$ to be free.

\subsection{A Simple Illustrative Computation.}

Let us consider the graph which is represented in figure 7.

\ps{7}
\rput(50,0)
{\cnodeput(-10,0){n1}{1}
\cnodeput(10,-20){n7}{7}
\cnodeput(30,0){n5}{5}
\cnodeput(0,0){n2}{2}
\cnodeput(10,0){n3}{3}
\cnodeput(20,0){n4}{4}
\cnodeput(10,-10){n6}{6}
\ncline{-}{n1}{n2}
\ncline{-}{n2}{n3}
\ncline{-}{n3}{n4}
\ncline{-}{n4}{n5}
\ncline{-}{n6}{n3}
\ncline{-}{n6}{n7}}
\ps{24}
\noindent{\mayp Figure 7. }{\peqfont A sample tree, on which we shall
apply the DMRG algorithm.\par}
\pss

Let site $2$ be our free site for the first step. The system, as it is
seen from that site, consists of {\it two\/} blocks (besides
itself). The first contains only site $S(2)_1=\{1\}$, while the second
contains the sites $S(2)_2=\{3,4,5,6,7\}$. Figure 8 shows this
graphically.

\def\ov(#1)#2#3{\rput(#1){\ovalnode{#2}{#3}}}
\midinsert
\rput(60,-10){
\ov(-25,0){A}{\vbox{\hbox{Block $S(2)_1$.}\hbox{Site: 1}}}
\cnodeput(0,0){B}{2}
\ov(30,0){C}{\vbox{\hbox{Block $S(2)_2$.}\hbox{Sites: 3,4,5,6,7}}}
\ncline{-}{A}{B}
\ncline{-}{B}{C}
}
\ps{17}
\noindent{\mayp Figure 8.} {\peqfont Schematic representation of the
system viewed from site 2.\par}
\pss
\endinsert

Let us suppose that only the ground state of the system is to be found
and we associate a single normalized state to each block. For the
first one we have the delta state
$\ket|\psi(S(2)_1)>=\ket|\delta_1>$. On the other hand, the state for
the second block, $\ket|\psi(S(2)_2)>$, encapsulates $5$ degrees of
freedom (as many as sites) in a single one. Thus, we may estimate {\it
variationally\/} the ground state of the system using as an {\it
Ansatz\/}

$$\ket|\phi>=a_1\ket|\psi(S(2)_1)> + a_2\ket|\psi(S(2)_2)> +
a_c\ket|\delta_2>$$

The superblock matrix, which ``represents'' the full system, is a
$3\times 3$ matrix which takes this form the point of view of site
$2$:

$$H^\Sb=\pmatrix{ H\,[S(2)_1] & 0 & T\,[S(2)_1] \cr 
	          0 & H\,[S(2)_2] & T\,[S(2)_2] \cr 
	T^\dagger\,[S(2)_1] & T^\dagger\,[S(2)_2] & H_{22}\cr}$$ 

Since the three states are normalized, the hamiltonian matrix elements
between them yield the elements of $H^\Sb$.

$$\displaylines{ H\,[S(2)_1]=\elem<\psi(S(2)_1)|H|\psi(S(2)_1)> \qquad
H\,[S(2)_2]=\elem<\psi(S(2)_2)|H|\psi(S(2)_2)> \cr
T\,[S(2)_1]=\elem<\psi(S(2)_1)|H|\delta_2> \qquad
T\,[S(2)_2]=\elem<\psi(S(2)_2)|H|\delta_2>\cr}$$

Once we have obtained the ground state we may carry out the blocks
fusion procedure

$$S(2)_1 + \bullet \to B\qquad \{1\} + \bullet \to \{1,2\}$$

Graphically, this is shown in figure 9.

\ps{8}
\rput(50,0){
\ov(0,0){A}{$S(2)_1$}
\cnodeput(20,0){B}{2}
\ncline{-}{A}{B}
\psframe(-8,-5)(25,5)
\psline{->}(25,0)(40,0)
\ov(53,0){C}{$B$}}
\ps{6}
\noindent{\mayp Figure 9.} {\peqfont Block renormalization: the old
block $S(2)_1$ plus the free site $2$ get renormalized to the new
block $B$.\par}
\pss

If the following free site is $3$, then the block $B$ corresponds to
one of its constituents, let us suppose it is $S(3)_1$. Let us see how
to calculate its block matrix element and its hook. We eliminate from
state $\ket|\phi>$ any mention of the block $S(2)_2$ and renormalize:

$$\ket|\psi(S(3)_1)> = A_1\ket|\psi(S(2)_1)> + A_c\ket|\delta_2>$$

\nind where $A_1=N\cdot a_1$ and $A_c=N\cdot a_c$, where $N$ is an
appropriate normalization constant. Now,

$$H\,[S(3)_1]=\elem<\psi(S(3)_1)|H|\psi(S(3)_1)>=|A_1|^2\; H\,[S(2)_1] +
|A_c|^2\; H_{22} + 2\,\hbox{Re}\,(A_1 A_c T\,[S(2)_1])$$

\nind and now only the ``hook'' is missing

$$T\,[S(3)_1]=\elem<\psi(S(3)_1)|H|\delta_3>=A_c H_{23}$$

Once we have completed the process, site $3$ may be set free, which
will have blocks $S(3)_1=\{1,2\}$, $S(3)_2=\{4,5\}$ and
$S(3)_3=\{6,7\}$. The pack for the first of those blocks has been
determined at the previous step, and the other two packs must be taken
either from a former sweep (RG cycle) or from the warmup.

\subsection{Blocks Composition.}

The major difference between 1D DMRG and its adaptation to trees is
the blocks fusion technique. We now give the general rule.

Let us suppose that, at a given RG step, the free site is $p$, while
the following one shall be $q$. Obviously, it must be true that $q\in
N(p)$.

Let $S(p)_x$ be the block from $p$ which contains its neighbour
$q$. And let $S(q)_y$ be the block from $q$ which contains $p$, as it
is reflected in figure 10.

\midinsert
\pss
\rput(50,-5){
\cnodeput*[fillstyle=solid,fillcolor=lightgray](-15,0){A}{Block}
\cnodeput(0,0){B}{$p$}
\cnodeput(18,0){C}{$q$}
\cnodeput*[fillstyle=solid,fillcolor=lightgray](33,0){D}{Block}
\cnodeput*[fillstyle=solid,fillcolor=lightgray](0,-15){E}{Block}
\cnodeput*[fillstyle=solid,fillcolor=lightgray](18,-15){F}{Block}
\ncline{-}{A}{B}
\ncline{-}{B}{C}
\ncline{-}{C}{D}
\ncline{-}{B}{E}
\ncline{-}{C}{F}
\psframe(-23,8)(8,-23)
\psframe(10,8)(41,-23)
\rput(-38,-10){$S(q)_y$}
\psline{->}(-32,-10)(-23,-10)
\rput[l](53,-10){$S(p)_x$}
\psline{->}(51,-10)(41,-10)
}
\ps{30}
\noindent{\mayp Figure 10.} {\peqfont Graphical description of blocks
$S(p)_x$ and $S(q)_y$.\par}
\pss
\endinsert

The objective, therefore, is to carry out the formal prescription

$$\sum_{i\in [1\ldots d(p)]-x} S(p)_i + \bullet_p \to S(q)_y$$

I.e.: the ``addition'' of all the blocks corresponding to site $p$,
{\it but\/} the one labelled $x$, along with the site $p$ itself to
``refresh'' the block from site $q$ which contains $p$.
\pss

Each block possesses a series of $m$ states which constitute an
orthonormal set\footn{This number might in principle vary from one
block to another, although we have considered it to be uniform
for the sake of simplicity.}. Let us call

$$\ket|\psi(S(p)_k)^l>$$

\nind the $l$-th state ($l\in [1\ldots m]$) of the $k$-th block when
splitting the graph according to site $p$. With these data we shall
constitute the variational {\it Ansatz\/} (notice how this expression
generalizes [\DMRGansatz]).

$$\ket|\phi>=\sum_{k=1}^{d(p)} \sum_{i=1}^m
c^{m(k-1)+i}\ket|\psi(S(p)_k)^i> + c^{md(p)+1} \ket|\delta_p>
\eq{\DMRGtreeansatz}$$

The data pack is now formed by the set

$$\displaylines{
H\,[S(p)_k]_{ij}\equiv \elem<\psi(S(p)_k)^i|H|\psi(S(p)_k)^j>\cr
T\,[S(p)_k]_i\equiv \elem<\psi(S(p)_k)^i|H|\delta_p>\cr}$$

With these elements the superblock hamiltonian is easy to construct,
and has a very characteristic structure

\newtoks\matH
\matH={\vcenter{\offinterlineskip\halign{&\quad#\quad&\vrule#\cr
\phantom{()}&&&\omit&&&\cr
$H\,[S(p)_1]$&&&\omit&&&$T\,[S(p)_1]$\cr
&height 6pt&&\omit&&&\cr
\omit{\hrulefill}&&\omit{\hrulefill}&&\omit&&\omit{\hrulefill}\cr
&height 6pt&&&&&\cr
&&$H\,[S(p)_2]$&&&&$T\,[S(p)_2]$\cr
&height 6pt&&&&&\cr
\omit&&\omit{\hrulefill}&&\omit{\hrulefill}&&\omit{\hrulefill}\cr
&\omit&&height 6pt&&&\cr
&\omit&&&$\ddots$&&$\vdots$\cr
&\omit&&height 6pt&&&\cr
\omit{\hrulefill}&&\omit{\hrulefill}&&\omit{\hrulefill}&&\omit{\hrulefill}\cr
&height 6pt&&&&&\cr
$T^\dagger\,[S(p)_1]$&&$T^\dagger\,[S(p)_2]$&&$\cdots$&&$H_{p,p}$\cr
\phantom{()}&&&&&&\cr
}}}

$$H^\Sb=\(\the\matH\)\eq{\treehsb}$$

This superblock hamiltonian needs to be diagonalized. The eigenvalues are the
actual estimate to the energies of the full system. The $m$ lowest
eigenstates, making up the matrix $B^i_j$ with $i\in [1\ldots m]$ and
$j\in [1\ldots md(p)+1]$ undergo the following process.
\ps2

\bul Making the components corresponding to block $S(p)_x$ zero, i.e.:
from $(x-1)m+1$ to $xm$.

\bul Re-orthonormalization of the vectors, through a Gram-Schmidt
process.
\ps2

We shall call $\hat B^i_j$ the same matrix after having finished both
processes. Now we may assert that the new states which correspond to
block $S(q)_y$ are

$$\ket|\hat\psi(S(q)_y)^i>=\sum_{k=1}^{d(p)} \sum_{i=1}^m
\hat B^{m(k-1)+i}\ket|\psi(S(p)_k)^i> + \hat B^{md(p)+1}
\ket|\delta_p>$$

The fact that the summation extends also to $k=x$ must not bother us,
since the corresponding values of $\hat B$ are null.

With the new block states ready, we proceed to renormalize the block
matrix and the hook vector of $S(q)_y$:

$$\hat H\,[S(q)_y]_{ij}=
\elem<\hat\psi(S(q)_y)^i|H|\hat\psi(S(q)_y)^j>=
\sum_{k,l=1}^{md(p)+1} \hat B^i_k \hat B^j_l H^\Sb_{kl}$$

The hook requires the matrix element between sites $p$ and $q$. The
probability of presence in site $p$ is given just by $\hat
B^i_{md(p)+1}$. Thus,

$$\hat T\,[S(q)_y]_i = \hat B^i_{md(p)+1} H_{pq}$$

\vfill\eject

\subsection{The Warmup.}

Once we have analyzed the tools, we tackle the concrete algorithm. The
first step we must take is the obtention of a sufficient number of
packs so as to start the process.

The only condition on the initial packs is their {\sl coherence}. We
mean that it is necessary that there exist real states whose matrix
elements are given by their values. If the initial packs are coherent,
the distance of the initial states to the true ones does not matter:
convergence is ensured (faster or slower).
\pss

It is impossible to give an universal algorithm for the warmup without
lowering the efficience in particular cases. The last section of this
chapter deals in detail with a process for the analysis of dendrimeric
molecules. For the moment we state some general rules.
\pss

If the tree is not infinite, there must exist a set of sites which
only have one neighbour. Let $D_1\equiv\{p\in {\cal G} |
d(p)=1\}$. Once the computation of those sites is finished, they are
{\it removed\/} from the graph. Let ${\cal G}_1$ be the resulting
graph after that elimination. Now we define $D_2\equiv\{p\in {\cal
G}_1 | d(p)=1\}$. This process yields a sequence of sets $D_k$ which
gives the different {\sl onion skins} of the graph. The number of
steps to take until there are no more sites in the tree shall be known
as its {\sl depth}.

The warmup algorithm must proceed {\it inwards\/}, i.e.: shall start
by grouping into blocks the sites belonging to the lower order onion
skins, which shall absorb the inner sites in an iterative way until
they reach the center. At that moment, the sweeping cycles may start.
\pss

An example may be useful in this case. Figure 11 shows the first step
on a tree. The sites of the first and second onion skins are put into
blocks. If these blocks contain more than $m$ states, it is necessary
to truncate. Otherwise, all the states are conserved.

\midinsert
\rput(30,-28){
\cnodeput(0,20){a}{1}
\cnodeput(-10,12){b}{2}
\cnodeput(10,12){c}{3}
\cnodeput(-20,5){e}{4}
\cnodeput(-30,0){f}{5}
\cnodeput(-5,3){g}{6}
\cnodeput(-10,-5){h}{7}
\cnodeput(0,-5){i}{8}
\ncline{-}{a}{b}\ncline{-}{a}{c}
\ncline{-}{b}{e}\ncline{-}{e}{f}\ncline{-}{b}{g}
\ncline{-}{g}{h}\ncline{-}{g}{i}
\psline{->}(0,22.5)(0,30)
\cnodeput(12,0){j}{10}
\cnodeput(22,5){k}{11}
\ncline{-}{c}{j}\ncline{-}{c}{k}
\psccurve(-5,7)(-13,-9)(3,-9)
\psccurve(-15,7)(-20,9)(-34,-2)
\psccurve(8,16)(8,-3)(28,2)}
\rput(65,-13){$\longrightarrow$}
\rput(100,-28){
\cnodeput(0,20){a}{1}
\cnodeput(-10,12){b}{2}
\cnodeput(15,10){c}{B(3)}
\cnodeput(-20,0){e}{B(2)}
\cnodeput(-5,-3){f}{B(3)}
\psline{->}(0,22.5)(0,30)
\ncline{-}{a}{b}\ncline{-}{a}{c}
\ncline{-}{b}{e}\ncline{-}{b}{f}}
\ps{40}
\noindent{\mayp Figure 11.} {\peqfont Warmup of a part of a tree. The
arrow on site number $1$ shows that the graph goes on in that
direction. The sites bagged in the left side belong to the first and
second onion skins. On the right, some blocks appear which have
substituted them, given in parenthesis the number of states they
contain. If this number is $\leq m$, it remains like that.  If it is
greater, it is truncated to $m$.\par}
\pss
\endinsert

The warmup is complete when all the sites belong, at least, to one
block. At this moment the sweeping may start, and at its first step
the free site must be one of the ones which belong to the innermost
onion skin.

\subsection{The Sweeping Cycle.}

The strategy starts by the establishment of a {\it sequence of free
sites\/} along the tree which we shall call the {\sl path}. Generally
speaking, the path may visit each site more than once.

This sequence does not need to cover the whole graph. It is not
necessary to {\it enter\/} into a block when the number of sites which
it contains is $\leq m$ (i.e.: there is no truncation). For example,
if $m=3$, it would not be required to enter at figure 11 into any of
the blocks which are marked.
\pss

The election of the path, and that of the warmup, must be built taking
into account the concrete problem, so as to take profit of the
available symmetries. The case of the dendrimers shall provide a clear
example.


\section{3.4. Physics and Chemistry of Dendrimers.}

A practical application of the DMRG technique applied to trees is the
study of the excitonic spectrum of the polymeric molecules called {\sl
dendrimers}. The main reference for that analysis is the work of
M.A. Mart\'{\i}n-Delgado, G. Sierra and the present author
\ct[MRS-00].

This section is dedicated only to provide a good physical and chemical
basis to the dendrimers' analysis. A great amount of basic information
may be obtained from \ct[V\"OG-98] and the references therein. Along
this section a part of the classical works on the subject are cited,
along with some of the recent developments which are relevant for our
work.

\subsection{Dendrimers.}

Dendrimers\footn{The word {\it dendrimer\/} comes from Greek
$\delta\acute\epsilon\nu\delta\rho o\nu$, which means {\it tree\/} and
$\mu\acute\epsilon\rho o\varsigma$, which means {\it part\/}. The term
was coined after its resemblance to {\it dendrites\/}.} are highly
branched polymers whose size in some cases exceed $10$
nanometers. They are synthesized molecules with a rather low
polydispersity (i.e.: the molecular masses distribution is highly
peaked around a value) which have quite peculiar architectural
features. These are:
\pss

\bul They have a ``core'', called the {\sl focal point}.

\bul Out of this core stem a number of branches, called {\sl
wedges}.

\bul After the addition of certain number of monomers, equal at all
the branches, these are forced to branch again. A {\sl generation} has
been completed.

\bul Successive generations are formed by adding a fixed number of
monomers to each branch until they are forced to branch again (see
figure 12).

\figura{6cm}{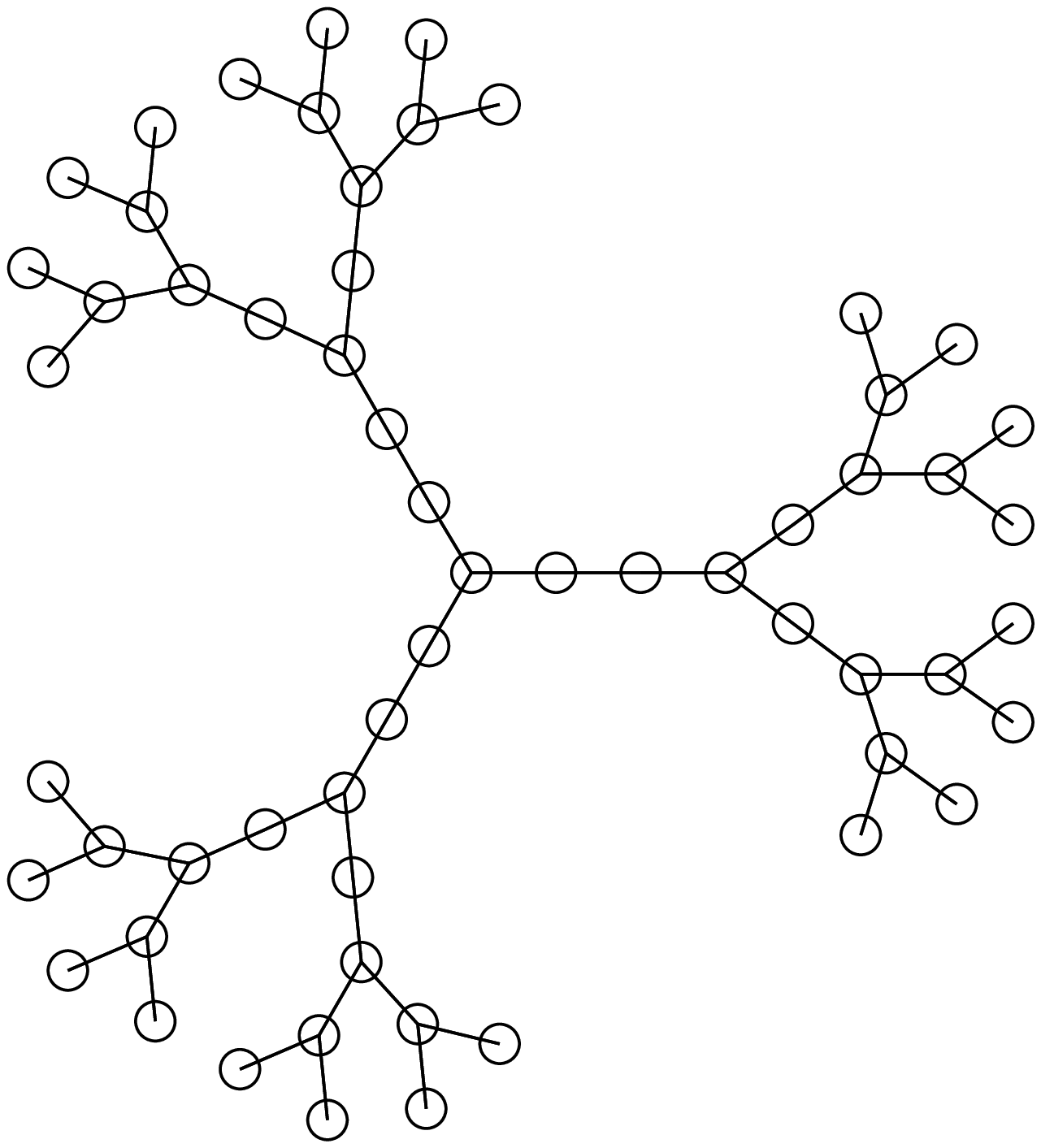}{Figure 12.}{A dendrimer with 3 wedges, 4
generations and connectivity 3 at the branching points.}

The number of generations is not unlimited. The molecules fold up in
three-dimensional space (they are not flat), but steric repulsion
(exclusion volume effects) inhibits high generation numbers. The exact
saturation generation depends highly on the nature of the monomers and
the molecular architecture.

So as to simplify the analysis it is a common practice to ignore the
chemical properties of the monomers, which are now considered to be
merely ``sites'' or ``nodes''. We might say that we {\it embed\/} the
molecule into a lattice or a graph for our purposes.

\subsection{Chemistry of Dendrimers.}

The synthesis of dendrimeric molecules was under very active search by
the end of the 70's, but it became realistic when Fritz V\"ogtle's
concept of {\sl molecular chain reaction} \ct[BWV-78] was developed.

V\"ogtle's idea, known nowadays as the {\sl divergent method} of
synthesis, works by adding iteratively shells from the central
core. Chemists may control the process through the addition and
removal of molecules which allow or inhibit polymerization and
branching. Through this process a high control on the molecular
architecture is gained, but is rather difficult to carry out in
practice \ct[TNG-90].

A new method was developed by Hawker and Frechet \ct[HF-90], which is
called the {\sl convergent method}, which makes dendrimers grow from
the surface inwards. The small pieces which are formed are afterwards
``self--assembled'' until the molecule is complete. The great
advantage of this second technique is its mechanizability.
\pss

Dendrimeric molecules may employ a great variety of monomers. Typical
elements are amino\-acids, polyamidoamines, DNA bases or
glucids. Recently also organo--silicic and other organo--inorganic
hybrids have been used.

Of special interest to our work are dendrimers whose monomers are
phenyl rings, which are linked by acetylene molecules. This link is
usually termed {\sl diphenylacetylene}. The molecules are named as
$Dn$, where $n$ is the number of phenyl rings (e.g.: $D4$, $D127$...).
\pss

Which are the good properties of dendrimeric molecules which made
them so actively pursued by the scientific community?

The geometrical configuration of the molecule makes it have many
``free ends'', which may be functional. Usual polymers with a linear
chain have only two such terminal monomers, and traditional branched
polymers have a number of them which is a small fraction of the
total. In dendrimers, on the other hand, they suppose a
non-negligeable fraction of the total (even beyond one half).

Its {\it self--similar\/} structure makes it have a large surface to
volume ratio \ct[GEN-79]. Concretely, they have been considered for
their usage as ``recipients'', where guest molecules may travel
through hostile media (e.g.: drugs deliverers). These hosting systems
are based on the compact packing of monomers on their surface if the
saturation generation has been achieved. It has been experimentally
tested \ct[JBBM-94] that trapped molecules have a negligeable trend to
diffuse out of the box, but the interaction between recipient and
guest is still under discussion.

Another application which is also based upon their fractal structure
is the possibility of building artificial molecular antennas for the
infra-red range \ct[KOP-97]. The wavelengths which an antenna may
detect depend on the lengths of the ``corridors'' that may be
travelled by the electrons undisturbed. Dendrimeric molecules may be
built so as their corridors range a wide spectrum of lengths. We have
focused on this application.

Let us cite at last some applications which have either been already
carried out or are being actively researched: gene therapy,
nanotechnology (as building blocks to construct bigger molecules),
magnetic resonance imaging (as contrast agents), molecule recognition,
antiviral treatments, energy converters (of light into electric
currents) \ct[V\"OG-98]. A host of applications which range from
materials science to medicine.

\subsection{Excitons in Dendrimers.}

As it was previously mentioned, one of the applications of dendrimers
which may attract the attention of physicists is the possibility of
building antennas for harvesting infra-red light. It is necessary,
therefore, to model the behaviour of the electrons.

The Frenkel exciton model is a development of the tight binding model
(TBM) for an electron in a lattice of similar atoms or molecules,
which we shall merely term {\sl sites}. The approximation takes
orbitals at each site as its starting point.

Let us suppose that there are only two orbitals per site which
contribute to the physics of the problem. We shall call $\ket|0_i>$
the orbital of site $i$ with smaller energy and $\ket|1_i>$ the
excited one at the same site.

\def\vac{\hbox{\mathtext vac}}\def\dg{\dagger}
Let us introduce a vacuum state per site $\ket|\vac_i>$, meaning that
there is no electron at all at the site. A couple of creation and
annihilation operators is suitable:

$$c^\dagger_{i0}\ket|\vac_i>=\ket|0_i> \qquad 
c^\dagger_{i1}\ket|\vac_i>=\ket|1_i>$$

As it was expected, $c_{0i}$ and $c_{1i}$ shall denote the
corresponding annihilation operators. Let us define now {\sl
excitation} operators: $b^\dg_i=c^\dg_{1i}c_{0i}$ and
$b_i=c^\dg_{0i}c_{1i}$. These operators fulfill 

$$\matrix{
b^\dg_i\ket|0_i>=\ket|1_i> \qquad& b^\dg_i\ket|1_i>=0\cr
b_i\ket|0_i>=0 \qquad& b_i\ket|1_i>=\ket|0_i>}$$

I.e.: the excitation operators act as creation and annihilation of a
new ``quasi-particle'': the {\sl exciton}.

The most simple hamiltonian which may be written for the dynamics of
these excitons is 

$$H=\sum_{i\in S} E_i b^\dg_i b_i + \sum_{\<i,j\>} J_{i,j} \(
b^\dg_i b_j + b^\dg_j b_i \) \eq{\excitonham}$$

\nind where $S$ represents the set of sites of the molecule, $E_i$ are
the energy gaps at each site (they may be different) and play the
r\^ole of chemical potentials. The sum extended over the $\<i,j\>$
runs, as it is usual, over the graph links. The origin of these terms
is the overlapping of the orbitals at different sites, which may make
the electron ``jump'' from one site to another.
\pss

An electrostatic interaction might be added between different
electrons to obtain a much more complex interacting hamiltonian such
as, for example, the {\sl Pariser--Parr--Pople} hamiltoinian (which
has been analyzed with DMRG \ct[RT-99] and similar techniques
\ct[MSPJ-99] for 1D chains).

Since the model formerly introduced [\excitonham] is {\it free\/}, it
may be converted into a problem in quantum mechanics just by writing
down a wavefunction

$$\ket|\psi>=\sum_{i\in S} C_i \ket|\delta_i>$$

And, therefore, in the one-particle sector, the hamiltonian gets
reduced to

\def\unex{{\hbox{\mathsubtext 1ex}}}
$$H_\unex=\sum_{i\in S} E_i
\ket|\delta_i>\bra<\delta_i| +
\sum_{\<i,j\>} \( J_{i,j} \ket|\delta_i>\bra<\delta_j| + {\rm h.c.}
\)\eq{\unexciton}$$

Schr\"odinger's equation is

$$H_\unex\ket|\psi>=E\ket|\psi>$$

\nind and the problem has been reduced to another one analyzable with
DMRG techniques for quantum mechanics. The eigenvalues of $H_\unex$,
anyway, must be taken carefully. The lowest eigenvalue does {\it
not\/} mean the energy of the ground state, but the minimum energy
which is needed to create an exciton. In the terminology of this
research field, it is called the {\sl optical absorption edge}.


\section{3.5. DMRG Algorithm for the Dynamics of Excitons in
Dendrimers.}

The physical problem which we shall study is the prediction of the
optical absorption edge for the series of dendrimers of the
diphenylacetylene family. There are experimental results, obtained by
Kopelman et al. \ct[KOP-97] to which we shall compare our results and
those of other authors.

The theoretical model we shall employ was developed by Harigaya
\ct[HAR-98] \ct[HAR-99]. It is a modification of the excitonic model
given in the previous section, with disorder added at the nondiagonal
terms.

Our target shall be, therefore, the analysis of the hamiltonian
[\unexciton] through DMRG. A set of realizations for the set of values
$\{J_{ij}\}$ shall be considered and the optical absorption edge shall
be obtained statistically.
\pss

Even though it is well known, it is appropriate to remark that
quantum-mechanical models with disorder are equivalent to many-body
interacting models \ct[ITZD-89]. The process consists (briefly
summarized) in the conversion of the summation on realizations into a
path integral for a new fermionic field.

\ps 7
\subsectionp{Compact and Extended Dendrimers. Combinatorial
Properties.}

A {\sl Cayley tree} (also known as {\sl Bethe lattice}) is a tree with
a uniform degree $d(i)=c$. It is impossible to make a finite model of
such a tree, so we may define a {\sl finite Cayley tree} as that which
has only two kind of sites: {\sl terminal} with $d(i)=1$ and normal
with $d(i)=c$.

A dendrimer whose sites are linked describing a finite Cayley tree is
called a {\sl compact dendrimer}. The connectivity is usually $c=3$
(see figure 13), since it is the smallest non-trivial value. We shall
only consider dendrimers with complete generations.

\figura{5cm}{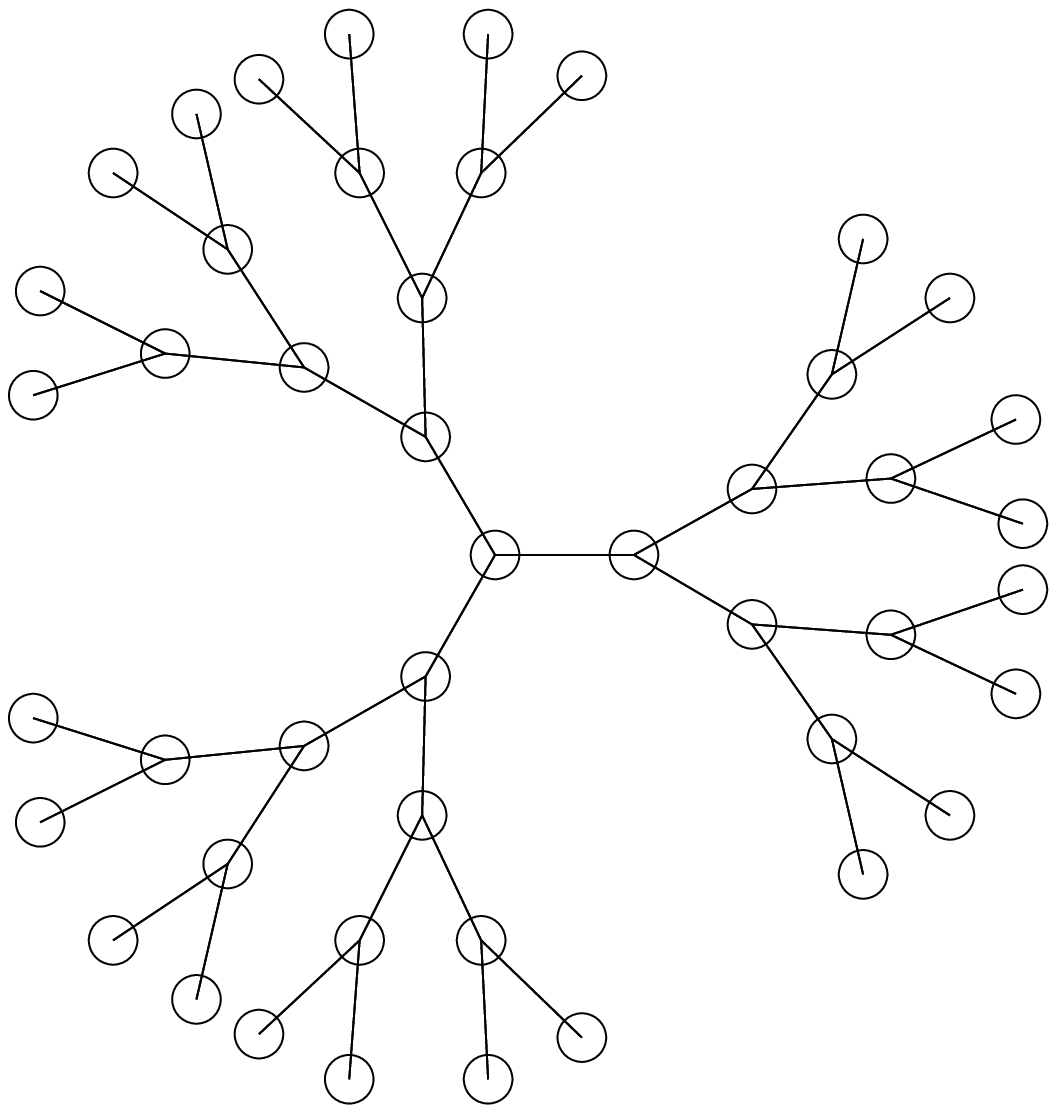}{Figure 13.}{A compact dendrimer, i.e.: a
finite Cayley tree. A further generation would force some sites to
overlap (or the molecule to fold up in 3D), which exemplifies steric
repulsion.}

The compact dendrimer of figure 13 has $4$ generations and shall be
denoted by $D_C(4)$. The number of sites of such a graph (translating:
the number of monomers of such a dendrimer) with $L$ generations is
given by

$$N_C(L)=1+3\sum_{n=1}^L 2^{n-1}=3\cdot 2^L -2$$

\nind They are also termed with the notation $Dn$, where $n$ is the
number $N(L)$. Thus, the first molecules of the compact series are
$D4$, $D10$, $D22$, $D46$,...
\pss

An {\sl extended dendrimer} is something more complex to
describe. Generations have different lengths. The branches which are
close to the core are longer than the ones which are near the
surface. If $L$ is the number of generations and $n$ is the generation
index (counting outwards), the length of the branches is $L-n$. A
``termination shell'' is added, which would correspond to $n=L$.

As an illustrative example, figure 12 may be good, which has $L=4$
generations. The innermost one, $n=1$, has length $L-n=3$, and at each
branching it decays by one site.
\pss

The number of sites of an extended dendrimer may be analytically
found. The computation is as follows:
\ps2

\bul A central site.

\bul The three first branches: in total $3(L-1)$ sites.

\bul The second level branches: in total $3\cdot 2\cdot (L-2)$ sites.

\bul The $n$-th shell gives $3\cdot 2^{n-1}\cdot (L-n)$. The shells
are considered from $1$ upto $n-1$, since there is no sense in adding
a null shell.

\bul The terminating shell: $3\cdot 2^{L-1}$.
\ps2

Summing up:

$$N_E(L)=1+3\sum_{n=1}^L (L-n)2^{n-1} + 3\cdot 2^{L-1}$$

\nind The final result is (see \ct[GKP-89], where techniques of
discrete analysis are discussed):

$$N_E(L)=1 + 3\cdot\(3\cdot 2^{L-1}-L-1\) \eq{\sizeext}$$

The advantage of this concrete form is that it makes explicit the
contribution of each wedge plus the central site. Thus, the first
dendrimers of the extended series are $D4$, $D10$, $D25$, $D58$,
$D127$,... Notice that the first two elements coincide in both series.

We remark also the fact that the function $N(L)$ presents an
exponential growth in both cases. For relatively high values of $L$,
the exact diagonalization of the hamiltonian is impracticable.

\subsection{Dendrimers and Computation.}

So as to make the numerical computations easily reproducible, we now
give a series of useful technical details.

The first practical problem is to write down a program which
implements computationally the neighbourhood structure. Once this has
been computed, the calculation of the hamiltonian matrix is
straightforward.
\pss

Although the programs are written in {\tt C++}, a typical structure of
the LOGO language was quite useful \ct[AdS-81]. It is the {\sl
turtle}: an object which is able to advance and rotate and, as it
moves, trace a drawing. Our turtles are also able to reproduce and,
thus, with the collaboration of the whole family, draw the graph of
the hamiltonian.

We give now in pseudocode the instructions which we have used, both to
draw compact and extended dendrimers. The routine of pseudocode 1,
called {\tt Launch}, only receives a parameter, which is a {\sl
turtle}. This has two inner variables: the generation number and the
direction it is pointing at.

The program invokes itself using as parameters each of its child
turtles. It is, therefore, {\sl recursive}, and the finalization
condition is to have reached the last generation.

\pss
\boxit{5pt}{\vbox{\advance\hsize by -10pt\tt Advance.

Make new site.

$\lin{level} \ot \lin{level}-1$.

if $(\lin{level}>0)$,

\quad Make new turtle with level $\lin{level}$ here.

\quad Rotate clockwise $\alpha$.

\quad Launch.

\quad Make new turtle with level $\lin{level}$ here.

\quad Rotate anticlockwise $\alpha$.

\quad Launch.}}
\ps2
\nind{\mayp PseudoCode 1.} {\peqfont Instructions for a turtle to
create a compact dendrimer.\par}
\pss

The complete program just creates three turtles (with angles $0$,
$2\pi/3$ and $4\pi/3$) and launch them. The {\may PostScript} graphics
which are shown in this thesis were obtained with this
program. Furthermore, the differences with the necessary code to
create the neighborhood structure are minimum.
\pss

In the case of the extended dendrimers a small modification is
necessary: the instruction {\tt Advance} must be carried out not just
once but as many times as the length of the branch we are at.
Pseudocode 2 gives the sequence of orders to substitute {\tt Advance}.

\pss
\boxit{5pt}{\vbox{\advance\hsize by -10pt
\tt For $i=1$ to $\lin{max\_level} - \lin{level}$\par
\quad Advance \ps0
\quad Make new site}}
\ps2
\nind{\mayp PseudoCode 2.} {\peqfont Modification of the former
program to create extended dendrimers.\par}
\pss

The instruction {\tt Make new site} adds a vertex to the graph,
assigns it the first available index and links it to the previous
site. This automatic manner to generate the numbering of the sites of
a dendrimer keeps an interesting relation to number systems.

The key concept is the {\sl chain of decisions} which must be taken so
as to localize the site in the dendrimer. We shall consider the
compact case for simplicity.

At the central site four decisions are possible: one for each of the
three branches plus the one corresponding to finish there. At each of
the other sites there are only three possible decisions: ``left'',
``right'' or, again, terminate. Thus, the specification of any site
may be given by a {\sl word}. The first letter belongs to the set
$\{0,1,2,T\}$, and the following ones to $\{0,1,T\}$. Of course, the
symbol $T$ must be the last one in each word. Valid examples are
$210010T$, $10T$ or, simply, $T$.

Let us define now a {\sl bush} of level $n$ as the set of sites which
would be obtained by launching a turtle at that level. The number of
sites of a bush of level $n$ is $2^{L-n+1}-1$. With this rule, the
conversion of a word into a site index is quite simple: each numerical
symbol contributes its value multiplied by the size of the bush of
that level. The symbol for termination contributes a single $1$.

In a single formula:

$$\{s_1s_2\ldots s_kT\} \to \sum_{i=1}^k s_i\cdot 2^{L-i+1} +1$$

\nind It is easy to convince oneself that this formula expresses a
particular number system.
\pss

The determination of a minimum path along the dendrimer is also an
interesting problem. Of course, it is impossible to find a hamiltonian
path (i.e.: one which runs over all the sites without repeating any
one). The following algorithm returns a list with the order of the
minimum length path which spans the whole tree.

We shall start at the central site. The rule is ``each time an option
appears, choose the site of smallest index among the ones which have
not been visited yet. If all the sites have already been visited,
return the way you came from.''
\pss

Given any tree, the number of steps of the minimum path is the double
of the number of links of that tree. Denoting this number by $|P|$ we
have

$$|P|=\sum_{i=1}^N d(i)$$

This sum may be determined analytically for compact and extended
dendrimers. In the first case we only have two kinds of sites
regarding the number of neighbours: ``inner'', which have $3$
neighbours, and ``terminals'', which have only $1$. Therefore, the
length of the path would be:

$$|P|=\sum_{\hbox{\mathsubtext Terminals}} 1 + 
\sum_{\hbox{\mathsubtext Inner}} 3$$

The number of inner sites is nothing but the number of sites of the
dendrimer with one generation less. Thus,

$$|P_C|=3N_C(L-1)+(N_C(L)-N_C(L-1))=3\cdot 2^{L+1}-6$$

\nind (let us remind that $N_C(L)$ is the number of sites of a compact
dendrimer of $L$ generations). In the case of the extended dendrimer
the computation is slightly longer. Sites may be classified according
to the following criterion:
\ps2

\bul ``Branching'' sites, with $3$ neighbours.

\bul ``Bridge'' sites, with $2$ neighbours.

\bul ``Terminal'' sites, with $1$ neighbour.
\ps2

The number of branching sites is equal to the size of the compact
dendrimer with the same number of levels: $3\cdot 2^{L-1}-2$, and the
number of terminal sites is the same as in the compact case: $3\cdot
2^{L-1}$. The number of ``bridge'' sites may be obtained by
substracting from [\sizeext] or either by direct computation. Choosing
the last option we have the summation

$$\hbox{\# ``Bridge'' sites} = 3\cdot \sum_{i=1}^{L-2} (L-i-1)\cdot
2^{i-1}= 3\cdot (2^{L-1}-L)$$

\nind (notice that, although the formula was designed for $L\geq 2$,
the final result takes into account correctly that for $L=1$ and $L=2$
there are no bridge sites). Putting all pieces together, the final
result is

$$|P_E|=9\cdot 2^L-6\cdot (L+1)$$

\subsection{The Laplacian on a Dendrimer.}

Regarding only the connectivity properties, a dendrimer has a very
large symmetry group. This group is $\hbox{\amsmath Z}_3\otimes
(\hbox{\amsmath Z}_2)^n$, where $n$ is the number of branching points
of the molecule (excluding the focal point). It is possible to
classify the eigenstates of the laplacian on a dendrimer, either
extended or compact, according to the group representations.

The ground state is, logically, uniform all over the graph and
corresponds to the trivial representation of the group. Figures
14{\mayp A} and 14{\mayp B} show other eigenfunctions of the laplacian
with free b.c. for an extended dendrimer with $6$ generations which
correspond to higher dimensional representations.

\centerline{\fig{5cm}{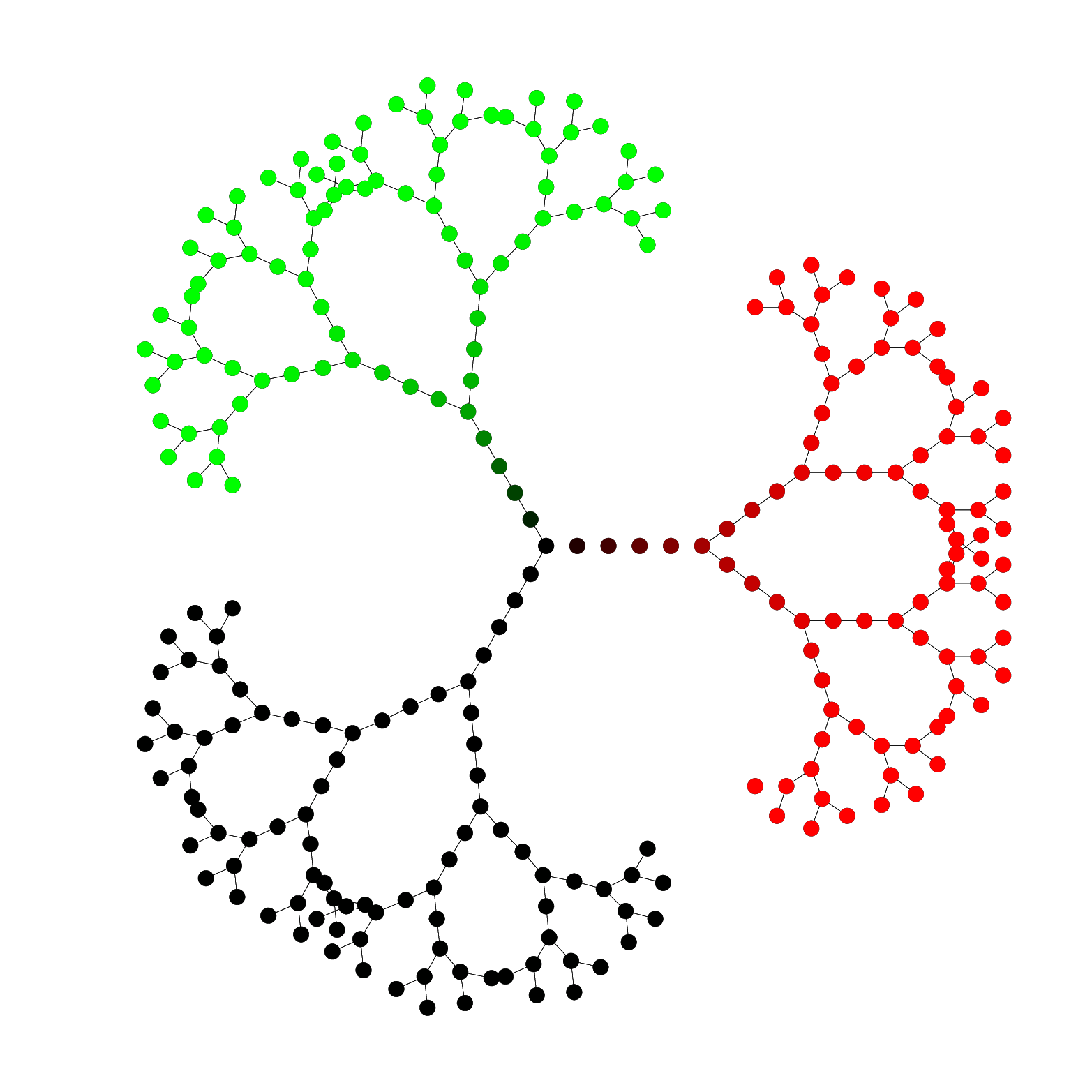}\hskip 1cm
\fig{5cm}{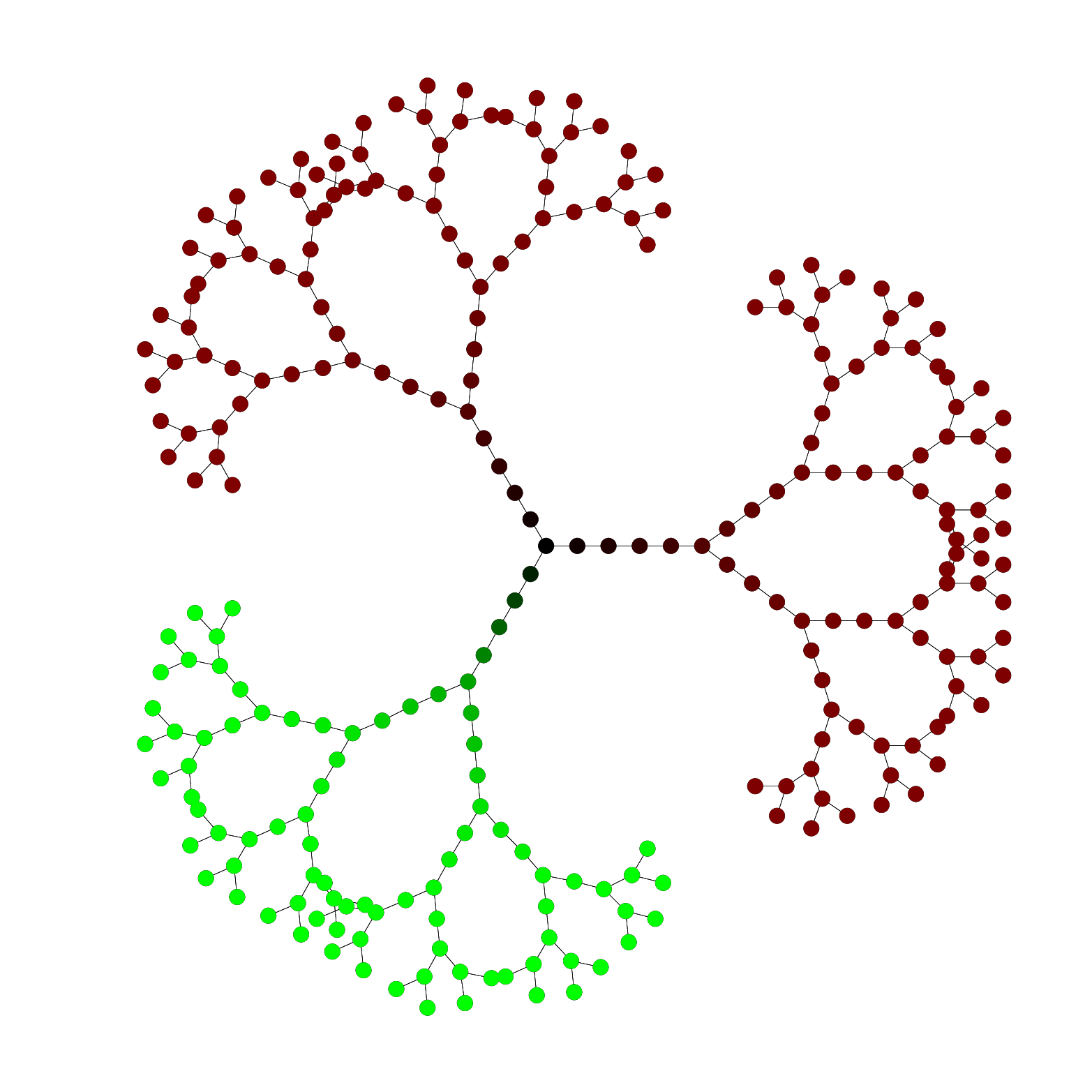}}
\nind{\mayp Figure 14a. }{\peqfont The two first excited states of the
laplacian with free b.c. on an extended dendrimer with 6
generations. Both of them have the same energy and correspond to a
bidimensional representation of the molecular symmetry group.\par}

\centerline{\fig{5cm}{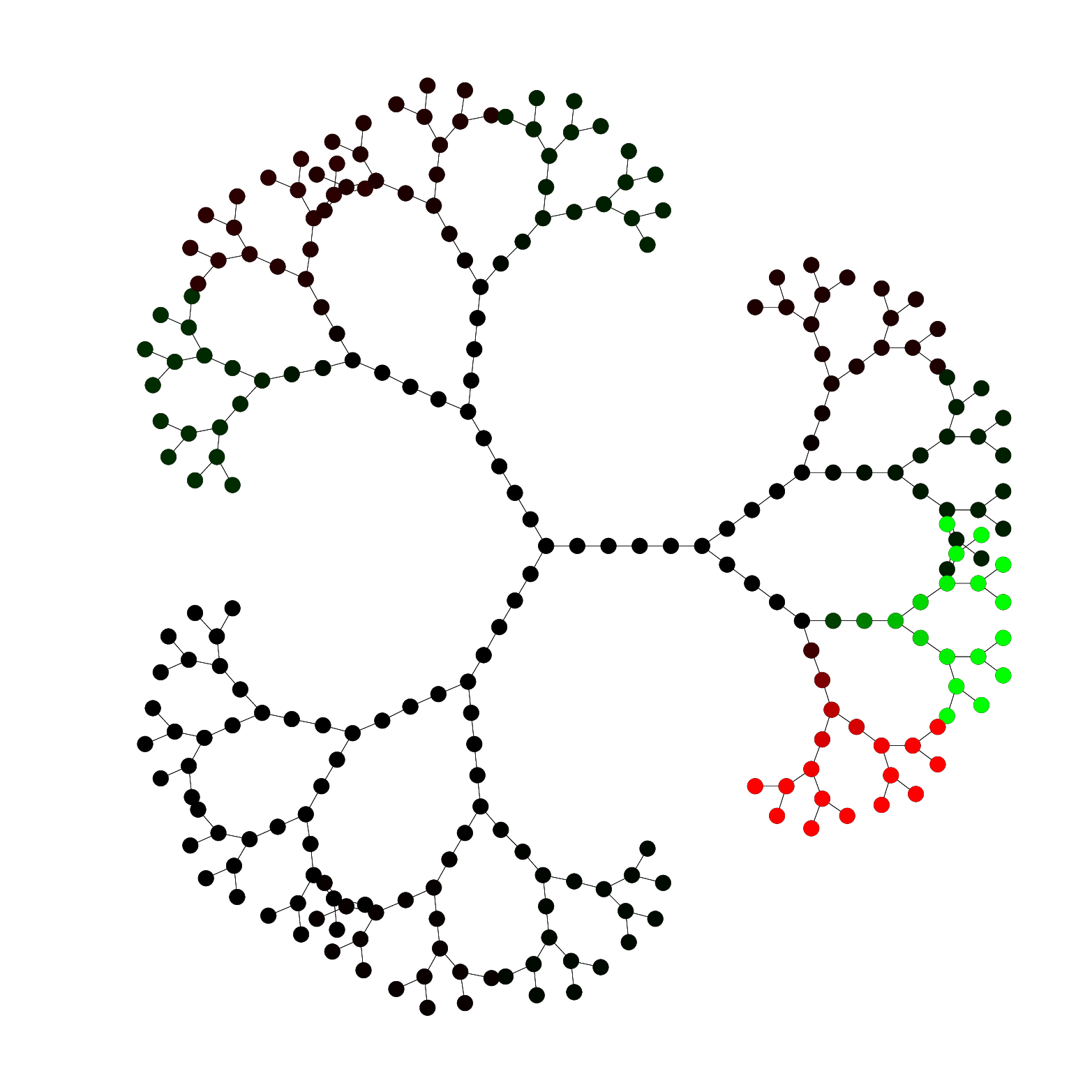}\hskip 1cm
\fig{5cm}{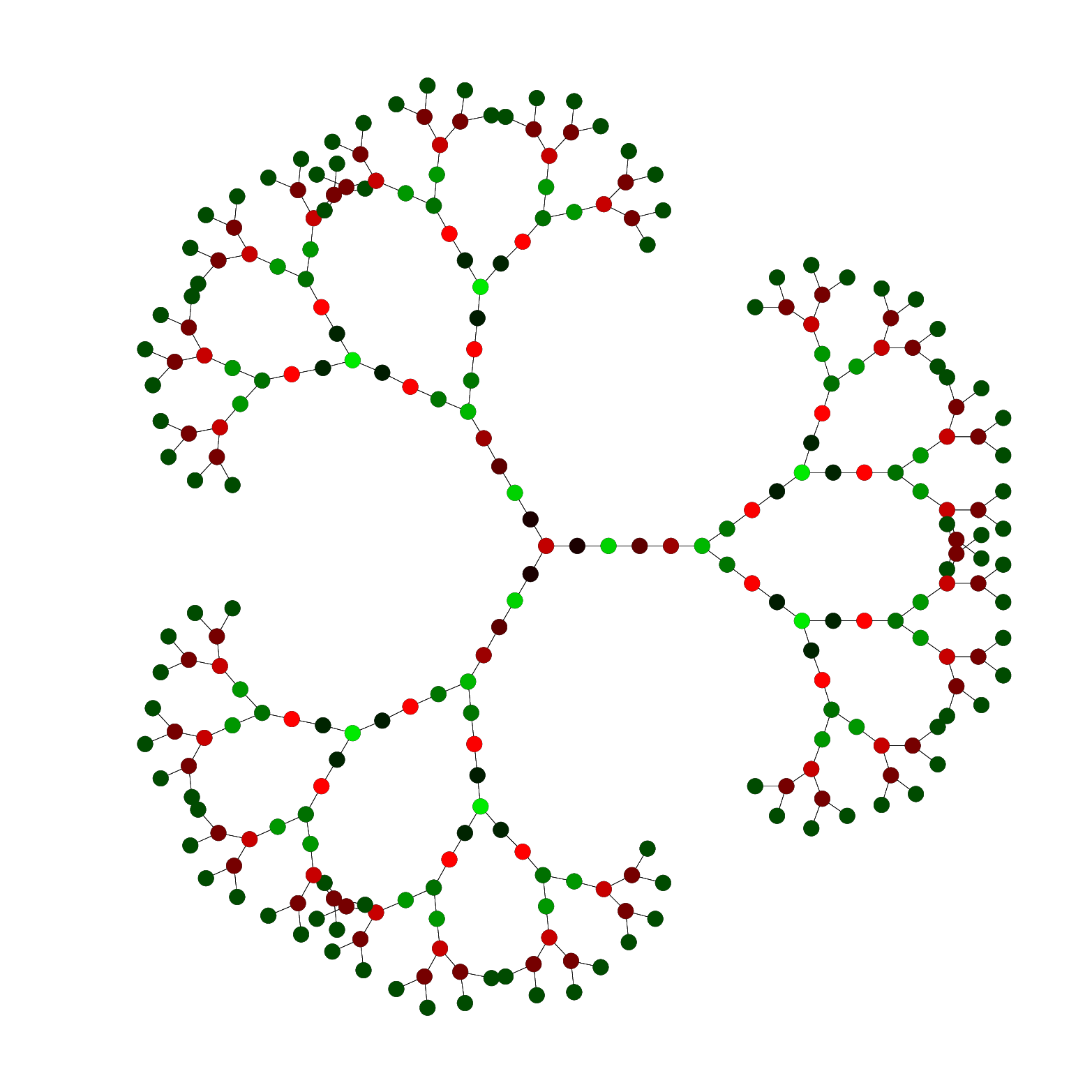}}
\nind{\mayp Figure 14b. }{\peqfont On the left, the seventh
eigenstate, with sixfold degeneration. On the right, the eigenstate
number 166, which corresponds to an unidimensional
representation.\par}

\subsection{Harigaya's Model.}

The model proposed by K.~Harigaya for the computation of the optical
absorption edge of extended dendrimers of the diphenylacetylene family
\ct[HAR-98], \ct[HAR-99] has the following features:
\ps2

\bul Each site (molecule) is represented by a single pair of molecular
orbitals.

\bul It is a quantum-mechanical model of a single exciton.

\bul It has uniform chemical potential $E$.

\bul The hopping terms between sites are random, drawn from a
gaussian distribution with mean zero and standard deviation $J$.
\pss

Analytically, the hamiltonian is given by

$$H=\sum_{i=1}^N E\ket|\delta_i>\bra<\delta_i| + \sum_{\<i,j\>}
(J_{i,j} \ket|\delta_i>\bra<\delta_j| + h.c.)$$

\nind and the empirical constants are given by
$E=37.200\,\lin{cm}^{-1}$ y $J=3.552\,\lin{cm}^{-1}$.

\subsection{Application of the DMRG Algorithm.}

The general algorithm exposed at section 3.3 is directly applicable to
this problem. The warmup establishes the division into onion skins,
which are translated in this case to the division into generations of
the compact dendrimers. In the case of the extended dendrimers,
generations are subdivided into different shells because of the
corridor sites. Basically, there are only two schemes of blocks
fusion:

$$\displaylines{
\hbox{Branching site:}\qquad 
\lin{Block}+\lin{Block}+\bullet \to \lin{Block}\cr
\hbox{Bridge site:}\qquad
\lin{Block}+\bullet \to \lin{Block}\cr}$$

For our concrete application we chose $m=3$, so our last two
generations may be included in the first warmup step. Figure 15 shows
the typical scheme.

\midinsert
\rput(50,-27)
{\cnodeput(0,20){a}{1}
\psline{->}(0,22.5)(0,27)
\cnodeput(0,10){b}{2}
\ncline{-}{a}{b}
\cnodeput(-10,0){c}{3}
\ncline{-}{b}{c}
\cnodeput(10,0){d}{4}
\ncline{-}{b}{d}}
\psline{->}(55,-17)(75,-17)
\rput(90,-27)
{\cnodeput(0,20){a}{1}
\psline{->}(0,22.5)(0,27)
\cnodeput(0,5){b}{Block}
\ncline{-}{a}{b}}
\ps{30}
\noindent {\mayp Figure 15.} {\peqfont The last two generations are
encapsulated into blocks.\par}
\pss
\endinsert

As the block goes ``creeping'' along a corridor, it ``eats'' the sites
as DMRG does in 1D. When two blocks which are going upwards arrive at
the same time to a branching point, they are forced to get fused and
keep ascending together. Figure 16 shows these two processes.

\midinsert
\rput(50,-25){
\psline{->}(0,20.5)(0,25)
\cnodeput(0,18){a}{1}
\cnodeput(-10,5){b}{Block 1}
\cnodeput(10,5){c}{Block 2}
\ncline{-}{a}{b}
\ncline{-}{a}{c}}
\rput(100,-25){
\psline{->}(0,22.5)(0,27)
\cnodeput(0,20){a}{1}
\cnodeput(0,13){b}{2}
\cnodeput(0,1){c}{Block}
\ncline{-}{a}{b}
\ncline{-}{b}{c}}
\ps{32}
\noindent{\mayp Figure 16.} {\peqfont On the left side, a branching
site is going to be fused with its two children blocks. On the right
side, a ``bridge'' site (2) is going to be swallowed by the ascending
block.\par}
\pss
\endinsert

Once the central site is the only one which does not belong to any
block, the warmup may be said to be finished. The sweeping proceeds as
usual. Each block is represented by $m=3$ states, so the dimension of
the hamiltonian matrix shall be $3m+1=10$ at branching sites, and
$2m+1=7$ at bridge sites.

\subsection{Numerical and Experimental Results.}

The target of our calculations is the optical absorption edge for
extended dendrimers, as it was formerly stated. The procedure may
also be used to study compact dendrimers, with the results which shall
follow.

The following table compares the experimental results with two
theoretical predictions of different authors.
\pss
\centerline{\hfilneg
\sbtabla(4)
\li Generations && Experimental${}^1$ & Theor. Kopelman ${}^2$ & 
Theor. Harigaya ${}^3$ \cr
\lt \hfill 1 && 31500 $\pm$ 100 &       & 31511 \cr
\lt \hfill 2 && 31300 $\pm$ 100 & 31250 & 28933 \cr
\lt \hfill 3 && 27700 $\pm$ 200 & 27244 & 27750 \cr
\lt \hfill 4 && 25000 $\pm$ 300 & 25800 & 26000 \cr
\lt \hfill 5 && 25000 $\pm$ 600 & 25300 & 25022 \cr
\setabla}
\ps2
\noindent{\mayp Table 1.} {\peqfont Optical absorption edges measured
in $\linp{cm}^{-1}$. (1) Experimental measurements of Kopelman et
al. \ct[KOP-97]. (2) Theoretical predictions, also by Kopelman et
al. \ct[KOP-97]. (3) Theoretical results of Harigaya \ct[HAR-99].\par}
\pss

The predictions of Kopelman are based on a thoroughly different
theory, called {\it linear cluster theory\/}. This theory is based
on the analysis of ``corridors'' and, according to it, the optical
absorption edge is dominated by the longest length of a linear chain
of monomers. The energies of the excitons travelling through them are
calculated through Hoshen-Kopelman's theory \ct[HK-77].

Harigaya's model, which has already been exposed, is a model for a
single particle. One of the goals of our study is to know whether it
would be appropriate to undertake a more sophisticated study based on
a many-body theory, such as the Pariser--Parr--Pople hamiltonian. The
DMRG does {\it not\/} provide a new theory about excitons in
dendrimers. It just makes it easier to obtain the numerical
predictions of a given theory, if this theory fulfills some
pre-requisites (as Harigaya's model does).
\pss

\midinsert
\hbox{
\vtop{\kern0pt\hsize=7.5cm
\setbox\figurarot=\hbox{\epsfysize 8cm\epsfbox{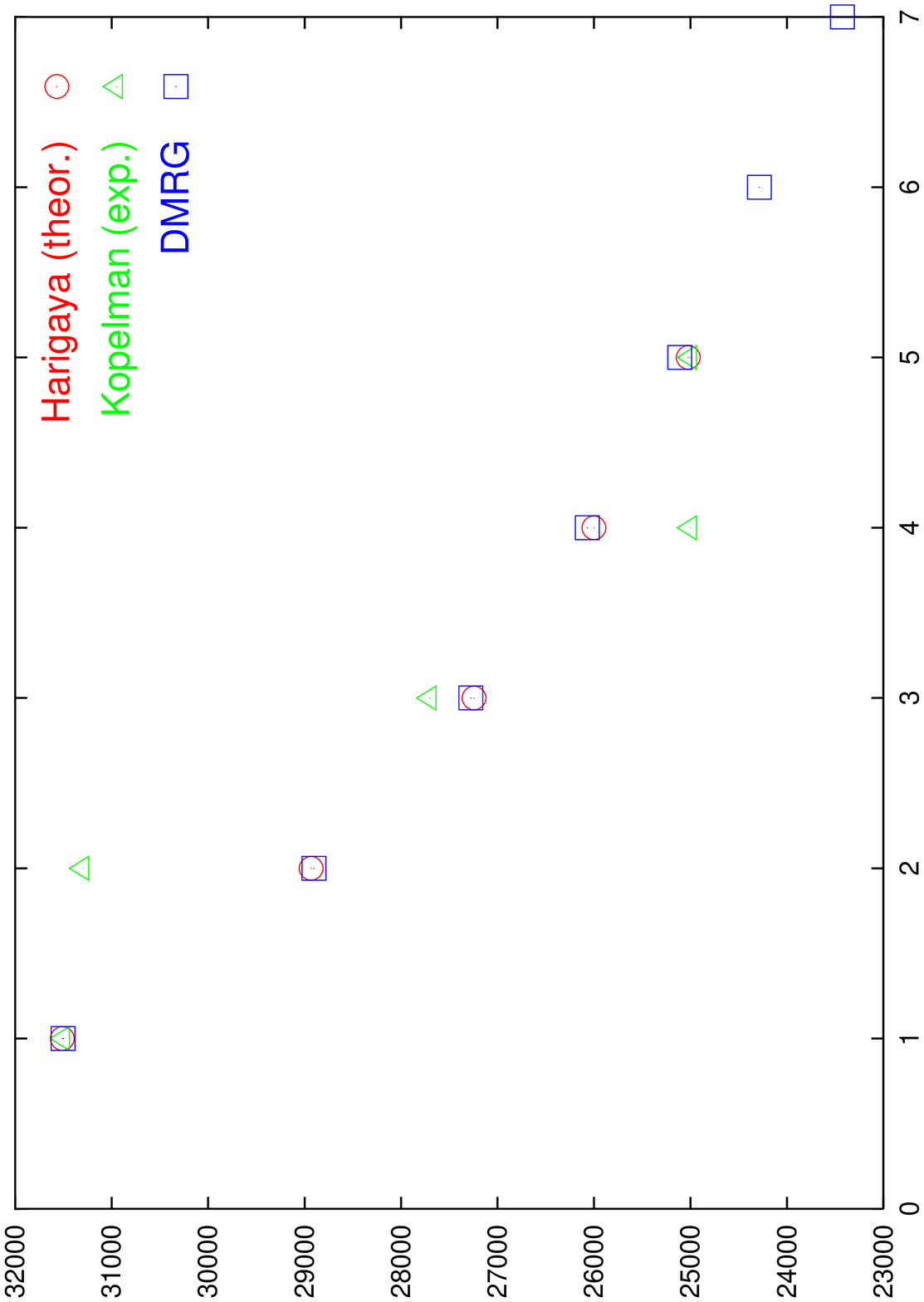}}
\rotr\figurarot
\ps2
\nind{\mayp Figure 17. }{\peqfont Optical absorption edge
experimentally obtained by Kopelman (as a function of the generation),
along with the theoretical results of Harigaya and those obtained with
DMRG.\par}}
\hskip 5mm
\vtop{\kern0pt\hsize=7.5cm 
\setbox\figurarot=\hbox{\epsfysize 8cm\epsfbox{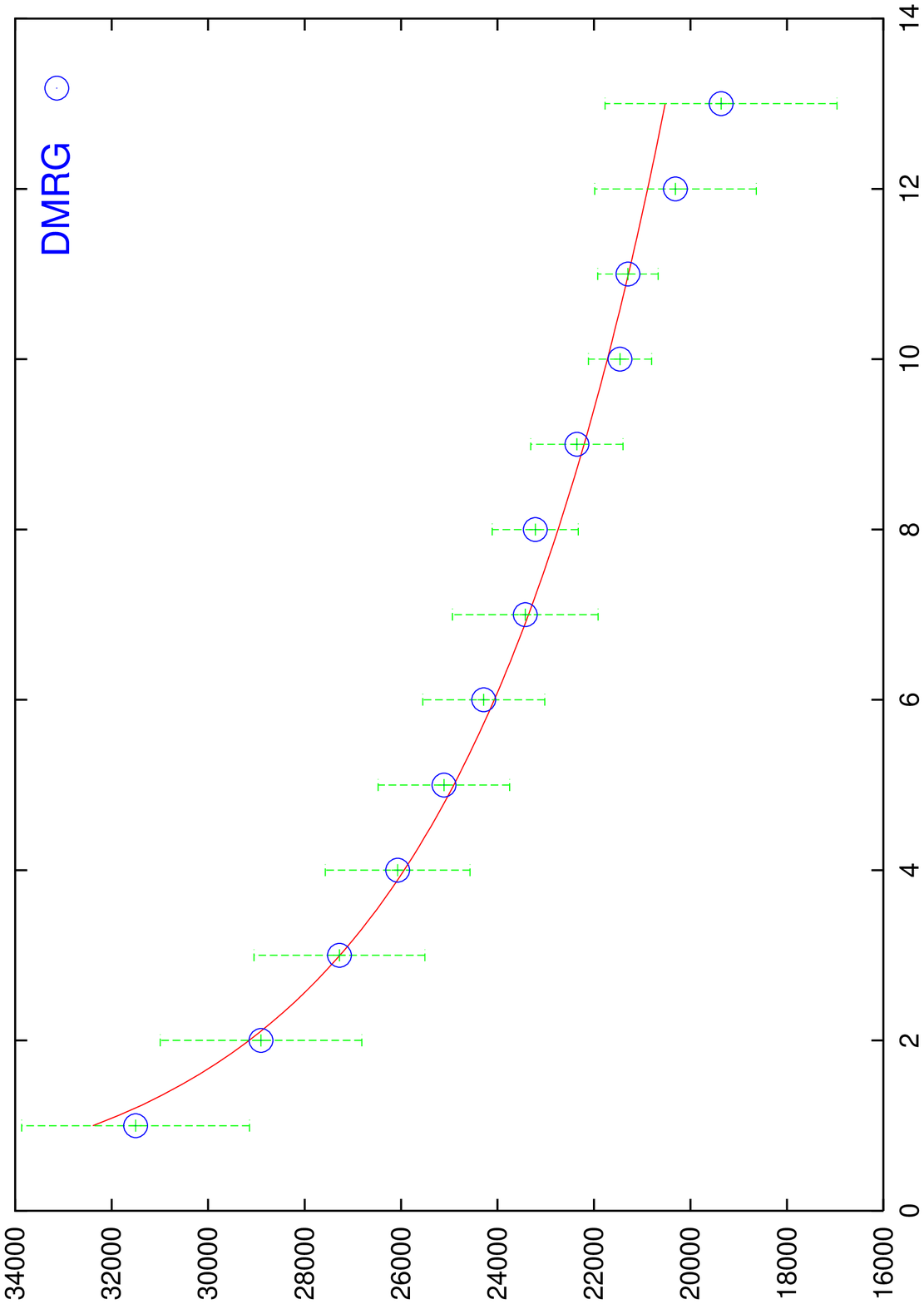}}
\rotr\figurarot
\ps2
\nind{\mayp Figure 18. }{\peqfont Optical absorption edge with its
errorbars for Harigaya's model up to 13 generations obtained with
DMRG. The fit to a power law is shown.\par}}}
\pss
\endinsert

Figure 16 shows the optical absorption edge against the generation
number according to experiments, Harigaya's calculations (exact
diagonalization) and the results obtained with DMRG. As it was to be
expected, the last two coincide. The DMRG data were obtained with
$10.000$ samples.

Our calculations may go beyond, and figure 18 shows the optical
absorption edge up to 13 generations along with their error bars. In
the last two cases only $1.000$ samples were drawn. The data fit
reasonably well to a power law

$$E_c(n)\approx 32000-C_e n^{\alpha_e}$$

\nind with $C_e\approx (2020\pm 200)\lin{cm}^{-1}$ and $\alpha_e\approx
0.71\pm 0.04$. Therefore, it may be asserted that the system does not
have a ``gap'' in the thermodynamic limit.

Since the DMRG calculations were carried out with $m=3$, the next two
states of the exciton were also calculated. Figure 19 shows
them. Figure 20 makes us watch the effect of the addition of wedges,
which is a usual technical process, from $w=2$ to $w=6$.

\midinsert
\hbox{
\vtop{\hsize=7.5cm
\setbox\figurarot=\hbox{\epsfysize 8cm\epsfbox{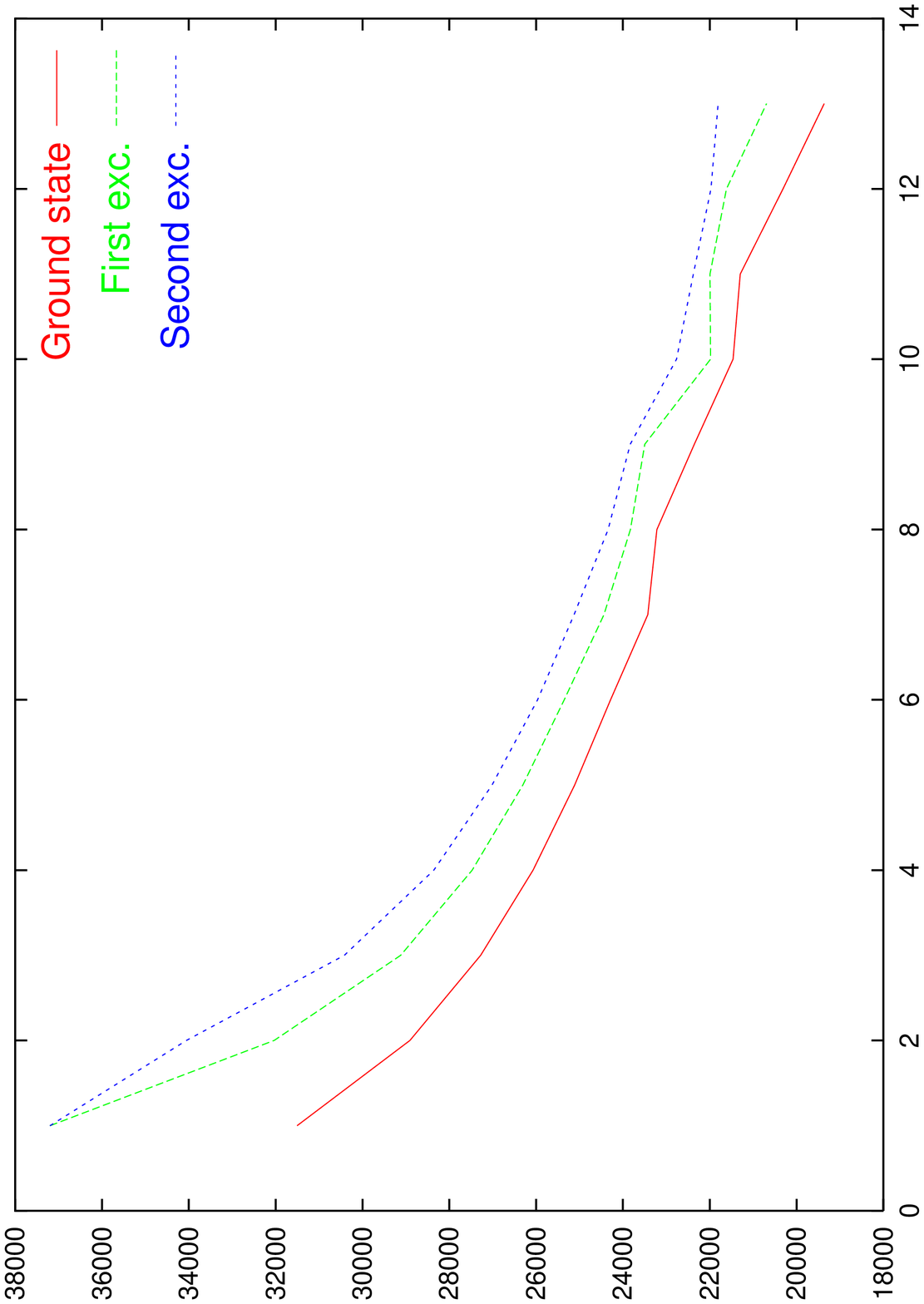}}
\rotr\figurarot
\ps2
\nind {\mayp Figure 19. }{\peqfont First three eigenvalues of the
hamiltonian of Harigaya's model as a function of the generation.\par}}
\hskip 5mm
\vtop{\hsize=7.5cm
\setbox\figurarot=\hbox{\epsfysize 8cm\epsfbox{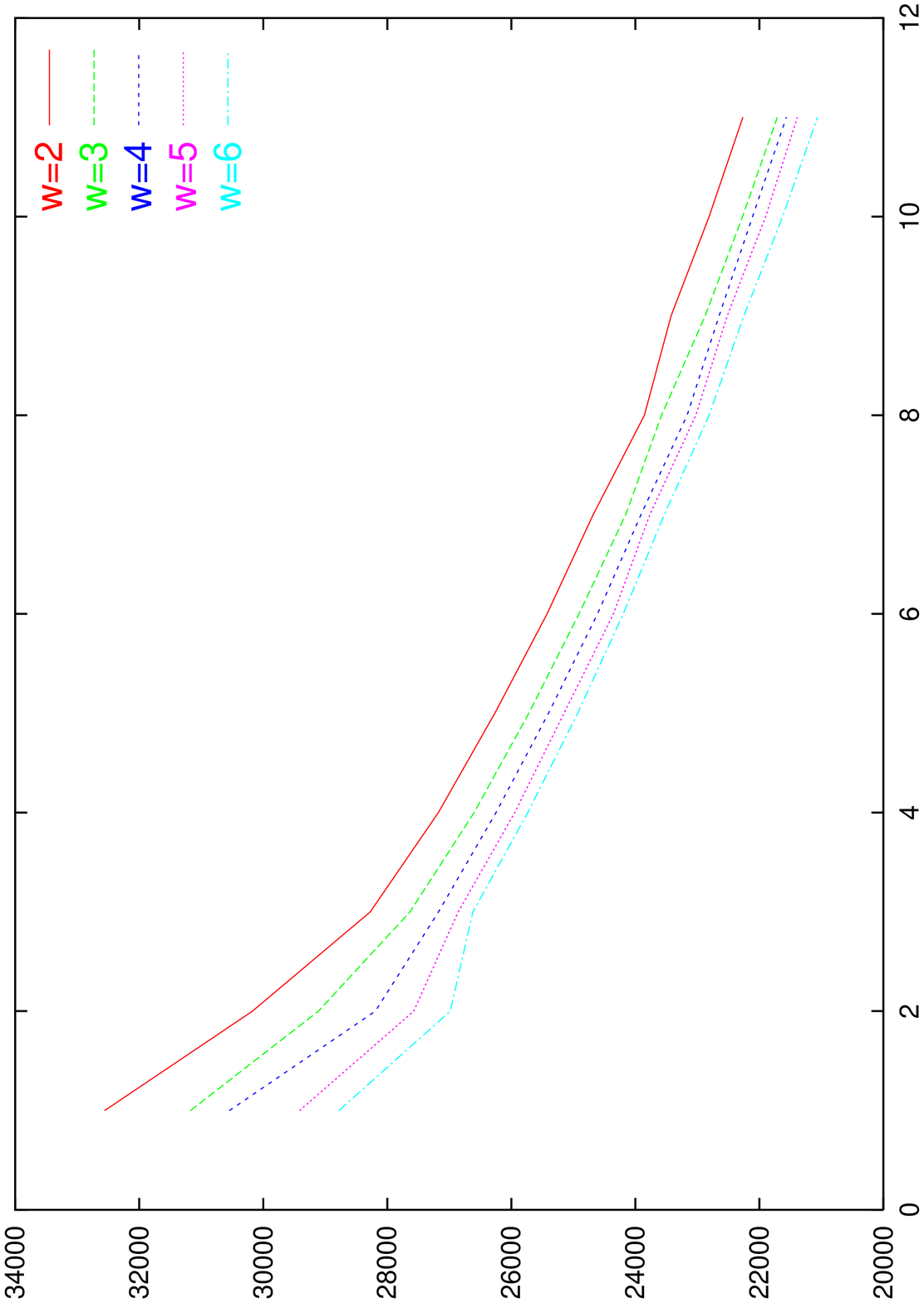}}
\rotr\figurarot
\ps2
\nind {\mayp Figure 20. }{\peqfont Optical absorption edge for
extended dendrimers with different numbers of wedges ($w=2$ up to
$w=6$) according to Harigaya's model with DMRG.\par}}}
\pss
\endinsert

When we consider compact dendrimers, the results diverge greatly from
the experiments. According to the data of Kopelman et al.
\ct[KOP-97] for the compact dendrimers series, the optical absorption
edge hardly decreases when the number of generations
increases. Concretely, $E_c(n)\approx 31500\pm 200$ for $n\in [1\ldots
5]$. I.e.: it is probable that the system has a finite gap in the
limit $n\to\infty$.

Notwithstanding, when analyzing Harigaya's model, the fit to a power
law of the results is also rather accurate:

$$E_c(n)\approx E_0 - C_c n^{\alpha_c}$$

\nind and it is obtained that $C_c\approx (3190\pm 120)\lin{cm}^{-1}$ and
$\alpha_c\approx (0.54\pm 0.02)$.
\pss

As a matter of fact, Harigaya's model of a single exciton is proposed
only for extended dendrimers, although no reason is presented for
which it might not be applied to compact ones. For the latter
case, linear cluster theory predicts successfully the constancy of the
data. The failure of the one-exciton theory leaves a single feasible
possibility: the higher links concentration makes the effects of
interaction between excitons much bigger than in the case of extended
dendrimers, rendering the approximation incorrect.


\section{3.6. Bibliography.}

\cx[AdS-81] {\may H. Abelson, A.A. diSessa,} {\sl Turtle geometry,}
The MIT Press (1981).

\cx[BAL-90] {\may L.E. Ballentine,} {\sl Quantum mechanics,} Prentice
Hall (1990).

\cx[BOL-98] {\may B. Bollob\'as}, {\sl Modern graph theory}, Springer (1998).

\cx[BWV-78] {\may E. Buhleier, W. Wehner, F. V\"ogtle}, Synthesis {\bf
1978}, 155 (1978).

\cx[CDL-73] {\may C. Cohen-Tannoudji, B. Diu, F. Lalo\"e}, {\sl M\'ecanique
quantique,} Hermann (1973).

\cx[FEY-72] {\may R.P. Feynman}, {\sl Statistical mechanics: 
a set of lectures}, Addison-Wesley (1972).

\cx[GEN-79] {\may P.G. de Gennes}, {\sl Scaling concepts in polymer
physics}, Cornell U.P. (1979).

\cx[GMSV-95] {\may J. Gonz\'alez, M.A. Mart\'{\i}n-Delgado, G. Sierra,
A.H. Vozmediano}, {\sl Quantum electron liquids and High-$T_c$
superconductivity}, Springer (1995).

\cx[GP-89] {\may A. Galindo, P. Pascual,} {\sl Mec\'anica cu\'antica}, Eudema
(1989).

\cx[HAL-99] {\may K. Hallberg,} {\sl Density matrix renormalization,}
available in {\tt cond-mat/9910082} (1999).

\cx[HAR-98] {\may K. Harigaya}, {\sl Coupled Exciton Model with
Off-Diagonal Disorder for Optical Excitations in Extended Dendrimers},
available in {\tt cond-mat/9810049} (revised en 1999).

\cx[HAR-99] {\may K. Harigaya}, {\sl Optical Excitations in
Diphenylacetylene Based Dendrimers Studied by a Coupled Exciton Model
with Off-Diagonal Disorder}, Int. J. Mod. Phys. B {\bf 13}, 2531-44
(1999).

\cx[HF-90] {\may C.J. Hawker, J.M. Frechet}, J. Am. Chem. Soc. {\bf 112},
7638-7647 (1990).

\cx[JBBM-94] {\may J.F.G.A. Jansen, E.M.M. de Brabander-van der
Berg, E.W. Meijer}, {\sl Encapsulation of Guest Molecules into a
Dendritic Box}, Science {\bf 226}, 1226-29 (1994).

\cx[KOP-97] {\may R. Kopelman, M. Shortreed, Z.Y. Shi, W. Tan, Z. Xu,
J.S. Moore}, {\sl Spectroscopic Evidence for Excitonic Localization in
Fractal Antenna Supermolecules}, Phys. Rev. Lett. {\bf 78}, 7, 1239-42
(1997).

\cx[LL-71] {\may L.D. Landau, E.M. Lifchitz}, {\sl M\'ecanique quantique},
\'Editions Mir (1989). First ed. in russian by Nauka (1971).

\cx[MRS-00] {\may M.A.~Mart\'{\i}n-Delgado,
J.~Rodr\'{\i}guez-Laguna, G.~Sierra}, {\sl A density matrix
renormalization group study of excitons in dendrimers}, available at
{\tt cond-mat/0012382}, and Phys.~Rev.~B {\bf 65}, 155116 (2002).

\cx[MSN-99] {\may M.A. Mart\'{\i}n-Delgado, R.M. Noack, G. Sierra,},
{\sl The density matrix renormalization group applied to
single-particle quantum mechanics}, J.~Phys.~A: Math. Gen. {\bf 32}
6079-6090 (1999).

\cx[MSPJ-99] {\may M.A. Mart\'{\i}n-Delgado, G. Sierra, S. Pleutin, 
E. Jeckelmann}, {\sl Matrix product approach to conjugated polymers},
Phys. Rev. B {\bf 61}, 1841-46 (2000) and available {\tt
cond-mat/9908066}.

\cx[NWH-92] {\may R.M. Noack, S.R. White,} {\sl Real-space quantum
renormalization groups}, Phys. Rev. Lett. {\bf 68}, 3487 (1992).

\cx[NWH-93] {\may R.M. Noack, S.R. White,} {\sl Real-space quantum
renormalization group and Anderson localization}, Phys. Rev. B {\bf
47}, 9243 (1993).

\cx[OMN-94] {\may R. Omn\`es,} {\sl The interpretation of quantum
mechanics,} Princeton U.P. (1994).

\cx[RT-99] {\may S. Ramasesha, K. Tandon}, {\sl Symmetrized DMRG
method for conjugated polymers}, in {\sl Den\-si\-ty-matrix
renormalization}, ed. by I. Peschel et al. (Notes from the Dresden
{\it workshop\/} in 1998), Springer (1999).

\cx[TNG-90] {\may D.A. Tomalia, A.M. Naylor, W.A. Goddard III},
Angew. Chem. (Int. Ed. Engl.) {\bf 29}, 138-175 (1990).

\cx[V\"OG-98] {\may F. V\"ogtle} (ed.), {\sl Dendrimers}, Springer (1998).

\cx[WHI-92] {\may S.R. White}, {\sl Density matrix formulation for
quantum renormalization groups}, Phys. Rev. Lett. {\bf 69}, 2863 (1992).

\cx[WHI-93] {\may S.R. White}, {\sl Density-matrix algorithms for
quantum renormalization groups}, Phys. Rev. B {\bf 48}, 10345 (1993).

\cx[WHI-98] {\may S.R. White}, {\sl Strongly correlated electron
systems and the density matrix renormalization group}, Phys. Rep. {\bf
301}, 187-204 (1998).

\cx[WHI-99] {\may S.R. White}, {\sl How it all began}, en
{\sl Density-matrix renormalization} ed. by I. Peschel et al. (Notes
of the Dresden {\it workshop\/} in 1998), Springer (1999).

\pasacap
\chapter{4. Multidimensional Formulations of the DMRG.}

\subsection{Synopsis.}

4.1. Long range formulation of DMRG.

4.2. Punctures renormalization group.

4.3. 1D implementation and numerical results.

4.4. PRG analysis of 2D and 3D lattices.

4.5. Warmups for the PRG.

4.6. Blocks algebra.

4.7. Towards a PRG for many-body theory.

4.8. Application of PRG to a model of excitons with disorder.

4.9. Bibliography.


\section{4.1. Long Range Formulation of DMRG.}

In the previous chapter some ``implicit'' RSRG algorithms were
developed for quantum mechanics (QM). In these algorithms the
wave-function is not explicitly stored and is not used in the
calculations, so the RG step requires only $O(m^k)$ operations, where
$m$ is the number of states we wish to obtain and $k$ is a certain
power.

But the DMRG algorithm for trees --the most developed of the
algorithms developed for QM-- {\it may not\/} be generalized to
various dimensions. This does not imply that RSRG may not be applied
successfully to multidimensional problems, but that methods must be
``explicit'': the full wave-function must be stored and the RG step
shall take, at the best possible case, $O(N)$ steps, with $N$ the
total number of sites to be considered. Even so, RSRG methods may
result quite profitable both theoritically and numerically for these
problems.

\subsection{In Search for an Implicit Multidimensional Algorithm.}

The question which we shall succinctly review in this section is ``Why
is it necessary to store the full wave-function when the problem is
2D?'' The answer is, rather briefly, {\it because adding is easy, but
substracting is not\/}. The present paragraph elaborates this
argument.

DMRG is based upon the {\sl addition} process or {\it modus ponens\/}:
it was said in the former chapter that at each step one of the blocks
grows and the other (or others) decrease, but this assertion should be
carefully restated. Indeed, there is a block which increases size; but
the information for the shrunken block is {\it taken from the previous
cycle\/}. The reason is that we do not know how to {\sl remove} sites:
we do not have a substraction or {\it modus tollens\/}\footn{The terms
{\it modus ponens\/} and {\it modus tollens\/}, inspired by the
terminology of formal logic, mean in latin {\sl adding} and {\sl
removing} respectively.} available.

How does this fact influence our problem? DMRG is applicable to trees
because, given any site, it is possible to split the system into
blocks such that:
\ps2

\bul {\sl The blocks are disconnected among themselves}. Only in this
case may the {\it modus tollens\/} be simulated by taking the
information from a previous step. The reason is that, in this case,
the modification of the states of a block {\sl does not alter} the
matrix elements of another one, which conserve their validity from a
sweeping cycle to the following one.

\bul {\sl Each neighbour of the site lies in a different block}. This
must be true so as all sites may become ``free sites''. At any RG
step, all neighbouring blocks of the site {\it but\/} the one which
contains the new free site shall get fused. Thus, this block should
not share sites with the others.
\pss

Once we have accepted the fact that the {\it modus tollens\/} or
substraction process may not be simulated in non--tree graphs, the
following question remains: ``Why can not an authentic implicit {\it
modus tollens\/} be built?''

The answer to this question is the following: the removal of a site
from a block forces the modification of matrix elements of some
operators. Such a modification requires the knowledge of the real
value of the wave-function at the site. But since the process shall be
iterated, it shall be necessary to know, along with it, all the other
values.
\pss

The explicit algorithm is, therefore, unavoidable. It is then possible
to write down a DMRG process for {\sl long range} quantum mechanical
problems, i.e.: where there is no underlying graph structure
whatsoever. Afterwards we shall see that, if it exists, such a
structure may be used to reduce drastically the computation time (even
when it is not a tree).

The first algorithm of this kind was developed in the work of
M.A. Mart\'{\i}n-Delgado and G. Sierra \ct[MDS-99].

\subsection{An Asymptotically Free Model in Quantum Mechanics.}

One of the physical systems which results more difficult to tackle is
probably Quantum Chromodynamics (QCD), which is the standard theory
for the behaviour of hadronic matter (quarks and gluons). This theory
is {\sl asymptotically free}, i.e.: at short distances the interaction
is so weak that perturbation theory may be employed. The problem
appears at the energy scales at which bound states (hadrons) exist
\ct[DGMP-97]. The analysis is rather difficult since all scales are
{\it strongly coupled\/}.

The simplest analogue in quantum mechanics was proposed by K.G.~Wilson
and St.D.~G\l a\-zek \ct[GW-97] and it is the momentum space analysis
of the behaviour of a bidimensional particle bound by a delta
potential.

The regularization employed for the problem required a certain {\sl IR
cutoff} and {\sl UV cutoff}, given by two integer numbers $M$ and
$N$. Just like the shell models of turbulence, the states have momenta
which follow a certain geometric progression. Given a momentum scaling
factor $b$ (chosen to be $\sqrt{2}$), the self-energy of the $n$-th
state is $b^{2n}$ and the coupling between two states $n$ and $m$ is
$-g\sqrt{E_nE_m}$. The discretized hamiltonian therefore takes the
form

$$H_{n,m}=\delta_{nm}b^{2n}-gb^{n+m} \qquad M\leq n,m\leq N
\eq{\potdelta}$$

\nind The full hamiltonian has discrete scale invariance $H_{n+1,m+1}=
b^2 H_{n,m}$, which is broken by the IR and UV cutoffs. We say that it
represents an asymptotically free model since the coupling between
neighbour momenta are arbitrarily small in one of the extremes of the
spectrum (IR or UV depending on the value of $b$).
\pss

The analysis of G\l azek and Wilson is based on the method they
developed, called the {\sl Similarity Renormalization Group} (SRG)
\ct[GW-93] \ct[GW-94], which works by applying similarity
transformations to the hamiltonian until it becomes band diagonal. It
``decouples'' in momentum space the scales which are far one from the
other in a perturbative way, by integrating {\sl Wegner's equation}

$${dH(s)\over ds}=\Bigl[\bigl[H_d(s),H(s)\bigr],H(s)\Bigr]$$

\nind (where $H_d(s)$ is the diagonal part of $H(s)$) with the initial
condition $H(0)=H$ and then taking the limit $s\to\infty$. The
parameter $s$ may be identified with the inverse square of the
``width'' in energies of the matrix $H(s)$ and, thus, in the
aforementioned limit, the hamiltonian is diagonal.
\pss

The DMRG analysis of this problem, which is non-perturbative, makes up
the first incursion of this technique into the field of asymptotically
free theories.

\subsection{DMRG for Long Range Problems.}

We shall now describe the DMRG technique which was employed to deal
with the given problem. It is a technical modification of the method
introduced in \ct[MDS-99], which fully respects the general idea.

First and foremost, the system is divided into ``left'' and ``right'',
which in this case means {\sl low momentum states} and {\sl high
momentum states}. In any case, it is important to observe that there
is a strong interaction between states which belong to different
blocks, in contrast to what we saw in the previous chapter.
\pss

Two blocks (left and right, $L$ and $R$) plus two sites between them
shall be considered. Block $L$ is formed by the sites $[1,\ldots,
p-1]$. Let $\{\ket|\psi^{Li}>\}$ and $\{\ket|\psi^{Ri}>\}$ with
$i\in[1,\ldots, m]$ be the series of $m$ orthonormal states for the
left and right sides. At each step the full wave-functions should be
available:

$$\{\<\delta_p\right|\left.\psi^{Li}\>\} \qquad
\{\<\delta_q\right|\left.\psi^{Ri}\>\}$$ 

\nind for all $p\in [1,\ldots, p-1]$ and $q\in[p+2,\ldots, N]$. I.e.:
in total $m(N-2)$ numbers.
\pss

The {\it Ansatz\/} is exactly the same as in the 1D-DMRG case:

$$\ket|\phi>=\sum_{i=1}^m c^i \ket|\psi^{Li}> + c^{m+1} \ket|\delta_p>
+ c^{m+2} \ket|\delta_{p+1}> + \sum_{i=1}^m c^{m+2+i}
\ket|\psi^{Ri}>$$

With this {\it Ansatz\/} we may build a superblock hamiltonian which
is different from the one we previously used:

\def\tablerule{&height 5pt&&&&&\cr\noalign{\hrule}&height 5pt&&&&&\cr}
$$H^\Sb=\left(
\vcenter{\offinterlineskip\halign{&\hfil$\;#\;$\hfil&\vrule#\cr
\strut &&&&&&\cr
\hbox to 1cm{\hfil $L$ \hfil} && T^{Ll\dagger} && 
T^{Lr\dagger} && H^{LR}\cr
\tablerule
T^{Ll} && H_{cl,cl} && H_{cl,cr} && T^{Rl}\cr
\tablerule
T^{Lr} && H_{cr,cl} && H_{cr,cr} && T^{Rr} \cr 
\tablerule
H^{LR\dagger} && T^{Rl\dagger} && T^{Rr\dagger} && \hbox to
1cm{\hfil $R$\hfil} \cr
\strut &&&&&&\cr
}}\right) \eq{\hamSBlong}$$

In this hamiltonian the classical elements of the DMRG are present,
which are:
\pss

\bul The intra-block elements $L_{ij}=\elem<\psi^{Li}|H|\psi^{Lj}>$,
$R_{ij}=\elem<\psi^{Ri}|H|\psi^{Rj}>$.
\ps2
\bul The hooks of the blocks to the free sites:
$T^{Ll}_i=\elem<\psi^{Li}|H|\delta_p>$,
$T^{Rr}_i=\elem<\psi^{Ri}|H|\delta_{p+1}>$.
\ps2
\bul The terms corresponding to the free sites: $H_{cl,cl}$, etc.
\pss

But there appear three new elements of the superblock
matrix which are non--zero {\it a priori\/}:
\pss

\bul The link between the left part and the right free site
$T^{Lr}_i=\elem<\psi^{Li}|H|\delta_{p+1}>$ and viceversa
$T^{Rl}_i=\elem<\psi^{Ri}|H|\delta_p>$.
\ps2

\bul The link between the left and right parts:
$H^{LR}_{ij}=\elem<\psi^{Li}|H|\psi^{Rj}>$.
\pss

All these matrix elements of the superblock hamiltonian may be fully
calculated if the states are available, but it is convenient to hold
in memory the superblock hamiltonian of the former step and modify it
consequently.

The superblock hamiltonian is diagonalized and its $m$ lowest energy
states are retained. We now give two options for the last step of the
procedure, which is the states updating. The first is more simple but
much less efficient. In both cases, we shall assume that the left
block is growing.
\pss

$\bullet$ {\it First procedure\/} Let us build the complete states
$\ket|\phi^i>$ by substituting the coefficients of the eigenstates of
$H^\Sb$ into the {\it Ansatz\/}. We write down these $m$ states
twice. On the first copy we make all components from the $p+1$ onwards
vanish, while on the second one we put zeroes on all components up to
the number $p+2$.

These states correspond to the new $L$ and $R$ states respectively,
but they require re--ortho\-nor\-mal\-ization through a Gram-Schmidt
process. Once it is finished, the matrix elements of the new
superblock hamiltonian $\hat H^\Sb$ are computed.

The full process requires $O(N^2)$ operations.
\pss

$\bullet$ {\it Second procedure.\/} A better notation is required to
explain this method, which is much more complex (and efficient). Let
capital indices $I$, $J,\ldots$ run over the left half of the
components of the eigenstates of $H^\Sb$, in the range
$[1,\ldots,m+1]$. Let us define the original states of the left part
(with ``Block $+\ \bullet$'' structure) as

$$\ket|\phi^{LI}> = \cases{ \ket|\psi^{LI}> & if $I\leq m$ \cr
	                   \ket|\delta_p>  & if $I=m+1$\cr}$$

\nind I.e.: we include the state $\ket|\delta_p>$ into this
set. Let us consider now the basis change matrix obtained through the
eigenstates of $H^\Sb$: $B^i_J$ denotes the $J$-th component of the
$i$-th eigenstate (with $i\in [1,\ldots,m]$ and $J\in [1,\ldots,
m+1]$). By not allowing $J$ to run over its full natural range (which
would be $[1,\ldots,2m+2]$), we are losing the {\sl orthogonality} of
the transformation. Let us see how we may recover it.
\pss

Let $C_{ij}$ be the dot products matrix among the states of $B$,
defined by

$$C_{ij}=\sum_{K=1}^{m+1} B^i_K B^j_K$$

Appendix D shows how to obtain the Gram-Schmidt basis change rather
fast by using this information. Let $G$ be the matrix implementing
that basis change and let us define

$$B^{Li}_J = G^i_j B^j_J$$

\nind to be the orthogonal basis change matrix from the old states
$\ket|\phi^{LI}>$ to the new ones, given by

$$\ket|\hat\psi^{Li}>=B^{Li}_J\ket|\phi^{LJ}>=\sum_{j=1}^m B^{Li}_j
\ket|\psi^{Lj}> + B^{Li}_{m+1} \ket|\delta_p>$$

\ps2
It is possible to write an equivalent basis change for the right side
states (with structure ``\hbox{Block $-\ \bullet$}'', i.e.: block {\it
minus\/} site). But there is an added difficulty: the state to remove
($\ket|\delta_{p+2}>$) is {\it not\/} one of the old states of the
basis of $H^\Sb$. Let us see how to approach the new problem.

Let $\psi^{Ri}_{p+2}$ be the component of the $i$-th state of the
right block on site $p+2$. Such a block undergoes the transformation

$$\ket|\psi^{Ri}> \to \ket|\psi^{Ri}> - \psi^{Ri}_{p+2}
\ket|\delta_{p+2}>$$

\nind which we shall call substraction algorithm (or {\it modus
tollens\/}). These states are not orthonormal. So as they can be, the
fast Gram-Schmidt procedure is applied (see appendix D) on its dot
products matrix:

$$\delta_{ij}-\psi^{Ri}_{p+2}\psi^{Rj}_{p+2}$$

\nind and let us call $Q^i_j$ the basis change matrix which ensures
orthonormality. Defining the right side states as

$$\ket|\phi^{RI}>=\cases{\ket|\psi^{RI}> & if $I\leq m$\cr
                      \ket|\delta_{p+2}> & if $I=m+1$\cr}$$

\nind and the transformation matrix 

$$\hat B^j_K \equiv \delta^j_K - \psi^{Rj}_{p+2}\delta_{K,m+1}$$

\nind the new right side states may be expressed as

$$\ket|\hat\psi^{Ri}>=Q^i_j \hat B^j_K \ket|\phi^{RK}> \equiv B^{Ri}_K
\ket|\phi^{RK}>$$

Therefore, we have formally the same type of transformation for both
blocks, even though we know that there are rather different operations
involved. The following step is the computation of the new matrix
elements between those renormalized states. Let us see how the
calculation may be done for each box the matrix $H^\Sb$.
\pss

$\bullet$ {\it Intra--block left part\/}. The computation is
straightforward.

$$\hat L_{ij}=\elem<\hat\psi^{Li}|H|\hat\psi^{Lj}>=\sum_{I,J=1}^{m+1}
B^{Li}_I B^{Lj}_J \elem<\phi^{LI}|H|\phi^{LJ}>$$

\nind But since all the states appearing in the sum are states of the 
old $H^\Sb$, we have only to transform the upper-left corner of the
old matrix:

$$\hat L_{ij}=\sum_{I,J=1}^{m+1} B^{Li}_I B^{Lj}_J H^\Sb_{IJ}$$

$\bullet$ {\it Left hook to the left free site\/}. The new free left
site is $p+1$, which beforehand was the right free site. Thus,

$$\hat T^{Ll}_i =\elem<\hat\psi^{Li}|H|\delta_{p+1}>=\sum_{I=1}^{m+1}
B^{Li}_I \elem<\phi^{LI}|H|\delta_{p+1}>$$

In the same fashion as before we obtain

$$\hat T^{Ll}_i = \sum_{I=1}^{m+1} B^{Li}_I H^\Sb_{I,m+2}$$

$\bullet$ {\it Link between the left block and the right free
site\/}. The new right free site did not exist before as an
independent entity, so there are no shortcuts for the computation:

$$\hat T^{Lr}_i = \elem<\hat\psi^{Li}|H|\delta_{p+2}>$$

The only possible ``trick'' to reduce the computational effort comes
up if the hamiltonian respects some kind of neighborhood structure
(even if it is not linear).
\ps2

$\bullet$ {\it Left--right inter-blocks part\/}. The calculation
provides some terms which may be simplified and other ones which may
not.

$$\displaylines{\hat H^{LR}_{ij}=\elem<\hat\psi^{Li}|H|\hat\psi^{Ri}>=
\sum_{I,J=1}^{m+1} B^{Li}_I B^{Rj}_J \elem<\phi^{LI}|H|\phi^{RJ}> = \cr
=\sum_{k,l=1}^m  B^{Li}_k B^{Lj}_l \elem<\psi^{Lk}|H|\psi^{Rl}>
+ \sum_{k=1}^m B^{Li}_k B^{Rj}_{m+1} \elem<\psi^{Lk}|H|\delta_{p+2}>+
\sum_{l=1}^m B^{Li}_{m+1} B^{Rj}_l \elem<\delta_p|H|\psi^{Rl}> +\cr
\hfill+ B^{Li}_{m+1}B^{Rj}_{m+1} \elem<\delta_p|H|\delta_{p+2}>}$$

Fortunately, only the second term requires a full computation (because
it makes interact the site $p+2$ with the left block). Thus, we have

$$\hat H^{LR}_{ij}=\sum_{k,l=1}^m B^{Li}_k B^{Rj}_l H^{LR}_{kl} +
\sum_{k=1}^m B^{Li}_{m+1} B^{Rj}_k T^{Rl}_k + B^{Li}_{m+1}B^{Rj}_{m+1}
H_{p,p+2} + \sum_{k=1}^m B^{Li}_k
B^{Rj}_{m+1}\elem<\psi^{Lk}|H|\delta_{p+2}>$$

$\bullet$ {\it Right intra--blocks part\/}. This part has also got
some term which must be calculated in full (of course, it involves
site $p+2$).

$$\displaylines{
\hat R_{ij}= \elem<\hat\psi^{Ri}|H|\hat\psi^{Rj}>=
\sum_{k,l=1}^m B^{Ri}_k B^{Rj}_l \elem<\psi^{Rk}|H|\psi^{Rl}> + \cr + 
\sum_{k=1}^m B^{Ri}_k B^{Rj}_{m+1} \elem<\psi^{Rk}|H|\delta_{p+2}> + 
(i\leftrightarrow j, \lin{hc}) + 
B^{Ri}_{m+1} B^{Rj}_{m+1} \elem<\delta_{p+2}|H|\delta_{p+2}> =\cr
\sum_{k,l=1}^m B^{Ri}_k B^{Rj}_l R_{kl} + 
\sum_{k=1}^m \left(B^{Ri}_k B^{Rj}_{m+1} + B^{Rj} B^{Ri}_{m+1} \right)
\lin{Re}\elem<\psi^{Rk}|H|\delta_{p+2}> + 
B^{Ri}_{m+1}B^{Rj}_{m+1} H_{p+2,p+2}}$$

$\bullet$ {\it Right link to the left free site\/}. This is
straightforward, since the site $\ket|\delta_{p+1}>$ belonged to the
old superblock.

$$\displaylines{
\hat T^{Rl}_i = \elem<\hat\psi^{Ri}|H|\delta_{p+1}> = 
\sum_{k=1}^m B^{Ri}_k \elem<\psi^{Rk}|H|\delta_{p+1}> +
B^{Ri}_{m+1}\elem<\delta_{p+2}|H|\delta_{p+1}> = \cr
\sum_{k=1}^m B^{Ri}_k T^{Rr}_k + B^{Ri}_{m+1} H_{p+2,p+1}}$$

$\bullet$ {\it Right hook to the right free site\/}. This new free
site is, of course, $\ket|\delta_{p+2}>$, so we must compute fully some
matrix elements:

$$\hat T^{Rr}_i = \elem<\hat\psi^{Ri}|H|\delta_{p+2}> =
\sum_{k=1}^m B^{Ri}_k \elem<\psi^{Rk}|H|\delta_{p+2}>+
B^{Ri}_{m+1} H_{p+2,p+2}$$
\pss

We may observe that the computations we have named as full only
require the knowledge of $2m$ numbers:

$$\elem<\psi^{Li}|H|\delta_{p+2}> \qquad
\elem<\psi^{Ri}|H|\delta_{p+2}>$$

Thus, the calculation of the matrix elements (and the updating of the
wave-functions) is performed in $O(N)$ steps, while the {\it naive\/}
version was carried out in $O(N^2)$ operations\footn{Notice that in
the short range 1D-DMRG the first of these matrix elements is null,
while the second one would have been taken from a previous cycle,
leaving the RG step with $O(m^2)$ operations.}. It is the computation
of the second matrix element $\elem<\psi^{Ri}|H|\delta_{p+2}>$ which
forces the storage of the full wave-functions.

\subsection{Application to the Asymptotically Free Model.}

The implementation of the original long range DMRG to the 2D delta
potential in momentum space (considered to be a quantum-mechanical
model of an asymptotically free system), which not only lacks an
underlying graph structure, but also combines matrix elements of very
different orders of magnitude, had a rather good precision.

As a test, system [\potdelta] was considered with $n=38$
``shells''\footn{The reader may ask herself whether $n=38$ is a
respectable number. The answer is that, in these models where the
shell momenta follow a geometric progression, one runs the risk of
overloading the machine precision. $38$ jumps of a factor $b=\sqrt{2}$
is already a very interesting computation.} with a scale factor
$b=\sqrt{2}$, IR cutoff $M=-21$ and UV cutoff $N=16$. The coupling
constant is fixed to a precise value\footn{$g=0.0606060003210886$} so
as to obtain a ground state with energy $-1$. The obtained
wave-function appears in figure 1.

\figrot{5cm}{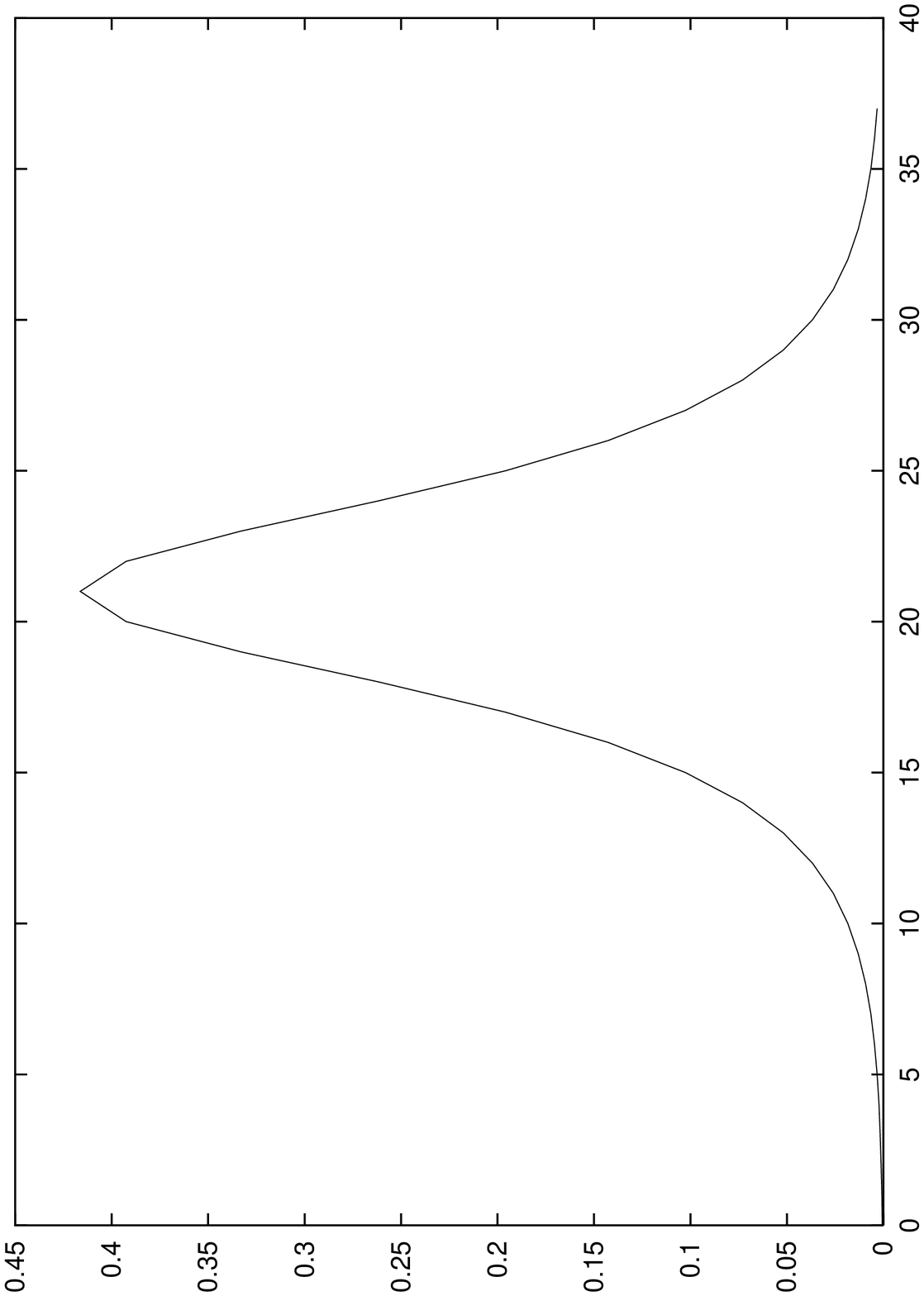}{Figure 1.}{Ground state of the asymptotically
free model under study. The abscissa represents the momentum
``shell''.}

As a result it yields, in two sweeping cycles, a precision of $15$
orders of magnitude for the ground state and, if $m=4$, $6$ orders of
magnitude for the three first excited states. For more details,
calculated with the first version of the long range DMRG (which gives
exactly the same results albeit more slowly), see \ct[MDS-99].

\subsection{Application to Short Range 2D Hamiltonians.}

The case of a bidimensional problem with short range terms is not
analyzable by the DMRG which was developed in the former chapter,
since the underlying graph has no tree structure. With the formalism
exposed in this section, it is possible to study such systems with an
arbitrary potential, and we may take profit from the underlying graph
structure so as to compute quickly the necessary matrix elements of
the hamiltonian.

The method starts with a ``unidimensionalization'' of the
bidimensional system, converted into a {\it snake\/}, as it is shown
in figure 2.

\midinsert
\pss
\newgray{gris}{0.8}
\rput(20,0)
{\multirput{0}(0,0)(10,0){3}{\multirput{0}(0,0)(0,-10){3}{
\pscircle[fillstyle=solid,fillcolor=gris](0,0){1.5}}}
\psline(-2,-3)(-2,3)(23,3)(23,-13)(4,-13)(4,-17)(23,-17)(23,-23)
(-2,-23)(-2,-7)(18,-7)(18,-3)(-2,-3)
\psline[linestyle=dashed](0.5,-1.5)(0.5,-8.5)
\psline[linestyle=dashed](10.5,-1.5)(10.5,-8.5)
\psline[linestyle=dashed](10.5,-11.5)(10.5,-18.5)
\psline[linestyle=dashed](20.5,-11.5)(20.5,-18.5)}
\rput(60,-13){\psline(0,0)(80,0)
\multirput{0}(0,0)(10,0){9}
{\pscircle[fillstyle=solid,fillcolor=gris](0,0){1.5}}
\pccurve[angleA=90,angleB=90,linestyle=dashed](0,1.5)(50,1.5)
\pccurve[angleA=90,angleB=90,linestyle=dashed](10,1.5)(40,1.5)
\pccurve[angleA=90,angleB=90,linestyle=dashed](30,1.5)(80,1.5)
\pccurve[angleA=90,angleB=90,linestyle=dashed](40,1.5)(70,1.5)
}
\ps{27}
\nind{\mayp Figure 2. }{\peqfont Unidimensionalization of a $3\times
3$ bidimensional lattice. The ``snake'' gives the order of the sites
in the new structure, and the broken links appear in dotted lines. On
the right, the linear chain with its {\sl long range} structure.\par}
\pss
\endinsert

The links which, in a bidimensional lattice, are local and natural,
appear under this transformation to be long ranged. 2D--DMRG
computations for many-body problems yield a much lower precision than
in the 1D case \ct[WHI-99]. The left-right distinction, which forces
unidimensionalization, is {\it unnatural\/} and destroys the good
properties of DMRG.

The numerical results for this technique shall appear in section
4.4. The convergence of these results is (at least) as safe as those
in 1D, but the computation times are bigger.


\section{4.2. Punctures Renormalization Group.}

The long range DMRG algorithms are based on the left-right
distinction, which only makes sense in 1D. This distinction is
unnecessary and even harmful in many cases, as it shall be clear when
comparing the results of the 2D DMRG with the new method which is
about to be introduced in this section.

This new method, which has been called {\sl Punctures Renormalization
Group (PRG)}, was developed by M.A. Mart\'{\i}n-Delgado, G. Sierra and
the present author \ct[MRS-00]. It requires a {\sl single block}
instead of two. Some ``punctures'' are drilled into this unique block,
which allow us to study with detail the behaviour of the system at
individual sites.
\pss

In the first section the simplest version of the algorithm is
explained, in which a single state and a single puncture are studied.

\subsection{Single Block -- Single State -- Single Puncture.}

Let us consider a global state $\ket|\psi_0>$, defined over the sites
of a given graph ${\cal G}$. This state is a certain approximation to
the ground state of a hamiltonian which fulfills the connectivity
rules of ${\cal G}$. We may obtain a better approximation to the
ground state by following this process:
\pss

\bul We choose any site $p$ of the graph.

\bul We project on the subspace orthogonal to $\ket|\delta_p>$. Or, in
other words, we make the component $\psi_0(p)$ vanish. We
re--normalize the state. Let us call it now $\ket|\psi^*_0>$.

\bul We write an {\it Ansatz\/} of the following form:

$$\ket|\hat\psi_0>=a_B\ket|\psi^*_0> + a_p\ket|\delta_p>$$

\bul As the states entering the summation are orthogonal, we may
proceed to write an effective hamiltonian (for the {\sl superblock}):

$$H_\Sb=\pmatrix{ \elem<\psi^*_0|H|\psi^*_0> &
\elem<\psi^*_0|H|\delta_p> \cr
\elem<\delta_p|H|\psi^*_0> & \elem<\delta_p|H|\delta_p>\cr}$$

\bul We diagonalize the hamiltonian and retain only the ground
state. This has only two components: the {\it weight\/} of the block
and the {\it weight\/} of the state $\ket|\delta_p>$. The lowest
eigenvalue of that matrix is an approximation to the ground state
energy.

\bul We recompose the state $\ket|\hat\psi_0>$ according to those weights.
\pss

It is convenient to emply a graphical representation. Let us consider
a bidimensional lattice such as that of figure 3.

\def\pscr{\pscircle[fillstyle=solid,fillcolor=gray]}
\midinsert
\newgray{lightgray}{.8}
\rput(30,-25){
\psline[fillstyle=solid, fillcolor=lightgray](-2.5,-2.5)(23.5,-2.5)
(23.5,23.5)(-2.5,23.5)(-2.5,-2.5)%
\psframe[fillstyle=solid, fillcolor=white](13,12.5)(18,17.5)
\multirput(0,0)(5,0){5}{
\multirput(0,0)(0,5){5}{\pscr(0,0){1}}}}
\ps{29}
\nind{\mayp Figure 3. }{\peqfont Representation of the ``single block +
single puncture'' scheme.\par}
\endinsert

Thus, the block contains the whole system but a site, which we shall
call the puncture. Figure 3 shows that the block (shadowed) is
connected, so it makes no sense to split it. This is the ``single
block'' philosophy, essential to the method.

When we diagonalize the superblock hamiltonian we are looking for the
best approximation of the ground state (and, incidentally, of the
first excited state) in the subspace spanned by the punctured block
and the delta state at the site. An appropriate way of thinking
about the process is that the wavefunction improves at the site,
meanwhile the rest of the system is merely adapted to the norm changes
which are introduced.

After a RG step is finished, the puncture may translate to any other site
of the system. The best results, notwithstanding, are obtained when the sites
through which the ``puncture'' travels make up a connected path on the
graph. Moreover, let us state that the movement from a site $p$ to a neighbour
$q$ {\sl covers the link} $\<pq\>$; it is then convenient to cover all the
links of the graph in at least one sense. Let us call a PRG-``sweeping'' to
a sequence of steps such as the former one which passes through all the
sites, leaving the covering of all the links as a {\it desideratum\/}. If
this last requirement is fulfilled we shall call it a ``sewing''.

\subsection{Single Block -- Various States -- Various Punctures.}

Once the basic idea of the method has been exposed, we give the
technical details for the process with many states and punctures, such
as it is performed in practice. The system takes profit of any graph
structure, but it is also efficient under its absence (as in the case
of the delta potential formerly described).

\bul {\it Warmup.}

Any orthonormal set of states $\ket|\psi_i>$ is suitable for the beginning
of the computation. Obviously, a right guess speeds the convergence up, as
it is always true for the variational methods. We shall discuss in section
4.5 various appropriate warmup techniques, but for the moment we shall
undertake the most simple one: the orthonormalization of a set of states
whose components are randomly selected.

\bul {\it Punctures and Sewing.}

The set of punctures (or {\sl patch}) will be denoted by $P$. Its
size, $N_p$, may change throughout the process. The strategy for the
election of a patch and its movement through the system is a subject
which, inasmuch as the warmup, may only affect the convergence speed.

Just so as not to leave everything living in an abstract realm, let us
propose a concrete example for 1D. At the $t$-th PRG step, the patch
may be given by the set $P(t)=\{2t,2t+1\}$. Thus, the patches would
all be made up of two punctures. Notice that the set $P(t)=\{t,t+1\}$
is very similar but not equal to the former one. In the second case,
patches overlap. Therefore, more PRG steps are required in order to
complete a sewing. It may be worth if the number of necessary sewings
is substantially smaller.

For 2D lattices a customary patch is the $2\times 2$ square:

$$P=\{(i,j),(i+1,j),(i,j+1),(i+1,j+1)\}$$

All the example patches which have been given are formed by connected
punctures, although it is not a {\it sine qua non\/} condition.

\pss

\bul {\it States, Ansatz and Superblock.}

Once the patch $P$ has been chosen, the $N_p$ delta states
$\{\ket|\delta_p>\}_{p\in P}$ become part of the {\it Ansatz\/}. The
block states shall be obtained from the states which were achieved in
the last step, denoted by $\{\ket|\psi_i>\}_{i=1}^m$. Of course, $m$
is the number of states we wish to obtain.

Let us call $Q_P$ the projector on the states which are orthogonal to
the set of delta states on the patch sites:

$$Q_P\equiv I - \sum_{p\in P} \ket|\delta_p>\bra<\delta_p|$$

The action of this operator is easy to describe. It simply removes the
components which correspond to the patch sites:
$\elem<\delta_p|Q_P|\psi_i>=0$ for all the $p\in P$.
\pss

Maybe the most important technical problem of the procedure appears at
this point: the states $\{Q_P\ket|\psi_i>\}$ are not an orthogonal
set. The re--orthonormalization may be carried out by the standard
Gram-Schmidt technique, but it is more appropriate to use the fast
Gram-Schmidt procedure described in appendix D. It is thus important
to know the scalar products matrix
$C_{ij}=\elem<\psi_i|Q_P|\psi_j>$. Let $G^i_j$ be the basis-change
matrix obtained through the process. Now the states

$$\ket|\psi^*_i>\equiv \sum_{j=1}^m G^i_j Q_P \ket|\psi_j> \qquad
i\in[1,\ldots, m]$$

\nind make up an orthonormal set. These states, along with the delta
states of the patch, make up the {\it Ansatz\/}:

$$\ket|\phi>=\sum_{i=1}^m a_i \ket|\psi^*_i> + \sum_{j=1}^{N_p}
a_{m+j} \ket|\delta_{P(j)}> \eq{\prgansatz}$$

\nind where $P(j)$ denote each of the patch punctures. The
$\{a_i\}_{i=1}^{m+N_p}$ are the variational parameters.
\pss

The superblock hamiltonian $H_\Sb$ which corresponds to the {\it
Ansatz\/} [\prgansatz] is now written straightforwardly:

$$\pmatrix{\vcenter{%
\offinterlineskip\halign{\hfil\ $#$\ \hfil&\hfil\ $#$\ \hfil&\hfil\ $#$\ \hfil%
&\vrule#&&\hfil\ $#$\ \hfil\cr
\elem<\psi^*_1|H|\psi^*_1>&\cdots&\elem<\psi^*_m|H|\psi^*_1> &&
\elem<\delta_{P(1)}|H|\psi^*_1>&\cdots
&\elem<\delta_{P(N_p)}|H|\psi^*_1>\cr
&&&height 6pt&&&\cr
\vdots &\ddots &\vdots&&\vdots&\ddots&\vdots\cr
&&&height 6pt&&&\cr
\elem<\psi^*_1|H|\psi^*_m> & \cdots& \elem<\psi^*_m|H|\psi^*_m> &&
\elem<\delta_{P(1)}|H|\psi^*_m> & \cdots &
\elem<\delta_{P(N_p)}|H|\psi^*_m> \cr
&&&height 6pt&&&\cr
\noalign{\hrule}
&&&height 6pt&&&\cr
\elem<\psi^*_1|H|\delta_{P(1)}> & \cdots &\elem<\psi^*_m|H|\delta_{P(1)}>
&& \elem<\delta_{P(1)}|H|\delta_{P(1)}> & \cdots &
\elem<\delta_{P(N_p)}|H|\delta_{P(1)}>\cr
&&&height 6pt&&&\cr
\vdots &\ddots &\vdots&&\vdots&\ddots&\vdots\cr
&&&height 6pt&&&\cr
\elem<\psi^*_1|H|\delta_{P(N_p)}> &\cdots& \elem<\psi^*_m|H|\delta_{P(N_p)}> &&
\elem<\delta_{P(1)}|H|\delta_{P(N_p)}> & \cdots & 
\elem<\delta_{P(N_p)}|H|\delta_{P(N_p)}>\cr}}}\eq{\hamsbprg}$$

The structure of the matrix [\hamsbprg] is quite clear:

$$H_\Sb=\left(\vcenter{\offinterlineskip
\halign{$#$\ &\vrule#&\ $#$\cr
H_B && H_{BP} \cr
&height 3pt&\cr
\noalign{\hrule}
&height 3pt&\cr
H^\dagger_{BP} && H_P \cr}}\right)$$

\nind where $H_B$ is the single block hamiltonian (the
$\ket|\psi^*_i>$ among themselves), $H_P$ is the set of elements
related to the patch, directly taken from the total
hamiltonian. $H_{BP}$ are the mixed elements, between block and patch
states.
\pss

The aforementioned matrix elements may, of course, be computed {\it ab
initio\/}, i.e.: from the real states components. But, again, this
involves an innecessary waste of resources since:
\ps2

\bul We know the superblock hamiltonian at the former PRG step, and it
may be ``merely'' adapted to the new circumstances.

\bul An underlying graph structure boosts the computation of the
matrix elements.
\ps2

If it is naively computed, $H_\Sb$ requires $(m+N_p)^2N^2$
operations. We shall assume, therefore, that the former step $H_\Sb$
is known:

$$h_{ij}=\elem<\psi_i|H|\psi_j>$$

From this expression\footn{For a general PRG step,
$h_{ij}=E_i\delta_{ij}$, i.e.: matrix $h$ is diagonal and its elements
are the estimates for the energy at the former step. Nevertheless,
matrix $h$ may be ``complete'' after the warmup, so we conserve the
general form.} we obtain $H_\Sb$ in the straightest possible
way. Foremost, we define the intermediate matrix

$$h^*_{ij}\equiv \elem<\psi_i|Q_PHQ_P|\psi_j>$$

\nind Of course, these numbers are not matrix elements of the
hamiltonian for any set of {\it real\/} states, since the
$\{Q_P\ket|\psi_i>\}$ are not orthonormalized. Nonetheless, they are
easy to find:

$$h^*_{ij}=\sum_{k,l\in S-P} \psi_i(k)H_{kl}\psi_j(l)
=\left( \sum_{k,l\in S} - \sum_{k \in P \atop l\in S} - \sum_{k\in S
\atop l \in P} + \sum_{k,l \in P} \right) \psi_i(k)H_{kl}\psi_j(l)$$

\nind where $S$ denotes again the set of sites of the system. The
first summation runs over the sites of the block. Using the simplest
version of the ``{\sl inclusion--exclusion theorem}'', it is split
into four easier partial sums:
\ps2

\bul The first summation is, of course, $h_{ij}$.

\bul The fourth one is rather fast to compute, since it only involves
matrix elements among states of $P$. The number of operations is
$O(N_p^2)$.

\bul The second and third summations, equal in structure, must be
computed in full if there is no underlying graph structure. Otherwise,
they are fast to calculate:

$$\sum_{k\in P \atop l\in S} \psi_i(k)H_{kl}\psi_j(l)
=\sum_{k\in P}\sum_{l\in N(k)} \psi_i(k) H_{kl} \psi_j(l)$$

\nind where $N(k)$ denotes the set of neighbours of $k$. These
summations require $O(z\times N_p)$ operations, where $z$ is an upper
bound for the size of $N(p)$ for all $p$ (maximum coordination index).
\pss

After the computation of $h^*_{ij}$, it is easy to obtain $H_B$:

$$(H_B)_{ij}=\elem<\psi^*_i|H|\psi^*_j>=\sum_{k,l=1}^m G^i_k G^j_l
\elem<\psi_k|Q_PHQ_P|\psi_l>=\sum_{k,l=1}^m G^i_k G^j_l h^*_{kl}$$

\nind The elements of $H_{BP}$ are also quick to compute under a
neighbourhood structure:

$$(H_{BP})_{ij}=\elem<\psi^*_i|H|\delta_{P(j)}>=
\sum_{k=1}^m G^i_k\elem<\psi_k|Q_PH|\delta_{P(j)}>=
\sum_{k=1}^m \sum_{s\in N(P(j))-P} G^i_k \psi_k(s) H_{sP(j)}$$

\nind where the last summation over $N(P(j))-P$ means ``the neighbours
of the $j$-th puncture which {\it are not\/} punctures
themselves''. This computation takes also less than $z\times m$
operations.
\pss

After all these operations, all the elements of $H_\Sb$ have been
computed without using in any case all the components of the
states. The neighbourhood structure reduces the timing drastically. If
it is absent, the computation takes $O(N)$ operations (again, $N\gg m$
is assumed).
\pss

\bul {\it Truncation and States Recomposition.}

The superblock hamiltonian represents the whole of the system. When it
is diagonalized, its $m$ lowest energy states provide us with the
variational parameters $\{a^j_i\}$ (with $j\in [1,\ldots,m]$
--denoting which excited state-- and $i\in[1,\ldots,m+N_p]$ --denoting
which component) which must be inserted into [\prgansatz]. These
values yield the best approximation, within the subspace spanned by
our states, to the lowest energy states of the system. The eigenvalues
of $H_\Sb$ give, as it was assumed, the best estimates for the lowest
energies of the full system.
\pss

Using these values we reconstruct the wave-functions:

$$\ket|\hat\psi_k>=\sum_{i=1}^m a^k_i\ket|\psi^*_i> + \sum_{j=1}^{N_p}
a^k_{m+j} \ket|\delta_{P(j)}>$$

But the stored states are the $\ket|\psi_i>$ and not the
$\ket|\psi^*_i>$. Therefore,

$$\ket|\hat\psi_k>=\sum_{i,j=1}^m a^k_iG^i_jQ_P\ket|\psi_j> + \sum_{j=1}^{N_p}
a^k_{m+j} \ket|\delta_{P(j)}> \eq{\recompprg}$$

The action of the operator $Q_P$ is really simple: make all the
components associated with $P$ vanish. We may build just a matrix
which encapsulates the re--orthogonalization and the renormalization:

$$B^k_j\equiv \sum_{i=1}^m a^k_i (G^*)^i_j$$

\nind where the matrix $G^*$ is defined by

$$(G^*)^i_j = \cases{ G^i_j & if $i,j\leq m$ \cr
	\delta^i_j & if $i,j > m$\cr
	0 & otherwise \cr}$$

The reconstruction procedure yields, therefore:

$$\hat\psi_k(q)=\cases{ \sum_{j=1}^m B^k_j \psi_j(q) & if $q\notin P$\cr
	a^k_{m+j} & if $q=P(j)$\cr}$$

And, under a neighbourhood structure, {\it this\/} is the only step
which scales with $N$. Thus, in the worst case scenario, the number of
operations for the PRG step is $O(N)$.
\pss

There is only one more detail to be closed: the matrix

$$\hat h_{ij}=\elem<\hat\psi_i|H|\hat\psi_j>$$

\nind is necessary to consider at the next step. Due to the
construction rule of the states $\ket|\hat\psi_i>$ this matrix is
{\it diagonal\/}, and its elements are the energies of the states:

$$\hat h_{ij}=\cases{E_i & if $i=j$\cr 0 & otherwise\cr}$$

\nind We have mantained the notation $h_{ij}$ for the sake of
generality (e.g.: the states coming out of the warmup need not have
the same structure).

\ps7
\subsectionp{Summary.}

$\bullet$ {\it Stored Data\/}. The $N$ components of the $m$ states
$\ket|\psi_i>$ and the elements of $h_{ij}$ (which usually make up a
diagonal matrix).
\ps2

$\bullet$ {\it System States and Ansatz\/}. The punctures components
are removed with the operator $Q_P$. The states are re--orthonormalized
with a Gram-Schmidt matrix $G$. New states are defined

$$\ket|\phi_i>=\cases{ 
\sum_{j=1}^m G^i_j Q_P \ket|\psi_j> & if $i\leq m$\cr
\ket|\delta_{P(i-m)}> & if $i>m$ \cr}$$

The final states are created after the {\it Ansatz\/}

$$\ket|\phi>=\sum_{i=1}^{m+N_p} a_i\ket|\phi_i>$$ 

\nind where the $\{a_i\}$ with $i\in [1,\ldots, m+N_p]$ are the
variational parameters.
\ps2

$\bullet$ {\it Superblock Hamiltonian Construction\/}. Given by
$H_\Sb=\elem<\phi_i|H|\phi_j>$. Its construction, using the
neighbourhood structure and the knowledge of $h_{ij}$ is rather fast.
\ps2

$\bullet$ {\it Diagonalization of the Superblock and Updating\/}. The
$m$ lowest eigenstates of the superblock hamiltonian are retained and
the $a_i$ weights are inserted into the {\it Ansatz\/} according to
the expression [\recompprg]. The diagonal matrix elements $\hat
h_{ij}$ are the obtained energies.

\subsection{General Features of the PRG.}

The Real Space Renormalization Group method known as PRG ({\it
Punctures Renormalization Group\/}) has the following general
features:
\ps2

\bul {\it Ensured Convergence\/}. Section 4.5 discusses the conditions
under which such statement may be held.

\bul {\it Absolute Generality\/}. Any quantum--mechanical model in a
finite dimensional Hilbert space is analyzable through this technique.

\bul {\it Adaptative Efficience\/}. It takes profit if the hamiltonian
matrix is sparse, i.e.: the existence of an underlying neighbourhood
structure.

\bul {\it Explicit Method\/}. Wave--functions are stored in full. The
number of operations of a PRG step may be not lower than $O(N)$, the
inequality being saturated in the case of underlying graph structure.


\section{4.3. 1D Implementation and Numerical Results.}

For the unidimensional examples we have chosen the following models:
\ps2

\bul Free particle in a box divided into $N=200$ cells with fixed
boundary conditions. The eigenstates are extended throughout the
system and the boundary conditions are rather relevant.

\bul Particle in a harmonic potential, with space again divided into
$N=200$ cells. The boundary conditions are much less relevant and the
eigenstates are spatially localized.

\bul Free particle in a 1D box with periodic boundary conditions. The
underlying graph is {\it not\/} a tree.
\ps2

In all the cases the objective of the calculations were the $m=4$
lowest energy eigenstates and the convergence criterion was a
precision of one part in $10^{10}$ for the two first states compared
to the {\it exact values\/}\footn{Obviously, there are also
convergence criteria which are fully internal, as it is the imposition
that the variation through a full sweep to be smaller than a prefixed
quantity.}.
\pss

Table 1 shows the numerical results. The numbers given in parenthesis
after the PRG heading are the number of punctures for the
patch. Respectively, $4$ and $10$ punctures were used.

\midinsert
\centerline{\hfilneg
\sbtabla(4)
\li && Fixed B.C. & Harm. Pot. & Periodic B.C. \cr
\lt Exact && 2.13 & 2.14 & 2.09 \cr
\lt DMRG && 0.83 (3) & 0.6 (2) & 109 (11) \cr
\lt PRG (4) && 25.6 (198) & 9.8 (76) & 10.94 (87) \cr
\lt PRG (10) && 6.97 (65) & 3.22 (30) & 3.81 (34) \cr
\setabla}
\ps2
\nind{\mayp Table 1. }{\peqfont Benchmarking for the 1D-PRG
algorithm, DMRG and exact diagonalization for {\it a)\/} a free
particle with fixed b.c., $N=200$ and $m=4$, {\it b)\/} in presence of
an harmonic potential, {\it c)\/} free, but with periodic boundary
conditions. In all the cases a precision was reached of one part in
$10^{10}$ for the first two states (when the exact energy was null,
the precision was imposed in absolute value). The numbers show the CPU
time in arbitrary units (Pentium III at 450 MHz) and, in parenthesis,
the number of PRG or DMRG sweeps.\par}
\endinsert

The numerical results for 1D yield a rather clear advantage of DMRG
with respect to the other methods on graphs ``without cycles''. The
reason is obvious: DMRG is implicit and, in 1D, the partition into
left and right blocks {\it is natural\/}. PRG in 1D gathers in a
single block two regions which are {\it really\/} separated in space,
slowing down the convergence.

It must be remarked the case of periodic boundary conditions (third
column). The underlying graph structure is not a tree for this
system, since it contains a cycle. Therefore, the left--right
distinction is unnatural and DMRG leads to worse results than PRG.
\pss

Both methods take a worse result than the exact diagonalization for
$N=200$, but for bigger sizes (empirically up to $N=120.000$ sites),
due to the different scaling regimes, PRG becomes a rather suitable
method. CPU times scale for all methods (exact, DMRG and PRG) as a
power law:

$$t_{\linp{CPU}}\approx K\cdot N^\alpha$$

\nind Albeit the parameter $K$ may be rather relevant for practical
applications, it is usual to consider the exponent $\alpha$ to be the
key, due to its ``universality''. In other terms: an improvement in
the implementation or in {\it hardware\/} may lower $K$, but the
exponent $\alpha$ may only diminish with a deep change in the
computation algorithm. This statement is illustrated by the following
results:
\ps2

\bul Exact diagonalization (periodic b.c.): $\alpha\approx 3.1$.

\bul Implicit DMRG (fixed b.c.): $\alpha\approx 1.2$.

\bul Explicit DMRG (periodic b.c.): $\alpha\approx 3.7$.

\bul PRG (periodic b.c.): $\alpha\approx 2.2$.
\ps2

It may be observed that the explicit DMRG for non-tree graphs scales
worse than the exact diagonalization. DMRG in its own scope is, no
doubt, the one which presents the best performance. PRG is an
alternative when DMRG can not tackle the problem in an appropriate
way: for long range and/or multidimensional systems.
\pss

Some practical questions on implementation:
\ps2

\bul When the number of punctures does not divide exactly the lattice
size, we may choose between these options for the analysis of the
last patch. {\it a)\/} The last patch contains less punctures than the
rest: $N'_p=\; \lin{mod}\;N_p$, or {\it b)\/} the movement backwards
starts when there are less than $N_p$ sites left to reach the extreme.

\bul The movement of the patch may be carried out {\sl with} or {\sl
without} punctures overlapping. In 1D the first case appears to be
slightly more efficient for big patches ($\approx 10$). Figure 4 shows
graphically the meaning of advance with overlapping.

\midinsert
\ps6
\rput(30,0)
{\psline(0,0)(100,0)
\multirput(0,0)(10,0){10}{\pscr(0,0){1}}
\psframe(-2,-2)(32,2)\rput(12,-6){\peqfont First patch}
\psframe(28,-3)(62,3)\rput(45,6){\peqfont Second patch}
\psframe(58,-2)(92,2)\rput(80,-6){\peqfont Third patch}
}
\ps{10}
\nind{\mayp Figure 4. }{\peqfont Movement of a $4$ sites patch with
overlapping in a 1D lattice. On the first PRG step, the patch includes
sites $1$ upto $4$. On the second step, sites $4$ to $8$. There is a
common site between successive patches.\par}
\pss
\endinsert

\subsection{Convergence in DMRG and PRG.}

Focusing on the DMRG and PRG techniques for 1D systems ({\it
without\/} periodic b.c.), figures 5 and 6 show the different
approaches to convergence for two free systems with fixed b.c. and
$N=100$ sites.

\midinsert
\hbox to \hsize{\vtop{\kern 0pt\hsize=8cm\figr{5.5cm}{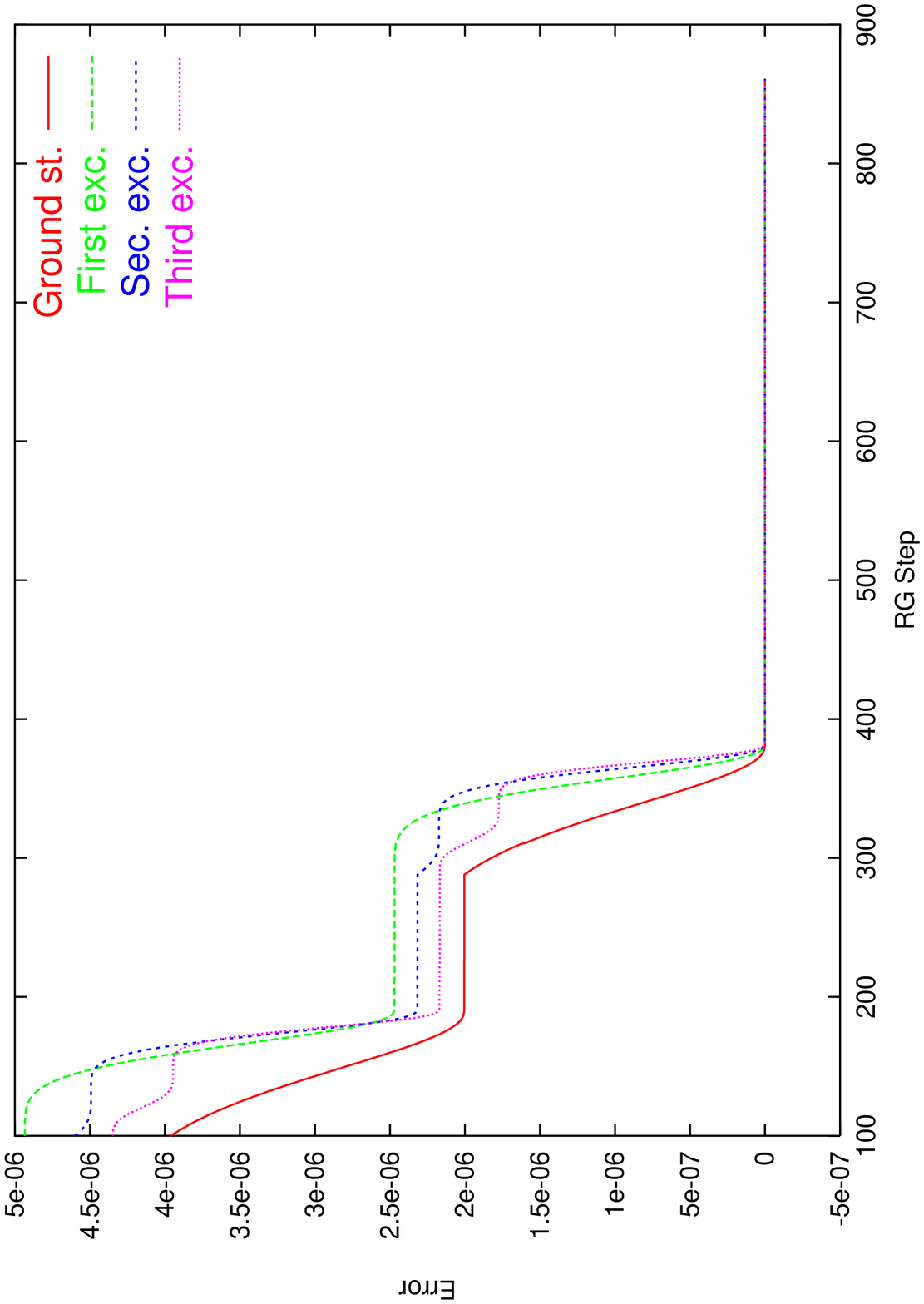}
\nind{\mayp Figure 5. }{\peqfont DMRG Convergence. Relative error for the
first four states as a function of the RG step. Notice how convergence
is led by fast downwards slopes and long plateaux.\par}}\hfil
\vtop{\kern 0pt\hsize=8cm\figr{5.5cm}{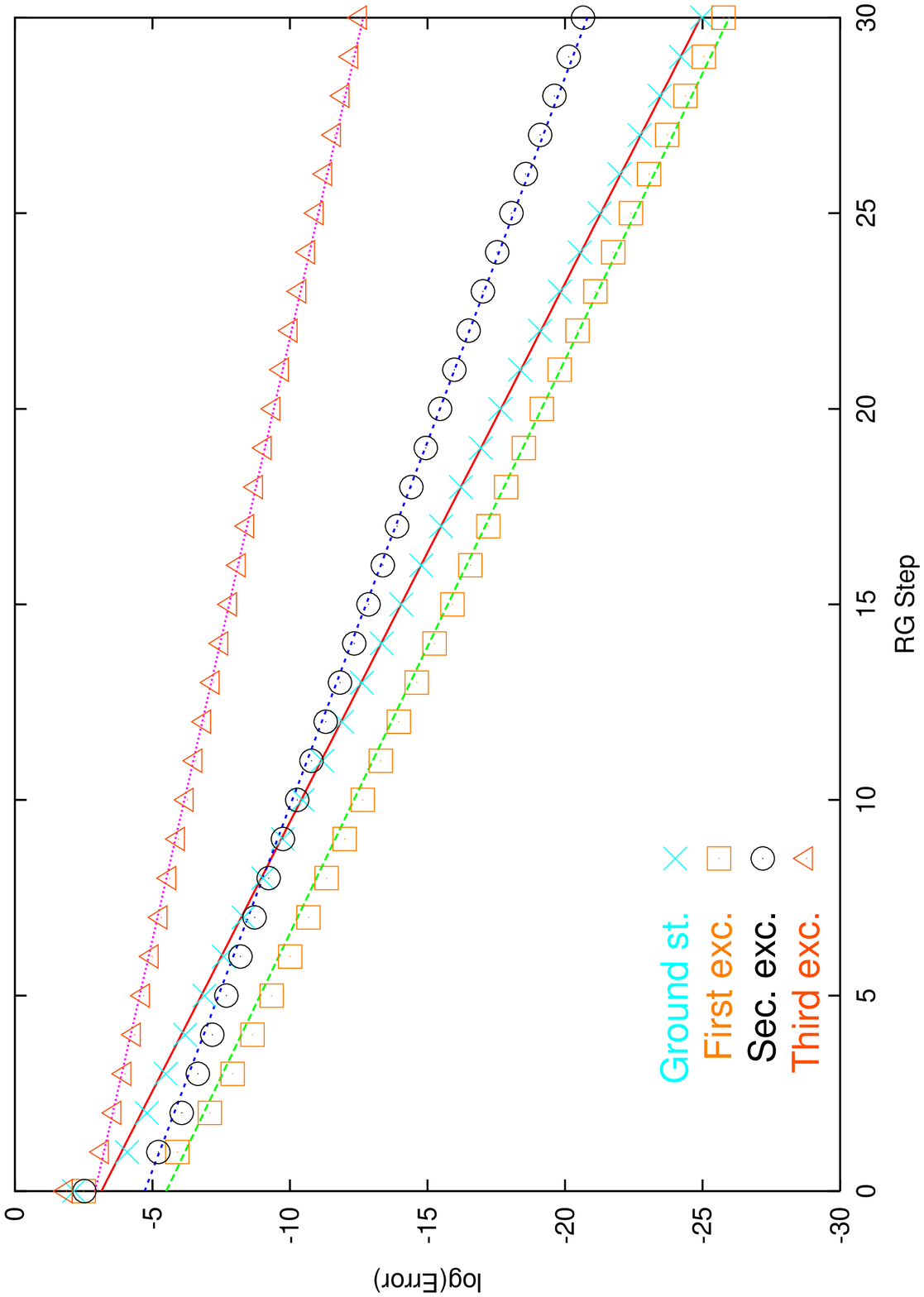}\nind{\mayp Figure
6. }{\peqfont PRG Convergence. The logarithm of the error is presented
as a function of the RG step. The straight lines show the fit to an
exponential decay.\par}}}
\endinsert

DMRG advances by very fast strides of the error and rather long
plateaux. On the other hand, PRG is a ``long-distance runner'', as it
is shown by the good fit to a law of the type

$$\Delta E \approx \exp(-K\cdot n)$$

\nind (where $n$ is the RG step index) in the whole range of values
until machine precision is reached.

This difference may be explained in qualitative terms. DMRG uses for
the decreasing block the information of a former step. This implies
that at the start of a ``half sweep'' (e.g. a sweep left $\to$ right)
we are employing a very big block which has not been recently
updated. The biggest advances are made when the decreasing block is
{\it very small\/} (in figure 5, the high slopes are just before steps
$200$ and $400$, which correspond to a sense change).

On the other hand, PRG has all its information updated. At every step,
the change in the wave--functions is {\it proportional\/} to its
distance to the desired state. This explains, at least qualitatively,
the exponential convergence.
\pss

It is interesting to study the PRG ``residual'' state. Let us consider
the case when only the ground state is retained ($m=1$), whose exact
value we shall assume to be known. We substract at each RG step the
exact state from the approximate one and normalize the result. This
residual state, after a number of RG steps, becomes the first excited
state. The transient is longer when both states have different
symmetries (even vs. odd). In this case, the residual state
corresponds to the second excited state for some time.

When $m$ states are conserved, the residual of the $i$-th state
converges to the $(m+i)$-th state after a transient, which may be
retarded for symmetry reasons (even vs. odd).


\section{4.4. PRG Analysis for 2D and 3D Lattices.}

The title for this chapter reminds us that the fundamental reason for
the improvements on the DMRG was the study of 2D and 3D systems. This
section deals with a series of fundamental technical questions when
applying the PRG in 2D and 3D, along with a series of applications and
numerical results.

\subsection{Sewing Lattices.}

Let us consider a 2D rectangular lattice $L_x\times L_y$ and a
puncture formed by a single site. For DMRG the only possible path is
the one we shall call ``DMRG sweep'', which traverses the system in an
anisotropic zig-zag (see figure 2).

PRG, dealing with a single block, allows an isotropic path in which
not only all sites are visited, but also all links. In the terminology
introduced in section 4.2 we may talk of a ``sewing''.

The change from the anisotropic to the isotropic sweeping obeys the
following reason: the derivatives of the wave--functions along a {\sl
link} between two sites are updated more appropriately when such a
link is traversed by the puncture. The DMRG sweeping leaves almost
half of the links without covering, slowing down the convergence.
\pss

The algorithm for drawing a sewing depends on the parity of $L_x$ and
$L_y$. Figure 7 shows a sewing for the even-even lattice.

\figura{2.5cm}{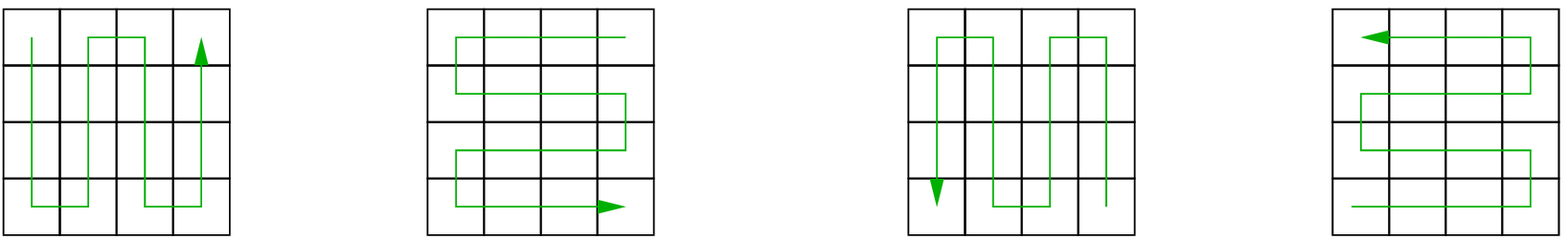}{Figure 7.}{A possible sewing for a
$4\times 4$ lattice.}

The structure of the sewings is exposed more rigorously in an
algebraic notation, which on the other hand describes the code of the
employed programs. Let us define the four operators $L$, $R$, $U$ and
$D$ as the moves of the puncture leftwards, rightwards, upwards and
downwards respectively. The path in figure 7 is algebraically
described in pseudocode 1.

\def\ot{\leftarrow}

\midinsert
\ps3
\boxit{7pt}{\vbox{\hsize=15cm\tt
$Z_\linp{hor}\ \ot\ U^{L_y-1}RD^{L_y-1}\ [RU^{L_y-1}RD^{L_y-1}]^{L_x/2-1}$
\ps2
$Z_\linp{vert}\ \ot\ R^{L_x-1}DL^{L_x-1}\ [DR^{L_x-1}DL^{L_x-1}]^{L_y/2-1}$
\ps2
Path\ $\ot\ Z_\linp{hor}^{-1}\ Z_\linp{vert}^{-1}\ 
Z_\linp{hor}\ Z_\linp{vert}$}}
\ps3
\nind{\mayp PseudoCode 1. }{\peqfont Algorithm for the sewing of an
even-even lattice.\par}
\ps2
\endinsert

The odd-odd sewing, exemplified in figure 8 for a $5\times 5$ lattice,
is described in pseudocode 2 in a more rigorous way\footn{The last
expression of pseudocode 1, describing the sewing of an even-even
lattice in terms of two operators and their inverses, reminds the
topological description of a torus. On the other hand, the description
of the odd-odd sewing (pseudocode 2) would be topologically trivial
\ct[BLA-82].}.

\figura{2.8cm}{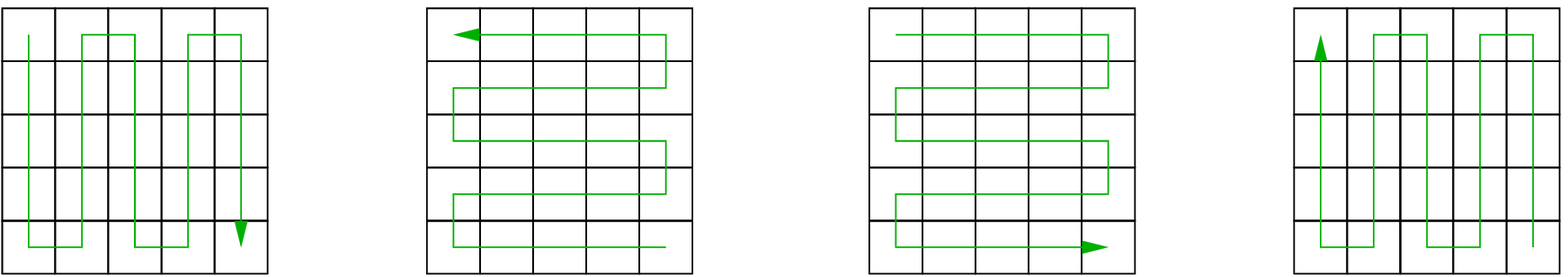}{Figure 8.}{Sewing for a $5\times 5$
lattice. Notice that, in this case, each link is traversed in both
senses.}

\midinsert
\ps3
\boxit{7pt}{\vbox{\hsize=15cm\tt
$Z_\linp{hor}\ \ot\ D^{L_y-1}\ [RU^{L_y-1}RD^{L_y-1}]^{(L_x-1)/2}$
\ps2
$Z_\linp{vert}\ \ot\ L^{L_x-1}\ [UR^{L_x-1}UL^{L_x-1}]^{(L_y-1)/2}$
\ps2
Path\ $\ot\ Z^{-1}_\linp{hor}\ Z^{-1}_\linp{vert}\ Z_\linp{vert}\
Z_\linp{hor}$}}
\ps3
\nind{\mayp PseudoCode 2. }{\peqfont Algorithm for the sewing of an
odd-odd lattice.\par}
\ps2
\endinsert

As a matter of fact, the {\it sewing\/} is one of the most important
features of the bidimensional and tridimensional PRG. 2D DMRG, due to
its own unidirectional structure, is forced to carry out the sweeping
in a necessarily anisotropic way.
\pss

Results improve significantly when patches composed of various
punctures are employed. Having tried many alternatives, we have
checked that the square patch of $L_p\times L_p$ sites is the one
leading to best results. The algorithm exposed previously for the
sewing is suitable when $L_p$ is commensurate both with $L_x$ and
$L_y$. In this case, it is possible to split the system into blocks of
size $L_p\times L_p$ for the patch to cover. Figure 9 illustrates the
process.

\midinsert
\ps{30}
\newgray{gris}{.8}
\rput(40,0){
\multirput(0,0)(5,0){6}{
\multirput(0,0)(0,5){6}{\pscr(0,0){1}}}
\psframe(-2,-2)(28,27)
\multirput(8,0)(10,0){2}{\psline(0,-2)(0,27)}
\multirput(0,7.5)(0,10){2}{\psline(-2,0)(28,0)}
\psframe[fillstyle=solid,fillcolor=gris,framearc=.3](-1,8)(7.25,17)
\psline{->}(3,17)(3,23)(13,23)(13,2.5)(24,2.5)}
\ps4
\nind{\mayp Figure 9. }{\peqfont A path through a $6\times 6$ lattice
by a $2\times 2$ patch. The dimensions of the patch must be
commensurate with the ones of the lattice.\par}
\ps2
\endinsert

\subsection{Numerical Results in 2D.}

Two quantum--mechanical problems have been chosen so as to check the
performance of the 2D-PRG: the free particle in a box (with fixed
b.c.) and the bidimensional hydrogen atom.

The first one is fully analogous to the 1D problem: the
diagonalization of the laplacian matrix of the 2D graph with fixed
b.c. The eigenstates are analytically obtainable. If $L_x=L_y$, the
$D_4$ group of symmetries guarantees the degeneration of the first
excited state (among many others): there are two {\it twin\/} states
with nodes lines which may point in any couple of orthogonal
directions.

The hydrogen atom on a 2D lattice is exactly solvable in the continuum
limit. If $L_x$ and $L_y$ are not both odd, then the atomic nucleus
stays at the center of a ``plaquette'' or an ``edge''. Otherwise, a
{\it regularization\/} is required so as the potential energy is
finite throughout the system. In the continuum limit this distinction
is senseless and energies are

$$E_n=-Z^2/n^2 \qquad\qquad n=1,2,\ldots$$

The production of a graphical output in the form of a ``movie'' is
straightforward within PRG. Figure 10 shows a typical picture from
that film.

\midinsert
\fig{3.6cm}{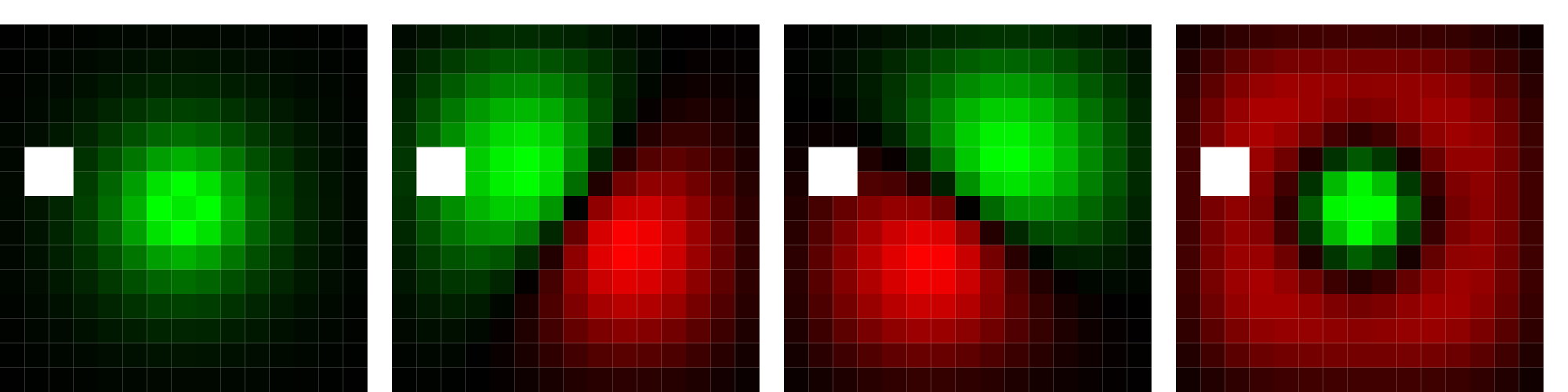}\ps2
\fig{3.6cm}{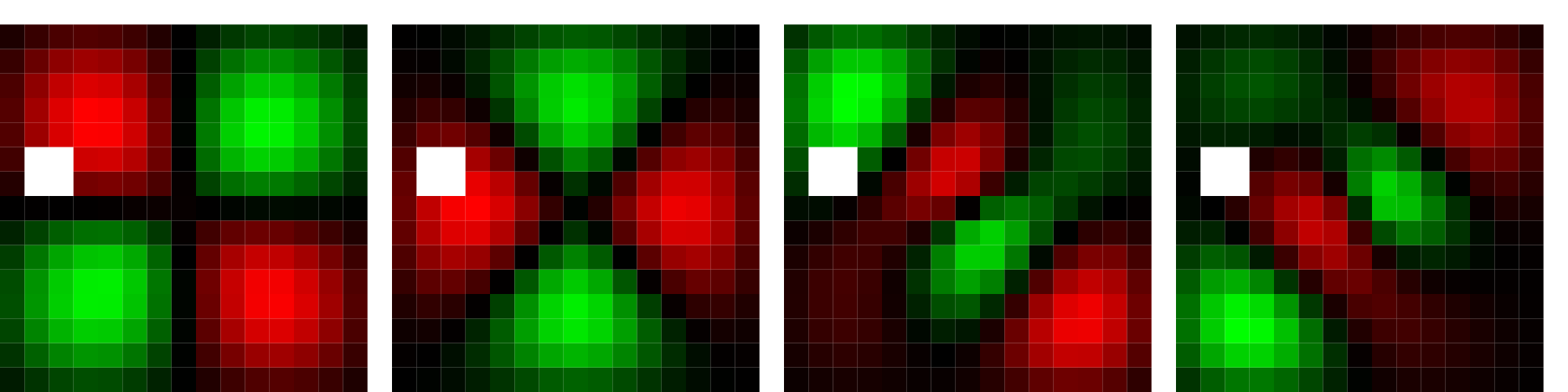}\ps2
\nind{\mayp Figure 10. }{\peqfont A density plot for the
wave--functions obtained with the 2D-PRG algorithm for the particle on
a $15\times 15$ lattice under the effect of a Coulomb potential,
before convergence has been reached. The states grow in energy in the
sense of reading. White pixels at the left side of the eight states
correspond to the position of the $2\times 2$ patch.\par}
\endinsert

The computation time has been obtained for the hydrogen atom through
three methods: 2D-DMRG, PRG and exact diagonalization. The results for
different lattices are shown in figure 11. The three data series fit
to power laws:

$$t_\linp{CPU}\propto L^\theta$$

Figure 11 shows that the lowest slope corresponds to PRG. In table 2
we may observe the corresponding scaling exponents. These results
prove that PRG is the best choice for big lattices, since its
convergence is assured and it is faster than the others.

\figrot{6.5cm}{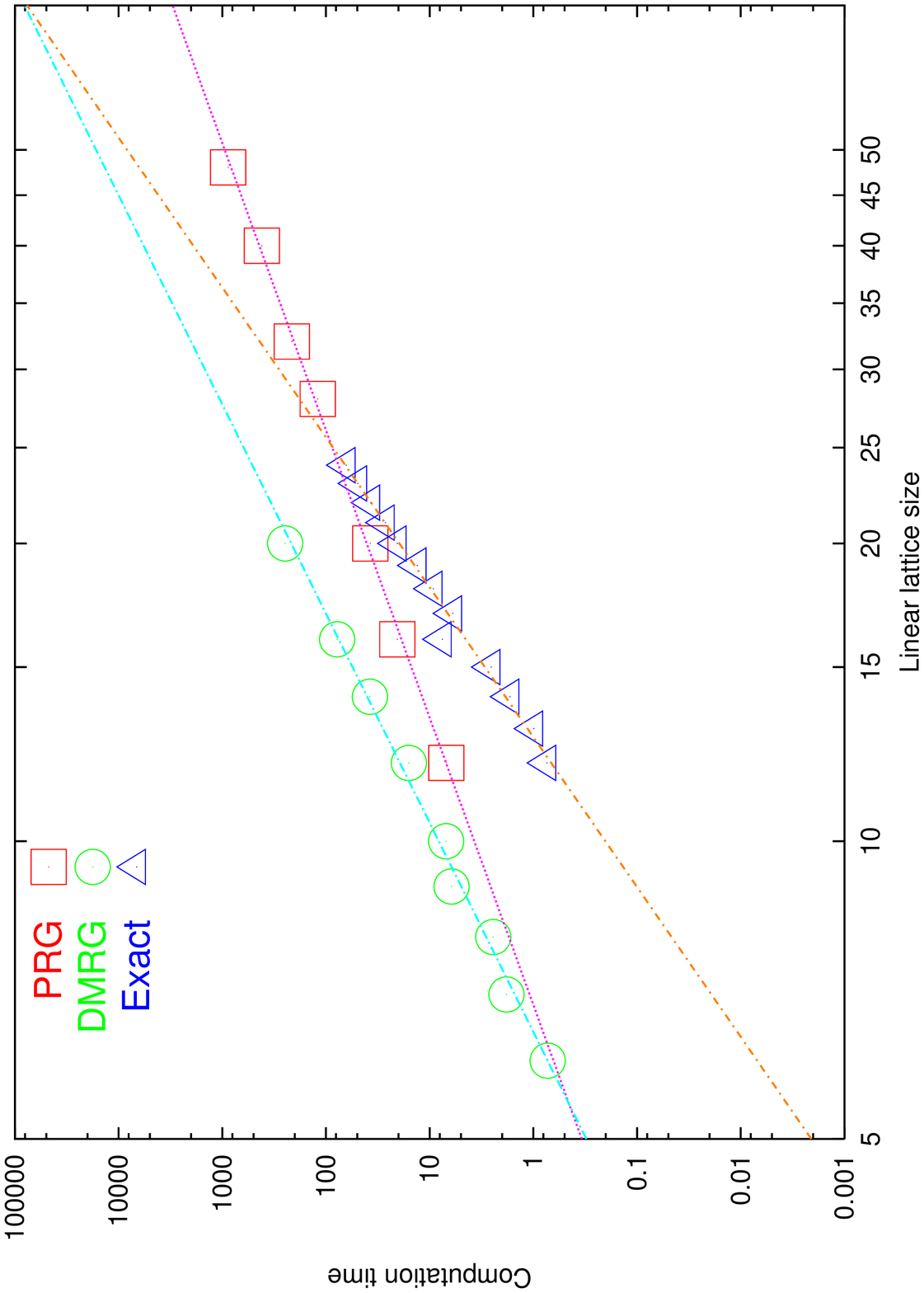}{Figure 11.}{Log-log graph of the CPU time
versus the length $L$ of a square lattice in a Coulomb
potential. Four states are stored, a precision of $10^{-5}$ is
required for 2D-DMRG and $10^{-10}$ for PRG with a $2\times 2$
patch.\par}

\midinsert
\centerline{
\sbtabla(2)
\li && Free Part. & 2D Hydrogen \cr
\lt Exact && $6.4\pm 0.3$ & $6.61\pm0.27$\cr
\lt DMRG && $5.5\pm 0.1$ & $5.21\pm 0.05$\cr
\lt PRG && $3.92\pm 0.1$ & $3.85\pm 0.07$\cr
\setabla}
\ps2
\nind{\mayp Table 2. }{\peqfont Scaling exponents for the free
particle with fixed b.c. and for the particle in a Coulomb
potential.\par}
\ps2
\endinsert

\subsection{PRG in 3 Dimensions.}

The PRG algorithm is generalizable to any dimension. Even more,
results improve comparatively with it. We have prepared programs for a
3D lattice, although tests have not been so exhaustive.

The 2D sewing algorithms have their 3D analogues, which are naturally
more complex. Thus, for the sake of simplicity, we shall only show the
path for the odd-odd-odd case\footn{Such a solution was initially
developed by Silvia N. Santalla, who put it kindly at our
disposal.}, shown in figure 12 and pseudocode 3.

\midinsert
\ps{10}
{
\newrgbcolor{rojo}{1 0 0} \newrgbcolor{verde}{0 1 0} \newrgbcolor{azul}{0 0 1}
\psset{linewidth=.2mm, arrowsize=5pt 3, unit=7mm}
\def\ly#1#2{\psline[linecolor=verde]{->}(#1)(#2)}
\def\lx#1#2{\psline[linecolor=rojo]{->}(#1)(#2)}
\def\lz#1#2{\psline[linecolor=azul]{->}(#1)(#2)} 
\def\punto#1{\pscircle*(#1){0.08}}
\newgray{grisclaro}{.8}
\def\pscr#1{\pscircle[fillstyle=solid, fillcolor=grisclaro](#1){0.2}}
\def\tut{0,-2}
\def\tud{0,-4} 
\def\tuu{0,-6} 
\def\tdt{2,-2} 
\def\tdd{2,-4}
\def\tdu{2,-6} 
\def\ttt{4,-2} 
\def\ttd{4,-4}
\def\ttu{4,-6}
\def\dut{0.866,-1.5} 
\def\dud{0.866,-3.5}
\def\duu{0.866,-5.5} 
\def\ddt{2.866,-1.5} 
\def\ddd{2.866,-3.5}
\def\ddu{2.866,-5.5} 
\def\dtt{4.866,-1.5} 
\def\dtd{4.866,-3.5}
\def\dtu{4.866,-5.5} 
\def\uut{1.732,-1}
\def\uud{1.732,-3} 
\def\uuu{1.732,-5} 
\def\udt{3.732,-1}
\def\udd{3.732,-3} 
\def\udu{3.732,-5} 
\def\utt{5.732,-1}
\def\utd{5.732,-3} 
\def\utu{5.732,-5} 

\def\reja{\punto\uuu \punto\uud \punto\uut \punto\udt \punto\udd 
\punto\udu \punto\utt \punto\utd \punto\utu \punto\duu \punto\dud 
\punto\dut \punto\ddt \punto\ddd \punto\ddu \punto\dtt \punto\dtd 
\punto\dtu \punto\tuu \punto\tud \punto\tut \punto\tdt \punto\tdd 
\punto\tdu \punto\ttt \punto\ttd \punto\ttu 
\psline[linecolor=grisclaro](\uuu)(\uut)
\psline[linecolor=grisclaro](\uuu)(\utu)
\psline[linecolor=grisclaro](\uuu)(\tuu)
\psline[linecolor=grisclaro](\ttt)(\utt)
\psline[linecolor=grisclaro](\ttt)(\ttu)
\psline[linecolor=grisclaro](\ttt)(\tut)
\psline[linecolor=grisclaro](\tuu)(\tut)
\psline[linecolor=grisclaro](\tut)(\uut)
\psline[linecolor=grisclaro](\utu)(\utt)
\psline[linecolor=grisclaro](\utt)(\uut)
\psline[linecolor=grisclaro](\ttu)(\tuu)
\psline[linecolor=grisclaro](\ttu)(\utu)}

\rput(0,0){\hbox{
$P_z$
\rput(0,1){\pscr\uuu \reja
\ly\uuu\utu \lx\utu\dtu \ly\dtu\duu \lx\duu\tuu \ly\tuu\ttu 
\lz\ttu\ttd
\lx\ttd\utd \ly\utd\udd \lx\udd\tdd \ly\tdd\tud \lx\tud\uud 
\lz\uud\uut
\ly\uut\utt \lx\utt\dtt \ly\dtt\dut \lx\dut\tut \ly\tut\ttt
}}}
\rput(0,-7){\hbox{
$P_x$
\rput(0,1){\pscr\ttt \reja
\ly\utu\uuu \lz\utd\utu \ly\uud\utd \lz\uut\uud \ly\utt\uut
\lx\dtt\utt
\lz\dtu\dtt \ly\ddu\dtu \lz\ddt\ddu \ly\dut\ddt \lz\duu\dut
\lx\tuu\duu
\ly\ttu\tuu \lz\ttd\ttu \ly\tud\ttd \lz\tut\tud \ly\ttt\tut
%
}}}
\rput(0,-14){\hbox{
$P_y$
\rput(0,1){\pscr\uuu \reja
\lx\uuu\tuu \lz\tuu\tud \lx\tud\uud \lz\uud\uut \lx\uut\tut
\ly\tut\tdt
\lz\tdt\tdu \lx\tdu\ddu \lz\ddu\ddt \lx\ddt\udt \lz\udt\udu
\ly\udu\utu
\lx\utu\ttu \lz\ttu\ttd \lx\ttd\utd \lz\utd\utt \lx\utt\ttt
}}}
\rput(11,0){\hbox{
$Q_z(P^{-1}_z)$
\rput(0,1){\pscr\ttt \reja
\lx\tuu\uuu \ly\tdu\tuu \lx\udu\tdu \ly\utu\udu \lx\ttu\utu
\lz\ttd\ttu
\ly\tud\ttd \lx\dud\tud \ly\dtd\dud \lx\utd\dtd \ly\uud\utd
\lz\uut\uud
\lx\tut\uut \ly\tdt\tut \lx\udt\tdt \ly\utt\udt \lx\ttt\utt
%
}}}
\rput(11,-7){\hbox{
$Q_x(P^{-1}_x)$
\rput(0,1){\pscr\uuu \reja
\lz\uuu\uut \ly\uut\udt \lz\udt\udu \ly\udu\utu \lz\utu\utt
\lx\utt\dtt
\ly\dtt\dut \lz\dut\dud \ly\dud\dtd \lz\dtd\dtu \ly\dtu\duu
\lx\duu\tuu
\lz\tuu\tut \ly\tut\tdt \lz\tdt\tdu \ly\tdu\ttu \lz\ttu\ttt
}}}
\rput(11,-14){\hbox{
$Q_y(P^{-1}_y)$
\rput(0,1){\pscr\ttt \reja
\lz\uut\uuu \lx\dut\uut \lz\duu\dut \lx\tuu\duu \lz\tut\tuu
\ly\tdt\tut
\lx\udt\tdt \lz\udd\udt \lx\tdd\udd \lz\tdu\tdd \lx\udu\tdu
\ly\utu\udu
\lz\utt\utu \lx\dtt\utt \lz\dtu\dtt \lx\ttu\dtu \lz\ttt\ttu
%
}}}


 } 
\vskip 14cm
\nind{\mayp Figure 12. }{\peqfont Three-dimensional sewing for an
odd-odd-odd lattice, exemplified with a $3\times 3\times 3$ one. Red
lines are parallel to the $x$-axis, green ones to the $y$-axis and
blue ones to the $z$-axis. Notice that each link is traversed at least
once, and inner links are covered just once. The notation $P_{x_i}$
and $Q_{x_i}(P^{-1}_{x_i})$ is defined in pseudocode 3.\par}
\endinsert

\def\tot{\leftrightarrow}
\midinsert
\ps2
\boxit{7pt}{\vbox{\hsize=15cm\tt
define $Z(P,Q,a,b)\ \equiv\
P^{a-1}[Q(P^{-1})^{a-1}QP^{a-1}]^{(b-1)/2}$
\ps2
define $Q_z\ \equiv\ (B\tot L; F\tot R)$
\ps2
define $Q_x\ \equiv\ (R\tot U; L\tot D)$
\ps2
define $Q_y\ \equiv\ (D\tot B; U\tot F)$
\ps2
$P_z\ \ot\ Z(R,F,L_y,L_x)\ [U\ Z(B,L,L_x,L_y)\ U\ Z(R,F,L_y,L_x)]^{(L_z-1)/2}$
\ps2
$P_x\ \ot\ Z(D,L,L_z,L_y)\ [B\ Z(R,U,L_y,L_z)\ B\ Z(D,L,L_z,L_y)]^{(L_x-1)/2}$
\ps2
$P_y\ \ot\ Z(F,U,L_x,L_z)\ [R\ Z(D,B,L_z,L_x)\ R\ Z(F,U,L_x,L_z)]^{(L_y-1)/2}$
\ps2
Path$\ \ot\ Q_y(P_y^{-1})\ Q_x(P_x^{-1})\ Q_z(P_z^{-1})\ P_y\ P_x\ P_z$}}
\ps2
\nind{\mayp PseudoCode 3. }{\peqfont Algorithm for the sewing of an
odd-odd-odd lattice shown in figure 12. Abbreviations $L-R$, $D-U$ and
$F-B$ correspond to movements in each axis ({\it left--right\/}, {\it
down--up\/}, {\it forward--backward\/}). Operators $Q_x$, $Q_y$ and
$Q_z$ denote appropriate rotations around each axis which swaps the
direction operators. Obviously, it is accepted that $U^{-1}=D$,
$R^{-1}=L$ and $B^{-1}=F$. Between the path $P_{x_i}$ and
$Q_{x_i}(P_{x_i}^{-1})$ a sewing of all planes perpendicular to axis
$x_i$ is carried out.\par}
\ps2
\endinsert

Pseudocode 3 uses the following movement operators: $B$ and $F$
($x$-axis), $L$ and $R$ ($y$-axis), $D$ and $U$ ($z$-axis). Values of
the exponent $\theta$ for the hydrogen atom are given in table 3.

\midinsert
\ps2
\centerline{\sbtabla(1)
\li Method && $\theta$ \cr
\lt Exact && $9.5\pm 0.6$\cr
\lt PRG (2) && $6.6\pm 0.4$\cr
\lt PRG (3) && $5.6\pm 0.3$\cr
\setabla}
\ps2
\nind{\mayp Table 3. }{\peqfont Scaling exponent $\theta$ for the CPU
time versus the linear size of the 3D lattice ($t_\linp{cpu}\propto
L^\theta$). The number in parenthesis shows the cubic patch size.\par}
\endinsert


\section{4.5. Warmups for the PRG.}

\subsectionp{Kadanoff Partition into Blocks.}

Although it does not affect the final convergence of the RSRG explicit
algorithms (long range DMRG and PRG), a good warmup may speed up the
process considerably. A technique which may provide the basis of a
full RSRG algorithm on its own (see following chapter) is the {\sl
partition into blocks} \'a la Kadanoff, also known as ``Kadanoff
blocking'' or ``Kadanoff coarse-graining'', which we shall describe in
this section.

The simplest presentation of this technique is for the 1D system with
$N$ sites and fixed or free b.c. Let $H$ be the hamiltonian $N\times
N$ matrix and $M$ be any divisor of $N$. Then, the complete system may
be split into $M$ cells of equal size $f\equiv N/M$.

\newgray{lgray}{.8}
\midinsert
\rput(30,-6){\psline(0,0)(95,0)
\multirput(0,0)(5,0){20}{
\pscircle[fillstyle=solid, fillcolor=lgray](0,0){1.5}}
\multirput(0,0)(25,0){4}{\psframe(-2,-3)(23,3)}}
\ps{10}
\nind{\mayp Figure 13. }{\peqfont Making blocks within a $N=20$ sites
1D graph.}
\endinsert

A characteristic wave--function $\ket|\chi_i>$ for each block is
defined, with $i\in [1,\ldots M]$.

$$\ket|\chi_i>={1\over\sqrt{f}}\sum_{j=(i-1)f+1}^{i\cdot f}
\ket|\delta_j>$$

In other words: normalized states which are uniform over each of the
blocks. After that, the effective hamiltonian is

$$H^R_{ij}=\elem<\chi_i|H|\chi_j>$$

\nind which is a $M\times M$ matrix. The computation of the matrix
elements is immediate. Let us define the sets
$S_i\equiv[(i-1)f+1,\ldots,if]$, for which every pair is disjoint and
whose union makes up the full lattice. Thus,

$$H^R_{ij}= {1\over f} \sum_{k\in S_i}\sum_{l\in S_j} H_{kl}$$

If the original hamiltonian $H$ is the laplacian on the graph, the
effective hamiltonian $H^R$ is called the {\sl collapsed laplacian}
in graph theory \ct[BOL-98].
\pss

The eigenstates of $H^R$ are posteriorly expanded to yield states for
the full graph. Let $a^k_j$ be the $j$-th component ($j\in[1,\ldots
M]$) of the $k$-th eigenstate of $H^R$. Then, the full states are

$$\ket|\phi^k>=\sum_{j=1}^M a^k_j \ket|\chi_j>$$

It is interesting to ask about the similarity between the functions so
obtained and the exact eigenstates of the hamiltonian. Let us see an
example in figure 14: the free particle in a box discretized into
$N=108$ sites, divided into $M=6$ blocks. The number of states to be
considered is $m=4$.

\figura{5.5cm}{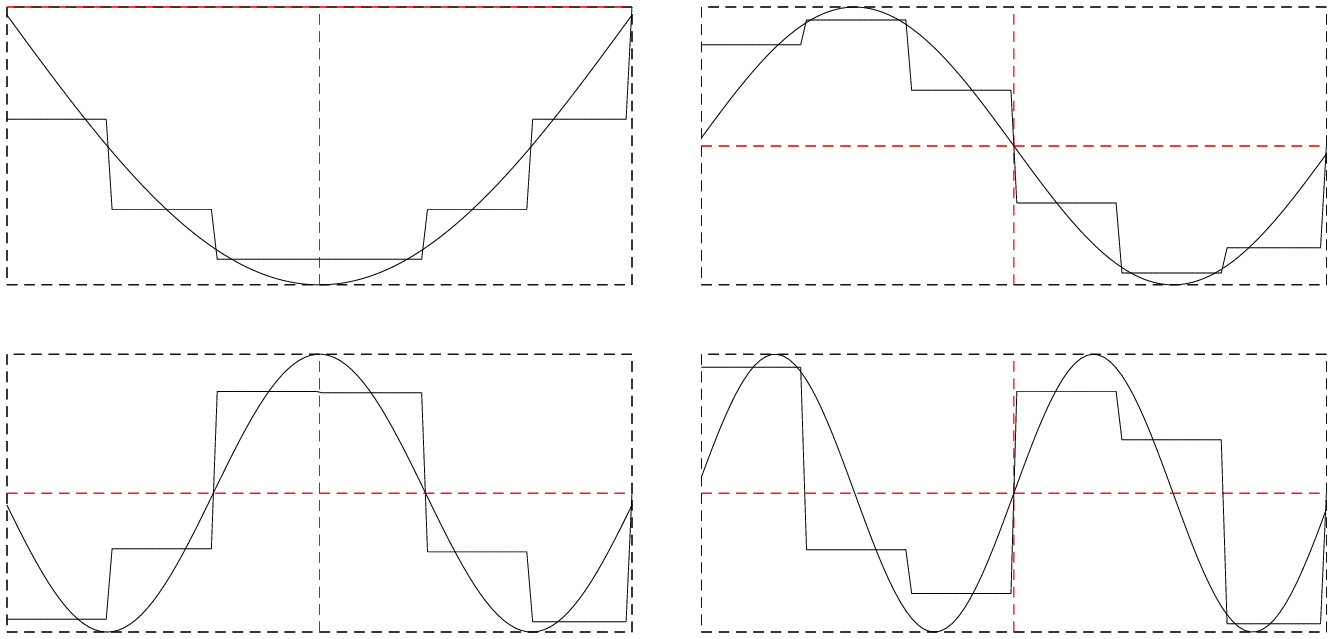}{Figure 14.}{Results of the partition 
of a 1D lattice with $108$ sites split into $6$ Kadanoff blocks. The
smooth curves represent the eigenstates.}

It may be observed in figure 14 that the ``coarse--grained''
wave--functions resemble the authentic ones as much as it is possible
within their own subspace, which is a consequence of the variational
nature of the computation. Nevertheless, the energies of the Kadanoff
states and the exact ones differ enormously, as it may be checked in
table 4.

\midinsert
\centerline{\hfilneg
\sbtabla(4)
\li && Ground state & $1^{\rm st}$ exc. & $2^{\rm nd}$ exc. & 
$3^{\rm rd}$ exc. \cr
\lt Kadanoff && 0.0110035 & 0.0418345 & 0.0863866 & 0.135836\cr
\lt Exact && 0.000830647 & 0.0033219 & 0.00747169 & 0.0132766 \cr
\setabla}
\ps2
\noindent{\mayp Table 4. }{\peqfont Energies of the wave--functions in
figure 14.\par}
\endinsert

The similarity between the warmup and the exact states may be measured
through the ${\cal L}^2$ norm of its difference. We obtain for the
states of figure 14, respectively, $3\%$, $11\%$, $24\%$ and $40\%$
error\footn{The mentioned error is obtained multiplying by $100$
the ${\cal L}^2$ norm of the difference between both functions, being
normalized both of them.}. The fact that a $3\%$ of error in ${\cal
L}^2$ norm may lead to an order of magnitude error in energy
($1375\%$) requires an explanation.

The reason is the inability of the ${\cal L}^2$ norm to apprehend
``all'' the aspects of the similarity between two functions. A Sobolev
norm \ct[TAY-97], which is also sensitive to derivatives (of arbitrary
order) may be more appropriate. As a matter of fact, the energy of the
free particle in a box may be considered to be a {\sl Sobolev-type}
norm:

$$E=\sum_{\<i,j\>} (\phi_i-\phi_j)^2\approx \int_\Omega
|\nabla\phi|^2$$

The Kadanoff wave--functions are {\it smooth\/} when considered in
their own ``blocked'' space, but they are not when extrapolated to a
refined space. These ideas may lead to further approximations, as it
is shown in the next chapter.
\pss

The Kadanoff blocking warmup technique is directly generalizable to
two or three dimensions. Figure 15 shows the wave--functions which
result from the warmup of the particle in a bidimensional box with
fixed b.c.

\figura{3.2cm}{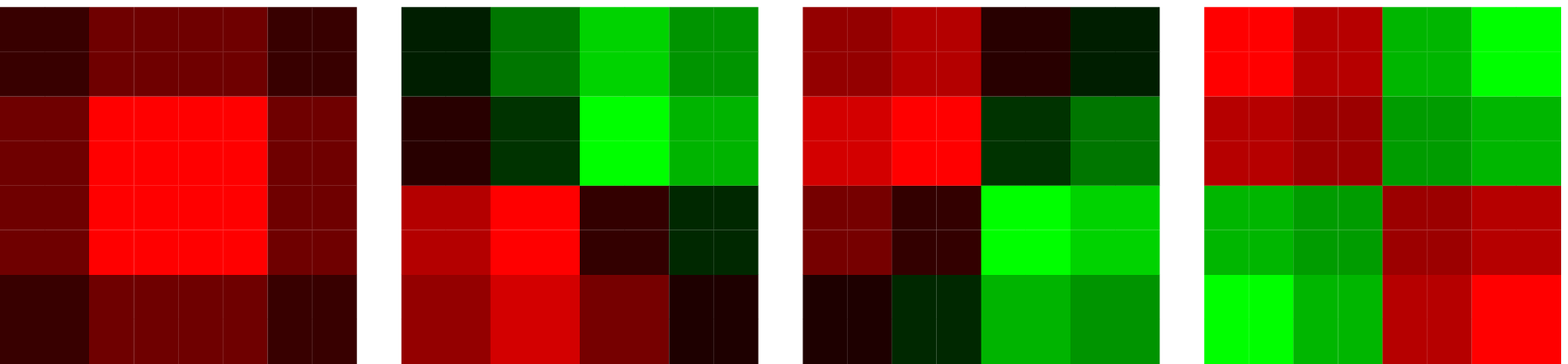}{Figure 15.}{2D Kadanoff warmup for a free
particle in a box with fixed b.c. The reduced Hilbert space has 16
degrees of freedom ($4\times 4$ blocks).}

Albeit the Kadanoff warmup has been employed in all the cases exposed
in the precedent sections, it is not the only possibility.

\subsection{Warmups inspired in Wavelets.}

Wavelet theory is a rather developed branch of applied mathematics
\ct[LEM-89] \ct[PTVF-97]. We have found in it inspiration for the
development of a warmup suitable for PRG. Similar computations have
been carried out by S.R.~White himself in his ``orthlets'' theory
\ct[WHI-99].
\pss

The general idea is to expand into a multi-resolution or {\sl
multi-scale} basis. This basis contains also $N$ states which are
organized in a hierarchical way. Figure 16 shows graphically a 1D
example with $N=8$.

\midinsert
\pss
\psset{arrowsize=2pt 3}
\def\eje{\psframe(0,-4)(16,4)\psline[linewidth=0.1pt](0,0)(16,0)}
\def\bl{\psframe[fillstyle=solid,fillcolor=lgray]}
\rput(50,0){
\rput(0,0){\eje\bl(0,0)(16,4)}
\rput[l](18,0){\peqfont $W(0,0)$}
\psline{->}(8,-5)(8,-11)
\rput(0,-17){\eje\bl(0,0)(8,4)\bl(8,0)(16,-4)}
\rput[l](18,-17){\peqfont $W(1,0)$}
\psline{->}(8,-22)(-8,-29)\psline{->}(8,-22)(24,-29)
\rput(-16,-34){\eje\bl(0,0)(4,4)\bl(4,0)(8,-4)}
\rput[l](2,-34){\peqfont $W(2,0)$}
\rput(16,-34){\eje\bl(8,0)(12,4)\bl(12,0)(16,-4)}
\rput[l](34,-34){\peqfont $W(2,1)$}
\psline{->}(-8,-40)(-24,-48)\psline{->}(-8,-40)(-2,-48)
\psline{->}(24,-40)(40,-48)\psline{->}(24,-40)(18,-48)
\rput(-32,-53){\eje\bl(0,0)(2,4)\bl(2,0)(4,-4)}
\rput[u](-24,-60){\peqfont $W(3,0)$}
\rput(-10,-53){\eje\bl(4,0)(6,4)\bl(6,0)(8,-4)}
\rput[u](-2,-60){\peqfont $W(3,1)$}
\rput(10,-53){\eje\bl(8,0)(10,4)\bl(10,0)(12,-4)}
\rput[u](18,-60){\peqfont $W(3,2)$}
\rput(32,-53){\eje\bl(12,0)(14,4)\bl(14,0)(16,-4)}
\rput[u](40,-60){\peqfont $W(3,3)$}}
\ps{63}
\nind{\mayp Figure 16. }{\peqfont Multi-resolution basis inspired on
wavelets for $N=8$ sites. It might also serve as the starting point for
a larger hierarchy. At each frame, the abscissa represents the
spatial position (1D), and the ordinate stands for the
wave--function.\par}
\pss
\endinsert

The numerical procedure consists of the following steps:
\pss

\bul The original lattice size is chosen, which must be a power of
$2$. The technique is, nonetheless, generalizable to other values.
\ps2

\bul {\it Establishment of the functional basis\/}. We choose a
representation such as that of figure 16. The set of functions
$W(l,i)$ is spanned by two indices: $l$ means the ``level'' and $i$
the position. Function $W(0,0)$ is uniform. All the rest have zero
``average'' and support of magnitude $N\cdot 2^{-l+1}$. Functions at
the same level have fully disjoint supports. Notice that the set
$\{W(l,i)\}$ with $l\in [0,\ldots,\log_2(N)]$ makes up an orthogonal
basis of the full original Hilbert space.
\ps2

\bul {\it Initialization\/}. We establish $m$ initial functions: the
first $m$ functions of lowest order in the set $\{W(l,i)\}$.
\ps2

\bul {\it Iteration\/}. At each step a ``superblock'' is formed with
the $m$ functions which make up the approximation so far and a new one
taken from the set $\{W(l,i)\}$. The superblock hamiltonian is
diagonalized and we retain just the first $m$ states. These states
make up the approximation for the next step.
\pss

The optimum strategy implies to start at the lowest level available
and changing from $l$ to $l+1$ only when convergence at level $l$ has
been reached.

The practical implementation of the method is slow because of the full
computation of the matrix elements. Even though, the results encourage
for further insight and the search for better techniques. A couple of
sweeps at each level is enough to assure convincing results. For
example, for the free particle in a box with $N=256$ sites and fixed
b.c., $3$ sweeps yield the energies shown in table 5.

\pss
\centerline{
\sbtabla(4)
\li && Ground state & $1^\linp{st}$ exc. & $2^\linp{nd}$ exc. &
$3^\linp{rd}$ exc. \cr
\lt Warmup && 0.000155129 & 0.000616833 & 0.0014313 & 0.00253865 \cr
\lt Exact && 0.000149427 & 0.000597684 & 0.00134471 & 0.00239038 \cr
\setabla}
\ps2
\nind{\mayp Table 5. }{\peqfont Energies obtained with the
wavelet-inspired warmup for the free particle in a box with fixed
b.c. and $N=256$ sites.\par}
\pss


\section{4.6 Blocks Algebra.}

The renewal of RSRG techniques is due in great extent to a much freer
usage of the block concept, even though K.G.~Wilson in his study of
the Kondo effect had already employed the site by site growth
pattern. This section is much more abstract than the previous ones,
and intends to provide a general view of the subject.

The methods employed in the whole of this thesis may be simply exposed
is terms of a {\sl blocks algebra}, which we now develop.
\pss

Let us consider a system with an underlying graph structure ${\cal
G}$. Let ${\cal B}({\cal G})$ be the set of the sub-graphs or {\it
blocks\/} of ${\cal G}$. Let $B_k$ and $B_l$ be elements of ${\cal
B}({\cal G})$. The links between sites of $B_k$ and $B_l$ which
do not belong to any of the two blocks shall be contained by
definition by the set $L_{kl}$. Formally, we may write

$$B_k+B_l\equiv B_k\cup B_l \cup L_{kl}$$

The condition for a set of blocks ${\cal P}$ to be a {\sl blocks
algebra} is to be closed under the operation $+$: if $B_k$ and $B_l$
belong to the set, then also does $B_k+B_l$\footn{Notice that the
operation $\cup$ is only set-oriented: $B_k\cup B_l$ means the union
of both blocks, but without restoring the links between them, which is
the key notion.}. Although it is not strictly required, we shall only
take non-overlapping blocks: addition shall only take place if
$B_k\cap B_l=\emptyset$.
\pss

A matricial representation of a blocks algebra is obtained by
assigning a ``block matrix'' to each $B\in {\cal P}$ and a ``hook
matrix'' to each pair of blocks $L\in {\cal P}\times {\cal
P}$. Abusing notation slightly we shall denote by $B_k$ both the
block and its associated matrix, and we shall do the same with hooks.

The dimensions of the matrices must fulfill the following
requirement. A block matrix $B_k$ is a square matrix of arbitrary
dimension $d(k)$, meanwhile $L_{kl}$ must be a rectangular matrix of
dimension $d(k)\times d(l)$. Even more, it shall be necessary that, if
$B_s=B_kk+B_l$, $d(s)\leq d(k)+d(l)$ (an analogue of the triangle
inequality).
\pss

A set $\{B_1,\ldots,B_n,L_{12},L_{13},\ldots,L_{n-1,n}\}$ shall be
termed a {\it representation\/} of a blocks algebra of a
quantum--mechanical system if there is an orthonormal set of system
states $\{\ket|\psi^k_i>\}$ (where $k\in [1,\ldots,n]$ denotes the
block index and $i\in[1,\ldots d(k)]$ denotes the state within that
block) such that

$$(B_k)_{ij}=\elem<\psi^k_i|H|\psi^k_j> \qquad
(L_{kl})_{ij}=\elem<\psi^k_i|H|\psi^l_j>$$

We shall now define the operation $+$ acting on the block matrices of
a system representation. If, in terms of blocks, $B_k+B_l=B_s$, then
operation $+$ acting on the matrices $B_k$ and $B_l$ consists of:
\pss

\bul The building of an effective superblock matrix

$$B_k \oplus B_l \equiv \pmatrix{ B_k & L_{k,l} \cr L_{k,l}^\dagger &
B_l\cr }$$

\item{}which is a square matrix $(d(k)+d(l))\times(d(k)+d(l))$.
\ps2

\bul The diagonalization of the aforementioned matrix and the
retention of the $d(s)$ lowest energy eigenstates (which is always
possible due to the {\it triangle\/} restriction $d(s)\leq
d(k)+d(l)$).
\ps2

\bul Truncation so as to retain an effective matrix for the lowest
states. If $T$ is the matrix whose columns are the weights of the new
global states on the old blocks states (first those of $B_k$ and later
those of $B_l$) then

$$B_s=B_k+B_l=T(B_k\oplus B_l)T^\dagger$$

\item{} Now the non-overlapping requirement may be understood: it
guarantees the orthonormality of the full set of test states:
$\{\ket|\psi^k_i>\}_{i=1}^{d(k)} \cup \{\ket|\psi^l_i>\}_{i=1}^{d(l)}$
\ps2

\bul We must still ``adapt'' the hooks. We may have to perform this
process in an explicit way (i.e.: having recourse to the real
hamiltonian matrix elements and the total wave--functions).
\pss

A representation of the blocks algebra may always be built if the set
{\sl minimum blocks} (i.e.: the sites) are known. In this case, a
series of additions among them provide us with a full representations.
\pss

Two blocks may be added in {\sl presence} of a third block which does
not overlap with them. We shall consider blocks $B_k$ and $B_l$ along
with an extra block named $B_p$. Then, the addition

$$B_k +_p B_l = T_p( B_k\oplus B_l \oplus B_p) T^\dagger_p$$

\nind where $+_p$ means ``addition in presence of $B_p$'', the direct
ternary sum means

$$B_k\oplus B_l \oplus B_p = \pmatrix{ B_k & L_{k,l} & L_{k,p} \cr
	L^\dagger_{k,l} & B_l & L_{l,p} \cr
	L^\dagger_{k,p} & L^\dagger_{l,p} & B_p \cr}$$

\nind and $T_p$ and $T^\dagger_p$ mean the first $d(B_k+B_l)$
eigenstates of the effective superblock matrix with the weights
corresponding to the states $\ket|\psi^p_k>$ removed (and consequently
re--orthonormalized).

The most useful case of ``sum in presence'' is that of {\sl sum in
total presence}: $B_k+_T B_l$, which means that the sum is extended in
such a way that $B_k+B_l+\sum_{p\in R}B_p={\cal G}$, where $R$ is any
partition into blocks of the rest of the system. Thus, the superblock
hamiltonian represents the {\sl complete system}. Of course, the
symbol $+_T$ is not uniquely defined: any partition of $R={\cal
G}-B_k-B_l$ may do.
\pss

We shall consider the following dynamics on a representation of the
system blocks algebra. At each step, the block matrix $B_s$, which
corresponds to $B_k+B_l$ is substituted for the matrix resulting from
$B_k+_TB_l$ for some valid representation of $+_T$.

Let $X_0$ be the original representation of the system, obtained by
any kind of warmup, and let $X_t$ be the representation after $t$
steps. Calling the described procedure an RSRG step and denoting the
set $\{X_t\}$ as an RG trajectory, we observe that the described
process is a variational finite-size RSRG (either implicit or
explicit). CBRG, DMRG and PRG are contained as particular instances. A
fixed point\footn{The only fixed point, as a matter of fact, if the
blocks algebra contains all the minimum blocks --those of a single
site.} of this RG is that in which the states of each block are the
(orthonormalized) projections over the corresponding sites of the
lowest energy eigenstates of the full hamiltonian. In this case, the
eigenvalues of the superblock matrix are the lowest energies of the
system.


\section{4.7. Towards a PRG Algorithm for Many Body Problems.}

Having acknowledged the interest of quantum--mechanical problems, it
is important to remark that the original aim of the effective RSRG
techniques was the application to many body problems in condensed
matter and quantum field theories. The initial development of the BRG
was mostly performed in the study of such problems (see sections 1.4
and 2.1). 1D-DMRG has been employed in a large amount of cases and
with great success to the analysis of problems of highly correlated
electrons (Heisenberg model, Hubbard, $t-J$,... see \ct[HAL-99]). The
natural question is: may the PRG be generalized so as to work on many
particle systems?

The answer is, at the same time, ``yes'' and ``in practice, not until
now''. In principle, being an explicit method, it should suffice to
choose a complete basis of the Hilbert space, write on it the total
hamiltonian and start to operate. The problem, obviously, is the {\it
combinatory explosion\/} in the number of states. For a simple model
such as ITF (see section 2.1), the number of states is $2^L$. A better
alternative is required.

Let us consider as an example the {\sl antiferromagnetic spin $1/2$
anisotropic Heisenberg model} (also known as {\it XXZ model\/}). It is
a quantum magnetism model quite well known \ct[MAN-91], which is
analytically solvable only in 1D via the {\it Bethe Ansatz\/}
\ct[GRS-96]. The model hamiltonian is given by:

$$H=\sum_{\<i,j\>}\[ \Delta S_i^z S_j^z +{1\over 2}\(S_i^+S_j^- +
S_j^+S_i^-\) \]$$

In order to carry out the calculations a full computational platform
for the analysis of many body problems was developed (see appendix B
for computation related questions).

\subsection{Sites Addition and Warmup.}

Let us consider the graph which represents the system and let us
establish a {\it building list\/}, i.e.: an order for the sites to be
added in such a way that the graph is never disconnected and, besides,
the number of created links at each step is minimum. Let us denote that
list by $\{B_i\}_{i=1}^N$.

Let $m$ be the number of states which we intend to obtain from the
hamiltonian. Let us take the first integer $n_0$ such that $2^{n_0}>m$
and let us build exactly the effective hamiltonian matrix for the
first $n_0$ sites of the list\footn{To be precise, because of the
natural division of the Hilbert space into sectors with defined $S^z$,
{\it boxed-matrices\/} were employed. Calling ``box'' $C_{a,b}$ to the
matrix elements set between states in the sectors $\<S^z\>=a$ and
$\<S^z\>=b$, these matrices store explicitly only the boxes which
contain non-null elements.}. This matrix is diagonalized and we retain
the $m$ lowest states. If it is observed that, when ``cutting'' at the
$m$-th state we have destroyed a multiplet, the necessary states are
added in order to complete it. Let $m_r$ be the ``real'' number of
retained states.
\pss

Now the system is prepared for the ``{\it modus ponens\/}'' or sites
addition process. The effective matrices are calculated\footn{Split
into boxes, of course.} for all the operators which may be needed to
keep on building, storing the matrix elements among the $m_r$ retained
states.

Let us call ``active edge'' of the system, ${\cal A}$, to the set of
sites which have not completed their links yet. We shall store,
therefore, the operators $S_p^z$, $S_p^+$ and $S_p^-$ for the sites
$p$ of that active edge, along with the total hamiltonian matrix
$h_\Sb$ for the system.
\pss

A new site to be added is chosen, which is given by $q=B_{n_0+1}$. We
call $G(n_0)$ the set of sites which already belong to the system. The
addition of $q$ is performed in the following way:
\ps2

\bul The stored operators are tensorially right--multiplied by the identity
on the new site:

$$\forall p\in{\cal A},\quad\forall \mu\in\{z,-,+\}\qquad
S^\mu_p \to S^\mu_p \otimes I_2$$

\nind where $I_2$ denotes the identity matrix of dimension $2\times
2$.
\ps2
\bul The Pauli matrices for the new site are tensorially
left--multiplied by the identity on the rest of the system:

$$\forall\mu\in\{z,-,+\}\qquad \hat S_q^\mu\equiv I_{m_r}\otimes
\sigma^\mu$$

\nind where $\sigma^\mu$ denotes, logically, the $2\times 2$ matrices
associated to each component.
\ps2

\bul The ``links'' are calculated between the site $p$ and those of
its neighbours which already belong to $G(n_0)$:

$$H_\linp{new}=\sum_{p\in N(q)\cap G(n_0)} \Delta S^z_p \hat S^z_q +
{1\over 2}\(S^+_p \hat S^-_q + \hat S^+_p S^-_q\)$$

\bul This operator shall be given by a matrix $(2m_r)\times (2m_r)$,
which must be summed to the old hamiltonian matrix tensorially
right--multiplied by the identity on the new site:

$$\hat H_\Sb= h_\Sb \otimes I_2 + H_\linp{new}$$

\bul $H_\Sb$ is diagonalized and the $m$ lowest states are retained
(again taking care not to break multiplets). The operators are
renormalized with the truncation operator built from these states and
the process has completed a RG cycle.
\pss

It is necessary to remark that for the anisotropic Heisenberg model
the previously stated method, being a mixture of the BRG and the site
by site growth techniques (used by Wilson on the Kondo problem and in
the DMRG), works qualitatively well, and with a reasonable degree of
precision.

As an example, table 6 shows some results for square lattices with free
boundary conditions of small sizes. Conserving only $m=8$ states
(thus, $n_0=3$), we obtain less than $5\%$ error. The first excited
states share the same degree of matching, both in their spin structure
and in their energy.

\midinsert
\centerline{
\sbtabla(2)
\li System && Exact && PRG \cr
\lt Linear $N=10$, $\Delta=0.125$ && $-3.5903$ && $-3.4995$\cr
\lt Linear $N=11$, $\Delta=1.0$ && $-10.736$ && $-10.655$\cr
\lt $3\times 3$, $\Delta=0.25$ && $-4.7443$ && $-4.5486$\cr
\lt $3\times 3$, $\Delta=0.5$ && $-6.8797$ && $-6.3389$\cr
\setabla\hfil}
\ps2
\nind{\mayp Table 6. }{\peqfont Results of the PRG warmup for a many
body problem on linear and square lattices, compared to the exact
diagonalization. The b.c. are always free.\par}
\endinsert

\ps7
\subsectionp{Absence of Site Substraction Algorithm.}

\newtoks\eiz \newtoks\ede
\eiz={\hbox{\raise8pt\hbox{\peqfont$\leftarrow$}\kern-6pt\hbox{$E$}}}
\ede={\hbox{\raise8pt\hbox{\peqfont$\rightarrow$}\kern-9pt\hbox{$E$}}}

Once the former cycle has been finished, the problem is to make
sweepings, so as to approach asymptotically the exact solution. The
addition of sites is a well understood variational process. The PRG
cycle works through the substraction ({\it modus tollens\/}) and
posterior addition of the site. But, how can we perform this removal?
We must start by saying that it {\it has not been carried
out yet\/}. The reasons for this failure are interesting and shall be
displayed in this paragraph.
\pss

Let us consider a family of operators which halve the dimension of the
Hilbert space by fully removing a site: $E_p:{\cal H}^N\mapsto {\cal
H}^{N-1}$. Let us choose now any orthonormal set of states
$\ket|\phi_i>$.

The set $\ket|\phi'_i>\equiv E_p\ket|\phi_i>$ {\it needs not\/} be
orthonormal. It would be if each state belonged to a sector with $S^z$
well defined. But in this case, the states of the set $\ket|\phi'_i>$
would mix different sectors up. These two complications are the gate
through which the difficulties enter.

Let us suppose that we lift the condition that every state has $S^z$
well defined, since it is only technically desirable, but not a required
condition for the computation (as it is the orthonormality of the
states). In this case, we must re--orthonormalize the states. It is
not too difficult to do, as the reader may observe, since the full
chapter 4 has successfully dealt with such troubles.

But the situation is now rather different. Let us consider the matrix

$$C^p_{ij}\equiv \<\phi'_i|\phi'_i\> = \elem<\phi_i|\the\eiz_p\;
\the\ede_p|\phi_j>$$

\nind (where $\the\eiz$ means that the operator acts on its left)
which serves to compute the basis change matrix $G^i_j$ which
re--orthonormalizes through the fast Gram-Schmidt method. Let us apply
this matrix on the states so as to obtain a new series
$\ket|\phi^{-p}_i>=G^i_j\ket|\phi'_j>$. For all the operators ${\cal
O}$ which act trivially on site $p$ we have:

$${\cal O}_{ij}=\elem<\phi^{-p}_i|{\cal O}|\phi^{-p}_j>=G^i_k
G^j_l\elem<\phi_k|{\cal O}|\phi_l>$$

\nind while the operators acting on the site must be calculated
anew. The removal process has successfully ended and we may add it
variationally again.
\pss

The problem appears in its whole magnitude when the moment to update
the operators comes before starting the next RG step. As it is
logical, the matrix $C^p_{ij}$ must be stored {\it for each value of
$p$\/} at every step, and they must be updated. How are we to do it?
Let us consider $C^q$ with $q\neq p$. Its matrix elements on the
states $\ket|\phi^{-p}_i>$ would be

$$\tilde C^q_{ij}=
\elem<\phi^{-p}_i|\the\eiz_q\; \the\ede_q|\smash{\phi^{-p}_j}>=
\elem<\phi_k| G^i_k \the\eiz_p\the\eiz_q\; \the\ede_q\the\ede_p G^j_l
|\phi_l>$$

The problem is that the scalar products between the states
$E_pE_q\ket|\phi_i>$, in which two sites have been removed instead of
one, can not be ``deduced'' from the knowledge of the matrices
$C^s_{ij}$ with $s\in S$ indexing the set of sites. It is required,
therefore, to store the matrices of {\it mixed\/} scalar products:

$$C^{pq}_{ij}=\elem<\phi_i|\the\eiz_p\the\eiz_q\;\the\ede_q\the\ede_p|\phi_j>$$

\nind for every couple $pq$ (without caring about order). The reader
may now suspect that the game goes on. Effectively, in order to
renormalize these two-sites operators the knowledge is required of the
three-points operators... and so on {\it ad nauseam\/}. Due to
elementary combinatorics we require

$$N + \( {N\atop 2}\) + \( {N\atop 3} \) + \ldots + \( {N \atop N} \)
= 2^N-1$$

\nind $m\times m$ matrices. I.e.: a quantity of information superior
to the storage of the full wave--functions. If this difficulty reminds
of the {\sl closure problem} in turbulence theory, the absence of
solution reinforces the analogy.
\pss

In the end, the reason of the failure of the many body PRG is the same
reason for which DMRG {\it needed\/} to become explicit when dealing
with long-range or multidimensional problems. The open question is:
``is it {\it really\/} necessary for an RSRG variational algorithm on
non-tree systems to be explicit?''


\section{4.8. Application of the PRG on a Model of Excitons with Disorder.}

\subsectionp{Localization, Delocalization and Disorder.}

It is well known since the late 50's \ct[AND-58] \ct[MT-61], that the
eigenstates of a 1D quantum--mechanical system with noise strongly
tend to be exponentially localized, {\it for any amount\/} of
noise. Anderson and other authors \ct[AALC-79] proved that, under very
general conditions, any decorrelated noise repeated the same situation
in 2D (see \ct[LR-85] and \ct[RMD-00] for further explanations and a
series of references).

The panorama changed drastically when it was discovered that the
absence of correlation was crucial. Moreover, the high conductivity of
certain materials where disorder is present was explained by proving
the existence of correlations.

Recently, F.~Dom\'{\i}nguez-Adame et al. have proved \ct[RMD-00] the
existence of delocalized states in 1D for systems with decorrelated
noise, {\it but with\/} long range order. In their computations they
studied the model

$${\cal H}=\sum_i \epsilon_i\ket|\delta_i>\bra<\delta_i| +
\sum_{i,j}J_{ij}\ket|\delta_i>\bra<\delta_j|$$

The hopping terms are chosen so as they decay with distance according
to a power law:

$$J_{ij}= J\;|i-j|^{-\alpha}$$

\nind where $J$ is the coupling between nearest neighbours. The noise
appears in the values $\epsilon_i$, which are taken from a uniform
probability distribution on the interval $[-\Delta/2,\Delta/2]$, where
$\Delta$ is called the ``disorder factor''. The relevant parameters
are, of course, $\alpha$ and $\Delta$.

Although this model was not proposed in order to explain any real
system, in \ct[RMD-00] some possibilities are considered whose physics
might be modelled by it (e.g., planar dipolar systems).

\subsection{PRG Application.}

In this case, as in that of the excitons on dendrimers, the numerical
experiments involve the obtention of the lowest energy spectrum of a
system with many degrees of freedom. Although in this example there is
no underlying graph structure, PRG is probably the best possible tool
to carry out this computation. The analytical--numerical development
of the bidimensional analogue of this problem was carried out by
F.~Dom\'{\i}nguez-Adame, J.P.~Lemaistre, V.A.~Malyshev,
M.A.~Mart\'{\i}n-Delgado, A.~Rodr\'{\i}guez, J.~Rodr\'{\i}guez-Laguna
and G.~Sierra \ct[DLMRS-01].

In this section we shall focus only on the adaptation of the PRG to
the mentioned calculations, leaving the physical conclusions and the
theoretical implications for later work.
\pss

Since the noise is diagonal and the non-diagonal elements follow a
scaling law, matrix elements of the hamiltonian in the canonical basis
are stored {\it by reference\/}. This means that, in the 1D case, a
vector $D$ of $N$ components is kept which contains the diagonal
elements $D_i=\epsilon_i$ and another vector $V$, whose $k$-th
component has the value $V_k=J k^{-\alpha}$. In the 2D case only the
vector $V$ changes, whose $k$-th component is now $V_k=J
k^{-\alpha/2}$.

In order to compute the matrix element $H_{ij}$ some steps must be
taken:
\ps2
\bul Convert indices $i$ and $j$ into lattice coordinates: $x_i$,
$y_i$, $x_j$ and $y_j$.
\ps2
\bul We compute the square of the distance among them:
$d^2=(y_j-y_i)^2+(x_j-x_i)^2$.
\ps2
\bul We obtain the $d^2$-th component of the vector: $V_{d^2}$.
\pss

The rest of the procedure fully coincides with the one discussed in
the previous sections. The warmup used is the Kadanoff blocking and
the computation of the hamiltonian matrix elements bears no shortcut
due to the absence of graph structure.
\pss

The 1D results from \ct[RMD-00] are checked by our own
computations. Figure 17 shows, for $N=1000$ sites\footn{Our reference
value for the computations is $N=10^4$, but for the sake of a proper
visualization, it has been reduced by an order of magnitude.} in a 1D
chain with $\alpha=3/2$, three ground states for the system, with
$\Delta=1$, $\Delta=8$ and $\Delta=30$.

\midinsert
\pss
\line{\fig{2.7cm}{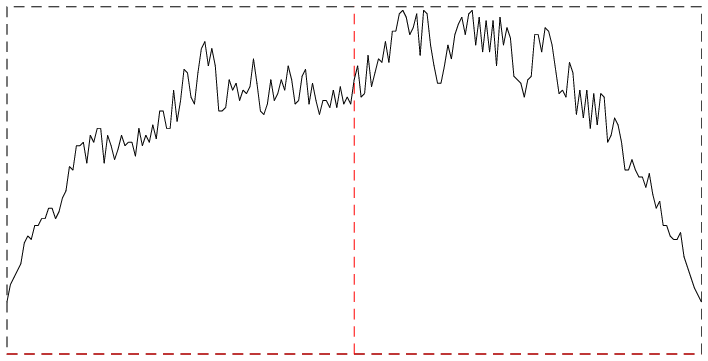}\hfil\fig{2.7cm}{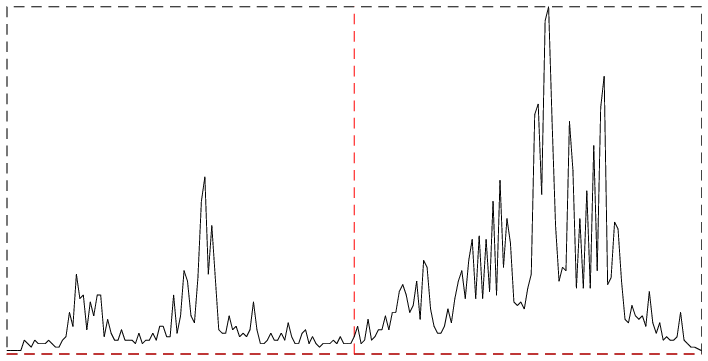}
\hfil\fig{2.7cm}{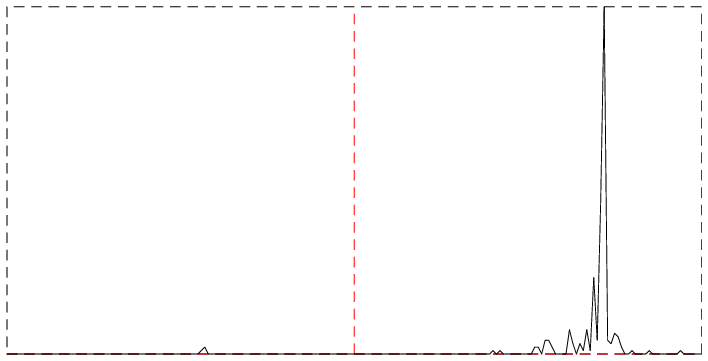}}
\ps2
\nind{\mayp Figure 17. }{\peqfont Ground states of three realizations
of the 1D system with $N=1000$ and $\alpha=3/2$, respectively with
$\Delta=1$, $\Delta=8$ and $\Delta=30$.\par}
\ps2
\endinsert

We define the {\it Inverse Participation Ratio\/} (IPR) as

$$I\equiv\sum_{i=1}^N |\psi_i|^4$$

\nind When it is computed on a normalized wave--function, it takes the
value $1$ only if it is a delta function (concentrated on a single
point), and it is of order $O(1/N)$ when the state is
delocalized. Figure 18 shows the value of the logarithm of the IPR
when we vary $\Delta$ for $N=1000$ and $\alpha=3/2$ in 1D. In the
right-hand plot we observe the number of sewings which were needed to
reach the desired precision of one part in $10^{10}$ for the energies.

\midinsert
\centerline{\figr{6cm}{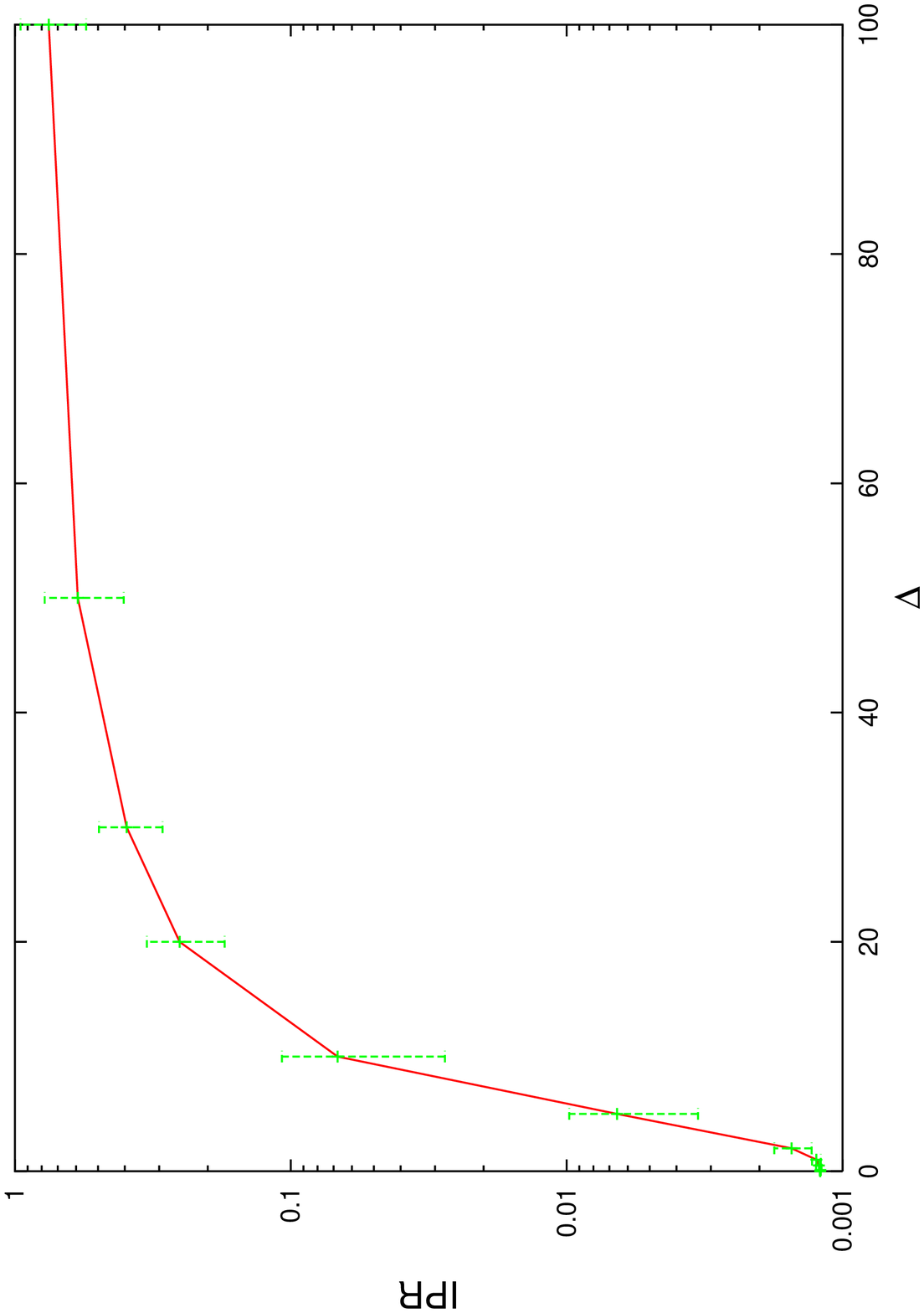}}%
\rput(90,32){\figr{3.5cm}{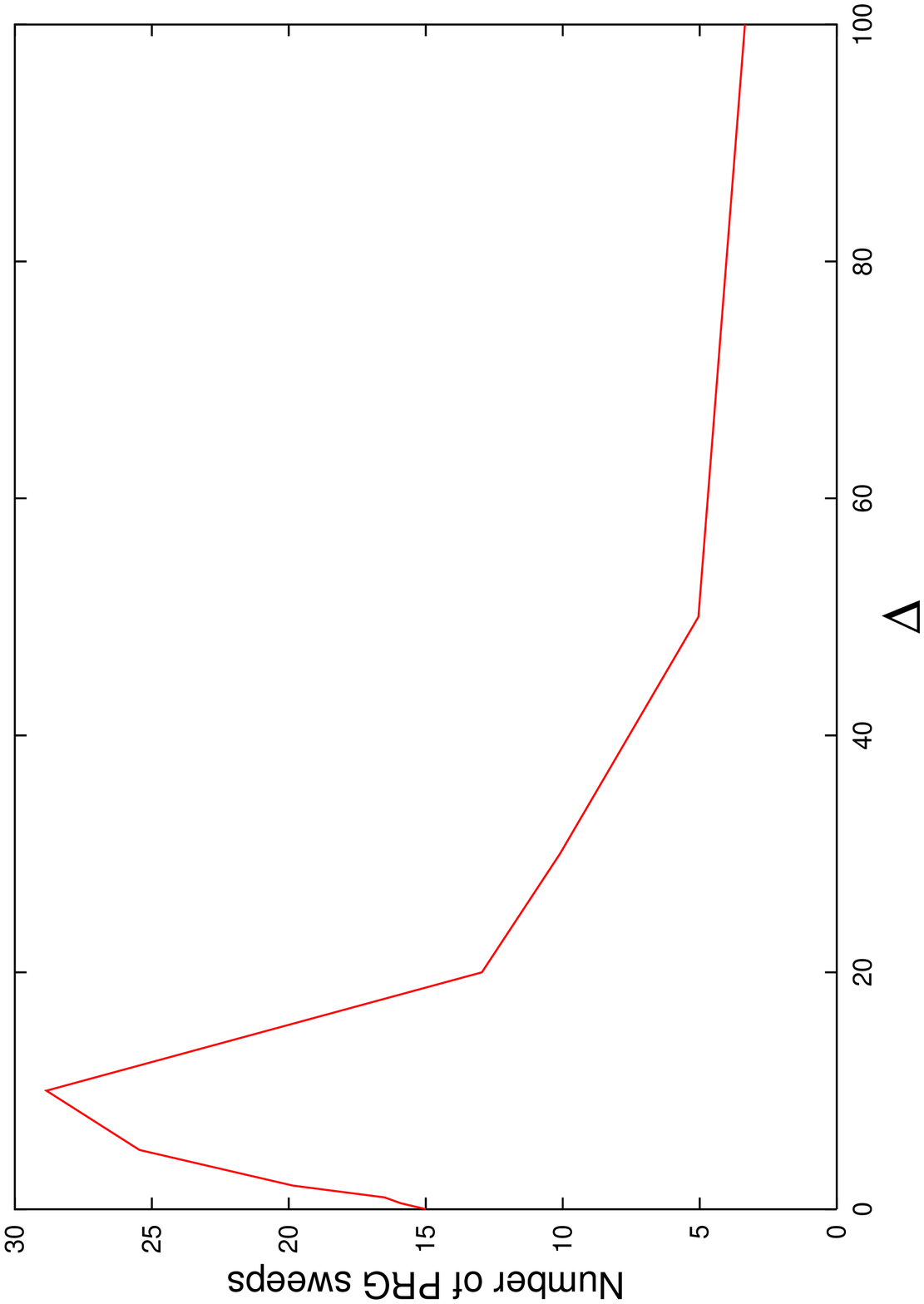}}%
\ps{-2}
\nind{\mayp Figure 18. }{\peqfont The outer plot shows the IPR (in a
logarithmic scale) against the parameter $\Delta$ for 1D chains with
$\alpha=3/2$ and $N=1000$. The inner shows the number of PRG sweeps
required to reach the desired convergence (one part in
$10^{10}$).\par}
\endinsert

The difficulty of the problem is seen to grow in the transition
region, which, as it may be readily checked in figure 17, corresponds
to wave--functions with the most complex structure. Part of the
following process, which shall not be considered in the present work,
shall be to carry out a {\it multifractal analysis\/} of these
objects.
\pss

The results of the PRG application for the bidimensional case are now
introduced. Figure 19 shows the dependence of the IPR on the linear
size of the lattice for $\alpha=3$ and $\Delta=5$. Figure 20 depicts a
particular realization of the ground state of the system, obtained
with $L=70$.

\figrot{5.5cm}{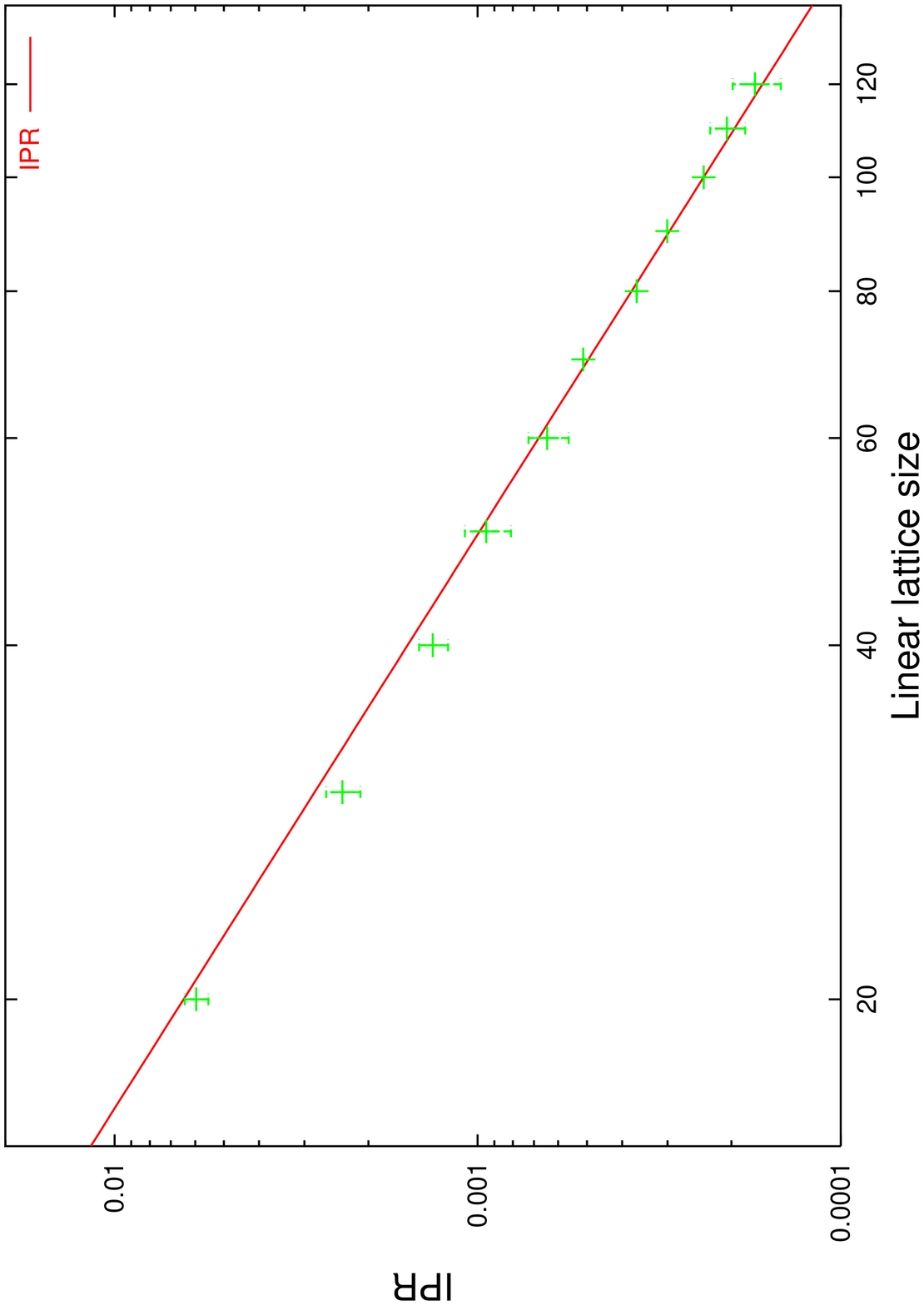}{Figure 19.}{Dependence of the IPR on the
length of the 2D lattice. 20 samples were obtained for each of the
values of $L$ but for the last two ones (10 samples), checking that
the statistics were robust for all the cases. The straight line has
slope $-2.05\pm 0.05$.}

\midinsert
\ps{-5}
\centerline{\figr{6.7cm}{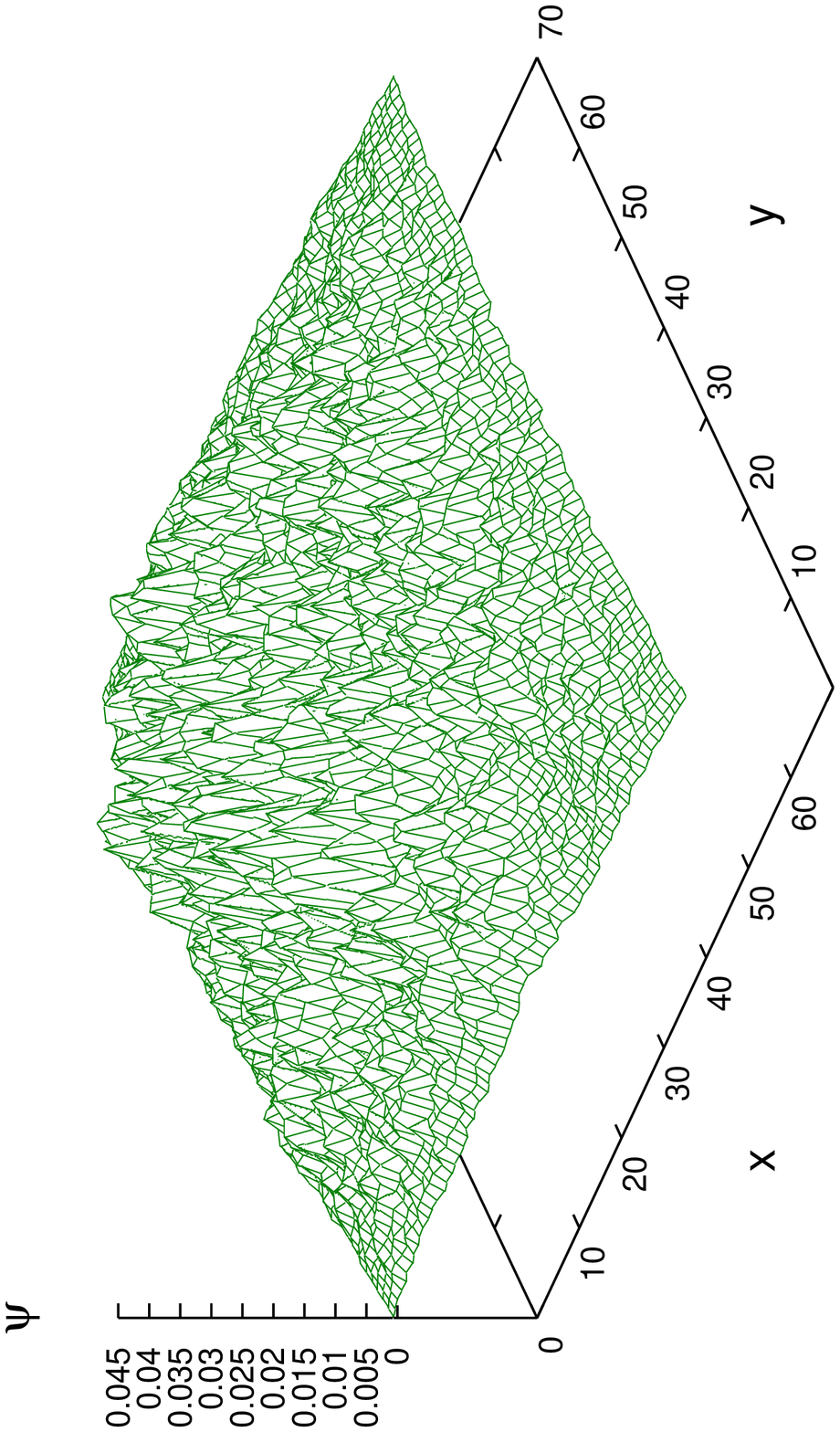}}
\nind{\mayp Figure 20. }{\peqfont Bidimensional wave--function in
a $70\times 70$ lattice for the ground state of a realization of the
system with $\alpha=3$ and $\Delta=5$.\par}
\endinsert


\section{4.9. Bibliography.}

\cx[AALC-79] {\may E. Abrahams, P.W. Anderson, D.C. Liciardello,
V. Ramakrishnan}, Phys.~Rev.~Lett. {\bf 42}, 673 (1979).

\cx[AND-58] {\may P.W. Anderson}, Phys.~Rev. {\bf 109}, 1492 (1958).

\cx[BLA-82] {\may D.W. Blackett}, {\sl Elementary topology. A
combinatorial and algebraic approach.} Academic Press (1982).

\cx[BOL-98] {\may B. Bollob\'as}, {\sl Modern graph theory}, Springer (1998).

\cx[DGMP-97] {\may A. Dobado, A. G\'omez-N\'{\i}cola, A. Maroto, 
J.R. Pel\'aez}, {\sl Effective lagragians for the standard model},
Springer (1997)

\cx[DLMRS-01] {\may F. Dom\'{\i}nguez-Adame, J.P. Lemaistre,
V.A. Malyshev, M.A. Mart\'{\i}n-Delgado, A. Rodr\'{\i}guez,
J.~Ro\-dr\'{\i}\-guez-Laguna, G.~Sierra}, {\sl Absence of weak
localization in two-dimen\-sional disordered Frenkel lattices},
available at {\tt cond-mat/0201535} and J.~Lumin. {\bf 94-95}, 359-363
(2001).

\cx[GRS-96] {\may C. G\'omez, M. Ruiz-Altaba, G. Sierra}, {\sl Quantum
groups in two-dimensional physics}, Cambridge U.P. (1996).

\cx[GW-93] {\may St.D. G\l azek, K.G. Wilson}, {\sl Renormalization of
hamiltonians}, Phys. Rev. D {\bf 48}, 12, 5863-72 (1993).

\cx[GW-94] {\may St.D. G\l azek, K.G. Wilson}, {\sl Perturbative
renormalization group for hamiltonians}, Phys. Rev. D {\bf 49}, 8,
4214-18. (1994).

\cx[GW-97] {\may St.D. G\l azek, K.G. Wilson}, {\tt hep-th/9707028} (1997).

\cx[HAL-99] {\may K. Hallberg}, {\sl Density matrix renormalization},
in {\tt cond-mat/9910082} (1999).

\cx[LEM-89] {\may  P.G. Lemari\'e} (ed.), {\sl Les Ondelettes en 1989},
Springer (1989).

\cx[LR-85] {\may P.A. Lee, T.V. Ramakrishnan}, {\sl Disordered
electronic systems}, Rev.~Mod.~Phys. {\bf 57}, 287 (1985).

\cx[MAN-91] {\may E. Manousakis}, {\sl The spin 1/2 Heisenberg
antiferromagnet on a square lattice and its application to the cuprous
oxides}, Rev.~Mod.~Phys. {\bf 63}, 1, 1-62 (1991).

\cx[MDS-99] {\may M.A. Mart\'{\i}n-Delgado, G. Sierra}, {\sl A Density Matrix
Renormalization Group Approach to an Asymptotically Free Model with
Bound States}, {\tt hep-th/9903188} and Phys.~Rev.~Lett.~{\bf 83},
1514 (1999).

\cx[MRS-00] {\may M.A. Mart\'{\i}n-Delgado, J. Rodr\'{\i}guez-Laguna,
G. Sierra}, {\sl Single block formulation of the DMRG in several
dimensions: quantum mechanical problems}, available at {\tt
cond-mat/0009474}, and Nucl.~Phys.~B {\bf 601}, 569-590 (2001).

\cx[MT-61] {\may N.F. Mott, W.D. Twose}, Adv.~Phys. {\bf 10}, 107 (1961).

\cx[PTVF-97] {\may W.H. Press, S.A. Teukolsky, W.T. Vetterling,
B.P. Flannery}, {\sl Numerical Recipes in C}, Cambridge U. P. (1997),
and freely available at {\tt http://www.nr.com}.

\cx[RMD-00] {\may A. Rodr\'{\i}guez, V.A. Malyshev, F. Dom\'{\i}nguez-Adame},
{\sl Quantum diffusion and lack of universal one-parameter scaling in
one-dimensional disordered lattices with long-range coupling},
J.~Phys. A (Math.~Gen.) {\bf 33} L161-66 (2000).

\cx[TAY-97] {\may M.E. Taylor}, {\sl Partial differential equations}, 
Springer (1997).

\cx[WHI-99] {\may S.R. White}, {\sl Electronic structure using DMRG},
in {\sl Density-matrix Renormalization}, notes from the Dresden {\it
workshop\/} in 1998, ed. by I.~Peschel et.~al., Springer
(1999).

\pasacap
\def\lin#1{{\hbox{\mathtext #1}}}
\def\linp#1{{\hbox{\mathsubtext #1}}}

\chapter{5. An RSRG approach to Field Evolution Equations.}

\subsection{Synopsis.}

5.1. Introduction.

5.2. RSRG Formalism for Field Evolution Prescriptions.

5.3. Cells Overlapping Truncators.

5.4. Applications and Numerical Results.

5.5. Towards a Physical Criterion for Truncation.

5.6. Bibliography.
\pss

The present chapter introduces the main lines of an RSRG approach to
classical field evolution equations which is being developed by the
author in collaboration with A.~Degenhard\footn{{\it Dept. f\"ur
Theoretische Physik --Universit\"at Bielefeld} (Germany) and {\it
Physics Dept. --Institute of Cancer Research} (London).}. Albeit some
interesting results have already been obtained \ct[DRL-01{\mayp A}]
\ct[DRL-01{\mayp B}], the exposition is mainly focused on the
fundamental ideas, remarking the close connection to the rest of the
RG techniques in this work.

\section{5.1. Introduction.}

Field evolution equations, whether deterministic or stochastic, are a
fundamental tool in many branches of physics, e.g. hydrodynamics,
surface growth phenomena, optics or self-gravitating media. These
equations are usually highly non linear, due to the complexity of the
studied phenomena.
\pss

A constant background topic during the previous chapters was the
consideration that physical measurements always take place within a
certain {\it observational frame\/}, which includes not only the
reference system and the set of observables to be measured, {\it but
also\/} the ``grain size'' (or ultraviolet (UV) cutoff) and the
``plate size'' (or infrared (IR) cutoff). As relativity theory was
born from the analysis of the changes of reference frames and quantum
mechanics from the study of the influence of the measurement of
different observables on a process, renormalization group theory stems
from the changes suffered by a physical system when modifying the
observational frame.

Thus, when modelling physical phenomena, {\sl partial differential
equations} (PDE)\footn{We shall adopt the term PDE with a rather wide
meaning, covering any field evolution equation on a continuous
space-time.} can have {\it no real physical meaning\/}. In other
words: space can not be assumed to constitute a mathematical continuum
as far as physical theories are concerned. Observational frames must
{\it always\/} have attached their IR- and UV-cutoffs. The continuous
space-time hypothesis is mathematically equivalent to taking the limit
in which UV-cutoff $\to\infty$, which is highly non-trivial.

As a matter of fact, it may be argued that the continuum hypothesis
has been a handicap in the conceptual development of physical theories
by spawning them with UV and/or IR catastrophes. Chaotic dynamics,
scaling anomalies, divergences in QFT... were poorly understood
pathologies before they received the appropriate treatment.
\pss

A field evolution {\sl prescription} must be an algorithm which
provides, once a set of observables has been given (along with their
uncertainty), the probability density for the same observables after
some time. Because of overusage of the term ``discretization'', we
have chosen the term {\sl aspect} to denote any such a set of
observables. A prescription, therefore, evolves a field aspect in
time.

PDEs may often be converted into field evolution prescriptions, but
not always. Models with high Lyapounov exponents (for which sometimes
not even an existence theorem for the solutions is available) yield
important problems in practice. For example, currently there is no
existence and uniqueness theorem for solutions of the 3D Navier-Stokes
equation (used to describe newtonian fluid mechanics). It might be not
a coincidence that there is {\it neither\/} a practical scheme for the
simulation of high Reynolds number flows.

The process through which PDE are converted into field evolution
prescriptions is usually called ``discretization''. This process
reduces the continuous field equations, which have infinite degrees of
freedom, into a finite-dimensional model. The regions removed from the
phase space are expected, in some sense, to be {\it irrelevant\/}. In
some cases there is a clear-cut algorithm to choose the relevant
degrees of freedom, but more often (e.g. almost all non-linear PDE) it
does not exist, and physical intuition is the only guide.
\pss

It is possible to reformulate the question in more general terms if an
infinite number of degrees of freedom is not assumed for the initial
equation. Let us consider the reduction problem from a field evolution
prescription into another one which has a smaller number of degrees of
freedom. Let us denote by $\phi$ an aspect associated to the original
prescription, and $\phi'$ a {\sl renormalized} aspect associated to
the new one. There must be an operator $R$ such that

$$\phi'=R(\phi)$$

\nind which we shall denote as {\sl coarse-graining} or {\sl
truncation} operator. Of course, if there is a {\it real\/} reduction
of the number of degrees of freedom, it shall be impossible to reconstruct
the original ones, so the inverse operator $R^{-1}$ does not exist.

The evolution prescription for the initial discretization may be given
formally by

$$\pl_t\phi=H\phi$$

\nind (where $H$ is the {\sl evolution generator}, for which 
linearity is {\it not\/} assumed) and the evolution of the
renormalized aspect shall be given by

$$\pl_t\phi'=H'\phi' \equiv RHR^p \phi'$$

\nind where $R^p$ is a suitably chosen {\sl pseudo-inverse}
operator. (Of course, $RHR^p$ is a formal operation, whose exact
implementation depends on the nature of $H$). The action of different
compatible $R$ operators constitute a {\sl renormalization
semi-group}. Usual discretizations from continuous space into any
finite-dimensional aspect are implemented by an $R$ operator with
an infinite degree of reduction.
\pss

To sum up, we have introduced the following concepts:
\ps2

\bul {\it Aspect of a field\/}: any finite set of observables,
preferably along with their uncertainties or probability distributions,
for an extended object (a field).
\ps2
\bul {\it Field evolution prescription\/}: a set of rules to
compute the time evolution of an aspect of a field.
\ps2
\bul {\it Truncation or coarse--graining operator\/}: operator which
reduces the number of degrees of freedom of an aspect of a field.
\ps2
\bul {\it Embedding operator\/}: Pseudoinverse of the former operator,
which ``refines'' an aspect.
\pss

Of course, we might have conserved the more classical terms
``discretization'' and ``discretized evolution equation'', but we have
chosen not to do since the change of meaning might make the statements
misleading.
\pss

Much interest has been devoted to this subject in the last years, and
this work is greatly influenced by the recent contributions of
N.~Goldenfeld and his group: \ct[GHM-98] and \ct[GHM-00]. These works
were themselves inspired ultimately by the {\sl perfect action}
concept of Hasenfrantz and Niedermayer \ct[HN-94], which intended to
remove ``lattice artifacts'' from lattice gauge theories, and that of
Katz and Wiese \ct[KW-97] about the application of the same concept to
fluid mechanics.

The former works employed a truncation or coarse--graining operator
$R$ created by a blocks fusion scheme \`a la Kadanoff. The most
important innovation of the present work is the consideration of a
wider set of possible $R$ operators. Sections 5.2, 5.3 and 5.4 deal in
some detail with the possibility of a geometric approach based on the
overlapping of cells \ct[DRL-01{\mayp A}]. Section 5.5 drafts the
possibility to use a {\sl physical} criterion (i.e.: problem
dependent) in order to choose $R$ \ct[DRL-01{\mayp B}].


\section{5.2. RSRG Formalism for Field Evolution Prescriptions.}

\subsection{Evolution Prescriptions on Partitions.}

Let us consider the following evolution prescription

$$\pl_t \phi_i = H_{ij}\phi_j \eq{\linearham}$$

\nind This equation may represent any discretization of a linear PDE,
whether the algorithm is explicit or implicit. The form [\linearham]
may also accommodate non-local linear equations and those of higher
order in time, along with some complex boundary conditions. Operator
$H$ shall be known as {\sl evolution generator} or {\sl hamiltonian}.
\pss

Some non-linear equations may enter this formalism rather easily. For
example, any quadratic evolution generator might be implemented as

$$\pl_t \phi_i = Q_{ijk}\phi_j\phi_k + H_{ij}\phi_j \eq{\quadham}$$

Surface growing phenomena as governed by the Kardar--Parisi--Zhang
(KPZ) equation \ct[KPZ-10] \ct[BS-95] or one-dimensional turbulence
as described by Burgers equation \ct[BUR-74] \ct[TAF-81] equation may
be studied this way. 

\subsection{Truncation Operator.}

The natural interest in the construction of $R$ maps can stem either
from limited computer resources or due to theoretical reasons,
since it may constitute a tool for successively integrating out the
irrelevant degrees of freedom.

Field aspects described by equation [\linearham] dwell naturally in a
vector space $E^N$. A truncation operator $R:\,E^N\mapsto E^M$ defines
a ``sub--discretization'', and we shall denote the effective field as
$\phi'\in E^M$. Therefore, the new aspect only has $M$ degrees of
freedom. Thus, $R$ must have a non-trivial kernel.

The election of the $R$ operator is the key problem. Ideally, it
should depend on the problem at hand, i.e. on the field equation and
the observables to be measured. In the next section a purely
geometrical approach is introduced which is independent of the physics
of the dynamical system, but which uses a quasi-static truncation
procedure for a careful selection of the relevant degrees of freedom.

The truncation operator shall be forced to be linear\footn{Linearity
on the truncation operator is a highly non-trivial imposition. It
forces the non-mixing nature of the evolution generators of different
``arity'', i.e.: a purely quadratic term can not develop a linear term
under such a RG construction. This is dubious, e.g., for Navier-Stokes
equation, where the nonlinear term is supposed to develop a
contribution to the viscous (linear) term.}. This enables us to write
its action as

$$\phi'_I = R_{Ii}\phi_i$$

\nind Here we denote the transformed aspect components with primes
and capital letter indices.

\subsection{Embedding Operator.}

If the $R$ operator had a trivial kernel, an inverse operator might be
written, $R^{-1}$ (the {\sl embedding} operator) and the following
equation would be exact

$$\pl_t \phi_i = H_{ij} R^{-1}_{jJ} \phi'_J$$ 

\nind I.e.: the evolution of the exact aspect might be found from
the values of the truncated one. Therefore, one might integrate a
prescription which has only got $M$ degrees of freedom through equation

$$\pl_t \phi'_I = R_{Ii} H_{ij} R^{-1}_{jJ} \phi'_J \equiv
H'_{IJ} \phi'_J  \eqno{[\linearham']}$$

\nind where $H'$ is the {\sl renormalized evolution generator}. Having
evolved the reduced aspect, the original aspect may be found:

$$\phi_i(t) = R^{-1}_{iI} \phi'_I(t)$$

We may express this situation by the commutative diagram:

\advance\eqnumber by 1
\global\edef\commatm{\the\eqnumber}
{\input eplain
$$\commdiag{E^N & \mapright^H & E^N \cr
           \adjmapdown\lft{R}\rt{R^{-1}} &  & \adjmapup\rt{R}\lft{R^{-1}}\cr
           E^M & \mapright^{H'} & E^M} \eqno{[\commatm]}$$
}
\eqnumber=\commatm

Equation [\linearham'] requires less CPU time and storage capacity
than equation [\linearham] when simulated on a computer, but this
situation is generally impossible: the truncation operator has a
non-trivial kernel, so there can be no real inverse. Nevertheless, it
is possible to find an ``optimum'' pseudoinverse: an operator $R^p$
which fulfills the Moore-Penrose conditions (see \ct[GvL-96]). 

$$ \displaylines{ RR^pR =R \qquad  R^pRR^p=R^p \cr
(R^pR)^\dagger=R^pR \qquad (RR^p)^\dagger=RR^p}$$

\nind These equations are fulfilled only if $R^p$ is the singular values
decomposition (SVD) pseudoinverse of $R$. $R^p$ is an
``extrapolation'' operator, which takes a (reduced) discretization
from $E^M$ and returns a full $E^N$ one.
\pss

Operators $R$ and $R^p$ play the same r\^ole as the truncation
operators we have met throughout this work, but some extra
considerations are needed.

Operator $R$ as a full matrix retains a big amount of spurious
information, inasmuch as the only relevant datum is {\sl its kernel},
i.e.: the degrees of freedom which are neglected. If $R^p$ is the SVD
pseudo-inverse, then $RR^p$ is the identity on $E^M$, and $R^pR$ is a
projector on the {\sl relevant degrees of freedom} subpsace of
$E^N$. This operator shall also be known as the {\sl reduction}
operator.

The information about these degrees of freedom is stored in the rows
of $R$, which may be read as vectors spanning the retained
subspace. These vectors need not form an orthonormal set (they may
even be non-independent). But an orthonormalization operation
guarantees that $R^p=R^\dagger$. Whenever we make use of an $R$
operator in this work, we shall mean the equivalence class of
operators of the same dimension sharing its kernel. Formally,

$$R\in {\cal GL}(M\ot N)/{\cal K}$$

\nind where ${\cal GL}(M\ot N)$ represents the set of matrices taking
$N$ degrees of freedom into $M$ and ${\cal K}$ stand for the set of
internal operations in that space which conserve the kernel.

\subsection{Evolution Prescription Transformations.}

With the pseudoinverse $R^p$, the diagram [\commatm] does not commute
in general and its ``curvature'' represents the error within the
procedure. The renormalized evolution generator is written as

$$H'_{IJ}=R_{Ii} H_{ij} R^p_{jJ} \eq{\renlinear}$$

\nind where indices are kept for clarity. A renormalized quadratic 
evolution generator might be written like this

$$Q'_{IJK} = R_{Ii} Q_{ijk} R^p_{jJ} R^p_{kK} \eq{\renquad}$$ 

\nind This expression shall be shorthanded as $Q'=RQR^p$ so as to unify
notation. Higher degree operators are possible, of course, but for the
sake of simplicity we shall restrict to the first
two\footn{Trascendental evolution generators, such as that of the
Sine-Gordon equation, do not fit in this formalism.}.
\pss

The schedule for all the simulations which shall be presented in the
rest of this work is
\pss

$\bullet$ Present an evolution generator $H$ (at most quadratic) and
an initial field $\phi(0)$.
\ps1

$\bullet$ Calculate the exact evolution and obtain $\phi(t)$.
\ps1

$\bullet$ Propose a truncation operator $R$ and obtain its pseudoinverse
$R^p$.
\ps1

$\bullet$ Compute the renormalized hamiltonian and the truncated
initial field: $H'=RHR^p$ and $\phi'(0)=R\phi(0)$.
\ps1

$\bullet$ Simulate the renormalized evolution on $\phi'(0)$ and obtain
$\phi'(t)$.
\ps1

$\bullet$ Compare $\phi(t)$ and $R^p\phi'(t)$.
\pss

We distinguish between the ``real space error'', which is given by the
${\cal L}^2$ norm of $[\phi(t)-R^p\phi'(t)]$ (a vector from $E^N$) and
the ``renormalized space error'', which is the ${\cal L}^2$ norm of
$[R\phi(t)-\phi'(t)]$ (a vector in $E^M$). These two errors need not
coincide. It is impossible for the first one to be zero for all the
functions in a discrete functional space, but not for the second
one. In that case, the retained degrees of freedom are exactly evolved
despite the information loss. We shall speak of {\sl perfect action}.


\section{5.3. Overlapping-Cells Truncators.}

In the former section an abstract formalism was introduced for the
truncation operators. In this section a series of concrete rules shall
be proposed to build up $R$ operators by geometric principles.

\subsection{Cells Overlapping.}

Let ${\cal P}$ be a partition of a given manifold ${\cal M}$ (possibly
with boundary) into the cells $\{C_i\}_{i=1}^N$. Let $\phi$ be a
scalar field on this region of space. As it was stated above, an {\sl
aspect} of that field shall be defined to be any discrete set of
values which attempts to represent the whole knowledge of a given
observer about the field. The aspect associated to a partition is
defined by

$$\phi_i \equiv \int_{\cal M} d\mu\, \phi(x)$$

\nind where $d\mu$ is any measure on ${\cal M}$. 
\pss

Let us choose the interval $[0,1]$ as our manifold (with boundary) and
let ${\cal P}_n$ denote a regular partition of that interval into $n$
equal cells, denoted by $C^n_i \equiv [(i-1)/n, i/n]$. The truncation
operator $R^{M\ot N}$ shall be defined by

$$R^{M\ot N}_{Ii} \equiv {\mu(C^M_I \cap C^N_i) \over
\mu(C^M_I)}\eq{\overlap}$$

\nind where $\mu(\cdot)$ denotes the measure on ${\cal M}$. In geometrical
terms $C_i$ and $C_I$ denote the respective cells in the original and
the destination partitions. The $R$ matrix elements are the overlap
fractions between the cells on the destination cell:

$$R^{M\ot N}_{Ii} = { \hbox{ Overlap between cells $C_i$ and $C_I$. }
\over \hbox{ Measure of $C_I$.} } \eqno{[\overlap]}$$

The rationale behind this expression may be described with a physical
analogy. Let us consider the $\phi_i$ to denote the density of a gas
in each cell of the source partition. The walls between cells are
impenetrable. Now a new set of new walls is established: the ones
corresponding to the new partition. The old walls are, after that,
removed. The gas molecules redistribute uniformly in each new
cell. The new densities are the values $\phi'_I$ which constitute the
transformed field aspect (see figure 1).

\figura{6cm}{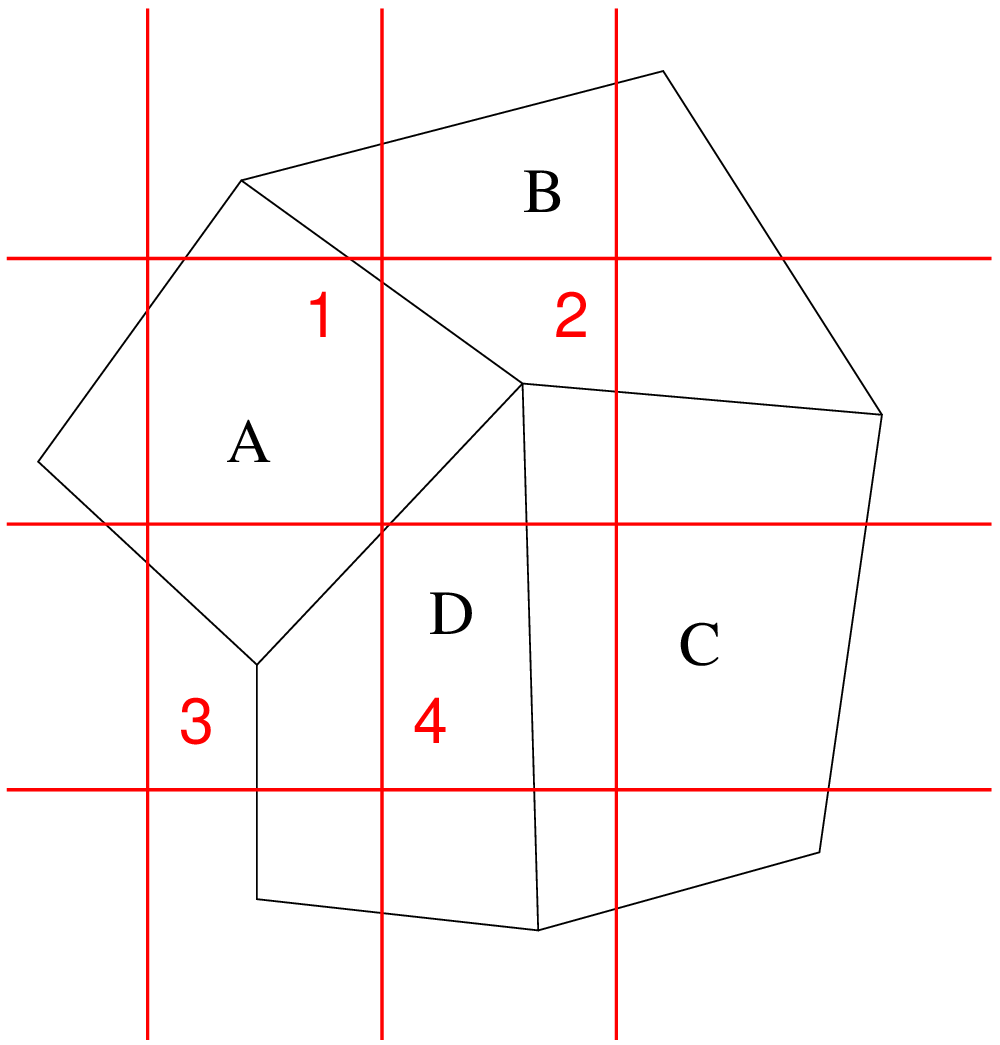}{Figure 1.}{A region of the plane on
which two partitions have been drawn. Black lines delimit the old
cells ($A$, $B$...), meanwhile the red ones correspond to the new ones
($1$, $2$,...). Thus, e.g., there would be no $R$ matrix element
between cells $1$ and $C$, since they do not overlap. On the other
hand, the element $R_{1A}$ must be near unity.}

In more mathematical terms the value of $\phi_I$ is a linear estimate
for

$$\phi_I = \int_{C_I} \phi(x) dx$$

\nind conserving total mass: $\sum_I \phi_I = \sum_i \phi_i$.

The resulting $R^{M\ot N}$ operators shall be called {\sl sudden
truncation operators}. Notice that the Kadanoff transformations, in
which an integer number of cells get fused into a bigger one, are
included.

One of the fundamental differences between usual (integer reduction
factor) truncation operators and the sudden truncation operators
defined in [\overlap] is the possibility of applying truncations which
remove a {\it single degree of freedom\/}. The situation is shown in
figure 2.

\figura{2cm}{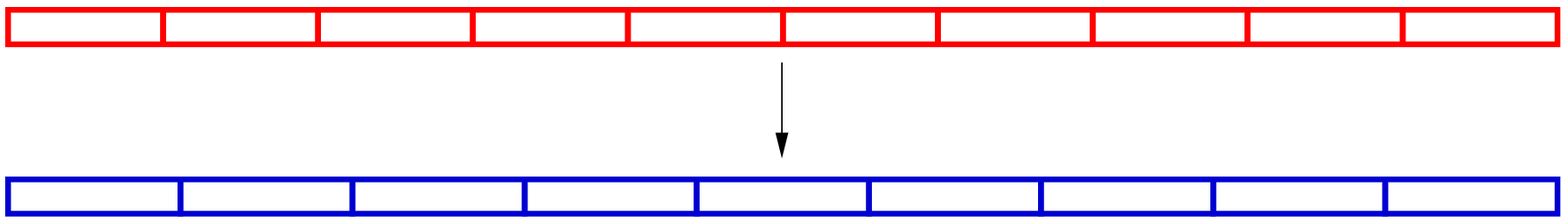}{Figure 2.}{The lower partition has just one 
degree of freedom less than the one above.}

\subsection{Quasistatic Transformations.}

The composition of sudden truncation operators takes us to the concept
of {\sl quasistatic} or {\sl adiabatic} truncation operators. These
are defined by

$$qR^{M\ot N} = R^{M\ot M+1} R^{M+1\ot M+2} \cdots R^{N-1\ot N}$$

\nind Of course, $qR^{M\ot N}$ differs greatly from $R^{M\ot N}$. The
``quasistatic'' term is suggested by the thermodynamical analogy which
was formerly introduced. Reversibility of a process is related to
quasistaticity, i.e.: proceeding through very small steps and waiting
for relaxation between these. In a certain sense, we might expect this
transformation to be {\it more reversible\/} and therefore better
suited to our purposes.

We formulate analytically a single step sudden transformation, upon
which the quasistatic operators are based, as

$$R^{N-1\ot N}_{Ii} = \delta_{I,i} {N-I\over N} + 
\delta_{I,i-1} {I\over N}$$

Iteration of this relation leads to a recursion relation fulfilled by
the quasistatic operators

$$qR^{M\ot N}_{Ii} = {M+1-I\over M+1} qR^{M+1\ot N}_{Ii} +
{I\over M+1} qR^{M+1\ot N}_{I+1,i}$$

\nind This recursion allows the calculation of the matrices using no 
matrix products, which therefore renders the approach much more
efficient in terms of computer time.
\pss

A question of special relevance is: which are the degrees of freedom
retained by this truncation? Figure 3 plots five rows of the matrix
$qR^{20\ot 80}$.

\figura{5cm}{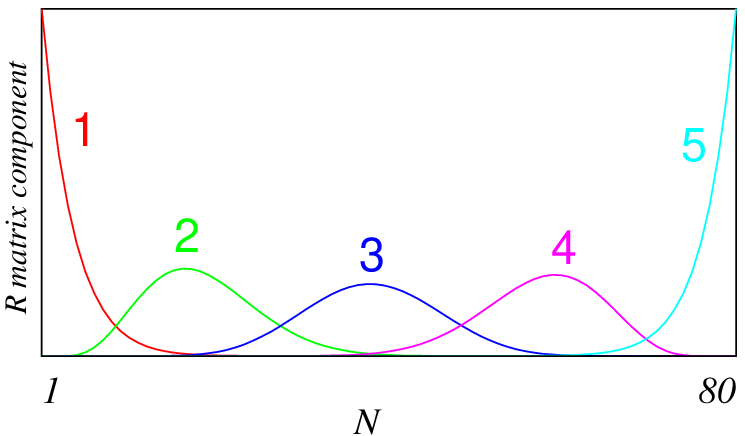}{Figure 3.}{Degrees of freedom which are
retained by the quasistatic truncation operator proceeding from $80$
upto $20$ sites. Rows $1$, $5$, $10$, $15$ and $20$ are
depicted. Notice that these ``cells'' are now overlapping and have
slightly gaussian nature.}

Each row of matrix $qR^{M\ot N}_{Ii}$, visualized in figure 3,
describes the discretization pattern for one of the final $20$ cells.
Although the functions representing the degrees of freedom are now
overlapping, they conserve a true geometric nature in real-space. It
must be noticed that the width of the extreme (leftmost and rightmost)
cells is smaller than that of the central one. The consequence is a
better representation of the boundary conditions.

The overlapping of cells is not a newcomer in RSRG applications
\ct[DEG-99]. In physical terms, cells overlapping takes into account 
inter--cells interaction in an improved way, which is the basic global
target of RSRG techniques.

\subsection{Other Possibilities.}

Even simpler truncation matrices are possible. According to the {\sl
decimation} scheme which is common in the literature, a cell out of
every $N/M\equiv f$ may be retained, where $f$ takes an integer
value. This truncation scheme is not well suited for our
formalism. The reason is that the chosen cells are disconnected. In
mathematical terms, the corresponding $R$ matrix would be

$$ R_{iI} = \delta_{i,fI} $$

\nind which, along with its SVD pseudoinverse, yield a trivial
dynamic. A way to hold linearity, though losing the Moore-Penrose
conditions, is to use an embedding operator $R^p$ specially designed.

$$R^p_{Ii}=\delta_{i\in\{f(I-1)+1,...,fI\}}=
\sum_{j=f(I-1)+1}^{fI} \delta_{i,j}$$

\nind Special care is required in this scheme to include the boundary
conditions appropriately. Of course, this is not the standard way to
implement this technique.
\pss

On the other hand, truncation processes based on the Fourier transform
with an UV cutoff enter the formalism perfectly, but they do not
correspond to an RSRG scheme.


\section{5.4. Applications and Numerical Results.}

This section discusses some concrete numerical applications, both to
linear and non-linear examples.

\subsection{Heat Equation.}

The heat equation is defined by declaring the laplacian to be the
evolution generator. It is known that the laplacian operator may be
sensibly defined on a great variety of spaces \ct[ROS-97], even on
discrete ones (see \ct[BOL-98] and appendix A).

Our 1D interval shall be $[0,1]$, split into $N$ cells of width
$\Delta x=1/N$. The discrete (or {\sl combinatorial}) laplacian on the
linear graph is given by:

$$L_{ij}=-2\delta_{i,j}+\delta_{|i-j|,1} \qquad\qquad\qquad
i,j\in\{1\ldots N\}$$

\nind with fixed boundary conditions: $L_{11}=L_{NN}=2$.
The evolution equation is therefore

$$\pl_t\phi_i = {\kappa\over \Delta x^2} L_{ij}\,\phi_j$$

The first check of the RSRG scheme shall be to take as initial
condition a random increments function, generated according to the
stochastic differences equation

$$\phi_{i+1}=\phi_i + w$$

\nind with $w$ a random variable with mean zero and uniformly distributed
on an interval of width $\Delta$. Using $N=200$, $M=20$ (a quite
severe reduction of a factor $10$) and $\Delta=1/4$ we obtain the
results depicted in figure 4.

\midinsert
\pss
\centerline{\fig{5cm}{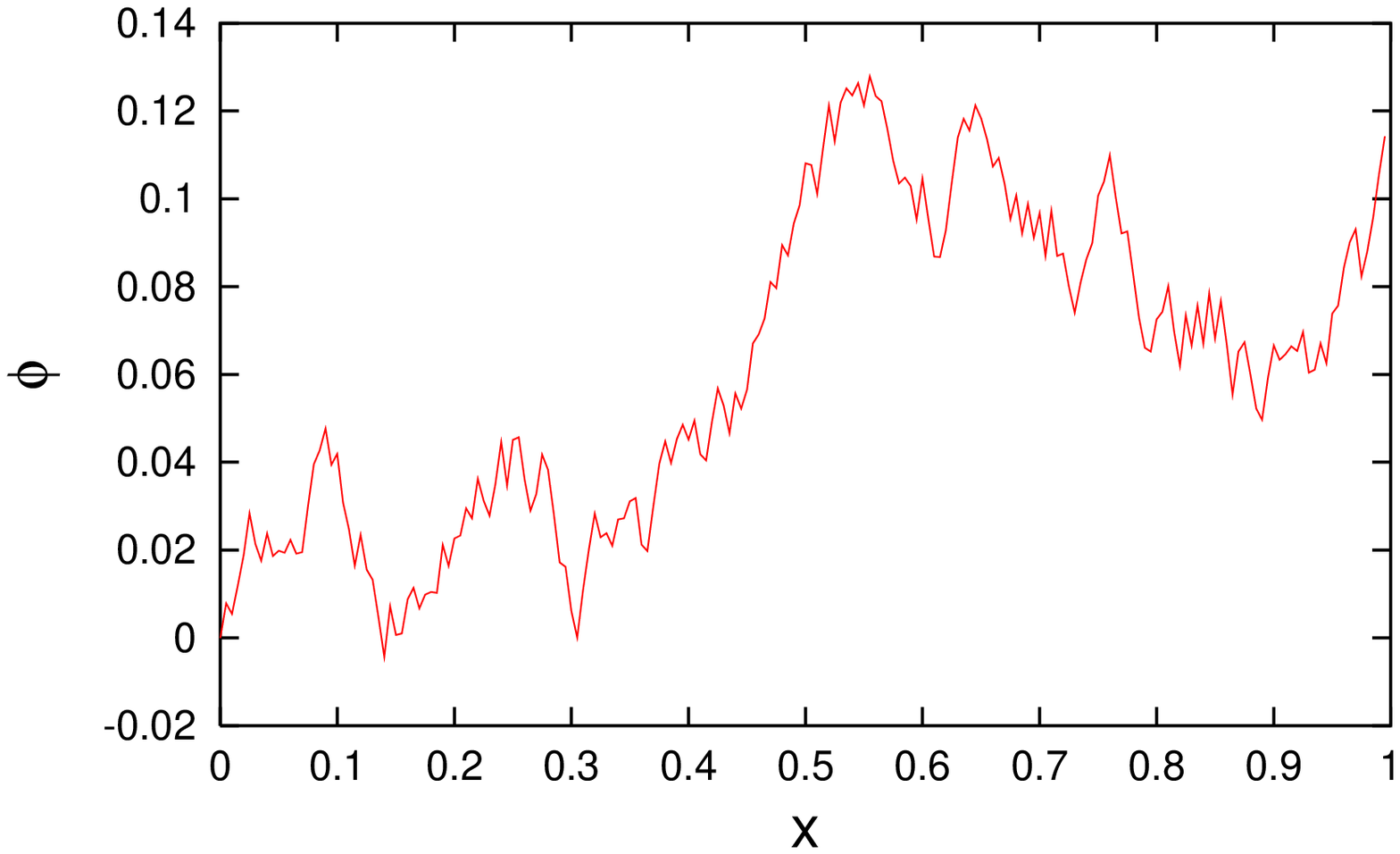}\fig{5cm}{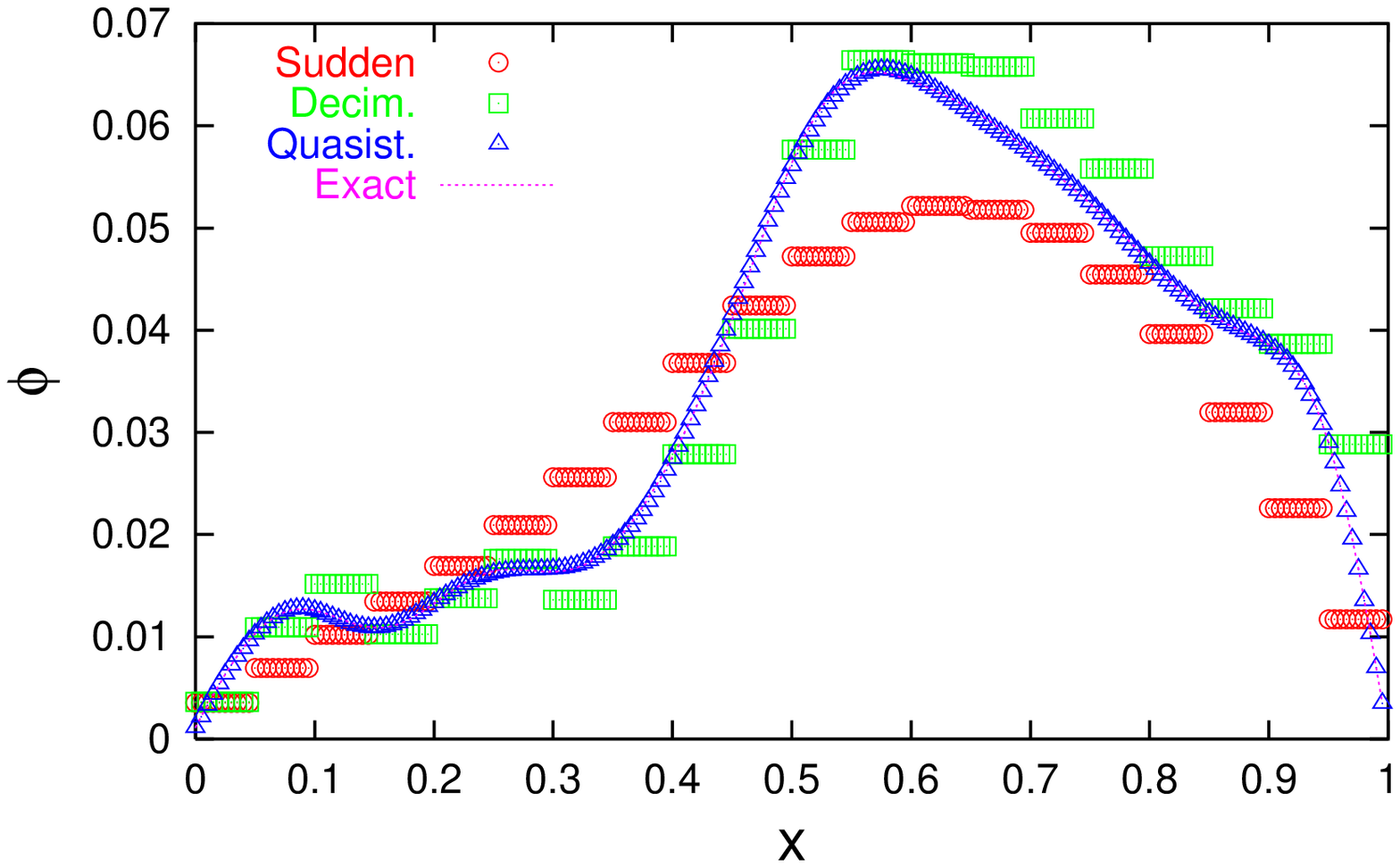}}
\ps2
\nind{\mayp Figure 4. }{\peqfont A random increments initial 
condition with $200$ cells (left) yields, under exact heat equation
evolution with $\kappa=1/2$ along $500$ time-steps with $\Delta
t=5\cdot 10^{-6}$, gives the full line on the right. The triangles
mark the approximation given by the quasistatic transformation with
$20$ degrees of freedom. The squares follow the decimation approach,
and the circles represent the sudden approximation (i.e. usual
coarse-graining).\par}
\pss
\endinsert

According to the explanations given in section 2, table 1 shows the
errors\footn{See end of section 5.2 for the specifications on the
calculation of errors.} for the curves plotted in figure 4.

\midinsert
\pss
\centerline{
\sbtabla(2)
\li && Real Space Error & Renormalized Space Error \cr
\lt Quasistatic && $0.53\%$ & $0.29\%$ \cr
\lt Sudden && $20\%$ & $19\%$ \cr
\lt Decimation && $13\%$ & $4.7\%$ \cr
\setabla}
\ps2
\nind{\mayp Table 1. }{\peqfont Errors associated to the functions of
figure 4.\par} 
\pss
\endinsert

Errors are usually smaller in renormalized space. The reason is that
in real space two sources of error are combined, which we shall know
as {\sl geometric error} and {\sl dynamical error}. The geometric
error is due to the transformation $R$ itself. If, e.g., the initial
condition belongs to the kernel of $R$, we shall evolve the null
function and the error shall be fully geometric. On the other hand,
the dynamical error is completely given by the ``non-closure'' of the
commutative diagram [\commatm]. Thus, a function which is initially
orthogonal to the kernel of $R$ may generate immediately components on
this kernel and be subsequently poorly represented.
\pss

In order to examine the relevant scaling laws \ct[BAR-96] a
discretization of $\delta(x-1/2)$ is used on the $200$ cells
partition, and is normalized by the condition

$$\sum_{i=1}^N \phi_i = 1$$

\nind Under time evolution this peak initial condition becomes a 
gaussian function whose width follows the law

$$w(t)\approx t^{1/2}$$ 

\nind which can be proved to be {\it valid\/} also in the discrete
case (see appendix C).

Using the same parameters as in the previous case, we have performed a
quasistatic simulation of the problem, showing in figure 5 a log-log
plot of the width against time.

\figura{7cm}{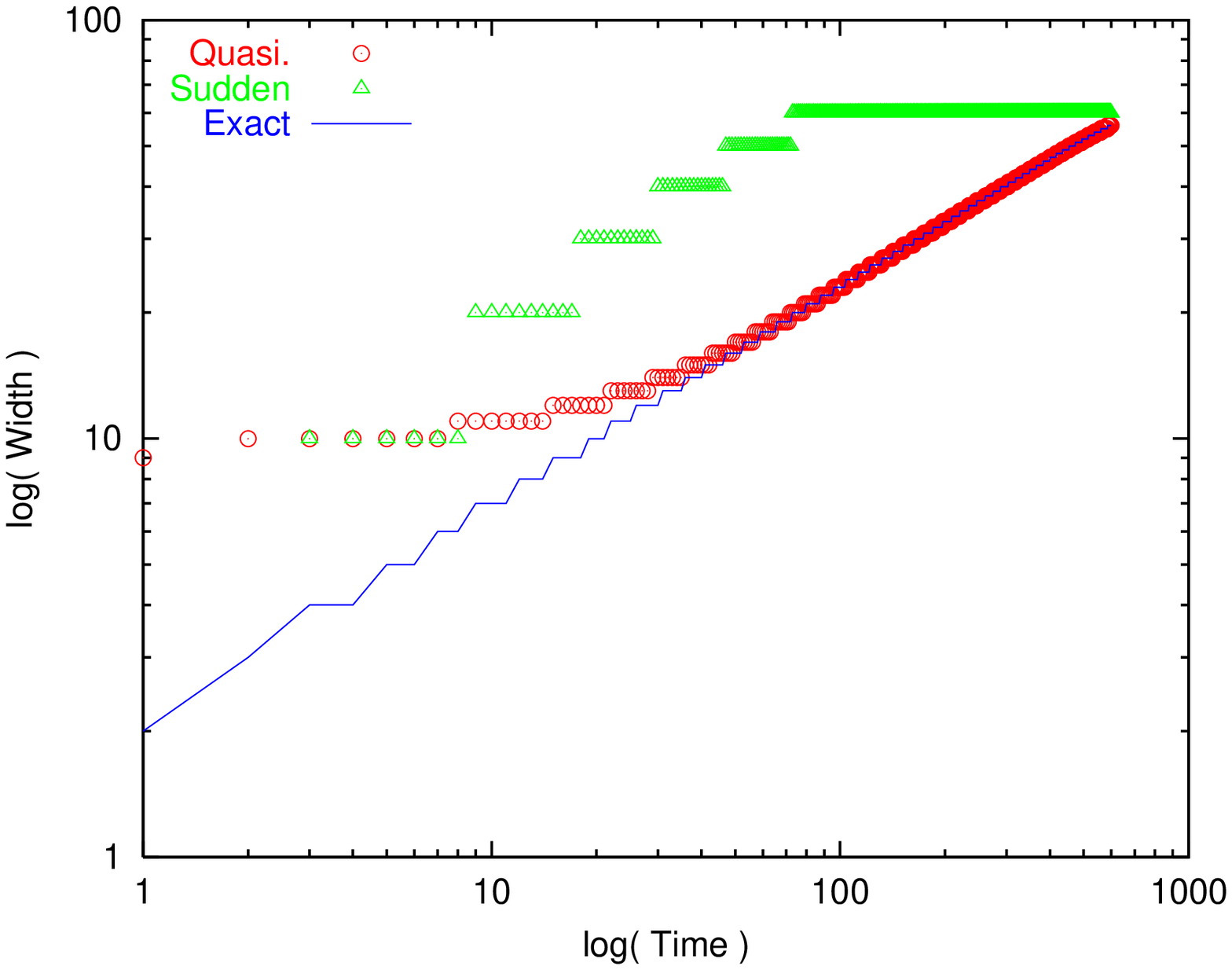}{Figure 5.}{Log-log plot of the width of 
the quasistatic simulation of the delta-function against time. The
straight line has slope $0.5$.}

The fit of the data for the quasistatic truncation in figure 5, after
a brief transient, fit a straight line with slope $0.4990\pm
0.0001$. The data for the exact evolution of the initial aspect yield
the same accuracy, but the validity range is complete. The sudden
approximation saturates at long times and gives a result of $0.66\pm
0.01$ for the slope just before that happens. In this case, as it is
proved in appendix C, decimation gives again the correct result.

\subsection{Low Energy States in Quantum Mechanics.}

The formalism presented here allows us to obtain a very accurate
approximation to the low energy spectrum of any unidimensional quantum
mechanical system, which was the main benchmark problem for the
previous chapters.

The transformation $H\to RHR^p$ may also give an effective
transformation for a hamiltonian, and its eigenvalues shall be upper
bounds to the lowest eigenvalues of the real spectrum\footn{And also,
logically, lower bounds for the highest ones.}, provided that the $R$
and $R^p$ are both orthogonal matrices.

The diagonalization of the quasistatically truncated laplacian yields
very precise values. For example, if we choose $N=100$ and $M=10$, we
obtain for the spectrum of $-L$ the values shown in table 2.

\midinsert
\pss
\centerline{\vbox{\offinterlineskip\halign{\quad #&\vrule #&&\quad #\cr
\noalign{\hrule height 1pt}
&height 5pt&\cr
Exact && 0.000967435 & 0.00386881 & 0.0087013 & 0.0154603 & 0.0241391
& 0.0347295 \cr
&height 10pt&\cr
Quasist. && 0.000967435 & 0.00386881 & 0.00870151 & 0.0154632 &
0.0247183 & 0.0363134 \cr
&height 10pt&\cr
Sudden && 0.00810141 & 0.0317493 & 0.0690279 & 0.116917 &
0.171537 & 0.228463 \cr
&height 5pt&\cr
\noalign{\hrule height 1pt}}}}
\ps2
\nind{\mayp Table 2. }{\peqfont Low energy spectrum of a particle in a
box split into $100$ discrete cells, calculated through exact
diagonalization and two effective variational RSRG techniques. The
first is based on a quasistatic transformation and the second on a
sudden one.\par}
\pss
\endinsert

The bad results for the sudden approximation are a bit misleading. For
example, the real space error (in $L^2$ norm) for the ground state is
only an $11\%$. But the energy is given by the derivatives of the
wave--function, so a small difference in norm may yield a big
error in energy. 

As a matter of fact, the sudden technique for the case of an integer
reduction factor coincides with the {\sl Kadanoff warmup} (see section
4.6). When analyzing it, the inability of the $L^2$ norm to
distinguish between two functions was discussed. The energy is in this
case a {\it Sobolev-like\/} norm \ct[TAY-97], which may be much more
appropriate\footn{Physics for two functions of almost equal $L^2$ norm
may be quite different. One of them may have, e.g., a high frequency
oscillation or {\it Zitterbewegung\/} superimposed on it.}.
\pss 

The procedure returns as many eigenvalues as degrees of freedom we
have retained, $10$ in our case. But only the $6$ first ones are
reasonable. The reason for the great accuracy of the first states and
the error in the last ones is the precision with which the boundaries
are represented. Thus, {\it a)\/} the boundary conditions are fairly
taken into account, but {\it b)\/} there are not many degrees of
freedom left for the bulk as we would need in order to have $10$ full
states correctly represented.
\pss

The method works fine not only for the free particle. Some potentials,
such as the harmonic oscillator and those studied in the previous
chapters, have been analyzed and the results have similar accuracy
{\it as long as\/} the resulting wave--functions are smooth.

\subsection{Kardar-Parisi-Zhang Equation.}

The Kardar-Parisi-Zhang (KPZ) equation is widely used as a model of
stochastic and deterministic surface growth \ct[KPZ-86]
\ct[BS-95]. We shall use the following version:

$$\pl_t\phi=\lambda|\nabla\phi|^2 + \kappa\nabla^2\phi$$

\nind which represents a surface in which absorption/desorption 
phenomena take place, but without surface diffusion (which would imply
a $\nabla^4\phi$ term).

Although this equation is usually employed in a noisy environment
(i.e.: the associated Langevin equation is the one which is currently
solved), we shall only deal with the deterministic version. The
squared gradient term shall be implemented through the quadratic
operator\footn{Warning added in proof: This discretization is known
not to conserve the universality class of the continuum limit.}

$$K_{ijk}={1\over 4}\( \delta_{j,i+1}-\delta_{j,i-1}\)
\(\delta_{k,i+1}-\delta_{k,i-1}\)$$

\nind Boundary conditions are imposed which use forward and
backward derivatives in the extremes.
\pss

The first test evolves a sinusoidal initial condition
$\phi(x)=\sin(4\pi x)$ with $x\in[0,1]$. A slight lack of symmetry of
the initial discretization (a ``lattice artifact'') develops a large
asymmetry in the result. The resolution change is $40\to 20$ and
$2000$ $\Delta t=5\cdot 10^{-6}$ timesteps were taken. Figure 6 shows
the results for $\lambda=2$ and $\kappa=1/2$. Errors are reported in
figure 3.

\midinsert
\centerline{\fig{5cm}{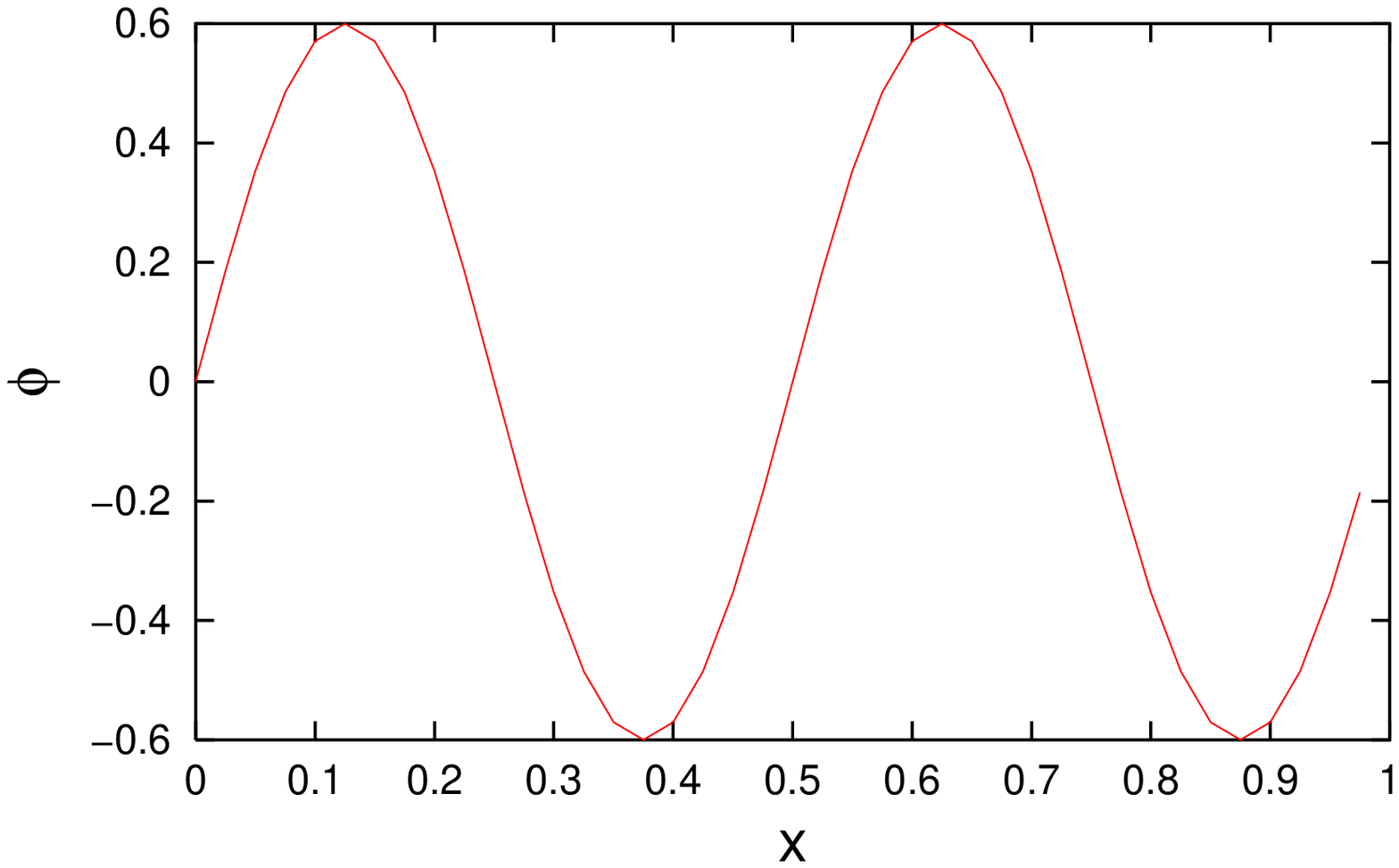}\hfil\fig{5cm}{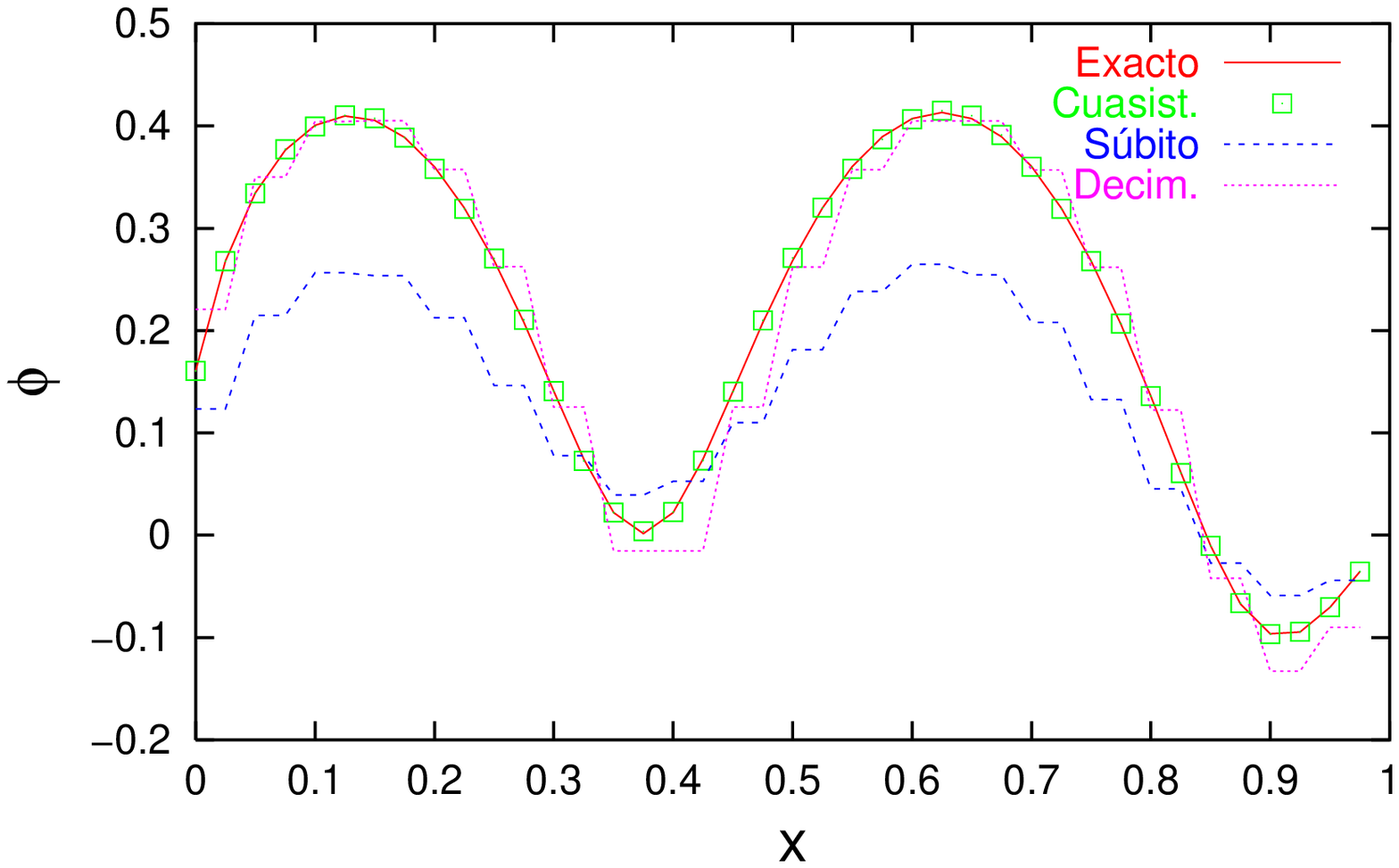}}
\par
\nind{\mayp Figure 6. }{\peqfont A sinusoidal initial condition (left)
evolves under KPZ equation (right) with the parameters specified in
the text.\par}
\endinsert

\midinsert
\pss
\centerline{
\sbtabla(2)
\li && Real Space Error & Renormalized Space Error \cr
\lt Quasist. && $0.5\%$ & $0.2\%$ \cr
\lt Sudden && $39\%$ & $38\%$ \cr
\lt Decim. && $15\%$ & $8\%$ \cr
\setabla}
\ps2
\nind{\mayp Table 3. }{\peqfont Errors in the evolution of a
sinusoidal initial condition under KPZ, corresponding to figure
7.\par}
\pss
\endinsert

A different test is obtained using a random increments function, as it
was done with the heat equation. The chosen parameters are the same as
the ones for the previous evolution and the results are shown in
figure 7, with the errors in table 4.

\midinsert
\pss
\centerline{\fig{5.2cm}{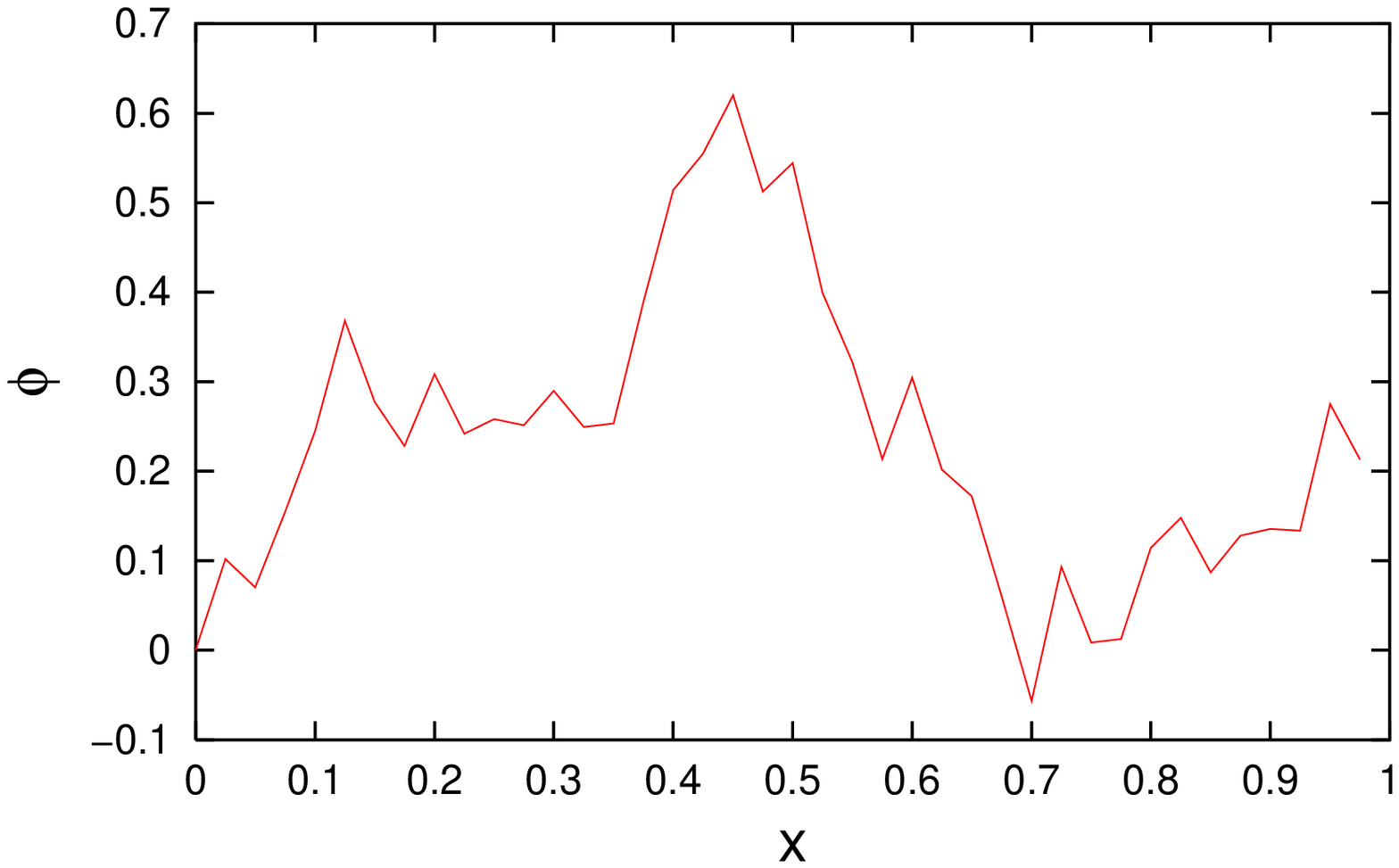}\fig{5.2cm}{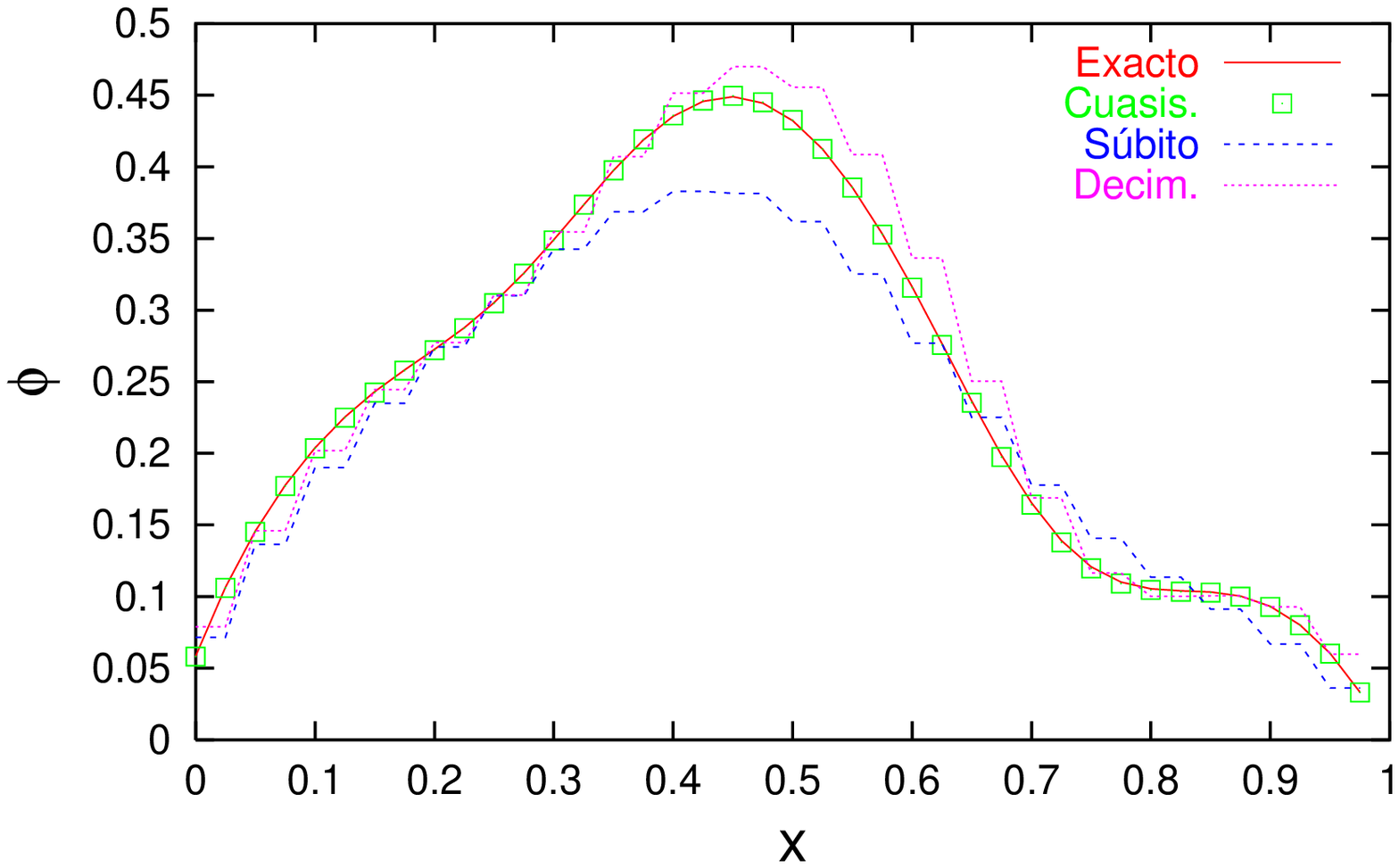}}
\ps2
\nind{\mayp Figure 7. }{\peqfont Random increments function evolved under
the same conditions as those for figure 6.\par}
\pss
\endinsert

\midinsert
\pss
\centerline{
\sbtabla(2)
\li && Real Space Error & Renormalized Space Error \cr
\lt Quasist. && $0.23\%$ & $0.23\%$ \cr
\lt Sudden && $12\%$ & $11\%$ \cr
\lt Decim. && $9.4\%$ & $6.5\%$ \cr
\setabla}
\ps2
\nind{\mayp Table 4. }{\peqfont Errors corresponding to plots in
figure 7.\par}
\pss
\endinsert

Numerical experiments were also carried out on Burgers equation (which
is analytically related to KPZ) and others, with similar results.


\section{4.5. Towards a Physical Criterion for Truncation.}

The success of the geometrically based truncation operators described
in the previous section lead us to think that the reduction of degrees
of freedom is feasible. Thus, the application of truncation operators
which were {\it specific\/} to a given physical model might improve
considerably the quality of the simulations. This approach shall be
termed the {\sl physical criterion}, as opposed to the {\sl geometric
criterion} which was our guide in the previous sections.

The schedule for the development of such physically oriented
truncation matrices is still at their first steps. In \ct[DRL-01{\mayp
B}] a couple of techniques were proposed which shall be briefly
commented in this section.

\subsection{Short--Time Dynamics.}

A first simple idea is the election of a set of states dependent both
on the initial condition and its initial ``movements''. Specifically,
if the equation is formally given by equation [\linearham] and
$\phi(t=0)=\phi_0$, then the set

$${\cal S}\equiv [\phi_0, H\phi_0, \ldots, H^{n-1}\phi_0]$$

\nind if duly orthonormalized, constitute the rows of an $R$ matrix (and
the columns of its pseudoinverse $R^p$) which is highly efficient at
{\sl short times}. 
\pss

The reason for the election of the set ${\cal S}$ should be
apparent. For a number of timesteps smaller than $n$, the state
$\phi(t)$ is fully contained in that subspace, which can not be
ensured for $t>n\Delta t$, although the approximation is generally
still correct for a longer time interval.

The short-time approach is strongly based on the discrete Taylor
series and is analogous to the Lanczos diagonalization technique
\ct[GvL-96]. This idea has not much utility beyond providing an
example of a {\sl physical criterion}, since the long-time dynamics is
typically much more interesting. Quite often, the approximation
becomes unstable beyond its applicability region.

A curious suitable combination consists in taking degrees of freedom
from the quasistatic truncation along with those of the short-time
approximation. For the heat equation figure 5 shows that the fit is
perfect {\it after\/} a short transient. This transient is fully
removed with the combination of both approaches.

\subsection{Error Minimization}

An approach which appears to be more promising is the minimization of
a suitable magnitude which measures the ``openness'' of the commuting
diagram [\commatm]. The usage of such measures is not new in the
literature. The original formulation of S.R.~White's DMRG
\ct[WHI-92] used a rather similar technique to choose the useful
degrees of freedom. In a wide sense, this technique generalizes DMRG
to non-linear systems.

So as to make the computations feasible, a single degree of freedom
shall be removed at each step. We shall employ Dirac's notation from
quantum mechanics and call $\ket|v>$ the state to be removed. Thus,
operators $R$ and $R^p$ shall be any pair fulfilling

\def\f{{\ket|\phi>}}
\def\brf{{\bra<\phi|}}

$$R^pR=I-P_\f=I-\f\brf$$

\nind the lack of closure of the diagram [\commatm] for a single time
step is given by the operator

$${\cal E}=(I+\Delta t H)- R^p(I+\Delta t H)R =P_\f + \Delta
t\( HP_\f + P_\f H- \elem<\phi|H|\phi> P_\f\)
\eq{\eliminator}$$

Let us consider the Frobenius norm of this operator as the error to be
minimized:

$$\lin{Error}=\sum_{i,j} {\cal E}_{ij}^2$$

When $H$ is a selfadjoint operator, equation [\eliminator] is rather
easily simplified. Let $\{\ket|u_i>\}$ be the basis of eigenstates of
operator $H$, ordered from lowest to highest eigenvalue:

$$H=\sum_{i=1}^N E_i \f\brf$$

Let us now decompose the desired state into the basis $\{\ket|u_i>\}$.

$$\f=\sum_{i=1}^N \mu_i \ket|u_i>$$

\nind Substituting into [\eliminator] we obtain the following
relation:

$${\cal E}_{ij}=\left[1+\Delta t(E_i+E_j-\<H\>)\]\mu_i\mu_j$$

\nind where $\<H\>\equiv \elem<\phi|H|\phi>$. We compute the error and
minimize it with respect to variations of the $\{\mu_i\}$ with the
constraint $\sum_i \mu_i^2=1$. Taking previously a first order
development in $\Delta t$ we get

$$\lin{Error}=\sum_{ij} {\cal E}^2_{ij} \approx 1+ 2\Delta t \<H\>$$

The minimum value of that expression is obtained when $\<H\>$ is the
lowest eigenvalue of $H$, so $\f=\ket|u_1>$. To sum up, the
progressive removal of the eigenstates with lowest eigenvalues is the
optimum truncation, following a {\sl physical criterion}.
\pss

The case of the non-selfadjoint linear evolution generator still
admits an analytical development which reduces the problem to that of
finding the ground state of a different matrix. Let us consider the
singular values decomposition of the $H$ operator.

$$H=\sum_i E_i \ket|u_i>\bra<v_i|$$

\nind and let us represent the target vector $\f$ in both bases:

$$\f=\sum_i \chi_i \ket|u_i> \qquad \f=\sum_i \mu_i \ket|v_i>$$

Both sets of coefficients $\{\chi_i\}$ and $\{\mu_i\}$ are not
independent. If $C_{ij}=\<u_i|v_j\>$, $\chi_i=\sum_j
C_{ij}\mu_j$. Expanding expression [\eliminator] in the same way as
that for the selfadjoint case a parallel expression is found to first
order in $\Delta t$:

$$\lin{Error}=1+2\Delta t \<H\>$$

\nind The difference appears when computing the minimum value for
$\<H\>$:

$$\<H\>=\sum_i E_i \chi_i \mu_i = \sum_{i,j} C_{ij}E_i \chi_i\chi_j$$

\nind This shows that the lowest eigenstate of the selfadjoint matrix

$$K_{ij}\equiv {1\over 2}\( C_{ij}E_i + C_{ji}E_j\)$$

\nind is the required solution.
\pss

The case of selfadjoint operators has a greater physical meaning,
since it includes the heat equation. In this case the {\sl physical}
recipe just asks to remove the {\it lowest eigenvalue eigenstates\/}
of the laplacian. These are the states which oscillate most strongly
and, in the continuous limit, the highest Fourier modes.
\pss

The problem of the computation of the optimum state to be removed
forces the minimization of a function of degree $2n$ (where $n$ is the
order of the equation) in $N$ variables. Thus, only the linear
equations offer a panorama suitable for analytical calculations. In
the case of quadratic equations, the formerly cited work
\ct[DRL-01{\mayp B}] presents some results obtained with numerical
minimization for Burgers and KPZ equations, and the results of the
geometrical approach are clearly improved.


\section{5.6. Bibliography.}

\cx[BAR-96] {\may G.I. Barenblatt}, {\sl Scaling, self-similarity and
intermediate asymptotics}, Cambridge U.P. \break(1996).

\cx[BOL-98] {\may B. Bollob\'as}, {\sl Modern Graph Theory}, Springer
(1998).

\cx[BS-95] {\may A.L. Barab\'asi, H.E. Stanley}, {\sl Fractal concepts in
surface growth}, Cambridge U.P. (1995).

\cx[BUR-74] {\may J.M. Burgers}, {\sl The nonlinear diffusion
equation}, Riedel (1974).

\cx[DEG-99] {\may A. Degenhard}, {\sl A non-perturbative real space
renormalization group scheme}, part I in {\tt cond-mat/9910511} and
part II in {\tt cond-mat/9910512}. Also at J.~Phys.~A (Math.~Gen.)
{\bf 33}, 6173 (2000) and Phys.~Rev.~B {\bf 64}, 174408 (2001).

\cx[DRL-01{\mayp A}] {\may A. Degenhard, J. Rodr\'{\i}guez-Laguna}, 
{\sl A real space renormalization group approach to field evolution
equations}, {\tt cond-mat/0106155} and Phys.~Rev.~E {\bf 65}
036703 (2002).

\cx[DRL-01{\mayp B}] {\may A. Degenhard, J. Rodr\'{\i}guez-Laguna}, {\sl 
Towards the evaluation of the relevant degrees of freedom in nonlinear
partial differential equations}, {\tt cond-mat/0106156} and
J.~Stat.~Phys. {\bf 106} 5/6, 1093 (2002).

\cx[GHM-98] {\may N. Goldenfeld, A. McKane, Q. Hou}, {\sl Block spins
for partial differential equations}, {\tt cond-mat/9802103} and
J.~Stat.~Phys. {\bf 93}, 3/4, 699 (1998).

\cx[GHM-00] {\may Q. Hou, N. Goldenfeld, A. McKane}, {\sl
Renormalization group and perfect operators for stochastic
differential equations}, {\tt cond-mat/0009449} and Phys.~Rev.~E {\bf
63}, 36125 (2001).

\cx[GvL-96] {\may G.H. Golub, C.F. van Loan}, {\sl Matrix
computations}, The John Hopkins U.P. (1996).

\cx[HN-94] {\may P. Hasenfrantz, F. Niedermayer}, {\sl Perfect lattice
action for asymptotically free theories}, Nucl. Phys. B {\bf 414}, 785
(1994).

\cx[KPZ-86] {\may M. Kardar, G. Parisi, Y.C. Zhang}, {\sl Dynamic
scaling of growing interfaces}, Phys. Rev. Lett. {\bf 56}, 889 (1986).

\cx[KW-97] {\may E. Katz, U.-J. Wiese}, {\sl Lattice fluid dynamics
from perfect discretizations of continuum flows}, {\tt
comp-gas/9709001} (1997).

\cx[ROS-97] {\may S. Rosenberg}, {\sl The laplacian on a riemannian
manifold}, Cambridge U.P. (1997).

\cx[TAF-81] {\may E. Taflin}, {\sl Analytic linearization, hamiltonian
formalism and infinite sequences of constants of motion for the
Burgers equation}, Phys. Rev. Lett {\bf 47}, 1425 (1981).

\cx[TAY-97] {\may M.E. Taylor}, {\sl Partial differential equations}, 
Springer (1997).

\cx[WHI-92] {\may S.R. White}, {\sl Density matrix formulation for
quantum renormalization groups}, Phys. Rev. Lett. {\bf 69}, 2863 (1992).

\pasacap

\titulartrue\denuevas
\vbox
{\ps{10}
\line{\hfil\vbox{\hrule height .5mm\ps3
\hbox{\grandota\verdeosc Conclusions and Future Work.}
\ps3\hrule height .5mm}\hfil}
\ps{10}}

\bul Insight has been gained about the cause of the initial failure of
the Blocks Renormalization Group (BRG), expressing the conclusions in
the development of a rather similar technique where the defects were
removed: the {\it Correlated\/} Blocks Renormalization Group (CBRG)
for problems in quantum mechanics both in 1D and 2D in the presence of
an arbitrary potential.
\pss

\bul The Density Matrix Renormalization Group (DMRG) has been extended
so as to work on trees, and it was applied to the analysis of the
excitonic spectrum of a family of polymeric molecules with fractal
character: the dendrimers.
\pss

\bul The required modifications of the DMRG algorithm in order to be
applied to multidimensional systems have been studied, arriving at the
Punctures Renormalization Group (PRG), which uses a single block
---more natural than the left-right distinction in $>1D$. The
application to the analysis of excitons in disordered systems with
long range interaction has been described. The difficulties on the
path towards a PRG for many body problems were analyzed in some
detail.
\pss

\bul Techniques based on real space renormalization group have 
been employed for the effective reduction of degrees of freedom in
partial differential equations. Concretely, a theoretical framework
has been exposed for the sub-discretization process and it has been
applied to both linear and non-linear equations in 1D.
\pss\pss

The Real Space Renormalization Group (RSRG) overflows with theoretical
questions which require more insight and practical applications which
wait for development. The number of publications on the field,
specially since the DMRG was developed, grows ever larger. In the near
future, we intend to undertake the following research lines:
\pss

\bul Application of the PRG, in parallel with analytical RG
techniques, to various problems of excitons in disordered media.
\pss

\bul Development of the PRG algorithm for many particles.
\pss

\bul Development of RSRG techniques for partial differential equations
in $>$1D, specially for fluid mechanical problems.

\vfill\eject

\blancotrue
\vbox{}
\vfill\eject
\blancofalse
\temaizq={Ap\'endices.}
\titulartrue
\denuevas
\vbox{\ps{10}
\nind\verdeosc\grandota Appendices.}
\ps{15}
\hrule\pss
{\parindent=0pt
A. The Laplacian on a Graph.
\ps2
B. Computational Issues.
\ps2
C. Asymptotics of the Discrete Heat Equation.
\ps2
D. Fast Gram-Schmidt Technique.
\ps2
E. Scale Dimension Associated to Energy.}
\pss\hrule


\def\titleap#1{\denuevas\vbox{\ps{10}\nind\azulosc\mediota #1}
\pss\pss\temader={#1}}

\titleap{Appendix A. The Laplacian on a Graph.}

In this appendix some elements of graph theory and their connection to
physics are exposed. The book of Bollob\'as \ct[BOL-98] is an
excellent introduction to the subject.

\subsection{Incidence, Adjacence and Laplacian.}

Let ${\cal G}$ be a simple graph, where $V({\cal G})$ denotes the set
of vertices and $E({\cal G})$ the set of edges, with respective sizes
$N_v$ and $N_e$. The graph structure is given by a neighbourhood
structure: $N: V\mapsto {\cal P}(V)$, which assigns to each vertex the
set of neighbouring vertices. The only constraint for that structure
is symmetry: $i\in N(j) \iff j\in N(i)$.
\pss

Let us assign an arbitrary {\sl orientation} to the graph by giving a
travelling order to each edge. The {\sl incidence matrix} is a
$N_v\times N_e$ matrix with elements $B_{ij}$ in the set $\{-1,0,+1\}$
such that

$$B_{ij}=\cases{+1 & if vertex $i$ is the origin of edge $j$\cr
		-1 & if vertex $i$ is the end of edge $j$\cr
		0 & otherwise}$$

A fundamental matrix related to $B$ is the {\sl adjacency matrix},
defined to be the $N_v\times N_v$ matrix whose elements $A_{ij}$ are

$$A_{ij}=\cases{+1 & if vertex $i$ is connected through an edge to
vertex $j$\cr 
	0 & otherwise}$$

The adjacency matrix is relevant for paths combinatorics on a
graph, since the number of different ways of traveling from site $i$
to site $j$ in $n$ steps is given by $(A^n)_{ij}$.
\pss

The most important matrix associated to a graph may be the {\sl
combinatorial laplacian}, which is defined as the $N_v\times N_v$
matrix $L=B^tB$. We define the {\sl degree} of a vertex $d_i$ to be
its number of neighbours (in physics it is usually called the {\sl
coordination number}), and $D$ to be the diagonal matrix $N_v\times N_v$
whose $(i,i)$-th entry is just $d(i)$. Then it is easy to prove the
following identity:

$$L=D-A$$ 

\nind where $A$ is the adjacency matrix. Therefore, the elements
of the combinatorial laplacian are just the number of neighbours in
the diagonal and $-1$ in all $(i,j)$ entries where vertex $i$ is
directly connected to vertex $j$. From now on, we shall drop the
adjective {\sl combinatorial}\footn{Notice that the sign is opposed to
the usual one, so as to generate a positive semidefinite operator. We
hope that this notation, usual in graph theory, does not mislead the
reader.}.
\pss

It is usual to consider {\sl fields} on graphs for the study of the
laplacian. We shall consider a field on the graph to be any function
$\phi:V\mapsto K$, where $K$ may be any numerical field (usually real
or complex numbers). The functional space to which $\phi$ belongs is a
vector space of finite dimension, so we may oscillate between the
terms ``field'' and ``vector''.

\subsection{Spectrum of the Adjacency Matrix.}

Let us denote by $\{\mu_i\}$ the set of eigenvalues of the adjacency
matrix. Let $d_\linp{max}$ the maximum degree of the graph (the
highest coordination number) and $d_\linp{min}$ the minimum degree. We
state a series of theorems whose proof is not complicated.
\pss

\item{\it a)} The highest eigenvalue fulfills $d_\linp{min}\leq
\mu_\linp{max}\leq d_\linp{max}$.
\ps2

\item{\it b)} The maximal degree $d_\linp{max}$ is an eigenvalue 
iff ${\cal G}$ is regular (i.e.: iff all the vertices have the same
degree). In that case, it is non--degenerate.
\ps2

\item{\it c)} If $-d_\linp{max}$ is an eigenvalue, then the graph is
regular and bipartite.
\ps2

\item{\it d)} If ${\cal H}$ is a subgraph of ${\cal G}$, then the
whole spectrum of its adjacency matrix lies within $\mu_\linp{min}$
and $\mu_\linp{max}$.

\subsection{Spectrum of the Laplacian}

The laplacian is always a selfadjoint positive semidefinite
matrix. The proof is straightforward from its definition or, as well,
from the fact that the quadratic form $\elem<\phi|L|\phi>$ (usually
called the {\sl lagrangian} form) may be written as a sum of squares.

Let us denote by $\ket|\delta_i>$ the field concentrated on the $i$-th
vertex. The field $\ket|\phi>$ may be expanded in that basis as
$\ket|\phi>=\sum_i \phi_i\ket|\delta_i>$ and, therefore:

$$\elem<\phi|L|\phi>= \sum_{i\in V} d_i \phi_i^2 
- \sum_{\<i,j\>\in E} \phi_i \phi_j = \sum_{\<i,j\>\in E}
(\phi_i-\phi_j)^2$$

\nind where the summation over $\<i,j\>\in E$ denotes all edges of the
graph. The following facts are easy to prove:
\pss

\item{\it a)} All the eigenvalues of the laplacian are either positive
or zero.
\ps2

\item{\it b)} If the graph is regular with uniform degree $r$, the
spectrum of the laplacian and that of the adjacency matrix are easily
related. If $\lambda$ is an eigenvalue of $L$, then $\mu=r-\lambda$ is
an eigenvalue of $A$ (and viceversa).
\ps2

\item{\it c)} The lowest eigenvalue of the laplacian is always zero. The
associated eigenvector is the uniform field. The reason is that each
row of the matrix has as many off--diagonal $-1$ elements as the
number in the diagonal marks. Therefore, the field given by
$(1,1,1,\ldots,1)$ has always zero lagrangian.
\ps2

\item{\it d)} The second smallest eigenvalue is of outmost importance. Its
magnitude refers to the {\sl global connectivity}. We define the
vertex connectivity to be the minimum fraction of the vertices which
it is necessary to remove for the graph to become disconnected. Then,
we get the following
\pss

{\may Theorem.} {\it The vertex connectivity of an incomplete graph
${\cal G}$ is never smaller than the second lowest eigenvalue of the
laplacian.}
\ps2

We shall not give a complete proof of this theorem, but only an
heuristic argument. The eigenfunction associated to the second lowest
eigenvalue must be orthogonal to the uniform field. Therefore, it must
have at least two domains of different signs. But, at the same time,
it must take the minimum possible value for the lagrangian, which
grows with nonuniformity (since it is the sum of the squares of the
field ``jumps''). So, the desired field should have just a single
``wall'', which should be as small as possible. The minimum set of
vertices which, when removed, separate the graph into two parts would
be a nice place to locate it.


\subsection{Boundary conditions.}

Probably, the most important constraint with physical applications
which may be imposed on the functional space of fields on a
graph is the presence of non--trivial boundary conditions. We shall
focus on the Dirichlet (fixed) and Neumann (free) types.
\pss

The discrete analogues of boundary conditions is, of course, highly
dependent on our concept of ``boundary''. We shall mark a subset of
$V({\cal G})$ to be the {\sl border set} of the graph (the rest being
called the {\sl bulk}). 

So as to make things concrete, we shall consider {\sl quasi--regular}
graphs, which consist of many vertices with a common degree $r$ but
for a few, which have a smaller degree. By adding a certain number of
extra vertices we may get all the authentic ones to have a homogeneous
degree $r$, while all the new vertices have degree $1$, making up what
we shall denote to be the {\sl closure} of the graph. The set of added
vertices shall be known as {\sl tack vertices}\footn{The word ``tack''
denotes, among other things, the loose stitches which fasten a piece
of cloth to a frame. The term ``auxiliar sites'', which has already
been used \ct[DEG-99], has a similar meaning, but some technical
details prevent us from using it here.} and the vertices which are
directly connected to them shall be the {\sl border vertices}.

At this point we may already consider various possibilities for the
boundary conditions.
\pss

a) {\it Free boundary conditions.\/} The acceptable fields take the
same value at the tack vertices and at the border ones.

Let us consider any of the border vertices, with index $i$ and degree
$d_i=r-v_i$, which is completed up to $r$ through the addition of tack
vertices. Denoting by $N(i)$ the set of bulk neighbours and by $T(i)$
the set of neighbouring tack vertices, the condition would be read as
$\forall k\in H(i), \phi_k=\phi_i$. The action of the laplacian on
such a field would be

$$(L_\linp{free}\ket|\phi>)_i = r\phi_i - \sum_{j\in N(i)} \phi_j -
\underbrace{\sum_{k\in T(i)}\phi_k}_{v_i\phi_i}= (r-v_i)\phi_i -
\sum_{j\in N(i)} \phi_j$$

\nind In other words: the contribution of the tack vertices is
exactly cancelled out. Therefore, the laplacian {\it does not change
form\/}. We might say that the meaning of the free boundary conditions
is the {\it absence of an exterior world\/} linked to the system: we
do not need to consider external elements to the graph.
\pss

b) {\it Fixed boundary conditions.\/} The acceptable fields take the
value zero on the tack vertices. Let $i$ be again the index of a
border vertex and, with the notation of the former paragraph, the
laplacian would be

$$(L_\linp{fixed}\ket|\phi>)_i=r\phi_i - \sum_{j\in N(i)} \phi_j -
\underbrace{\sum_{k\in T(i)} \phi_i}_{0} = r\phi_i - \sum_{j\in
N(i)}\phi_j = r-(A\ket|\phi>)_i$$

\nind thus we find that it may be rewritten as

$$L_\linp{fixed} = rI-A$$ 

\nind The physical meaning of the fixed boundary conditions is that
{\it there is an exterior world\/}, but it {\it is trivial\/}. Notice
also that the laplacian spectrum with fixed b.c. is closely related to
that of the adjacency matrix.
\pss

c) {\it Mixed boundary conditions.\/} If the value of the field at the
tack vertices is neither zero nor the same as that at the border
vertices, then the laplacian gets the form:

$$(L_\linp{mixed}\ket|\phi>)_i = r\phi_i - \sum_{j\in N(i)} \phi_j -
\underbrace{\sum_{k\in H(i)} \phi_k}_\linp{constant}$$

\nind The term marked as ``constant'' is not linear on the set
$\{\phi_i\}_{i\in V}$, so the laplacian receives an inhomogeneous term
when the tack vertices are removed.

This more general type of boundary conditions can take into account
the immersion of a system into a larger one.

\ps7
\subsectionp{Random walks, Diffusion and Quantum Mechanics on Graphs.}

Random walks on different kinds of spaces constitute a very important
part of the basis of theoretical physics. In particular, quantum
mechanics of spinless particles may be formulated as the statistical
theory of random walkers in {\it imaginary time\/} \ct[FH-65]
\ct[ITZD-89].

Let us consider a large number of non--interacting random walkers in a
graph. The number of particles at each vertex constitutes a
field\footn{To be strict, the interpretation ``number of particles''
implies positivity. The requirement may be relaxed accepting
``anti--particles''.}. Its time evolution is dictated by the master
equation:

$${\pl\phi_i\over \pl t} \propto 
\hbox{\# Particles which enter} - \hbox{\# Particles which exit}
\propto \sum_{j\in N(i)} \phi_j - d_i\phi_i = -(L\ket|\phi>)_i$$

\nind whenever vertex $i$ belongs to the bulk. Otherwise, then there is
a distinction:
\pss

\item{a)} {\it Fixed boundary conditions.\/} Tack vertices (walls)
absorb any incoming particle. Therefore, border vertices have sinks
linked to them. This justifies the ``zero'' values and the relation to
the adjacency matrix.
\ps2

\item{b)} {\it Free boundary conditions.\/} Tack vertices (walls)
reflect back any incoming particle. Therefore, they may be thought
{\it not to exist\/}. In mathematical terms, they take the same value
as on the border vertices, making the gradient across the walls
vanish.
\pss

The combinatorial laplacian {\it should not\/} be considered as a mere
discrete approximation to the continuous operator used for the
Schr\"odinger equation. The formulation should be the other way round:
the continuum is just an idealization useful for mathematical physics,
meanwhile all the real data are discrete. Moreover: the structure of
space--time might be discrete itself or, more precisely, better
approximable by discrete structures than by the continous concepts
used nowadays.

\subsection{Bibliography.}

\cx[BOL-98] {\may B. Bollob\'as}, {\sl Modern graph theory}, Springer
(1998).

\cx[DEG-99] {\may A. Degenhard}, {\sl A non-perturbative real space
renormalization group scheme}, part I in {\tt cond-mat/9910511} and
part II in {\tt cond-mat/9910512}. Also at J.~Phys.~A (Math.~Gen.)
{\bf 33}, 6173 (2000) and Phys.~Rev.~B {\bf 64}, 174408 (2001).

\cx[FH-65] {\may R.P. Feynman, A.R. Hibbs}, {\sl Quantum Mechanics and
Path Integrals}, McGraw-Hill (1965).

\cx[ITZD-89] {\may C. Itzykson, J.M. Drouffe}, {\sl Statistical Field
Theory}, Cambridge U.P. (1989).

\vfill\eject

\titleap{Appendix B. Computational Issues.}

This appendix deals with certain computational aspects of the present
thesis work which may be usually neglected.
\pss

We wish to remark that, for the reasons exposed in detail at the
preface, all the results shown in this memory have been obtained with
software written by us and with free software. The programs were
written in C and C++ (compiled with {\tt g++} and {\tt gcc} from GNU)
and it was necessary to write a series of libraries, which are cited
in the rest of the appendix.
\pss

$\bullet$ {\it Linear Algebra Library.\/} ({\tt matrix.cc}) A C++
library which implements the classes {\tt Vector} and {\tt Matrix}, in
which sum and multiplication operators have a natural meaning.
Standard routines for the inversion of matrices, Gram-Schmidt
orthonormalization, etc. are implemented.

Especially important is the exact diagonalization routine. It operates
in two steps: obtention of a tridiagonal matrix similar to the
original one by the Householder method and obtention of the
eigenvalues through a QL algorithm with implicit displacements
\ct[PTVF-97] \ct[GvL-96].
\pss

$\bullet$ {\it Graphs Library.\/} ({\tt graph.cc}) The generation and
easy manipulation of neighbourhood structures was performed with a
library which implemented the class called {\tt Graph} with natural
operations for the access of the vertices, graphical output and
computation of adjacency and laplacian matrices, among other functions
\ct[BOL-98].
\pss

$\bullet$ {\it Operators Description Library.\/} ({\tt odl.cc}) The
operators used for many body problems are manipulated as descriptions
in a certain language, having the possibility to act on states,
adding, multiplying and obtaining their matrix elements on any basis
of the Hilbert space.
\pss

$\bullet$ {\it Easy X-Window Graphics Library.\/} ({\tt easyx.c}) The
X-Window graphical system is a very extended standard both for PCs and
workstations \ct[X-96]. Through the usage of this library of easy
management, based on the {\tt Xlib} standard, it was possible to make
up animations and to obtain real time graphical output.
\pss

$\bullet$ {\it PostScript Graphics Library.\/} ({\tt easyps.c}) On the
other hand, {\may PostScript} is the standard format for graphics
printing \ct[ADO-85]. Many of the pictures of the present memory were
produced by our own programs using this library.
\pss

These libraries, which are by-products of this thesis, shall be put
in the public domain under the GPL ({\it General Public License\/})
\ct[GNU-01] in a near future, which shall allow its free usage and
copying. At the same time, some of these works have already yielded
some innovations in the computational sciences field, as it may be
checked at \ct[GDL-01], where the techniques exposed in the {\tt
odl.cc} are being applied to the development of a computational
language of geometrical constructions with pedagogical applications.

\subsection{Bibliography.}

\cx[ADO-85] {\may Adobe Systems Inc.} {\sl PostScript language:
tutorial and cookbook}, Adobe Systems Inc. (1985), 19th ed. in
1991.

\cx[BOL-98] {\may B. Bollob\'as}, {\sl Modern graph theory}, Springer
(2001).

\cx[GDL-01] {\may J. Rodr\'{\i}guez-Laguna}, {\sl The geometric
description language}, text available at {\tt http://\break
gdlang.sourceforge.net} (2001).

\cx[GNU-01] {\may GNU Project}, {\sl General public license}, text
available at {\tt http://www.gnu.org}.

\cx[GvL-96] {\may G.H. Golub, C.F. van Loan}, {\sl Matrix
computations}, The John Hopkins U.P. (1996).

\cx[PTVF-97] {\may W.H. Press, S.A. Teukolsky, W.T. Vetterling,
B.P. Flannery}, {\sl Numerical recipes in C}, Cambridge U. P. (1997)
and freely available at {\tt http://www.nr.com}.

\cx[X-96] {\may X Consortium}, {\sl X11R6: Xlib Documentation},
available at {\tt http://www.xfree86.org} (1996).


\titleap{Appendix C. Asymptotics of the Discrete Heat Equation.}

It is interesting to notice that many classical results of continuous
analysis are also {\it exactly\/} true as results of discrete
analysis. On the other hand, in many cases it is necessary to make
some changes so as to preserve the analogy (see, e.g., \ct[GKP-89]).
\pss

In chapter 5 a result was used according to which a delta function
under the heat equation evolves to a gaussian function which fulfills

$$\<x^2\> \approx t^{1/2} \eq{\leyescala}$$ 

\nind Notwithstanding, the calculations made in that chapter referred
to the {\it discrete version\/} of the same equation, given in 1D
by\footn{We shall assume the values $\phi_0$ and $\phi_{N+1}$ (the
tack vertices, see appendix A) to be well defined and constant.}:

$$\phi_i(t+\Delta t)=\phi_i(t) + {\kappa\Delta t\over \Delta
x^2}\(\phi_{i-1}(t)-2\phi_i(t) + \phi_{i+1}(t)\)\eq{\calordisc}$$

\nind This appendix proves that the relation [\leyescala] is also {\it
exact\/} for the finite differences equation [\calordisc] under
certain constraints for the boundary conditions (b.c.) \ct[DRL-01].
\pss

Let us suppose the function $\phi(0)$ to be {\it normalized\/}
according to 

$$\sum_{i=1}^N \Delta x \phi_i(0)=1$$

\nind and that the b.c. allow for the exact conservation of that
magnitude: either $N\to\infty$, or b.c. are free or periodical. It is
not allowed to fix the b.c. to any set of values.
\pss

The second moment $\<x^2\>$ is defined in the continuum by

$$\<x^2(t)\>_c\equiv \int d\mu\,x^2\phi(x,t)$$

\nind The discrete analogue would be

$$\<x^2(t)\>\equiv \sum_{i=1}^N (\Delta x) (i\Delta x)^2 \phi_i(t)$$

\nind We now compute the value for the following time-step:

$$\<x^2(t+\Delta t)\>=\sum_{i=1}^N \Delta x (i\Delta x)^2
\phi_i(t+\Delta t)=$$

$$=\sum_{i=1}^N (\Delta x)^3 i^2 \[ \phi_i(t) +
{\kappa \Delta t\over \Delta x^2} \(\phi_{i-1}(t) - 2\phi_i(t)
+\phi_{i+1}(t)\)\]=$$

$$=\kappa\Delta t\Delta x\[ \sum_{i=1}^N i^2\phi_{i-1}(t) -
2\sum_{i=1}^N i^2\phi_i(t) + \sum_{i=1}^N i^2\phi_{i+1}(t)\] +
\underbrace{\sum_{i=1}^N i^2\Delta x^3\phi_i(t)}_{\<x^2(t)\>}$$

\nind The quantity in brackets may be rewritten as

$$\sum_{i=1}^N \{ (i+1)^2 -2i^2 + (i-1)^2 \} \phi_i(t)=\sum_{i=1}^N
\phi_i(t)$$

\nind Due to the conditions previously exposed, the quantity
$m=\sum\phi_i(t)$ may not depend on time, so we have

$$\<x^2(t+\Delta t)\>=\<x^2(t)\>+2\kappa m\Delta t$$

\nind Iterating this equation we arrive at

$$\<x^2(t)\>=\<x^2(0)\> + 2\kappa mt$$

\nind as we wanted to prove.

\subsection{Bibliography.}

\cx[DRL-01] {\may A. Degenhard, J. Rodr\'{\i}guez-Laguna}, 
{\sl A real space renormalization group approach to field evolution
equations}, {\tt cond-mat/0106155} and Phys.~Rev.~E {\bf 65}
036703 (2002). 

\cx[GKP-89] {\may R.L. Graham, D.E. Knuth, O. Patashnik}, {\sl
Concrete mathematics}, Addison-Wesley Publ. Co. (1994) (Primera ed.:
1989).

\vfill\eject

\titleap{Appendix D. Fast Gram-Schmidt Technique.}

In many parts of this work it is necessary to re-orthonormalize a set
of vectors which were orthonormal before some operation on them took
place. This operation is in many cases of ``local'' nature, i.e.: it
affects only a small fraction of the components of the vectors.
Therefore, a full Gram-Schmidt technique is an innecessary waste of
computational resources.

Let us suppose the set of vectors $\{\ket|\phi^0_i>\}_{i=1}^n$ to be
orthonormal. An operator ${\cal O}$ has acted on them, so we have a
new series of states:

$$\ket|\phi_i>={\cal O}\ket|\phi^0_i>$$

\nind such that the obtention of the dot products on them is a
computationally simple task:

$$C_{ij}\equiv\<\phi_i|\phi_j\>$$

In the typical case, the dot product of two vectors need $N$
multiplications. We shall assume that this is not needed, and matrix
$C_{ij}$ is given as an {\it input\/}.
\pss

Once matrix $C_{ij}$ is known, it is possible to obtain a Gram-Schmidt
matrix $G^i_j$ so that the states

$$\ket|\phi'_i>=G^i_j\ket|\phi_j>$$

\nind make up an othonormal set. We describe now the process to obtain
such a matrix.
\pss

The Gram-Schmidt (GS) operation consists, briefly, in the removal from
the $i$-th vector of all the ``contribution'' of the previous vectors
(indices $1$ up to $i-1$). This implies that the GS matrix has a
structure which may be exploited: $G^i_j=0$ whenever $j>i$. Thus, the
orthonormal $i$-th vector is computed only from the previous vectors.

Let us suppose that the orthonormal set
$\{\ket|\phi'_j>\}_{j=1}^{i-1}$ has already been built. Then the new
state given by

$$\ket|\phi^*_i>=\ket|\phi_i>-\sum_{j<i}
\<\phi_i|\phi'_j\>\ket|\phi'_j> \eq{\nonormalized}$$

\nind is orthogonal to the previous ones, but is not normalized. Using
matrix $G^i_j$ we may rewrite expression [\nonormalized] as

$$\ket|\phi^*_i>=\ket|\phi_i> - 
\sum_{j<i}\sum_{k,l\leq j}\elem<\phi_i|G^j_k|\phi_k>
G^j_l\ket|\phi_l>$$

\nind where the second sum only extends up to $k,l\leq j$ due to the
matrix structure of $G$. Naturally, $G^i_j$ is a number and not an
operator, so we may extract the dot product $\<\phi_i|\phi_k\>$

$$\ket|\phi^*_i>=\ket|\phi_i>-\sum_{j<i}\sum_{k,l\leq j}
C_{ik}G^j_kG^j_l \ket|\phi_l>=\ket|\phi_i>-\sum_{l<i}\sum_{j<i \atop k\leq j}
G^j_k G^j_l C_{ik} \ket|\phi_l>$$

\nind where the last expression was obtained reorganizing the
sums. This last expression may be rewritten as

$$\ket|\phi^*_i>=\sum_{l\leq i}g^i_l \ket|\phi_l>$$

\nind with the coefficients $g^i_l$ given by

$$g^i_l=\cases{ -\sum_{j,k<i} G^j_k G^j_l C_{ik} & if $l<i$ \cr
1 & if $l=i$\cr
0 & otherwise\cr} \eq{\gij}$$

Let us suppose that the $g^i_l$ are known for a given value of $i$,
providing therefore the state $\ket|\phi^*_i>$. In due time, this
state shall be normalized to yield the desired state
$\ket|\phi'_i>$. The norm of this state shall be:

$$\<\phi^*_i|\phi^*_i\>=\sum_{j,k\leq i}\sum_{k\leq i} g^i_j g^i_k
\<\phi_i|\phi_j\>= \sum_{j,k\leq i} g^i_j g^i_k C_{jk}$$

Therefore, the values of $G^i_j$ are obtained from those of $g^i_j$ in
a very simple way:

$$G^i_j = {g^i_j \over \sqrt{\sum_{j,k\leq i} g^i_j g^i_k C_{jk}}}
\eq{\normalization}$$

The calculation procedure for the $G^i_j$ may be now established in an
inductive way. The $g^1_j$ do not require previous values of $G^i_j$:
$g^1_j=\delta^1_j$. The normalization is direct:

$$G^1_1={1\over\sqrt{C_{11}}}$$

\nind and the rest $G^1_j=0$.

The values for the $g^i_j$ may always be calculated from those of the
$G^k_l$ with $k\leq l$, as it is shown in equation [\gij]. The usage
of this equation along with [\normalization] provides the full
algorithm, shown in pseudocode 1.

\midinsert
\pss
\def\ot{$\leftarrow$}
\boxit{5pt}{\vbox{\tt for all $i$, $j$: G($i$,$j$)=0

G(1,1) \ot\ 1/$\sqrt{\hbox{(C(1,1))}}$

for $i$:= $2$ upto $n$

\quad for $j$:= $1$ upto $i-1$

\qquad for $k$, $l$:= $1$ upto $j$

\qquad\quad G($i$,$l$) \ot\ G($i$,$l$) -
G($j$,$k$)*G($j$,$l$)*C($i$,$k$)

\quad G($i$,1) \ot\ 1

\quad norm \ot\ 0

\quad for $k$, $l$:= $1$ upto $i$

\qquad norm \ot\ norm + G($i$,$k$)*G($i$,$l$)*C($k$,$l$)

\quad for $j$:= $1$ upto $i$

\qquad G($i$,$j$) \ot\ G($i$,$j$)/$\sqrt{\hbox{norm}}$
 
}}
\ps2
\noindent{\mayp PseudoCode 1.} {\peqfont Algorithm for the obtention
of the matrix $G^i_j$.\par}
\pss
\endinsert

At the end of the previous program, matrix $G$ is complete.


\titleap{Appendix E. Scale Dimension associated to Energy.}

In section 2.7 the self-replicability properties of certain sets of
functions were studied, along with their relation to CBRG. In this
section some results were shown referring to a certain
``pseudo-fractal dimension'' or ``scaling dimension for the
energy''. It is a definition with full geometric sense, but which is
not {\it sensu strictu\/} a fractal dimension \ct[MAN-82] \ct[EDG-90].
\pss

Let $\{\psi_i\}$ be the $N$ components of a normalized
wave--function. Let $T: V^N\mapsto V^{N/2}$ be the RG (or
coarse-graining) transformation which consists of the following steps:

$$\hat\phi_i={1\over 2}(\phi_{2i-1}+\phi_{2i}) \qquad
T\phi={\hat\phi \over |\hat\phi|}$$

In other words: we consider the vector formed by the local averages of
two sites, and after that we normalize it. We may generate a family of
states in different vector spaces:

$$\{\phi,T\phi,T^2\phi\ldots T^n\phi\}$$

\nind where $n$ is $\log_2(N)-1$. In our applications (although it is
not required), $N$ shall always be a power of $2$.
\pss

Now we take the sequence of expected values of the kinetic energy for
these functions, always with free b.c.:

$$E_m\equiv \elem<\phi|(T^m)^\dagger\ H\ T^m|\phi>$$

\nind In the form of an ordinary sum,

$$E_m=\sum_{i=2}^{N\cdot 2^{-m}} \((T^m\phi)_i -
(T^m\phi)_{i-1}\)^2$$

\nind I.e.: the squared sum of the derivative. For ``smooth''
functions, this sequence approximately fulfills

$$\log(E_m) \approx c_0+2m$$

\nind The reason is the following: for a smooth function, jumps will
be increased in a factor $2$ when averaging ($4$ when squared). When
normalizing a smaller number of components, these shall be increased
by a further factor $\sqrt{2}$ ($2$ upon squaring, so a factor $8$ so
far). But the sum extends only to half the components, so we must
divide by $2$ and a factor $4$ remains: the scaling exponent is
$\log_2(4)=2$ as it was announced.
\pss

We shall denote by ``{\sl energetic scaling exponent}'' or ``{\sl
pseudo-fractal dimension associated to energy}'' the number $\epsilon$
which makes the best fit to

$$\log(E_m) \approx c_0 + \epsilon m$$

In practice we obtain, for example, for the first four normalized
polynomials $\{1,x,x^2,x^3\}$ on $2048$ sites the values of
$\epsilon$: $x\to 1.90 \pm 0.03$, $x^2\to 1.87 \pm 0.04$ and $x^3\to
1.87 \pm 0.04$\footn{Obviously, there is no sense in trying to find it
for a constant function, whose energy is always zero.}. As it was
shown in section 2.7, for the fixed point of the eigenstates of the
particle in a box with free and fixed b.c. the results of table 4 in
chapter 2 are obtained. In the first case they are rather near to $2$,
meanwhile in the second one they are clearly $\approx 0$.
\pss

What physical interpretation do these results have? Let us imagine
that these wave--functions which were obtained as a fixed point were
real wave--functions of a particle and let us make the inverse path:
from big scales to smaller ones. If $\epsilon\approx 2$ or, at least,
is clearly incompatible with zero, then the energies of the lowest
energy states get smaller and smaller when one takes the continuum
limit. If the exponent were null, that would imply that the energy of
the ground state might have a finite limit {\it without\/} being
localized. This generation of a ``gap'' might be interpreted as a
dynamical mass generation.

\vfill\eject
\titulartrue
\def\p#1--#2\par{{\may\rojoosc #1} #2\ps2}
\def\t{$\to$}
\def\ast{{\verdeosc(*)}}
\denuevas
\vbox{\ps{30}
\verdeosc\grandota\centerline{Glossary.}}
\pss\pss\pss
\temaizq={Glossary.}
\temader={Glossary.}

We present in the following pages an index of commented terms or
glossary. Some of these terms have been introduced in our work, and
in that case they are marked with an asterisk.
\pss

\p Aspect \ast-- (of a field) Discrete and finite set of observables
referred to an extended physical system which aspire to represent the
whole knowledge of the observer about it. For the \t quasistatic
transformation, e.g., these observables are the integral values on
the cells of a given partition of space (concept more refined than
that of {\sl discretization}). Sections 5.1 to 5.3.

\p Basic Scale \ast-- Scale at which the physical laws explaining a
given phenomenon are simple. Thus, e.g., ferromagnetism is explained
at atomic scale. Macroscopic physics is obtained from it by applying
\t RG transformations. Section 1.2.

\p Blocks Algebra \ast-- In the meaning used in this work, blocks are
just subsets of a graph, on which an internal operation known as
addition may be defined. It is possible to interpret RSRG as a
flow on a matrix representation of that algebra. Section 4.6.

\p BRG, Blocks Renormalization Group,-- Process for the variational
computation of the lowest energy spectrum of a system using as {\it
Ansatz\/} the lowest energy eigenfunctions corresponding to smaller
blocks. Section 1.4.III and chapter 2 (part I).

\p CBRG, Correlated Blocks Renormalization Group \ast-- Modification of
the \t BRG which takes into account the correlation between blocks
through the introduction of the \t influence and \t interaction
matrices. Chapter 2 (part II).

\p Dendrimers,-- Polymeric molecules in which from a monomer stem
various branches which branch at due time all at once, resulting a
system of great symmetry. They have lots of applications, from
materials science to medicine. Its excitonic spectrum is analyzed
through \t DMRG applied to \t trees. Chapter 3 (part II).

\p DMRG, Density Matrix Renormalization Group,-- Drastic modification
of the \t BRG following the idea that the lowest energy states of the
blocks need not be the best bricks for building the global state. The
system is divided into left and right blocks, and the ``most
probable'' states for each block are chosen by fitting to a given \t
target state, projecting with a density matrix. Esp. section 1.4.V and
chapter 3.

\p Embedding Operator,-- Operator which projects a renormalized state
into a real space one. In QM it is forced to be the adjoint of the \t
truncation operator, while in other applications it is only required
to be its \t SVD pseudoinverse. Since the renormalized state has fewer
degrees of freedom, the consecutive application of truncation and
embedding does not yield the identity, but a projector on the relevant
degrees of freedom. Esp. sections 2.1 and 5.2.

\p Evolution Prescription \ast-- (for fields) Discrete computational
rule for the evolution (either deterministic or stochastic) of an \t
aspect of a field. Sections 5.1 and 5.2.

\p Implicit RSRG Method \ast-- Variational \t RSRG algorithm for QM in
which the wave--functions are not stored, but a given number of
matrix elements of chosen operators on them. Otherwise, the method is
known as {\sl explicit}. Sections 3.1 and 4.1.

\p Influence Matrix \ast-- In \t CBRG, an additive modification on the
block hamiltonian provoked by the presence of a neighbouring block. A
given block receives one influence matrix per neighbour. Section 2.3.

\p Interaction Matrix \ast--In \t CBRG, part of the total hamiltonian
of a system which does not belong to the hamiltonian of any block, but
stays between a given pair of blocks. Section 2.3.

\p ITF, Ising Model in a Transverse Field,-- Quantum model for an
uniaxial ferromagnet in which the $z$ components of the spins tend to
align, while a magnetic field in a perpendicular axis ($x$) tend to
prevent that alignment. It has a phase transition even in 1D and has
also an interesting analogy with the classical 2D Ising model. It is
analyzed with \t BRG in section 2.1.

\p Laplacian,-- In a rather generic way, it is an operator which
allows to compute statistically the diffusion of a collective of
non-interacting particles on a given space. Especially interesting for
this work is the laplacian on a graph. Appendix A.

\p Observational Frame \ast-- Any physical observation process
requires a certain ``grain size'' (lower or UV cutoff) and a ``plate
size'' (higher or IR cutoff). A pack formed by a reference frame
(Lorentz, Galilei...) along with the ``IR cutoff + UV cutoff'' and any
other required features for observation makes up an {\sl observational
frame}. Section 1.2.

\p Patch \ast-- Set of \t punctures. In \t PRG, region of the system
which, at a given RG step, is fully well represented in the {\it
Ansatz\/}. Section 4.2.

\p PRG, Punctures Renormalization Group \ast-- Variational RSRG
calculation of the low energy spectrum of a quantum mechanical system
via the {\it Ansatz\/} of some functions which represent a given
region of the system (the \t punctures) and others which represent the
rest of it. Chapter 4.

\p Puncture \ast-- In \t PRG, each of the sites which are well
determined in the {\it Ansatz\/} at a given step, represented by the
corresponding delta state $\ket|\delta_p>$. Section 4.2.

\p Quadtree,-- Structure for the addressing of 2D points or the
storage of a bidimensional field (e.g. an image) through divisions of
a square into $2\times 2$ smaller squares in an interative way. Used
by 2D \t CBRG. Section 2.6.

\p Quasistatic Transformation \ast-- \t Truncation operator which
proceeds by cells overlapping in an iterative way, removing one degree
of freedom at each step. Sections 5.3 and 5.4.

\p Renormalized Space,-- Space of a smaller number of degrees of
freedom than real space, in which each component represents the weight
of a state considered to be relevant. Real states may travel to
renormalized space via the \t truncation operators. From renormalized
space one may jump to real space by using the \t embedding
operators. Esp. sections 2.1, 2.5, 5.1 and 5.2.

\p RG, Renormalization Group-- In physical terms, a displacement of
the \t observational frame along the scales axis. Esp. chapter 1.

\p RSRG, Real Space Renormalization Group-- RG implementation which
only uses blocks formed according to geometric criteria. The term is
used as opposed to the RG methods based on Fourier space. Esp. chapter
1.

\p Scaling Exponents,-- Many physical laws are ruled by {\sl scaling
laws} or {\sl power laws}, which require a situation with a certain
degree of invariance under scaling (aka \t RG) transformations. The
exponents are robust observables, since they tend to be \t
universal. {\sl Critical exponents}, according to which physical
magnitudes diverge near a critical point, are a particular
case. Esp. sections 1.2 and 2.1.

\p Self-Replicability \ast,-- Property of certain sets of functions on
an interval (in $d$ dimensions) to be correctly approximated using
linear combinations of its copies scaled on a factor $1/2^d$ and
situated on the different quadrants of that interval. Useful concept
for the analysis of \t CBRG. Section 2.7.

\p Sewing \ast,-- Path followed by the \t puncture (or \t patch)
through the system in the \t PRG, which must traverse all the
links. Esp. section 4.4.

\p SVD, Singular Values Decomposition-- Analogue of the
diagonalization for rectangular matrices. It allows the easy obtention
of a ``pseudoinverse'' of a given matrix which fulfills the
Moore-Penrose conditions. Section 5.2.

\p Sweep,-- Process in the second part of traditional \t DMRG, for
which one of the two blocks (left or right) grows at the expense of
the other until an extreme is reached; after that, the process
continues in the other sense. Analogously, a full path of the free
site in the \t trees DMRG and \t PRG. Esp. section 3.1.

\p Target State,-- In \t DMRG the block states are chosen as those
which fit better to a certain state which was obtained for the full
system, known as {\sl target state}. Esp. section 3.1 and 3.2.

\p Tree,-- Connected graph in which there is an only continuous path
which does not repeat sites between any couple of sites. Especially
useful for making calculations with \t DMRG, since it allows the use
of an \t implicit method. Section 3.3.

\p Truncation Operator,-- Operator which takes an state from an
arbitrary space and returns another on a space of smaller dimension,
known as the \t renormalized space, where each component represents
the ``weight'' on a series of states of the original space thought
(for some reason) to be especially relevant. Esp. sections 2.1 and
5.2.

\p Warmup,-- First step for many RSRG techniques, which initializes
the suitable operators before the \t sweeping or \t sewing cycles
start. The only requirement is that the matrix elements of the
operators correspond to real states, but an intelligent election
boosts the computation a lot. Esp. sections 3.1, 3.2, 3.3 and 4.5.

\p Universality Class-- Set of physical systems which share the same
scaling exponents. The ubiquity of the phenomenon is explained via the
\t Renormalization Group (RG). Section 1.2.

\bye